\documentclass[aps,pra,10pt,a4paper,superscriptaddress,onecolumn,preprintnumbers,floatfix,nofootinbib, showkeys]{revtex4-2}

\usepackage{graphicx}
\usepackage[utf8]{inputenc}
\usepackage{tocloft}
\usepackage{epsf}
\usepackage{bm}
\usepackage{amsmath}
\usepackage{amsfonts}
\usepackage{amssymb}
\usepackage{color}
\usepackage{caption}
\usepackage{subcaption}
\usepackage{booktabs} 
\usepackage{float}
\usepackage{empheq}

\usepackage[citecolor=blue,urlcolor=blue,unicode=true,pdfusetitle, bookmarks=true,bookmarksnumbered=true,bookmarksopen=true,bookmarksopenlevel=3, breaklinks=false,pdfborder={0 0 1},backref=false,colorlinks=true, linkcolor=blue]{hyperref}

\numberwithin{equation}{section}

\makeatletter

\renewcommand*{\p@subsection}{}

\renewcommand*{\p@subsubsection}{}

\renewcommand*{\p@paragraph}{}
\makeatother

\providecommand{\U}[1]{\protect\rule{.1in}{.1in}}

\newcommand{\be}{\begin{equation}}
\newcommand{\ee}{\end{equation}}

\newcommand{\mincir}{\raise
-3.truept\hbox{\rlap{\hbox{$\sim$}}\raise4.truept\hbox{$<$}\ }}
\newcommand{\magcir}{\raise
-3.truept\hbox{\rlap{\hbox{$\sim$}}\raise4.truept\hbox{$>$}\ }}

\begin{document}
\title{I. Linear Interacting Dark Energy: Analytical Solutions and Theoretical Pathologies}

\author{Marcel van der Westhuizen}
\email{marcelvdw007@gmail.com}
\affiliation{Centre for Space Research, North-West University, Potchefstroom 2520, South
Africa}
\author{Amare Abebe}
\email{amare.abebe@nithecs.ac.za}
\affiliation{Centre for Space Research, North-West University, Potchefstroom 2520, South
Africa}
\affiliation{National Institute for Theoretical and Computational Sciences (NITheCS),
South Africa}
\author{Eleonora Di Valentino}
\email{e.divalentino@sheffield.ac.uk}
\affiliation{School of Mathematical and Physical Sciences, University of Sheffield, Hounsfield Road, Sheffield S3 7RH, United Kingdom}

\begin{abstract}
Interacting dark energy (IDE) models, in which dark matter (DM) and dark energy (DE) exchange energy through a non-gravitational interaction, have long been proposed as candidates to address key challenges in modern cosmology. These include the coincidence problem, the $H_0$ and $S_8$ tensions, and, more recently, the hints of dynamical dark energy reported by the DESI collaboration. Given the renewed interest in IDE models, it is crucial to fully understand their parameter space when constraining them observationally, especially with regard to the often-neglected issues of negative energy densities and future big rip singularities. In this work, we present a comparative study of the general linear interaction $Q=3H(\delta_{\rm dm}\rho_{\rm dm} + \delta_{\rm de}\rho_{\rm de})$ and four special cases: $Q=3H\delta(\rho_{\rm dm}+\rho_{\rm de})$, $Q=3H\delta(\rho_{\rm dm}-\rho_{\rm de})$, $Q=3H\delta \rho_{\rm dm}$, and $Q=3H\delta \rho_{\rm de}$. For these five models,  we perform a dynamical system analysis and derive new conditions that ensure positive, real, and well-defined energy densities throughout cosmic evolution, as well as criteria to avoid future big rip singularities. We obtain exact analytical solutions for $\rho_{\rm{dm}}$, $\rho_{\rm{de}}$, the effective equations of state ($w_{\mathrm{eff}}^{\rm{dm}}$, $w_{\mathrm{eff}}^{\rm{de}}$, $w_{\mathrm{eff}}^{\rm{tot}}$), and a reconstructed dynamical DE equation of state $\tilde{w}$. Using these results, we examine phantom crossings, address the coincidence problem, and apply the statefinder diagnostic to distinguish between models. We show that energy transfer from DM to DE inevitably produces negative energy densities and make future singularities more likely, while transfer from DE to DM avoids these pathologies and is thus theoretically favored.
\end{abstract}
\date{\today}
\maketitle
\date{\today }

\section{Introduction}

The standard model of cosmology, the $\Lambda$CDM model, describes a universe largely dominated by two mysterious dark components. First, dark matter (DM), introduced to explain galaxy rotation curves and underpinning many other astronomical and cosmological observations. Second, dark energy (DE), invoked to account for the observed accelerated expansion of the universe. DE has traditionally been associated with the cosmological constant $\Lambda$, the suggested energy of the vacuum that exerts constant negative pressure throughout all of space, but has been plagued from early on by the mismatch between the predicted and observed values of its energy density, known as the \textit{cosmological constant problem}~\cite{Weinberg:1988cp}. 
Despite this, the $\Lambda$CDM model has been largely supported by the recent era of precision cosmology~\cite{Plack2018,SPT-3G:2025bzu,ACT:2025fju}, but some discrepancies have recently arisen and have only grown more severe as measurements have improved. Among these, the most prominent is the \textit{Hubble tension}~\cite{Verde:2019ivm,DiValentino:2020zio,DiValentino:2021izs,Perivolaropoulos:2021jda,Schoneberg:2021qvd,Shah:2021onj,Abdalla:2022yfr,DiValentino:2022fjm,Kamionkowski:2022pkx, Gao:2022ahg,Giare:2023xoc, Vagnozzi:2023nrq,Hu:2023jqc,Verde:2023lmm,DiValentino:2024yew,Perivolaropoulos:2024yxv}, which refers to the discrepancy in the measurement of the Hubble constant $H_0$ by early- and late-time probes~\cite{Freedman:2020dne,Birrer:2020tax,Anderson:2023aga,Scolnic:2023mrv,Jones:2022mvo,Anand:2021sum,Freedman:2021ahq,Uddin:2023iob,Huang:2023frr,Li:2024yoe,Pesce:2020xfe,Kourkchi:2020iyz,Schombert:2020pxm,Blakeslee:2021rqi,deJaeger:2022lit,Murakami:2023xuy,Breuval:2024lsv,Freedman:2024eph,Riess:2024vfa,Vogl:2024bum,Scolnic:2024hbh,Said:2024pwm,Boubel:2024cqw,Scolnic:2024oth,Li:2025ife,Jensen:2025aai}. 
Another tension between early- and late-universe observations is the \textit{$S_8$ discrepancy}, where the $S_8$ parameter describes the amplitude of matter fluctuations on cosmological scales~\cite{DiValentino:2020vvd,DiValentino:2018gcu,Nunes:2021ipq,DES:2021bvc,DES:2021vln,KiDS:2020suj,Asgari:2019fkq,Joudaki:2019pmv,DAmico:2019fhj,Kilo-DegreeSurvey:2023gfr,Troster:2019ean,Heymans:2020gsg,Dalal:2023olq,Chen:2024vvk,ACT:2024okh,DES:2024oud,Harnois-Deraps:2024ucb,Dvornik:2022xap,DES:2021wwk,Wright:2025xka}. A comprehensive review of these tensions and their possible resolutions can be found in~\cite{CosmoVerse:2025txj}. 
Moreover, the release of DESI DR2 Baryon Acoustic Oscillation (BAO) data has cast further doubt on a pure cosmological constant description of DE, showing a preference for dynamical DE models (within the CPL parametrization) over $\Lambda$CDM~\cite{DESI:2025zgx,DESI:2025fii,DESI:2025qqy,DESI:2025wyn} (see also~\cite{DESI:2024mwx,Cortes:2024lgw,Shlivko:2024llw,Luongo:2024fww,Yin:2024hba,Gialamas:2024lyw,Dinda:2024kjf,Najafi:2024qzm,Wang:2024dka,Ye:2024ywg,Tada:2024znt,Carloni:2024zpl,Chan-GyungPark:2024mlx,DESI:2024kob,Ramadan:2024kmn,Notari:2024rti,Orchard:2024bve,Hernandez-Almada:2024ost,Pourojaghi:2024tmw,Giare:2024gpk,Reboucas:2024smm,Giare:2024ocw,Chan-GyungPark:2024brx,Menci:2024hop,Li:2024qus,Li:2024hrv,Notari:2024zmi,Gao:2024ily,Fikri:2024klc,Jiang:2024xnu,Zheng:2024qzi,Gomez-Valent:2024ejh,RoyChoudhury:2024wri,Lewis:2024cqj,Wolf:2024eph,Wolf:2024stt,Wolf:2025jed, Wolf:2025jlc,Shajib:2025tpd,Giare:2025pzu,Chaussidon:2025npr,Kessler:2025kju,Pang:2025lvh,RoyChoudhury:2025dhe,Scherer:2025esj,Teixeira:2025czm,Specogna:2025guo,Cheng:2025lod,Cheng:2025hug,Ozulker:2025ehg,Gialamas:2025pwv,Lee:2025pzo}).

Due to the concerns mentioned above, alternative cosmological models that may address these and other long-standing issues should be considered. One prominent class of models is that of interacting dark energy (IDE) models, where non-gravitational interactions exist between DM and DE. Historically, these models were introduced to address another issue, known as the \textit{coincidence problem} (CP), which refers to the prediction that the DM and DE densities should differ by many orders of magnitude in both the past and future, yet coincidentally happen to be of the same order today, when we are able to observe them. Historically important papers on IDE include~\cite{Amendola_2000, Zimdahl_2001, Chimento_2003, Farrar_2004, Wang_2004, Olivares_2006}, while their potential to address the coincidence problem has been studied in those as well as in~\cite{Sadjadi_2006, Quartin_2008, Campo_2009, Caldera_Cabral_2009_DSA, He_2011, delcampo2015interactiondarksector, von_Marttens_2019}. 
In recent years, IDE models have regained popularity for their potential to address both the $H_0$ tension~\cite{Kumar:2016zpg, Murgia:2016ccp, Kumar:2017dnp, DiValentino:2017iww, Kumar:2021eev, Pan:2023mie, Benisty:2024lmj, Yang:2020uga, Forconi:2023hsj, Pourtsidou:2016ico, DiValentino:2020vnx, DiValentino:2020leo, Nunes:2021zzi, Yang:2018uae, vonMarttens:2019ixw, Lucca:2020zjb, Zhai:2023yny, Bernui:2023byc, Hoerning:2023hks, Giare:2024ytc, Escamilla:2023shf, vanderWesthuizen:2023hcl, Silva:2024ift, DiValentino:2019ffd, Li:2024qso, Pooya:2024wsq, Halder:2024uao, Castello:2023zjr, Yao:2023jau, Mishra:2023ueo, Nunes:2016dlj, Silva:2025hxw, Zheng_2017, Kumar_2019, Anchordoqui_2021, Pan_2019, Guo_2021,Gao:2021xnk, Di_Valentino_2021_H0_review, Gariazzo_2022, Wang_2022, Califano_2023, Pan_2024, Liu_2023, Liu_2024, Sabogal_2025} and the $S_8$ discrepancy~\cite{Kumar_2017, Kumar_2019, Di_Valentino_2020_rhode, Anchordoqui_2021, Kumar_2021, Gariazzo_2022, lucca2021darkenergydarkmatterinteractions, Sabogal_2024, Liu_2023, Liu_2024, Sabogal_2025, yang2025probingcoldnaturedark, Liu_2025,Yang:2025uyv}. 
The success of IDE models in alleviating these issues depends strongly on the type of interaction kernel and the datasets chosen, as they may alleviate one tension while worsening the other (for a recent review, see Section 4.2.3 in~\cite{CosmoVerse:2025txj}). Most recently, hints of dynamical DE~\cite{DESI:2024mwx,Cortes:2024lgw,Shlivko:2024llw,Luongo:2024fww,Yin:2024hba,Gialamas:2024lyw,Dinda:2024kjf,Najafi:2024qzm,Wang:2024dka,Ye:2024ywg,Tada:2024znt,Carloni:2024zpl,Chan-GyungPark:2024mlx,DESI:2024kob,Ramadan:2024kmn,Notari:2024rti,Orchard:2024bve,Hernandez-Almada:2024ost,Pourojaghi:2024tmw,Giare:2024gpk,Reboucas:2024smm,Giare:2024ocw,Chan-GyungPark:2024brx,Menci:2024hop,Li:2024qus,Li:2024hrv,Notari:2024zmi,Gao:2024ily,Fikri:2024klc,Jiang:2024xnu,Zheng:2024qzi,Gomez-Valent:2024ejh,RoyChoudhury:2024wri,Lewis:2024cqj,Wolf:2024eph,Wolf:2024stt,Wolf:2025jed, Wolf:2025jlc,Shajib:2025tpd,Giare:2025pzu,Chaussidon:2025npr,Kessler:2025kju,Pang:2025lvh,RoyChoudhury:2025dhe,Scherer:2025esj,Teixeira:2025czm,Specogna:2025guo,Cheng:2025lod,Cheng:2025hug,Ozulker:2025ehg,Lee:2025pzo}, as well as the possibility of a phantom crossing in the DE equation of state~\cite{DESI:2025fii}, claimed by the DESI Collaboration (see also~\cite{Ozulker:2025ehg}), have reignited interest in IDE models. This is because the energy transfer between DM and DE in IDE cosmology provides a natural mechanism for an effective dynamical DE equation of state, as described in \eqref{DSA.omega_eff_dm_de}, which can experience a phantom crossing without requiring $w < -1$. Some IDE models have recently been constrained in light of the new DESI DR2 data release~\cite{guedezounme2025phantomcrossingdarkinteraction, silva2025newconstraintsinteractingdark, pan2025interactingdarkenergydesi, shah2025interactingdarksectorslight, lee2025shapedarkenergyconstraining, vanderwesthuizen2025compartmentalizationdarksectoruniverse, Giare:2024smz}. 
Given the increasing relevance of IDE models in cosmology, it is worth revisiting the background expansion of these models, their pitfalls, and how to possibly avoid them.

In this paper, we assume general relativity in a flat, isotropic, and homogeneous universe described by the FLRW metric. IDE models are characterized by the assumption that neither the DM nor the DE components of the universe are individually conserved, but rather that there is an energy exchange between them, implying that only the total energy of the dark sector is conserved. This assumption leads to modifications of the DM and DE conservation equations, whereas the radiation and baryonic matter conservation equations \eqref{DSA.H} remain unchanged. 
The interaction can be introduced into the conservation equations through an interaction function $Q$, or its effect can be encapsulated by introducing effective equations of state, such that we have two equivalent sets of coupled equations:
\begin{gather} \label{eq:conservation}
\begin{split}
\dot{\rho}_{\text{dm}} + 3H \rho_{\text{dm}} = Q\;, \quad & \quad  \dot{\rho}_{\text{de}} + 3H (1 + w) \rho_{\text{de}} = -Q\;,\\
\dot{\rho}_{\text{dm}} + 3H \rho_{\text{dm}}(1 + w_{\rm{dm}}^{\rm{eff}}) = 0\;, \quad & \quad  \dot{\rho}_{\text{de}} + 3H (1 + w_{\rm{de}}^{\rm{eff}}) \rho_{\text{de}} = 0\;.
\end{split}
\end{gather}
In \eqref{eq:conservation}, $\rho_{\text{dm}}$ and $\rho_{\text{de}}$ are the energy densities of DM and DE, respectively, and $\dot{\rho}$ denotes differentiation with respect to cosmic time. The parameter $w$ is the DE equation of state, while $w_{\rm{dm}}^{\rm{eff}}$ and $w_{\rm{de}}^{\rm{eff}}$ are the effective equations of state for DM and DE, respectively, also described by \eqref{DSA.omega_eff_dm_de}. The Hubble parameter $H$ is given in \eqref{DSA.H}. The interaction kernel $Q$ determines the energy transfer between DM and DE, with its sign indicating the direction of energy flow:
\begin{equation}
Q =
\begin{cases} 
> 0 & \text{Dark Energy} \rightarrow \text{Dark Matter (iDEDM regime)},\\ 
< 0 & \text{Dark Matter} \rightarrow \text{Dark Energy (iDMDE regime)},\\ 
= 0 & \text{No interaction.}
\end{cases}
\label{Q_directions}
\end{equation}
Some general consequences that apply to all interaction kernels $Q$ can be found in Table 1 of~\cite{vanderWesthuizen:2023hcl}. The question of which direction the energy transfer should occur in remains open and should be determined observationally. Currently, it can be argued that, from a theoretical point of view, there is a preference for energy transfer from DE to DM, as this regime helps to alleviate the coincidence problem and reduces the likelihood of both negative energy densities and future big rip singularities~\cite{vanderWesthuizen:2023hcl}. 
Another line of reasoning relies on the second law of thermodynamics or the Le Ch$\hat{\text{a}}$telier-Braun principle, which may be more or less convincing depending on the assumed fluid or scalar field nature of the dark components, as well as the possible dynamical nature of the DE equation of state~\cite{Pav_n_2008}. It has also been suggested that energy flow from DE to DM resembles the particle creation mechanisms invoked during inflation~\cite{Berera_1995}. These arguments likewise favor energy flow from DE to DM, since the reverse would violate these long-established principles (given the assumed nature of the dark components). Further discussion on this topic can be found in the following IDE review papers~\cite{Wang:2016lxa, wang2024understandinginteractiondarkenergy}.

It is natural to ask what form the interaction kernel $Q$ should take. At present, there is no fundamental theory that predicts a specific functional form for $Q$; however, this remains an active area of research, and many field-theoretic proposals have appeared in the literature~\cite{Karwan_2017, Landim_2018, Mifsud_2017, van_de_Bruck_2018, Dutta_2018_maybe, Roy_2019, Nakamura_2019, Chibana_2019, B_gu__2019, An_2019_2, Kase_2020, Pan_2020_Field, van_de_Bruck_2020, Johnson_2021, S__2021, Dusoye_2021, chatzidakis2022interactingdarkenergycurved, Johnson_2022, _lvarez_Ortega_2022, Kaneta:2022kjj, akarsu2024equivalencemattertypemodifiedgravity, Mandal_2023, De_Arcia_2023, G_mez_2023, Patil_2023, Cardona_2024, Liu_2023, Liu_2024, Thipaksorn_2022, nari2025dynamicalfrictioncoupleddark, Kaneta:2025kcn, rahimy2025decipheringcoupledscalardark}. For a discussion of these models, see~\cite{Lucca_2020}. Recent attempts have also been made to reconstruct $Q$ directly from observations~\cite{yang2019reconstructingdarkmatterdark, Escamilla_2023}.
In practice, most interaction kernels are chosen phenomenologically. Typically, the form of $Q$ is taken to be proportional, either linearly or non-linearly, to the energy densities of one or both dark components, as illustrated in the eight interaction kernels shown in Table~\ref{tab:positivityQ}. These kernels also include a dimensionless interaction constant $\delta$ (commonly denoted as $\xi$ in the literature), which indicates the strength of the interaction, and the Hubble rate $H$, to ensure dimensional consistency.
At first glance, the dependence of the interaction on a global quantity such as the Hubble rate $H$ may seem unnatural. However, it has been argued that this can be interpreted as a temperature dependence of the interaction rate~\cite{Väliviita_2008}. A related argument based on the first law of thermodynamics suggests that the appearance of $H = \dot{a}/a$, which is linked to the change in the scale factor $a$, simply reflects the fact that a change in density must be accompanied by a change in volume~\cite{Valentino_2020_DE, Nunes_2022}. Nonetheless, a few authors have also explored interaction kernels that do not include any explicit dependence on $H$~\cite{Yang_2021_no_H}.

Accepting that the interaction kernel is chosen phenomenologically, it is natural to wonder which kernel should be selected. This question must ultimately be answered observationally, but from a theoretical point of view, different kernels come with various advantages and disadvantages. These considerations include mathematical simplicity, the number of additional parameters, the stability of the system, the presence of negative energy densities, the prediction of future big rip singularities, the potential to solve the coincidence problem, and the possibility of connecting the kernel to more fundamental physical theories.
To address the question of interaction-kernel choice observationally, we need a good understanding of the parameter space and theoretical predictions associated with the interaction kernels we choose to constrain. The stability of IDE models has been widely studied, and the existence of a doom factor $\textbf{d}$, further discussed in \eqref{DSA.doom}, has often been used to guide the choice of parameter priors that avoid instabilities when performing observational constraints~\cite{M.B.Gavela_2009, Honorez_2010, Salvatelli_2013, Costa_2014, Yang_2016, Costa_2017, Costa_2018, Yang_2018_DOOM, Bachega_2020, Yang_2020, Lucca_2020, Valentino_2020_DE, Di_Valentino_2020, Di_Valentino_2021, Di_Valentino_2021_closed, lucca2021darkenergydarkmatterinteractions, Gariazzo_2022, Yang_2021, Joseph_2022, Nunes_2022, Califano_2023, Ghodsi_Yengejeh_2023, Forconi_2024, Giar__2024, Sabogal_2024}.
Conversely, the issues of negative DM and DE energy densities and big rip singularities are rarely mentioned when observationally constraining these IDE models. For example, the following papers assume energy transfer from DM to DE, which should result in instances of negative energy densities, but do not mention them~\cite{Salvatelli_2013, Yang_2016, Caprini_2016, Pan_2017, Di_Valentino_2017, Zheng_2017, Zhang_2019, Valentino_2020_DE, Di_Valentino_2020, Di_Valentino_2021, Di_Valentino_2021_closed, Anchordoqui_2021, Lucca_2020, Li_2020, Mukhopadhyay_2021, Xiao_2021, Li_2014_2, Yang_2021, Nunes_2022, Gariazzo_2022, Joseph_2022, Bernui_2023, Zhai_2023, Zhao_2023, Forconi_2024, Giar__2024}.
To this end, we present a comparative study focusing on these often neglected aspects for five linear interaction kernels in this paper, while the three non-linear interaction kernels in Table~\ref{tab:positivityQ} will be studied in a companion paper \cite{vanderWesthuizen:2025II}. 

The main aim of this paper is to facilitate future efforts to observationally constrain IDE models, building on previous efforts in \cite{vanderWesthuizen:2023hcl} where two linear IDE models were studied. Here we extend the analysis to five linear interaction kernels. Readers interested in the main findings relevant to observational constraints will find new analytical solutions for the energy densities in Section~\ref{Background_cosmology}, which can be used to obtain exact expressions for the corresponding Hubble functions. The implications of the parameter space for each interaction kernel are summarised in Section~\ref{summary}. The main results for all eight interaction kernels studied are summarised in \cite{vanderWesthuizen:2025III}. These theoretical insights may also be used to interpret the results of previously obtained observational constraints. A full overview of the paper’s structure follows.

\begin{itemize}
    \item In Section~\ref{BG_lit}, we provide further background on IDE cosmology. Specifically, in Subsection~\ref{BG_neg}, we discuss why negative energy densities commonly appear in IDE models, and in Subsection~\ref{BG_rip}, we examine the conditions leading to big rip singularities in such models. We also present the background equations that will be used throughout the rest of the paper in Subsection~\ref{BG_equations}. In Subsection~\ref{BG_lit_Q}, we summarize the existing literature on specific aspects of each of the five linear interaction kernels considered in this study.
    
    \item In Section~\ref{Sec.DSA}, we set up a dynamical system and perform a full analysis for the most general linear interaction kernel, $Q = 3 H (\delta_{\text{dm}} \rho_{\text{dm}} + \delta_{\text{de}} \rho_{\text{de}})$. We identify the critical points of the system and compute their corresponding eigenvalues (where useful), and we plot the phase portraits in 3D (Figure~\ref{fig:3D_Q_general_Phase_portrait_boundaries}) and 2D (Figure~\ref{fig:2D_Q_general_Phase_portrait_boundaries}). 
    We derive new conditions that ensure cosmological trajectories in which both the DM and DE energy densities remain positive throughout the entire evolution, given in \eqref{DSA.Q.delta_dm+delta_de.PEC_BG} and summarized in Table~\ref{tab:Qddm+dde_energy_conditions}. Conditions to ensure future accelerated expansion and to avoid a big rip are provided in \eqref{DSA.Q.ddm+dde.ACC} and \eqref{DSA.Q.ddm+dde.BR}, respectively. 
    Additionally, we summarize the expressions for various cosmological parameters at each critical point in Table~\ref{tab:CP_B_ddm+dde}, and provide a brief summary of the dynamical-system results in Subsection~\ref{DSA.summary}. 

    \item In Section~\ref{Finding_analytical_solutions}, we provide a derivation for solving the conservation equations for $Q = 3 H (\delta_{\text{dm}} \rho_{\text{dm}} + \delta_{\text{de}} \rho_{\text{de}})$, based on a method previously outlined in~\cite{Pan_2015, Pan_2017}. We derive new integration constants that lead to a new formalism for $\rho_{\text{dm}}$ \eqref{eq:rhodm_general} and $\rho_{\text{de}}$ \eqref{eq:rhode_general}, applicable to all linear models considered in this study. In Subsection~\ref{More_analytical_relations}, we present additional analytic expressions for the asymptotic behavior of the DM and DE densities, the redshift at which DM or DE crosses the zero boundary before becoming negative, the redshift of DM-DE equality, the redshift at which phantom crossing occurs, expressions for the ratio $r$ of DM to DE, the conditions for a big rip, and the time at which a big rip would occur.

    \item In Section~\ref{Background_cosmology}, we apply the expressions derived in the previous section to each interaction model, and we provide graphs showing the evolution of the density parameters $\Omega$, the ratio $r$ (to address the coincidence problem), the effective equations of state for DM and DE ($w^{\rm{eff}}_{\rm{dm}}$ and $w^{\rm{eff}}_{\rm{de}}$), the total effective equation of state $w^{\rm{eff}}_{\rm{tot}}$, and the scale factor corresponding to a big rip. For each model, we give explicit expressions for $\rho_{\text{dm}}$ and $\rho_{\text{de}}$, and identify the redshifts or scale factors at which the energy densities become undefined, imaginary, or negative. These results are presented for:
    \begin{itemize}
        \item $Q = 3 H (\delta_{\text{dm}} \rho_{\text{dm}} + \delta_{\text{de}} \rho_{\text{de}})$ in Subsection~\ref{Q_General},
        \item $Q = 3 H \delta (\rho_{\text{dm}} + \rho_{\text{de}})$ in Subsection~\ref{Q_dm+de},
        \item $Q = 3 H \delta (\rho_{\text{dm}} - \rho_{\text{de}})$ in Subsection~\ref{Q_dm-de},
        \item $Q = 3 H \delta \rho_{\text{dm}}$ in Subsection~\ref{Q_dm},
        \item $Q = 3 H \delta \rho_{\text{de}}$ in Subsection~\ref{Q_de}.
    \end{itemize}
    The results from this section are consistent with the findings of the dynamical system analysis presented in Section~\ref{Sec.DSA}.  To the best of our knowledge, all expressions in Subsection~\ref{Q_General}, \ref{Q_dm+de} and \ref{Q_dm-de} are new, while the results in Subsection~\ref{Q_dm} and \ref{Q_de} were previously discussed in \cite{vanderWesthuizen:2023hcl} and are presented here to ease comparison between models, and more importantly, to show that the new results in Subsection~\ref{Q_General} for the general interaction $Q = 3 H (\delta_{\text{dm}} \rho_{\text{dm}} + \delta_{\text{de}} \rho_{\text{de}})$ reduce back to familiar results in the special cases where either $\delta_{\rm{dm}}=0$ or $\delta_{\rm{de}}=0$.
    
    \item In Section~\ref{reconstructed_w}, for each of the five interactions studied, we derive expressions for a reconstructed dynamical dark energy equation of state $\tilde{w}(z)$. This is useful when comparing IDE models to other models and parametrizations of DE or modified gravity. We also use this section to show the relationship between the different equations of state used throughout this paper, namely $w$, $w^{\rm eff}_{\rm de}$, $w^{\rm eff}_{\rm dm}$, $w^{\rm eff}_{\rm tot}$, and $\tilde{w}$, which are plotted for each interaction in Figure \ref{fig:w_all_Qdm+de},  \ref{fig:w_all_Qdm-de},  \ref{fig:w_all_Qdm} and \ref{fig:w_all_Qde}. 
    
    \item In Section~\ref{statefinder}, we consider the statefinder diagnostics to help differentiate between the four special cases of the linear interaction studied in Section~\ref{Background_cosmology}. We plot the evolution of the statefinder parameters $r_{\rm{sf}}$, $s_{\rm{sf}}$, and $q$ in Figure~\ref{fig:Statefinder_linear_iDEDM}. We compare our results with those in the literature and examine how each interaction differs from the $\Lambda$CDM case at early and late times by presenting the new expressions in Tables~\ref{tab:Statefinder_past_linear}, \ref{tab:Statefinder_present_linear}, and \ref{tab:Statefinder_future_linear}.

    \item Finally, in Section~\ref{summary}, we summarize all our main results in a set of tables. We provide conditions to avoid parameter-space regions where the solutions become undefined or imaginary (Table~\ref{tab:Com_real}), lead to negative DM or DE energy densities (Table~\ref{tab:Com_PEC}), or predict future big rip singularities (Table~\ref{tab:Com_AE_BR}). We also evaluate how each model addresses the coincidence problem in both the past and future (Table~\ref{tab:Com_CP}). We conclude with a discussion of our results and outline directions for future research.
    
\end{itemize}

\section{Background on IDE models} \label{BG_lit}
\subsection{Understanding negative energies in IDE cosmology} \label{BG_neg}
 
In chemical or nuclear reactions, if the density of the decaying component becomes zero, the interaction or decay stops. Conversely, in the dark sector, there is not always a mechanism to halt the energy transfer when either the DM or DE density becomes zero (i.e., $Q \neq 0$ in \eqref{eq:NEGEN_1} when $\rho_{\text{de/dm}} = 0$), which can lead to negative energy densities.  
In this study, we consider special cases of an interaction of the form $Q = 3H (\delta_{\text{dm}} \rho_{\rm{dm}} + \delta_{\text{de}} \rho_{\rm{de}})$, which serves to illustrate the negative energy problem. For this interaction, the conservation equations are: 
\begin{gather} \label{eq:NEGEN_1}
\begin{split}
\dot{\rho}_{\text{dm}} + 3H \rho_{\text{dm}} &= 3H (\delta_{\text{dm}} \rho_{\rm{dm}} + \delta_{\text{de}} \rho_{\rm{de}}), \\
\dot{\rho}_{\text{de}} + 3H (1 + w) \rho_{\text{de}} &= -3H (\delta_{\text{dm}} \rho_{\rm{dm}} + \delta_{\text{de}} \rho_{\rm{de}}).
\end{split}
\end{gather}
Negative energy densities can only arise from positive initial conditions if there is still non-zero energy transfer ($\dot{\rho} \ne 0$) at the point where the energy density reaches zero ($\rho = 0$). In such cases, there is no mechanism to brake the transfer of energy, leading to negative values.  
In the non-interacting case, where the right-hand side of \eqref{eq:NEGEN_1} is zero, we can clearly define:
\begin{gather} \label{eq:NEGEN_2}
\begin{split}
\text{Dark Matter Braking Mechanism (DMBM): if } \dot{\rho}_{\text{dm}} \ge 0 \text{ at } \rho_{\text{dm}} = 0 \quad &\Rightarrow \quad \rho_{\text{dm}} \ge 0 \quad \forall \; a, \\
\text{Dark Energy Braking Mechanism (DEBM): if } \dot{\rho}_{\text{de}} \ge 0 \text{ at } \rho_{\text{de}} = 0 \quad &\Rightarrow \quad \rho_{\text{de}} \ge 0 \quad \forall \; a,
\end{split}
\end{gather}
where $a$ is the scale factor describing the expansion of the universe. These braking mechanisms constrain both DM and DE to remain within the positive energy domain, as they halt the flow of energy at the zero-energy boundary. By applying the DMBM and DEBM to the special cases of the interaction in \eqref{eq:NEGEN_1} (as well as to the three non-linear interactions studied in a companion paper \cite{vanderWesthuizen:2025II}), we can determine \textit{a priori} from Table~\ref{tab:positivityQ} whether the DM or DE densities will remain positive throughout cosmic evolution.

\begin{table}[h!]
\renewcommand{\arraystretch}{1.2} 
\setlength{\tabcolsep}{2pt}
\centering
\begin{tabular}{|c|c|c|c|c|}
\hline 
$\textbf{Interaction Q}$ & $\dot{\rho}_{\text{dm}} \textbf{ at } \rho_{\text{dm}}=0$  & $\rho_{\text{dm}}\ge0 \;\forall  \;a$   &$ \dot{\rho}_{\text{de}} \textbf{ at } \rho_{\text{de}}=0$ & $ \rho_{\text{de}}\ge0 \;\forall  \;a$ \\ \hline \hline
$3H \delta \rho_{\text{dm}}$ & $0$ & $\checkmark$ &$-3H \delta \rho_{\text{dm}}$ & $X$ \\ \hline
$3H\delta \rho_{\text{de}}$ & $3H \delta \rho_{\text{de}}$ & $X$ & $0$ & $\checkmark$  \\ \hline
$3H\delta( \rho_{\text{dm}}+\rho_{\text{de}})$ & $3H\delta \rho_{\text{de}}$ & $X$ & $-3H\delta \rho_{\text{dm}}$ & $X$  \\ \hline
$3H\delta( \rho_{\text{dm}}-\rho_{\text{de}})$ & $-3H\delta \rho_{\text{de}}$ &$ X$ & $-3H\delta \rho_{\text{dm}}$ & $X$  \\ \hline
$3H(\delta_{\text{dm}} \rho_{\text{dm}}+\delta_{\text{de}} \rho_{\text{de}})$ &$ 3H\delta_{\text{de}} \rho_{\text{de}}$ & $X$ & $-3H\delta_{\text{dm}} \rho_{\text{dm}}$ & $X$  \\ \hline 
$3H\delta \left(\frac{\rho_{\text{dm}}\rho_{\text{de}} }{\rho_{\text{dm}}+\rho_{\text{de}}} \right)$  & $0$ & $\checkmark$ & $0$ & $ \checkmark$ \\ \hline
$3H\delta \left(\frac{\rho^2_{\text{dm}} }{\rho_{\text{dm}}+\rho_{\text{de}}} \right)$  & $0$ & $\checkmark$ &$-3H \delta \rho_{\text{dm}}$ & $X$ \\ \hline
$3H\delta \left(\frac{\rho^2_{\text{de}} }{\rho_{\text{dm}}+\rho_{\text{de}}} \right)$ & $3H \delta \rho_{\text{de}}$ & $X$ & $0$ & $\checkmark$  \\ \hline
\end{tabular}
\caption{Possible presence of negative energies for various interaction kernels.}
\label{tab:positivityQ}
\end{table}

The absence of these braking mechanisms is what often leads to negative energy densities in IDE models. Consider what happens at the boundary where $\rho_{\text{dm}} = 0$ in the IDE conservation equation \eqref{eq:NEGEN_1}, both in the future and the past, \textit{noting that the direction of energy flow reverses when we consider the past (i.e., evolving backwards in time)}:
\begin{gather} \label{eq:NEGEN_3}
\begin{split}
\text{At }&\rho_{\text{dm}} = 0: \quad  \dot{\rho}_{\text{dm}} = 3H \delta_{\text{de}} \rho_{\text{de}}, \\ 
\text{if } \delta_{\text{de}} > 0 \; &\begin{cases}
\text{Future: } \quad \dot{\rho}_{\text{dm}} > 0  \;(\text{DE} \rightarrow  \text{DM}) \quad \Rightarrow \quad \rho_{\text{dm}} \geq 0  \quad \checkmark \\
\text{Past: } \quad\quad \dot{\rho}_{\text{dm}} < 0 \; (\text{DM} \rightarrow  \text{DE}) \quad \Rightarrow \quad \rho_{\text{dm}} \leq 0  \quad \text{X (less likely)}  
\end{cases}  \\ 
\text{if } \delta_{\text{de}} < 0  \; &\begin{cases}
\text{Future: } \quad \dot{\rho}_{\text{dm}} < 0 \; (\text{DM} \rightarrow  \text{DE}) \quad \Rightarrow \quad \rho_{\text{dm}} \leq 0  \quad \text{X (more likely)} \\
\text{Past: } \quad\quad \dot{\rho}_{\text{dm}} > 0  \;(\text{DE} \rightarrow  \text{DM}) \quad \Rightarrow \quad \rho_{\text{dm}} \geq 0  \quad \checkmark
\end{cases} ,
\end{split}
\end{gather}
where the symbols "$\checkmark$" and "X" indicate regions where negative energy densities are avoided or present, respectively. Similarly, the "less likely" and "more likely" correspond  concretely to the lower and upper bounds in Table~\ref{tab:Com_PEC}). In \eqref{eq:NEGEN_3}, the most problematic regime is $\delta_{\text{de}} < 0$ in the future, where in uncoupled models we already expect $\rho_{\text{dm}} \rightarrow 0$ as $a \rightarrow \infty$. For this IDE model, once $\rho_{\text{dm}} = 0$, DM continues to transfer energy to DE because there is no DMBM to stop the interaction, resulting in negative DM densities.
The presence of negative DM energy densities for $\delta_{\text{de}} < 0$ will be a recurring theme in this paper, ultimately ruling out the iDMDE regime for all models considered. Moreover, $\rho_{\text{dm}}$ can also become negative in the past when $\delta_{\text{de}} > 0$. However, this situation is less likely, as DM should dominate in the early universe and would not easily approach the $\rho_{\text{dm}} = 0$ boundary. This scenario only arises for sufficiently large values of $\delta_{\text{de}}$, as shown by the derived upper bounds on $\delta_{\text{de}}$ throughout the paper.
We obtain a similar result for the DE conservation equation:
\begin{gather} \label{eq:NEGEN_4}
\begin{split}
\text{At } &\rho_{\text{de}} = 0: \quad  \dot{\rho}_{\text{de}} = -3H \delta_{\text{dm}} \rho_{\text{dm}}, \\ 
\text{if } \delta_{\text{dm}} > 0 \; &\begin{cases}
\text{Future: } \quad \dot{\rho}_{\text{de}} < 0  \;(\text{DE} \rightarrow  \text{DM}) \quad \Rightarrow \quad \rho_{\text{de}} \leq 0  \quad \text{X (less likely)} \\
\text{Past: } \quad\quad \dot{\rho}_{\text{de}} > 0 \; (\text{DM} \rightarrow  \text{DE}) \quad \Rightarrow \quad \rho_{\text{de}} \geq 0  \quad \checkmark
\end{cases}  \\ 
\text{if } \delta_{\text{dm}} < 0  \; &\begin{cases}
\text{Future: } \quad \dot{\rho}_{\text{de}} > 0 \; (\text{DM} \rightarrow  \text{DE}) \quad \Rightarrow \quad \rho_{\text{de}} \geq 0  \quad \checkmark \\
\text{Past: } \quad\quad \dot{\rho}_{\text{de}} < 0  \;(\text{DE} \rightarrow  \text{DM}) \quad \Rightarrow \quad \rho_{\text{de}} \leq 0  \quad \text{X (more likely)}
\end{cases} .
\end{split}
\end{gather}
In \eqref{eq:NEGEN_4}, the most problematic domain arises when $\delta_{\text{dm}} < 0$ in the past, since $\rho_{\rm{dm}}$ is typically large at early times, resulting in a strong energy transfer that can drive $\rho_{\text{de}}$ to zero. In the absence of a DEBM, the energy transfer continues beyond this point, causing $\rho_{\text{de}}$ to become negative. This observation, along with the argument made earlier regarding DM positivity, leads us to rule out the $\delta_{\text{dm}} < 0$ case (iDMDE regime) for all models considered in this study.
The other problematic domain occurs when $\rho_{\text{de}}$ becomes negative in the future for $\delta_{\text{dm}} > 0$. This situation is less likely, as both $\rho_{\text{dm}}$ and $\rho_{\text{de}}$ tend to decrease in the future, reducing the strength of the interaction before either component reaches the zero-energy boundary. As a result, the system often avoids crossing into negative energy. This scenario is only possible if $\delta_{\text{dm}}$ is very large, as reflected by the upper bounds on $\delta_{\text{dm}}$ derived throughout this paper.

\begin{figure}
    \centering
    \includegraphics[width=0.99\linewidth]{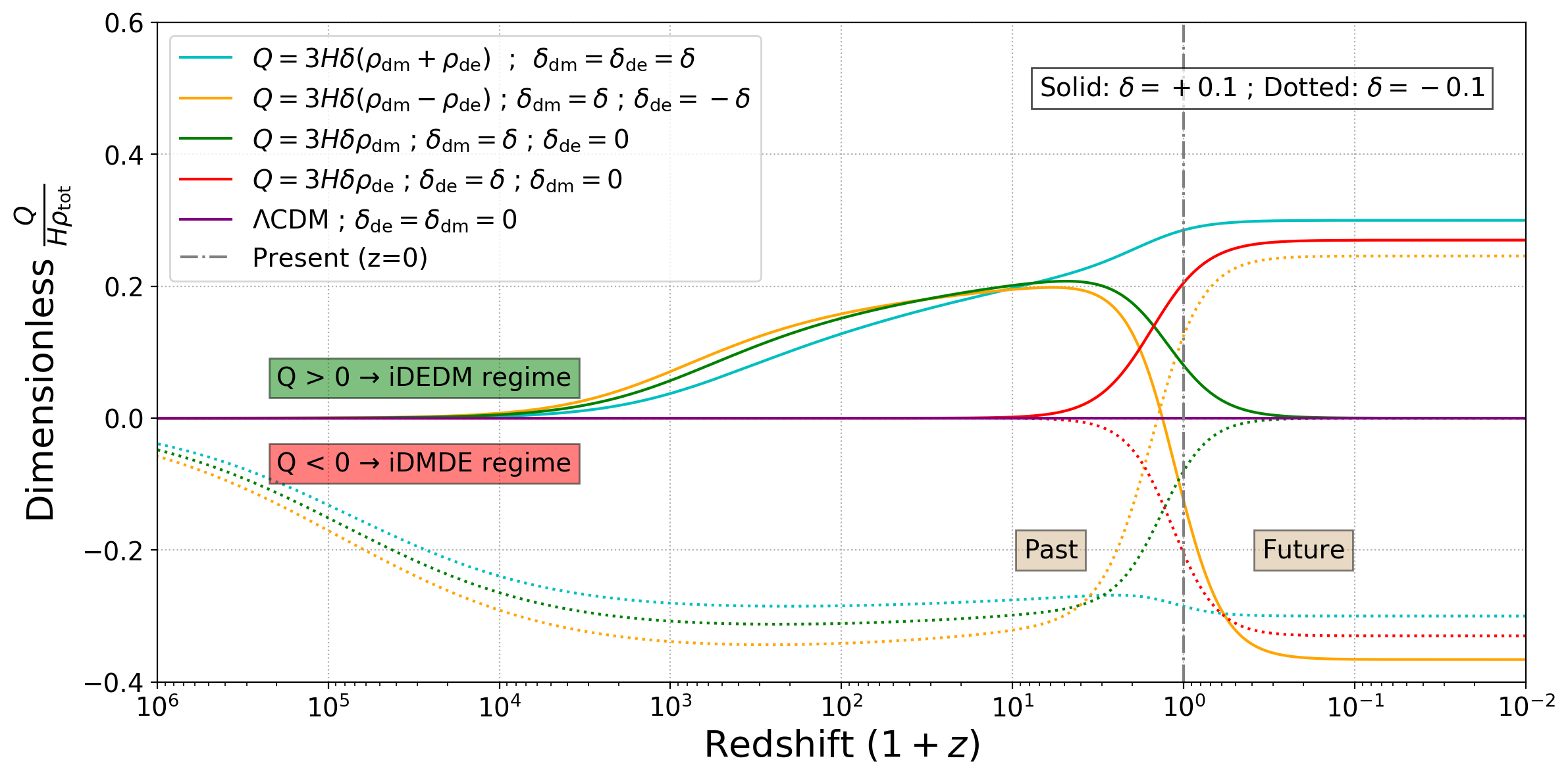}
\caption{Interaction strength $Q$ relative to $H \rho_{\rm{tot}}$ as a function of redshift for the interaction $Q = 3H (\delta_{\text{dm}} \rho_{\rm{dm}} + \delta_{\text{de}} \rho_{\rm{de}})$, shown for different special cases. The plot illustrates when the effect of the interaction becomes dominant during cosmic evolution. If $Q \propto \rho_{\rm{dm}}$, the interaction primarily affects the early-time dynamics, whereas if $Q \propto \rho_{\rm{de}}$, it mainly influences the late-time and future expansion. Sign-switching behavior occurs only when $\delta_{\text{dm}}$ and $\delta_{\text{de}}$ have opposite signs.}
    \label{fig:Q_Linear_dm+de}
\end{figure}

From these arguments, we may immediately conclude that the interaction $Q = 3H\delta\rho_{\text{dm}}$ can never lead to negative $\rho_{\text{dm}}$, while $Q = 3H\delta\rho_{\text{de}}$ can never lead to negative $\rho_{\text{de}}$, as in both cases the energy transfer is braked at the zero crossing (as illustrated in Figures~\ref{fig:Omega_Linear_dm} and~\ref{fig:Omega_Linear_de}). Conversely, if $Q \propto \pm \rho_{\text{dm}} \pm \rho_{\text{de}}$, then both $\rho_{\text{dm}}$ and $\rho_{\text{de}}$ may become negative (as seen in Figures~\ref{fig:Omega_Linear_dm+de} and~\ref{fig:Omega_Linear_dm-de}). 
Lastly, any interaction that is a product of the two energy densities, $Q \propto \rho_{\text{dm}} \rho_{\text{de}}$, will always yield positive energy densities for both components. An example of such a model, with $Q = 3H \delta \left( \frac{\rho_{\rm{dm}} \rho_{\rm{de}}}{\rho_{\rm{dm}} + \rho_{\rm{de}}} \right)$, will be studied in a companion paper \cite{vanderWesthuizen:2025II}. These results are summarized in Table~\ref{tab:positivityQ}, while the exact conditions for maintaining positive energy densities are given in Table~\ref{tab:Com_PEC}.
To visually interpret the arguments in \eqref{eq:NEGEN_3} and \eqref{eq:NEGEN_4}, Figure~\ref{fig:Q_Linear_dm+de} may be used, especially for special cases of $\delta_{\text{dm}}$ and $\delta_{\text{de}}$, where the interaction becomes prominent at different epochs. Figure~\ref{fig:Q_Linear_dm+de} shows the dimensionless variable $Q / H \rho_{\text{tot}}$, which highlights both the direction and relative magnitude of the energy transfer as a function of redshift.

It may be added that negative energy densities ($\rho < 0$) violate several key energy conditions in general relativity, including the weak energy condition (WEC) and the dominant energy condition (DEC). The strong energy condition (SEC) is also violated by any fluid that drives accelerated expansion, i.e., with $w < -\frac{1}{3}$. Moreover, for the IDE models considered in this study, stability arguments based on the doom factor (discussed later) may require phantom dark energy ($w < -1$), which additionally violates the null energy condition (NEC), alongside the other three conditions mentioned above. For a more comprehensive discussion on energy conditions and their connection to negative energy densities and phantom equations of state, see~\cite{ Carroll_2003, Santos_2007, Rubakov_2014, martinmoruno2017classicalsemiclassicalenergyconditions}. A discussion on WEC violation in the case of $Q = 3H\delta \rho_{\text{de}}$ can be found in~\cite{von_Marttens_2020}, which concluded that $\delta > 0$ (the iDEDM regime) is required to avoid this issue.
It has been argued that, since the nature of dark energy is still unknown, one might entertain the possibility of $\rho_{\text{de}} < 0$, as discussed in Appendix A3 of~\cite{V_liviita_2010}. There, it is shown that even if negative DE is allowed in the iDMDE regime with $Q = 3H\rho_{\text{dm}}$, the model necessarily involves a zero-energy crossing ($\rho_{\text{de}} = 0$) in the past. Furthermore, because the perturbation equations for this interaction (given in~\cite{V_liviita_2008}) contain $\rho_{\text{de}}$ in the denominator, they become singular at the crossing, effectively ruling out the model. Researchers considering other interacting models that permit negative energy densities are advised to be cautious of similar pathologies.
Recent discussions on the role of negative DE densities in alleviating cosmic tensions can be found in Section 4.2.2 of~\cite{divalentino2025cosmoversewhitepaperaddressing}. Hints of negative DE have also been suggested by observational reconstructions of the interaction kernel~\cite{Escamilla_2023}, and other reconstruction approaches show a preference for sign-switching behavior~\cite{guedezounme2025phantomcrossingdarkinteraction}. Other discussions on the possibility of negative DE are found in \cite{Poulin:2018zxs, Wang:2018fng, Visinelli:2019qqu, Calderon:2020hoc}. For now, we leave the question of the viability of negative DE densities to the discretion of the reader.

\subsection{Understanding future big rip singularities in IDE cosmology} \label{BG_rip}

The presence of big rip future singularities is well understood in the non-interacting regime when dark energy (DE) has an equation of state in the phantom regime ($w < -1$)~\cite{Caldwell_2003, Nojiri_2004, Nojiri_2005, Nojiri_2017, Elizalde_2019, Trivedi_2024}. In the asymptotic future, DE typically dominates the energy budget of the universe. For phantom DE, this implies that the total effective equation of state of the universe satisfies $w_{\text{tot}}^{\text{eff}} < -1$, where $w_{\text{tot}}^{\text{eff}}$ is defined in \eqref{DSA.omega_eff_tot}.
The largest physical distance over which causal processes can occur—i.e., the causally connected region of the universe—is given by the particle horizon $D_H$. The particle horizon is related to the Hubble parameter, which itself depends on the total effective equation of state, leading to the relation:
\begin{gather} \label{horizon.size}
\begin{split}
D_H = \frac{c}{H}\;, \quad H^2 \propto \rho_{\text{tot}} \propto a^{-3(1 + w_{\text{tot}}^{\text{eff}})} \quad \Rightarrow \quad D_H \propto a^{\frac{3}{2}(1 + w_{\text{tot}}^{\text{eff}})}.
\end{split}
\end{gather}
From \eqref{horizon.size}, it follows that when $w_{\text{tot}}^{\text{eff}} < -1$, the particle horizon $D_H$ decreases as the scale factor $a$ increases. Physically, this means that the number of galaxies within the horizon will decrease at an accelerating rate, as the horizon closes in on any observer. This shrinking continues until gravitationally bound systems are disrupted and the horizon eventually approaches the Planck scale, making communication between any two regions of the universe impossible~\cite{Caldwell_2003}.
The big rip singularity is characterized by both the DE density and the scale factor diverging ($\rho_{\text{de}} \rightarrow \infty$ and $a \rightarrow \infty$) within a finite time $t_{\text{rip}}$. This behavior, in the context of IDE cosmology, is illustrated in Figures~\ref{fig:eos_tot_BR_Qdm+de}, \ref{fig:eos_tot_BR_Qdm-de}, \ref{fig:eos_tot_BR_Qdm}, and~\ref{fig:eos_tot_BR_Qde}.
A big rip scenario within IDE cosmology differs from the non-interacting case~\cite{Nojiri_2005_IDE, Curbelo_2006, Pan_2020, de_Haro_2023, vanderWesthuizen:2023hcl}, as dark energy acquires an effective equation of state $w^{\text{eff}}_{\text{de}}$ (defined in \eqref{DSA.omega_eff_dm_de}) that differs from the intrinsic $w_{\text{de}}$. The situation is further complicated by the fact that, in many IDE models, there is hybrid dark matter and dark energy dominance in the asymptotic future. Therefore, the condition $w_{\text{de}} < -1$, which signals a big rip in non-interacting cosmology, is no longer sufficient in the interacting case.
Instead, the relevant condition must involve the total effective equation of state $w^{\rm{eff}}_{\rm{tot}}$, leading to:
\begin{equation} \label{omega_eff_tot_BG}
\boxed{
\text{Big rip condition for IDE cosmology:} \quad w^{\rm{eff}}_{\rm{tot}} < -1 \quad \text{in the asymptotic future.}
}
\end{equation}
This implies that a big rip may occur for $w$ in either the quintessence or phantom regimes, but it is more likely in the iDMDE regime, as this pushes $w^{\text{eff}}_{\text{de}}$ further into the phantom domain.
For the five interaction kernels studied in this paper, we derive the corresponding big rip conditions in equations~\eqref{omega_eff_tot_general_BG}, \eqref{omega_eff_tot_dm+de_BG}, \eqref{omega_eff_tot_dm-de_BG}, \eqref{omega_eff_tot_dm_BG}, and~\eqref{omega_eff_tot_de_BG}. The corresponding expressions for the time of the big rip, $t_{\text{rip}}$, are given by~\eqref{eq:Big_Rip_ddm+dde_BG}, \eqref{eq:Big_Rip_dm+de_BG}, \eqref{eq:Big_Rip_dm-de_BG}, \eqref{eq:Big_Rip_dm_BG}, and~\eqref{eq:Big_Rip_de_BG}.

\subsection{Background equations} \label{BG_equations}

The background equations that hold for any interaction $Q$, and which we aim to investigate, are:
\begin{gather} \label{DSA.H}
\begin{split}
H^2 = \left( \frac{\dot{a}}{a} \right)^2 = \frac{8\pi G}{3} \left( \rho_{\text{r}} + \rho_{\text{bm}} + \rho_{\text{dm}} + \rho_{\text{de}} \right), \quad
\rho_{\text{r}} = \rho_{\text{(r,0)}} a^{-4}, \quad
\rho_{\text{bm}} = \rho_{\text{(bm,0)}} a^{-3}.
\end{split}
\end{gather}
Equation~\eqref{DSA.H} is the Hubble equation for a flat universe containing dark matter (DM), dark energy (DE), radiation (r), and baryons (bm). Only the expressions for the DM and DE densities differ from those in non-interacting models, depending on the choice of interaction kernel $Q$. It should be noted that baryonic matter $\rho_{\text{bm}}$ and dark matter $\rho_{\text{dm}}$ are separately conserved and evolve independently. The energy densities can be expressed as density parameters via the relation $\frac{8\pi G}{3H^2} \rho_i = \Omega_i$.
\begin{gather} \label{DSA.omega_eff_dm_de}
\begin{split}
w^{\rm{eff}}_{\rm{dm}} = - \frac{Q}{3 H \rho_{\rm{dm}}}, \quad
w^{\rm{eff}}_{\rm{de}} = w_{\rm{de}} + \frac{Q}{3 H \rho_{\rm{de}}}.
\end{split}
\end{gather}
Equation~\eqref{DSA.omega_eff_dm_de} defines the effective equations of state for DM and DE. This provides an equivalent description of how each dark component would evolve in the absence of an interaction, as in \eqref{eq:conservation}, but instead with a time-varying equation of state.  
\begin{gather} \label{DSA.r}
\begin{split}
r \equiv \frac{\rho_{\rm{dm}}}{\rho_{\rm{de}}}
= \frac{\Omega_{\rm{dm}}}{\Omega_{\rm{de}}}
= \frac{\rho_{\text{dm,0}}\, a^{-3(1 + w^{\text{eff}}_{\text{dm}})}}{\rho_{\text{de,0}}\, a^{-3(1 + w^{\text{eff}}_{\text{de}})}}
= r_0\, a^{-3(w^{\text{eff}}_{\text{dm}} - w^{\text{eff}}_{\text{de}})}, 
\quad \zeta \equiv 3\left(w^{\text{eff}}_{\text{dm}} - w^{\text{eff}}_{\text{de}}\right).
\end{split}
\end{gather}
Equation~\eqref{DSA.r} defines the ratio $r$ of dark matter (DM) to dark energy (DE), which is a key quantity for discussing the coincidence problem. When $r$ remains constant over time, the coincidence problem is considered \emph{solved}. The parameter $\zeta$ quantifies the severity of the coincidence problem in interacting dark energy (IDE) models by measuring the deviation from $\zeta_{\rm IDE} = 0$. In the $\Lambda$CDM model, $\zeta = 3$.
\begin{gather} \label{DSA.zeta}
\begin{split}
\zeta_{\Lambda \text{CDM}} = 3 \; \Rightarrow \;
\begin{cases}
|\zeta_{\rm IDE}| > 3 & \text{\emph{worsens} the coincidence problem}, \\
|\zeta_{\rm IDE}| < 3 & \text{\emph{alleviates} the coincidence problem}, \\
\zeta_{\rm IDE} = 0 & \text{\emph{solves} the coincidence problem}.
\end{cases}
\end{split}
\end{gather}
From equation~\eqref{DSA.r}, we see that the closer the DM and DE effective equations of state are to each other, the more the coincidence problem is alleviated. In the special case where $w^{\text{eff}}_{\text{dm}} = w^{\text{eff}}_{\text{de}}$, the coincidence problem is completely solved, as the two fluids redshift at exactly the same rate, keeping their ratio fixed over cosmic time. In general, the coincidence problem is alleviated in the iDEDM regime, while it is worsened in the iDMDE regime.
\begin{gather} \label{DSA.omega_eff_tot}
\begin{split}
w^{\rm{eff}}_{\rm{tot}} 
= \frac{P_{\rm{tot}}}{\rho_{\rm{tot}}} 
= \frac{w_{\rm{r}} \Omega_{\rm{r}} + w_{\rm{bm}} \Omega_{\rm{bm}} + w_{\rm{dm}} \Omega_{\rm{dm}} + w_{\rm{de}} \Omega_{\rm{de}}}
{\Omega_{\rm{r}} + \Omega_{\rm{bm}} + \Omega_{\rm{dm}} + \Omega_{\rm{de}}} 
= \frac{1}{3} \Omega_{\rm{r}} + w\, \Omega_{\rm{de}}.
\end{split}
\end{gather}
Equation~\eqref{DSA.omega_eff_tot} gives the total effective equation of state $w^{\rm{eff}}_{\rm{tot}}$, treating the contents of the Universe as a single effective fluid whose pressure is related to its total energy density. The last equality holds for a flat universe with $w_{\rm{r}} = \frac{1}{3}$, $w_{\rm{bm}} = w_{\rm{dm}} = 0$, and $w_{\rm{de}} = w$.  
The sign of $w^{\rm{eff}}_{\rm{tot}}$ determines the expansion behavior: 
\begin{gather} \label{DSA.omega_eff_tot_ACC}
\begin{split}
\text{In the asymptotic future, if   }  \; &\begin{cases}
w^{\rm{eff}}_{\rm{tot}} > -\frac{1}{3} \quad \rightarrow \quad \text{decelerating expansion,} \\
w^{\rm{eff}}_{\rm{tot}} < -\frac{1}{3} \quad \rightarrow \quad \text{accelerated expansion}. \\
\end{cases}
\end{split}
\end{gather}

As discussed in Section~\ref{BG_rip}, a big rip occurs when the future attractor satisfies $w^{\rm{eff}}_{\rm{tot}} < -1$.
\begin{gather} \label{DSA.q}
\begin{split}
q &= -\frac{\ddot{a}a}{\dot{a}^2}   
= \Omega_{\rm{r}} + \frac{1}{2} \left( \Omega_{\rm{bm}} + \Omega_{\rm{dm}} \right)  
+ \frac{1}{2} \Omega_{\rm{de}} \left( 1 + 3w \right) 
= \frac{1}{2} \left( 1 + 3\Omega_{\rm{de}} w \right).
\end{split}
\end{gather}
Equation~\eqref{DSA.q} defines the deceleration parameter $q$, which measures whether the expansion of the Universe is slowing down ($q>0$) or speeding up ($q<0$). In the context of IDE models, a big rip singularity in the future may occur if $q < -1$.
\begin{gather} \label{DSA.r_st}
\begin{split}
r_{\rm sf} &= \frac{\dddot{a}}{aH^3} = 2q^2 + q + (1+z) \frac{dq}{dz} = 1 + \frac{9}{2} \, \Omega_{\rm de} \, w \left( 1 + w^{\rm{eff}}_{\rm de} \right), \\
s_{\rm sf} &= \frac{r_{\rm sf} - 1}{3\left(q - \frac{1}{2} \right)} = 1 + w^{\rm{eff}}_{\rm de}.
\end{split}
\end{gather}
Equation~\eqref{DSA.r_st} introduces higher-order time derivatives of the scale factor, which are used to distinguish between models that share similar background expansion histories. These include the statefinder diagnostics $r_{\rm sf}$ and $s_{\rm sf}$ proposed by~\cite{Sahni_2003, Alam_2003}. The subscripts are used here to avoid confusion with the ratio $r$ defined in equation~\eqref{DSA.r}.
In the context of IDE cosmology, where models are often analyzed using the effective equations of state in~\eqref{DSA.omega_eff_dm_de}, the statefinder parameters in  have been shown by~\cite{Zhang_2006} to be equivalent to the final equalities in ~\eqref{DSA.r_st}.

It is important to note that, for the $\Lambda$CDM model, both parameters approach a single fixed point in the late-time limit, $(r_{\rm sf}, s_{\rm sf}) = (1,0)$. The deviation from this point provides a useful diagnostic for distinguishing between different DE models.

Lastly, the stability of IDE models is usually determined by the sign of a doom factor $\textbf{d}$, first proposed in~\cite{M.B.Gavela_2009} and given for any interaction $Q$ as:
\begin{gather} \label{DSA.doom}
\begin{split}
\textbf{d} = \frac{Q}{3H\rho_{\rm{de}}(1+w)}.
\end{split}
\end{gather}
The sign of $\textbf{d}$ will determine if there are early-time instabilities, and more specifically, if there are non-adiabatic instabilities in the DE perturbation equations. If $\textbf{d}>1$, these DE perturbations will have a runaway unstable growth regime, while if $\textbf{d}<0$, the model should be free of these instabilities, giving an \textit{a priori} stable universe~\cite{M.B.Gavela_2009}. During observational constraints, the doom factor is generally used to define two stable regimes where $\textbf{d}$ is negative~\cite{M.B.Gavela_2009, Salvatelli_2013, Costa_2014, Yang_2016, Costa_2017, Costa_2018, Yang_2018_DOOM, Bachega_2020, Yang_2020, Lucca_2020, Valentino_2020_DE, Di_Valentino_2020, Di_Valentino_2021, Di_Valentino_2021_closed, lucca2021darkenergydarkmatterinteractions, Gariazzo_2022, Yang_2021, Joseph_2022, Nunes_2022, Califano_2023, Ghodsi_Yengejeh_2023, Forconi_2024, Giar__2024, Sabogal_2024}. These are the iDMDE regime combined with $w>-1$, and the iDEDM regime with $w<-1$. The vacuum scenario where $w=-1$ has also been shown to cause gravitational instabilities~\cite{Lucca_2020, lucca2021darkenergydarkmatterinteractions}.
It should be mentioned that, in recent years, it has been proposed that this issue can be resolved and that the whole parameter space may be opened up for observational constraints. This is achieved by considering perturbations in the Parametrized Post-Friedmann Framework~\cite{Li_2014, Li_2014_2, Skordis_2015, Zhang_2017, Feng_2018, Dai_2019, Li_2020_2, Li_2023}, which even has a CAMB implementation~\cite{Li_2023}. Further discussion on this matter falls outside the scope of this study, but it is worthy of future investigation.

For purely illustrative purposes, and for ease of comparison between IDE models and the $\Lambda$CDM model, we will plot all figures using the parameters $H_0=67.4$ km/s/Mpc, $\Omega_{\rm{(r,0)}}=9\times10^{-5}$, $\Omega_{\rm{(bm,0)}}=0.049$, $\Omega_{\rm{(dm,0)}}=0.266$, $\Omega_{\rm{(de,0)}}=0.685$, $w=-1$, and $\delta=\pm0.1$, unless otherwise stated.

\subsection{Literature on each interaction} \label{BG_lit_Q}
The five interaction kernels that we are studying have each been examined to varying degrees in the literature, although there remain gaps that we attempt to address in this study. In the following, we provide an overview of some of the main works associated with each interaction kernel. The groupings below are not definitive, as many papers overlap with other groups and either address multiple interaction kernels or discuss several theoretical and/or observational aspects of the interactions considered. The list presented here is not exhaustive, but is intended as a rough guide for researchers interested in specific aspects of certain interaction kernels. 

\begin{itemize}

\item \textbf{Literature on linear IDE model 1:} $\boldsymbol{Q= 3 H (\delta_{\text{dm}} \rho_{\text{dm}} + \delta_{\text{de}}  \rho_{\text{de}})}$ — 
Analytical solutions and background dynamics~\cite{Chimento_2012, Pan_2015, Pan_2017, von_Marttens_2019}; dynamical system analysis and discussions on negative energy densities~\cite{Quartin_2008, He_2008, Caldera_Cabral_2009_DSA, Caldera_Cabral_2009_structure, Pan_2020}; big rip future singularities~\cite{Nojiri_2005_IDE, Pan_2020, de_Haro_2023}; structure growth and instabilities~\cite{Caldera_Cabral_2009_structure, Väliviita_2008, He_2009, He_2011, Costa_2014, Eingorn_2015, Costa_2017}; coincidence problem~\cite{Sadjadi_2006}; statefinder diagnostics~\cite{carrasco2023discriminatinginteractingdarkenergy}; Bayesian comparison~\cite{Arevalo_2017, Cid_2019}; holographic modelling~\cite{Lepe_2016}; other~\cite{Costa_2019, An_2019, Mahata_2015}.

\item \textbf{Literature on linear IDE model 2:} $\boldsymbol{Q=3H\delta( \rho_{\text{dm}}+\rho_{\text{de}})}$ — 
Analytical solutions and background dynamics~\cite{Olivares_2005, Califano_2024}; dynamical system analysis and discussions on negative energy densities~\cite{V_liviita_2008, Pan_2020, Ar_valo_2022}; structure growth and instabilities~\cite{Olivares_2006, V_liviita_2008, Väliviita_2008, He_2009, He_2011}; coincidence problem~\cite{Campo_2009, Wang:2016lxa}; statefinder diagnostics~\cite{Ar_valo_2022}; Bayesian comparison~\cite{Arevalo_2017}; holographic modelling~\cite{Feng_2016, Huang_2019, AbdollahiZadeh:2019lsx}; field theory interpretation~\cite{Pan_2020_Field}; observational constraints~\cite{Costa_2014, Costa_2017, An_2017, An_2018, Bachega_2020, Li_2020, Halder_2021, Mukhopadhyay_2021, Xiao_2021, Aljaf_2021, Califano_2024}; N-body simulations~\cite{Zhang_2018, Zhang_2019, Liu_2022}; gravitational wave predictions~\cite{Yang_2019, Bachega_2020, Califano_2024}; other~\cite{Sharma_2021}.

\item \textbf{Literature on linear IDE model 3:} $\boldsymbol{Q=3H\delta( \rho_{\text{dm}}-\rho_{\text{de}})}$ or $\boldsymbol{Q=3H\delta( \rho_{\text{de}}-\rho_{\text{dm}})}$ — First proposal of the interaction kernel~\cite{Sun_2012}; dynamical system analysis~\cite{Zhang_2014, halder2024phasespaceanalysissignshifting}; addressing the Hubble tension~\cite{Pan_2019, Pan_2024}; observational constraints~\cite{Aljaf_2021}; Ghost dark energy modelling~\cite{Khurshudyan_2015}; other sign-changing interactions~\cite{Forte_2014, Guo_2018, Arevalo_2019, Pan_2020_Qsin, li2025probingsignchangeableinteractiondark}.

\item \textbf{Literature on linear IDE model 4:} $\boldsymbol{Q=3H\delta \rho_{\text{dm}}}$ — Analytical solutions and background cosmology~\cite{Wang_2004, von_Marttens_2019, Mishra_2023}; dynamical system analysis and discussions on negative energy densities~\cite{V_liviita_2008, He_2008, V_liviita_2010, Izquierdo_2017, Pan_2020, Ar_valo_2022, vanderWesthuizen:2023hcl, Rodriguez_Benites_2024, Paliathanasis:2024jxo}; big rip future singularities~\cite{Nojiri_2005_IDE, Pan_2020, vanderWesthuizen:2023hcl}; structure growth and instabilities~\cite{M.B.Gavela_2009, V_liviita_2008, Väliviita_2008, He_2009, Honorez_2010, He_2011, Khyllep_2022, pooya2024growthmatterperturbationsinteracting}; statefinder diagnostics~\cite{Ar_valo_2022, carrasco2023discriminatinginteractingdarkenergy}; $H_0$ and $\sigma_8$ tension~\cite{Kumar_2016, Li_2020_2, Guo_2021, Di_Valentino_2021_H0_review, Wang_2022}; observational constraints~\cite{Costa_2014, Costa_2017, Santos_2017, An_2017, An_2018, Grand_n_2019, von_Marttens_2019, Bachega_2020, Aljaf_2021, Califano_2024, Li_2024, Benisty:2024lmj, yan2025investigatinginteractingdarkenergy}; N-body simulations~\cite{Zhang_2018, Zhang_2019, Liu_2022}; gravitational wave predictions~\cite{Caprini_2016, Yang_2019, Bachega_2020, Califano_2024, wang2025prospectsconstraininginteractingdark}; 21cm cosmology~\cite{Costa_2018, Li_2020, Halder_2021, Mukhopadhyay_2021, Xiao_2021}; neutrino constraints~\cite{Kumar_2016, Guo_2017, Feng_2019_2, Feng_2020, Zhao_2020, Li_2020_2}; dynamical coupling constant~\cite{Guo_2018, li2025probingsignchangeableinteractiondark}; holographic modelling~\cite{Feng_2016, Feng_2018, Sadri_2019}; field theory and other interpretations~\cite{Guo_2018, B_gu__2019, Pan_2020_Field, Banerjee_2024, li2025probingsignchangeableinteractiondark, Guin:2025xki}; other~\cite{zhang2009crossingphantomdivide, Rezaei_2020, Zhao_2023, nagpal2025darksectorinteractionsprobing}.

\item \textbf{Literature on linear IDE model 5:} $\boldsymbol{Q=3H\delta \rho_{\text{de}}}$ — analytical solutions and background cosmology~\cite{V_liviita_2008, He_2009, von_Marttens_2019, vanderWesthuizen:2023hcl}; dynamical system analysis and discussions on negative energy densities~\cite{V_liviita_2008, He_2008, He_2009, Bahamonde_2018, Izquierdo_2018, Pan_2020, Panotopoulos_2020, Deogharia_2021, von_Marttens_2020, Ar_valo_2022, vanderWesthuizen:2023hcl}; big rip future singularities~\cite{Nojiri_2005_IDE, Pan_2020, de_Haro_2023, vanderWesthuizen:2023hcl}; structure growth and instabilities~\cite{M.B.Gavela_2009, V_liviita_2008, Väliviita_2008, He_2009, Honorez_2010, He_2011, Marcondes_2016, Yang_2016, Khyllep_2022, pooya2024growthmatterperturbationsinteracting, Silva_2024}; statefinder diagnostics~\cite{Panotopoulos_2020, Ar_valo_2022, carrasco2023discriminatinginteractingdarkenergy}; $H_0$ and $\sigma_8$ tension~\cite{Salvatelli_2013, Di_Valentino_2017, Kumar_2017, Kumar_2019, Li_2020_2, Yang_2020_2, Kumar_2021, Lucca_2020,  Di_Valentino_2020,  Di_Valentino_2020_rhode, Di_Valentino_2021_H0_review, lucca2021darkenergydarkmatterinteractions, Anchordoqui_2021, Wang_2022, Yang_2023, Bernui_2023, Califano_2023, Pan_2024, Sabogal_2024, Giare:2024smz, Sabogal_2025}; observational constraints~\cite{Clemson_2012, Costa_2014, Costa_2017, An_2017, Santos_2017, An_2018, Grand_n_2019, von_Marttens_2019, Aljaf_2021, Nunes_2022, Zhai_2023, Li_2024, yan2025investigatinginteractingdarkenergy}; N-body simulations~\cite{Zhang_2018, Zhang_2019, Liu_2022, Zhao_2023, Giar__2024, Silva_2024}; gravitational wave predictions~\cite{Caprini_2016, Bachega_2020, Yang_2020_3, Califano_2023, Califano_2024, wang2025prospectsconstraininginteractingdark}; 21cm cosmology~\cite{Costa_2018, Li_2020, Halder_2021, Mukhopadhyay_2021, Xiao_2021}; neutrino constraints~\cite{Guo_2017, Kumar_2017, Feng_2019, Feng_2019_2, Zhao_2020, Yang_2020, Yang_2020_2, Li_2020_2}; dynamical coupling constant~\cite{Guo_2018, Giar__2024, Sabogal_2025, silva2025newconstraintsinteractingdark, li2025probingsignchangeableinteractiondark}; holographic modelling~\cite{Feng_2016, Nayak_2020, Sinha_2020}; field theory and other interpretations~\cite{Pan_2020_Field, Ghodsi_Yengejeh_2023, Guin:2025xki}; other~\cite{Zheng_2017, Di_Valentino_2021_closed, Yang_2021, Joseph_2022, Zhao_2023, Forconi_2024, mbewe2024viscouscosmologicalfluidslargescale, yang2025probingcoldnaturedark}.

\end{itemize}

\section{Dynamical system analysis} \label{Sec.DSA}
\subsection{Setting up the dynamical system}  \label{Setup.DSA}
In this section, we apply standard techniques from dynamical system analysis to set up and analyze IDE models. For a review of dynamical system analysis applied to cosmology, see~\cite{Bahamonde_2018, phdthesis}. We define our dynamical system as a four-fluid system, including dark energy ($w_{\text{de}}=w$), dark matter ($w_{\text{dm}}=0$), radiation ($w_{\rm{r}}=1/3$) and baryonic matter ($w_{\rm{bm}}=0$). Starting from equation (2.16) derived in the previous work of some of us~\cite{vanderWesthuizen:2023hcl}, we have:
\begin{gather} \label{DSA.7}
\begin{split}
\dot{\Omega}_{\rm{de}} &= \Omega_{\rm{de}} H \left[ 2 \Omega_{\rm{r}}  + \Omega_{\rm{bm}}+ \Omega_{\rm{dm}} + \Omega_{\rm{de}}  \left(1 +3 w \right) -1 - 3w \right]  - \frac{8 \pi G }{3H^2} Q, \\
\dot{\Omega}_{\rm{dm}} &= \Omega_{\rm{dm}} H \left[ 2 \Omega_{\rm{r}}  + \Omega_{\rm{bm}}+ \Omega_{\rm{dm}} + \Omega_{\rm{de}}  \left(1 +3 w \right) -1  \right]  + \frac{8 \pi G }{3H^2} Q, \\
\dot{\Omega}_{\rm{bm}} &= \Omega_{\rm{bm}} H \left[ 2 \Omega_{\rm{r}}  + \Omega_{\rm{bm}}+ \Omega_{\rm{dm}} + \Omega_{\rm{de}}  \left(1 +3 w \right) -1  \right], \\
\dot{\Omega}_{\rm{r}} &= \Omega_{\rm{r}} H \left[ 2 \Omega_{\rm{r}}  + \Omega_{\rm{bm}}+ \Omega_{\rm{dm}} + \Omega_{\rm{de}}  \left(1 +3 w \right) -2 \right].  \\
\end{split}
\end{gather}
Equation~\eqref{DSA.7} reduces to the $\Lambda$CDM case if $Q=0$ and $w_{\rm{de}}=-1$, as shown in~\cite{GR_book}.
These equations can be made dimensionless by introducing the derivative with respect to the Hubble parameter, such that: 
\begin{gather} \label{DSA.8}
\begin{split}
{\Omega}'_{\rm{i}} =  \frac{d}{d \zeta} \Omega_{\rm{i}}=\frac{d}{H \, dt} \Omega_{\rm{i}} .
\end{split}
\end{gather}
The system of equations~\eqref{DSA.7} can be reduced to only three equations by introducing the assumption of a flat universe, such that $\Omega_{\rm{r}}+\Omega_{\rm{bm}}+\Omega_{\rm{dm}}+\Omega_{\rm{de}}=1$. Combining the flatness assumption with the new notation in~\eqref{DSA.8}, the dynamical system~\eqref{DSA.7} becomes: 
\begin{gather} \label{DSA.9}
\begin{split}
{\Omega}'_{\rm{de}} &= \Omega_{\rm{de}} \left[ 1 - \Omega_{\rm{bm}} - \Omega_{\rm{dm}} - \Omega_{\rm{de}}  \left(1 - 3 w \right) - 3w \right]  - \frac{8 \pi G }{3H^3} Q, \\
{\Omega}'_{\rm{dm}} &= \Omega_{\rm{dm}} \left[ 1 - \Omega_{\rm{bm}} - \Omega_{\rm{dm}} - \Omega_{\rm{de}}  \left(1 - 3 w \right) \right]  + \frac{8 \pi G }{3H^3} Q, \\
{\Omega}'_{\rm{bm}} &= \Omega_{\rm{bm}} \left[ 1 - \Omega_{\rm{bm}} - \Omega_{\rm{dm}} - \Omega_{\rm{de}}  \left(1 - 3 w \right) \right].
\end{split}
\end{gather}
We note that, from the flatness assumption, the radiation density obeys the relationship $\Omega_{\rm{r}} = 1 - \Omega_{\rm{bm}} - \Omega_{\rm{dm}} - \Omega_{\rm{de}}$. For these models, we will make only two assumptions regarding the parameter space: 
\begin{gather} \label{DSA.10}
\begin{split}
w < 0 \quad &\text{(DE has negative pressure)}, \\
\delta_{\text{dm}} + \delta_{\text{de}} < |w| \ \text{and} \ \delta < |w| \quad &\text{(interaction strength is not too strong)}.
\end{split}
\end{gather}
These assumptions arise from the expectation that the correct dynamics of the universe do not deviate too drastically from the current description of the relatively successful $\Lambda$CDM model, as used in~\cite{V_liviita_2008}. For our analysis, we determine the coordinates and stability of the critical points. The stability of the critical points relevant to our study can be classified into three types, depending on how the trajectories behave after the critical point is perturbed, and mathematically by the sign of the eigenvalues $\lambda$ of the corresponding Jacobian matrix of the system~\cite{Bahamonde_2018}. The three types are given below.

\begin{enumerate}
    \item \underline{Unstable node (source)}: All trajectories diverge away from the critical point as $t \rightarrow \infty$. This may be considered the origin. (Requirement: all eigenvalues are \textit{positive and real}, $\lambda > 0$);
    \item \underline{Saddle point}: Some trajectories converge towards the point, while others diverge away. This may be considered an asymptotically stable stop along the journey to the final attractor. (Requirement: some eigenvalues are \textit{positive}, $\lambda > 0$, while others are \textit{negative}, $\lambda < 0$);
    \item \underline{Stable node (sink)}: All trajectories converge at the critical point as $t \rightarrow \infty$. This may be considered a future attractor where the system ends. (Requirement: all eigenvalues are \textit{negative and real}, $\lambda < 0$).
\end{enumerate}
To determine positive energy conditions, we require the following to hold in the 2D projection of the system in the $(\Omega_{\rm{dm}}, \Omega_{\rm{de}})$ plane:
\begin{enumerate}
    \item \underline{Positive critical points}: We require conditions that ensure that the coordinates of each critical point are positive $(\Omega_{\rm{dm}} \ge 0 \;, \; \Omega_{\rm{de}} \ge 0)$.
    \item \underline{Positive trajectories}: The phase portraits we obtain will have a region bounded by three invariant submanifolds connecting critical points, where $(\Omega_{\rm{dm}} \ge 0 \;, \; \Omega_{\rm{de}} \ge 0)$ throughout the region. We require constraints on $(\Omega_{\rm{(dm,0)}} \;, \; \Omega_{\rm{(de,0)}})$ that ensure trajectories start and remain within this bounded region.
\end{enumerate}
In order to further understand the asymptotic behavior of the system (particularly with regard to big rip future singularities), we also determine the expressions for the equations given in Section~\ref{BG_equations} at each critical point.

\subsection{Dynamical system analysis of linear IDE model 1: $Q= 3 H (\delta_{\text{dm}} \rho_{\text{dm}} + \delta_{\text{de}}  \rho_{\text{de}})$} \label{DSA.Q.general}
We now consider the dynamical system behavior of the most general IDE model in Table~\ref{tab:positivityQ}, where the interaction strength can be coupled differently to DM and DE, as specified by the constants $\delta_{\text{dm}}$ and $\delta_{\text{de}}$, respectively. For this model, we adopt the assumption $\delta_{\text{dm}}+\delta_{\text{de}}< |w|$ stated in \eqref{DSA.10}, from which it follows that $w \pm\delta_{\text{dm}} \pm\delta_{\text{de}}<0$ and $-w \pm\delta_{\text{dm}} \pm\delta_{\text{de}}>0$. 

For the interaction $Q= 3 H (\delta_{\text{dm}} \rho_{\text{dm}} + \delta_{\text{de}}  \rho_{\text{de}})$, the dynamical system \eqref{DSA.9} becomes:
\begin{gather} \label{DSA.Q.ddm+dde.1}
\begin{split}
{\Omega}'_{\rm{de}} &= \Omega_{\rm{de}} \left[ 1 - \Omega_{\rm{bm}} - \Omega_{\rm{dm}} - \Omega_{\rm{de}} \left(1 - 3 w \right) - 3w \right] - 3 \delta_{\text{dm}} \Omega_{\rm{dm}} - 3 \delta_{\text{de}} \Omega_{\rm{de}}, \\
{\Omega}'_{\rm{dm}} &= \Omega_{\rm{dm}} \left[ 1 - \Omega_{\rm{bm}} - \Omega_{\rm{dm}} - \Omega_{\rm{de}} \left(1 - 3 w \right) \right] + 3 \delta_{\text{dm}} \Omega_{\rm{dm}} + 3 \delta_{\text{de}} \Omega_{\rm{de}}, \\
{\Omega}'_{\rm{bm}} &= \Omega_{\rm{bm}} \left[ 1 - \Omega_{\rm{bm}} - \Omega_{\rm{dm}} - \Omega_{\rm{de}} \left(1 - 3 w \right) \right], \\
\end{split}
\end{gather}
where we used the relation $\frac{8 \pi G}{3H^2} \rho_i = \Omega_i$. In this case, because the DM participates in the interaction while the baryonic matter does not, the two fluids evolve differently and cannot be grouped together. We find four solutions for the dynamical system \eqref{DSA.Q.ddm+dde.1}, corresponding to four critical points. \\ \\
\underline{Critical Point $P_{\text{r}}$: radiation-dominated phase.}
\begin{gather} \label{DSA.Q.ddm+dde.2}
\Omega_{\rm{bm}} = 0, \quad \Omega_{\rm{dm}} = 0, \quad \Omega_{\rm{de}} = 0, \quad \rightarrow \quad \Omega_{\rm{r}} = 1 \quad ; \quad \lambda = \begin{bmatrix}
1  \\
\frac{3}{2} \left( -\delta_{\text{de}} + \delta_{\text{dm}} - w - \Delta \right) + 1 \\
\frac{3}{2} \left( -\delta_{\text{de}} + \delta_{\text{dm}} - w + \Delta \right) + 1
\end{bmatrix}.
\end{gather}  
Here, the determinant $\Delta$ is given by:
\begin{gather} \label{DSA.Q.ddm+dde.4}
\Delta = \sqrt{ \left( \delta_{\text{dm}} + \delta_{\text{de}} + w \right)^2 - 4 \delta_{\text{de}} \delta_{\text{dm}} }
= \sqrt{ \delta_{\text{dm}}^2 + \delta_{\text{de}}^2 + w^2 - 2 \delta_{\text{dm}} \delta_{\text{de}} + 2 \delta_{\text{dm}} w + 2 \delta_{\text{de}} w }.
\end{gather}
The determinant \eqref{DSA.Q.ddm+dde.4} will appear throughout this analysis. From it, we obtain the reality condition, which ensures $\Delta$ is real for this model:
\begin{gather} \label{DSA.Q.ddm+dde.5}
\left( \delta_{\text{dm}} + \delta_{\text{de}} + w \right)^2 \ge 4 \delta_{\text{de}} \delta_{\text{dm}}. 
\end{gather}
Complex values of $\Delta$ will cause the flow lines to spiral around the critical point, leading to negative densities, which should be avoided. If we want this radiation phase to be an unstable node (source), as in the $\Lambda$CDM model, we need to ensure that all eigenvalues are real and positive. In \eqref{DSA.Q.ddm+dde.4}, the first eigenvalue is positive, while the third eigenvalue will also be positive if the second eigenvalue is positive, since it is more negative. We may obtain the conditions for a positive second eigenvalue, after some algebra, as:
\begin{gather} \label{DSA.Q.ddm+dde.6}
\begin{split}
\frac{3}{2} \left( -\delta_{\text{de}} + \delta_{\text{dm}} - w - \Delta \right) + 1 &> 0 
\quad \rightarrow \quad         
\delta_{\text{dm}} \left( \underbrace{w - \frac{1}{3}}_{<0} \right) + \underbrace{\frac{1}{3} \left( \delta_{\text{de}} + w \right)}_{<0} < \frac{1}{9}.
\end{split} 
\end{gather}
To square both sides in \eqref{DSA.Q.ddm+dde.6}, we require both sides to be positive, which is ensured by the initial assumption $-w \pm \delta_{\text{dm}} \pm \delta_{\text{de}} > 0$. From this assumption, we also find that both terms in brackets in the final equality are negative. Therefore, if $\delta_{\text{dm}} > 0$ (which corresponds to a positive energy density condition \eqref{DSA.Q.ddm+dde.15} for this model), the left-hand side will be negative, which is smaller than the positive right-hand side. Consequently, under our initial assumptions and positivity constraints, we will always have a past radiation-dominated phase.
\begin{gather} \label{DSA.Q.ddm+dde.7}
\begin{split}
\underline{\text{Conditions for:}} \quad \text{Radiation-dominated real unstable node (source)} \quad \begin{cases} 
w < 0,  \\
\delta_{\text{dm}} + \delta_{\text{de}} < |w|,  \\ 
\delta_{\text{dm}} > 0.
\end{cases}
\end{split}
\end{gather}
\underline{Critical Point $P_{\text{bm}}$: baryonic matter-dominated phase.}
\begin{gather} \label{DSA.Q.ddm+dde.8}
\Omega_{\rm{bm}} = 1, \quad \Omega_{\rm{dm}} = 0, \quad \Omega_{\rm{de}} = 0,  \quad \rightarrow \quad \Omega_{\rm{r}} = 0 \quad ; \quad  \lambda = \begin{bmatrix}
-1  \\
\frac{3}{2}(-\delta_{\text{de}} + \delta_{\text{dm}} - w - \Delta) \\
\frac{3}{2}(-\delta_{\text{de}} + \delta_{\text{dm}} - w + \Delta)
\end{bmatrix}.
\end{gather} \\

From the eigenvalues in \eqref{DSA.Q.ddm+dde.8}, we see that, since the same square root appears as in the previous critical point, all eigenvalues will be real if \eqref{DSA.Q.ddm+dde.5} holds.  
If we want this point to be a saddle, we require either the second or third eigenvalue to be positive.  
For the third eigenvalue to be positive, we need:
\begin{gather} \label{DSA.Q.ddm+dde.11}
\begin{split}
\frac{3}{2}(-\delta_{\text{de}} + \delta_{\text{dm}} - w + \Delta) &> 0 \quad \rightarrow \quad 
\underbrace{\Delta}_{>0} > \underbrace{\delta_{\text{de}} - \delta_{\text{dm}} + w}_{<0}.
\end{split} 
\end{gather}

\begin{figure}[H]
    \centering
    \includegraphics[width=0.95\linewidth]{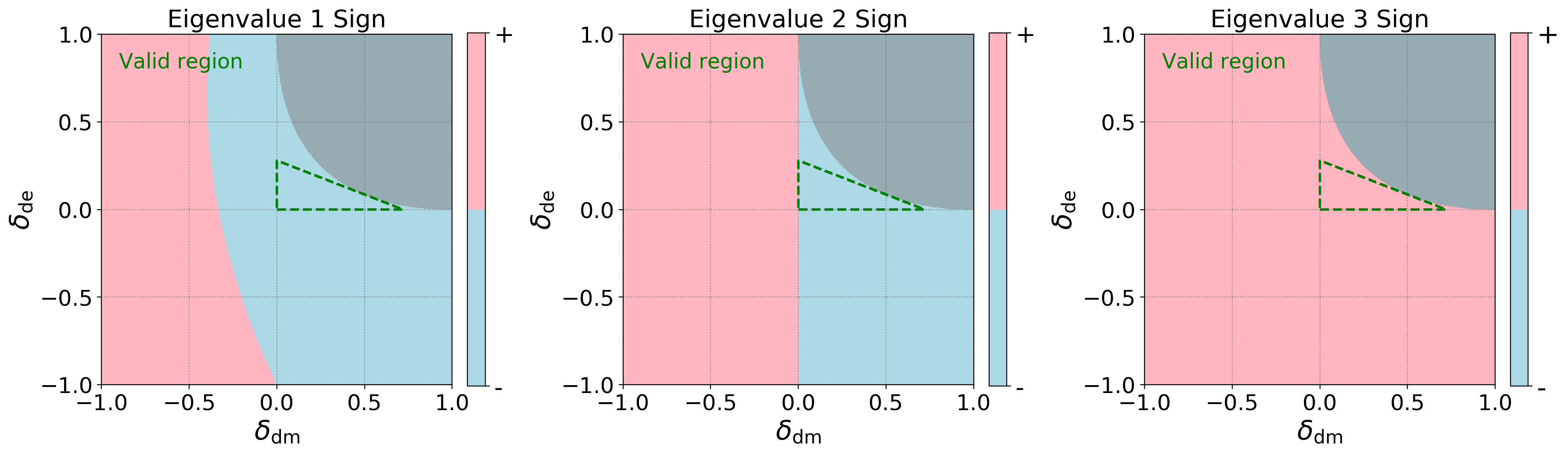}
    \caption{Parameter space of the eigenvalues for the critical point \eqref{DSA.Q.ddm+dde.13} - $Q= 3 H (\delta_{\text{dm}} \rho_{\text{dm}} + \delta_{\text{de}}  \rho_{\text{de}})$}
    \label{fig:Linear_general_CP3_eigenvalues}
\end{figure}

The R.H.S. of \eqref{DSA.Q.ddm+dde.11} is always negative due to our initial assumption $w \pm \delta_{\text{dm}} \pm \delta_{\text{de}} < 0$, while the L.H.S. is always positive from the reality condition.  
Therefore, this critical point will always act as a saddle point.
\begin{gather} \label{DSA.Q.ddm+dde.12}
\begin{split}
\underline{\text{Conditions for:}} \quad \text{Baryonic matter-dominated real saddle point} \quad \begin{cases} 
w < 0,  \\ 
\delta_{\text{dm}} + \delta_{\text{de}} < |w|.
\end{cases}
\end{split}
\end{gather} \\ 
\underline{Critical Point $P_{\text{\textbf{dm}+de}}$: dark matter--dark energy hybrid-dominated phase.}
\begin{gather} \label{DSA.Q.ddm+dde.13}
\Omega_{\rm{bm}} = 0, \quad 
\Omega_{\rm{dm}} = \frac{1}{2} + \frac{\delta_{\text{dm}} - \delta_{\text{de}} - \Delta}{2 w}, \quad 
\Omega_{\rm{de}} = \frac{1}{2} - \frac{\delta_{\text{dm}} - \delta_{\text{de}} - \Delta}{2 w},  
\quad \rightarrow \quad \Omega_{\rm{r}} = 0.
\end{gather} 
The eigenvalues for this critical point are too cumbersome to present in a useful analytic form (except in special cases such as $\delta_{\text{dm}} = 0$ or $\delta_{\text{de}} = 0$, where they become straightforward to analyze).  
Instead, we examine the parameter space by plotting the sign of each eigenvalue, as shown in Figure~\ref{fig:Linear_general_CP3_eigenvalues}.
\noindent  In Figure~\ref{fig:Linear_general_CP3_eigenvalues}, we show how the signs of the three eigenvalues depend on the coupling constants $\delta_{\text{dm}}$ and $\delta_{\text{de}}$, fixing $w=-1$ and $r_0=0.388$.  
Red regions correspond to positive eigenvalues, while blue regions correspond to negative eigenvalues.  
Areas where the eigenvalues have imaginary components are marked with a grey overlay.  
The region satisfying the constraints for both positive and real energy densities (derived later in \eqref{DSA.Q.delta_dm+delta_de.PEC_BG}) is enclosed within the green triangle.  
Within most of this viable region, the first eigenvalue is positive, while the second and third eigenvalues are negative.  
Since eigenvalues of opposite signs dominate across the parameter space, this critical point behaves as a saddle in most viable cases, consistent with the trajectory patterns around $P_{\text{\textbf{dm}+de}}$ shown in Figure~\ref{fig:2D_Q_general_Phase_portrait_boundaries}. \\ 
  
From the coordinates we find that $\Omega_{\rm{dm}}+\Omega_{\rm{de}}=1$ at this critical point, indicating a DM–DE hybrid dominated phase.  
Although both components contribute, we expect DM to constitute the larger fraction of the total density, since setting $\delta_{\text{dm}}=\delta_{\text{de}}=0$ in \eqref{DSA.Q.ddm+dde.13} yields $\Omega_{\rm{dm}}=1$ and $\Omega_{\rm{de}}=0$.  
The coordinates also involve the same determinant $\Delta$ defined in \eqref{DSA.Q.ddm+dde.4}, so the same reality condition \eqref{DSA.Q.ddm+dde.5} must be satisfied.  
We now determine the conditions ensuring positive energy densities $\Omega_{\rm{dm}}$ and $\Omega_{\rm{de}}$ at this critical point.  
For $\Omega_{\rm{dm}}>0$ we require:
\begin{gather} \label{DSA.Q.ddm+dde.14}
\begin{split}
 \Omega_{\rm{dm}}=\frac{1}{2} + \frac{ \delta_{\text{dm}}-\delta_{\text{de}} -\Delta}{2 w} > 0  
 \quad \rightarrow \quad  
 \underbrace{\delta_{\text{dm}}-\delta_{\text{de}} + w}_{<0} < \underbrace{\Delta}_{>0}.
\end{split} 
\end{gather}
From our assumption $w \pm\delta_{\text{dm}}\pm\delta_{\text{de}}<0$, the left-hand side of \eqref{DSA.Q.ddm+dde.14} is necessarily negative, while the determinant on the right-hand side is always positive.  
Consequently, the inequality in \eqref{DSA.Q.ddm+dde.14} is automatically satisfied, and $\Omega_{\rm{dm}}$ remains positive at this critical point under our initial assumptions.  
For $\Omega_{\rm{de}}>0$, some algebra (performed with careful attention to sign changes when multiplying inequalities) yields the condition:
\begin{gather} \label{DSA.Q.ddm+dde.15}
\begin{split}
 \Omega_{\rm{de}}=\frac{1}{2} - \frac{ \delta_{\text{dm}}-\delta_{\text{de}} -\Delta}{2 w} > 0  
 \quad \rightarrow \quad \delta_{\text{dm}} > 0. \\
\end{split} 
\end{gather}
Thus, DE will be positive at the past attractor if $\delta_{\text{dm}} > 0$, 
which effectively rules out any model with $\delta_{\text{dm}} < 0$ — a regime that typically corresponds to iDMDE scenarios. 

\begin{figure}[H]
    \centering
    \includegraphics[width=0.95\linewidth]{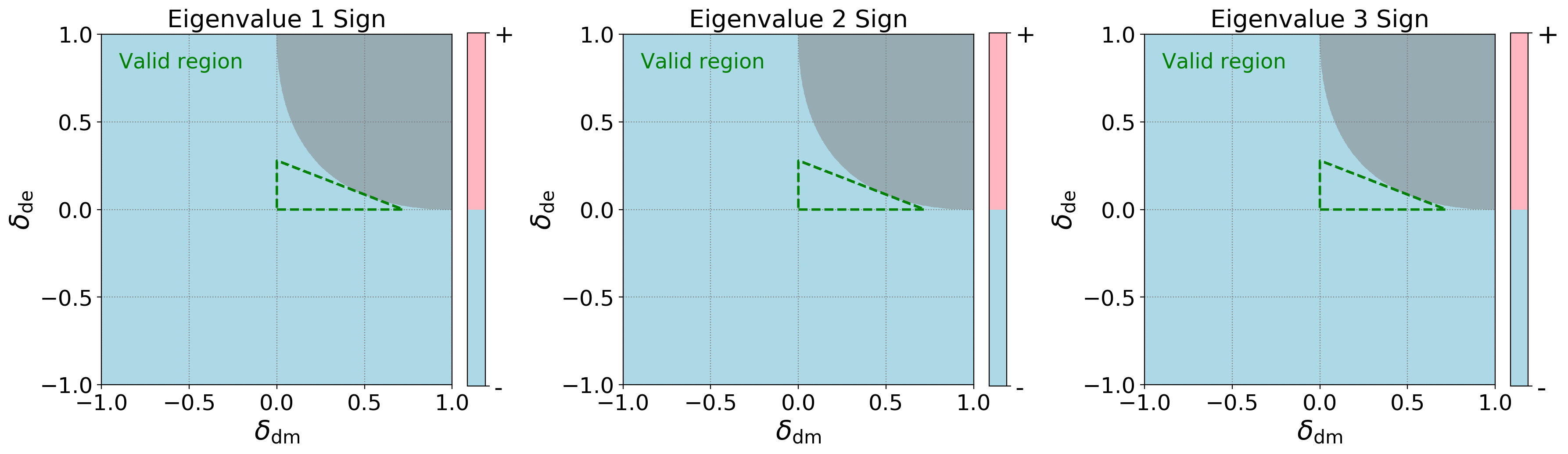}
    \caption{Parameter space of the eigenvalues for the critical point \eqref{DSA.Q.ddm+dde.16} - $Q= 3 H (\delta_{\text{dm}} \rho_{\text{dm}} + \delta_{\text{de}}  \rho_{\text{de}})$}
    \label{fig:Linear_general_CP4_eigenvalues}
\end{figure}

\begin{figure}[htbp]
    \centering
    \begin{subfigure}[b]{0.495\linewidth}
        \centering
        \includegraphics[width=\linewidth]{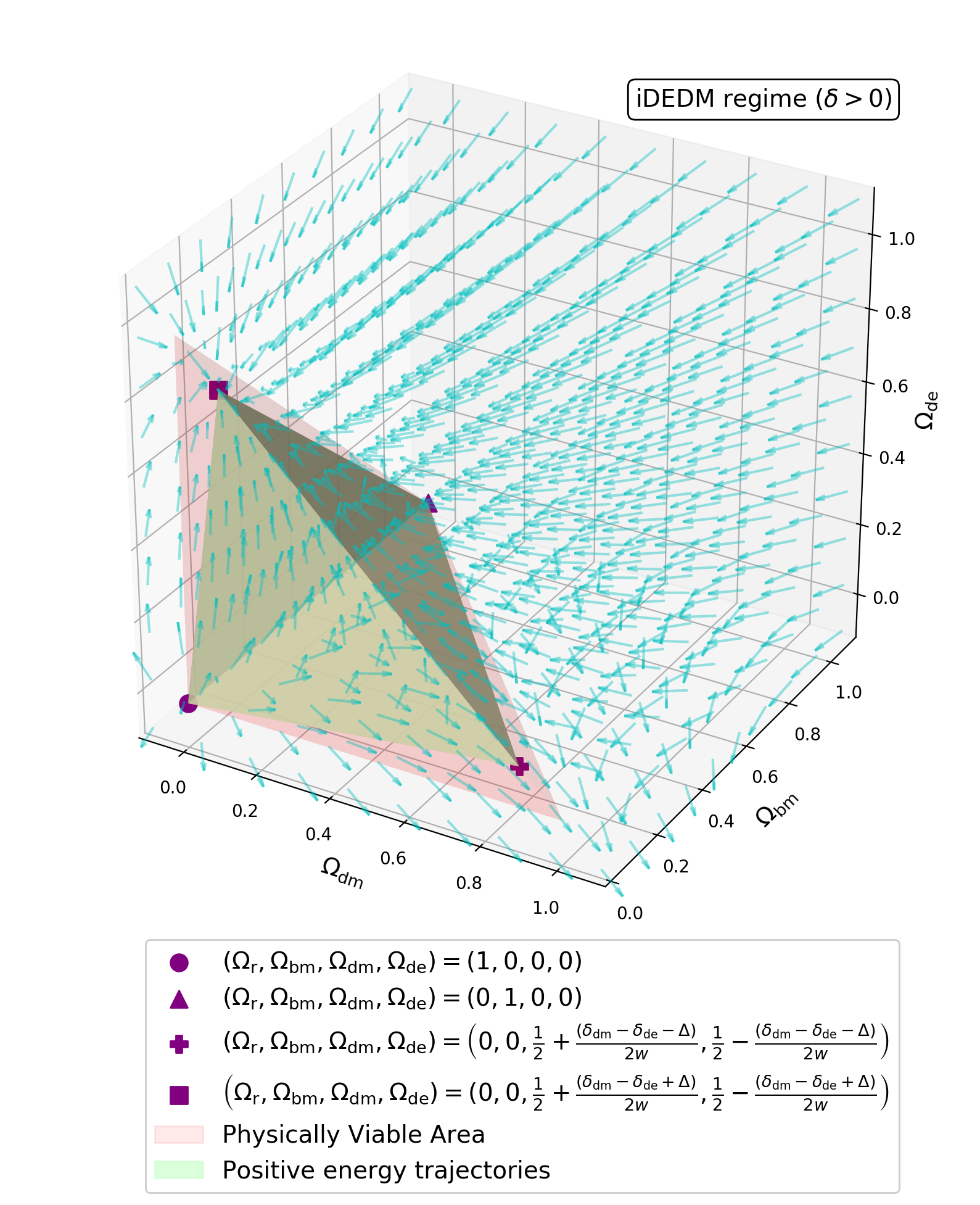}
        \label{fig:iDEDM3D}
    \end{subfigure}%
    \hspace{0pt} 
    \begin{subfigure}[b]{0.495\linewidth}
        \centering
        \includegraphics[width=\linewidth]{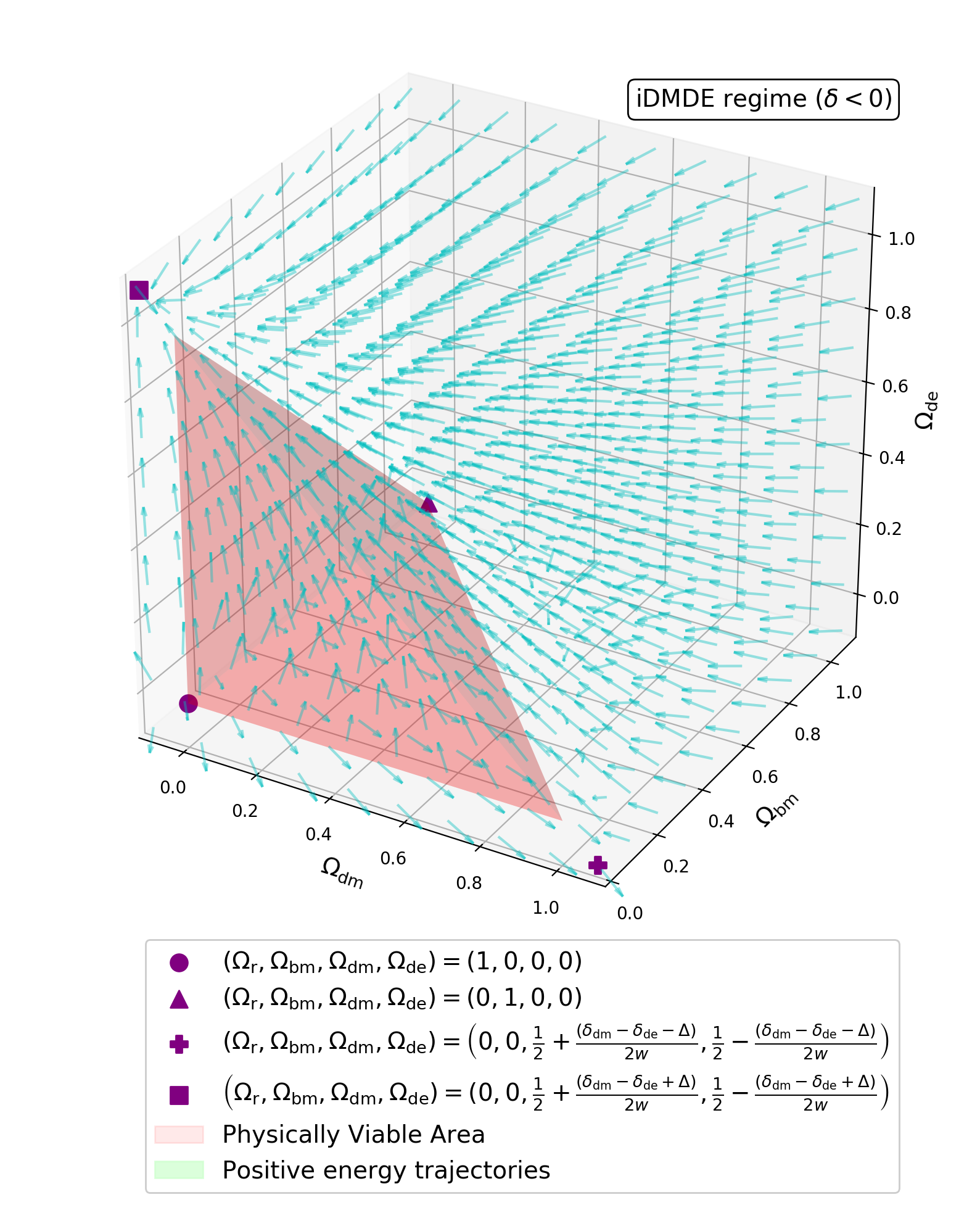}
        \label{fig:iDMDE3D}
    \end{subfigure}    
\caption{3D phase portrait for $Q= 3 H (\delta_{\text{dm}} \rho_{\text{dm}} + \delta_{\text{de}} \rho_{\text{de}})$, showing positive-energy trajectories in the iDEDM regime ($\delta_{\text{dm}}=\delta_{\text{de}}=+0.1$, left panel) and negative-energy trajectories in the iDMDE regime ($\delta_{\text{dm}}=\delta_{\text{de}}=-0.1$, right panel). The trajectories also show radiation, matter (DM+BM) and DE-dominated eras.}
    \label{fig:3D_Q_general_Phase_portrait_boundaries}
\end{figure}

\begin{figure}[htbp]
    \centering
    \begin{subfigure}[b]{0.49\linewidth}
        \centering
        \includegraphics[width=\linewidth]{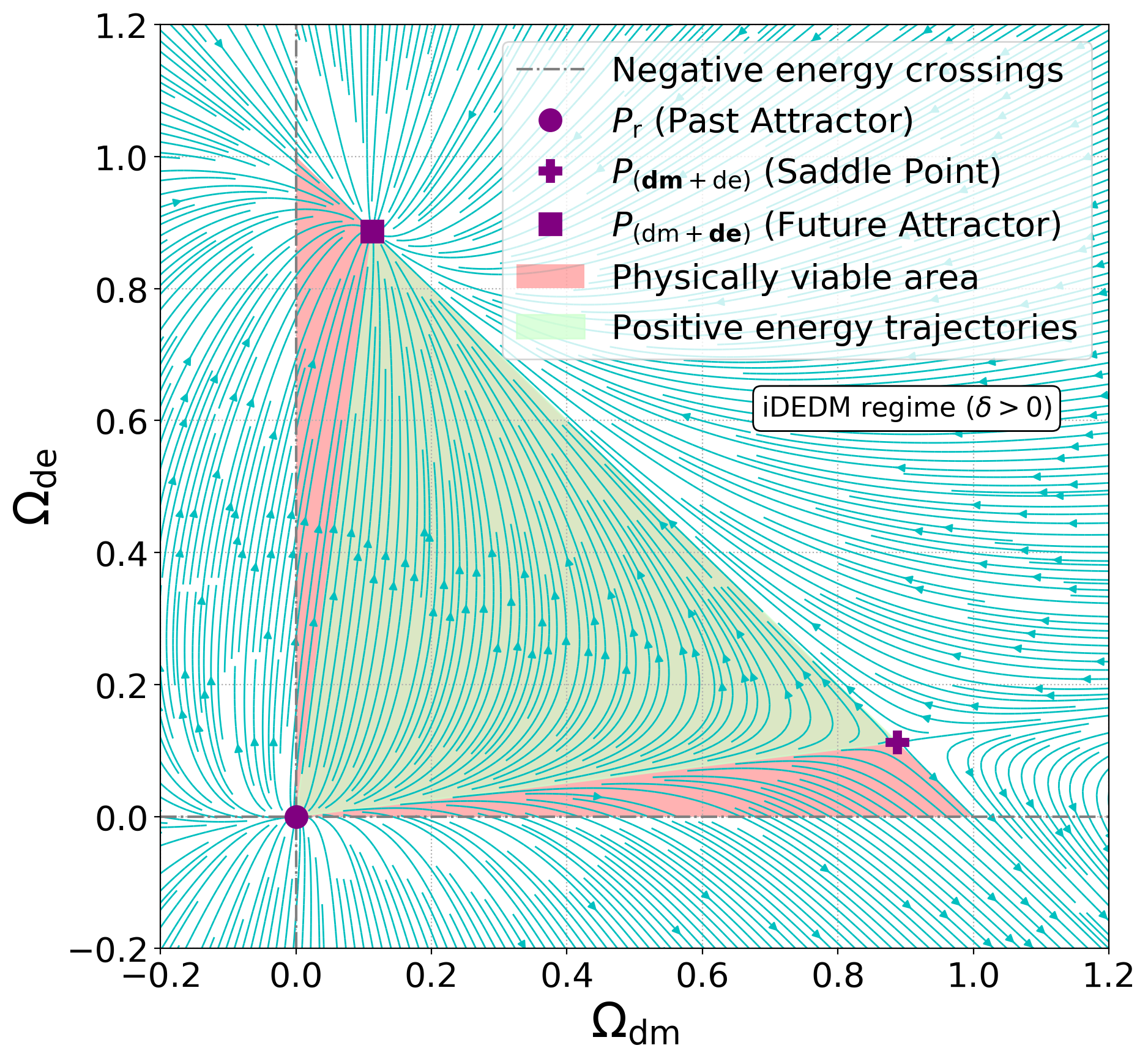}
        \label{fig:iDEDM2D}
    \end{subfigure}%
    \hspace{0pt} % No extra space between subfigures
    \begin{subfigure}[b]{0.49\linewidth}
        \centering
        \includegraphics[width=\linewidth]{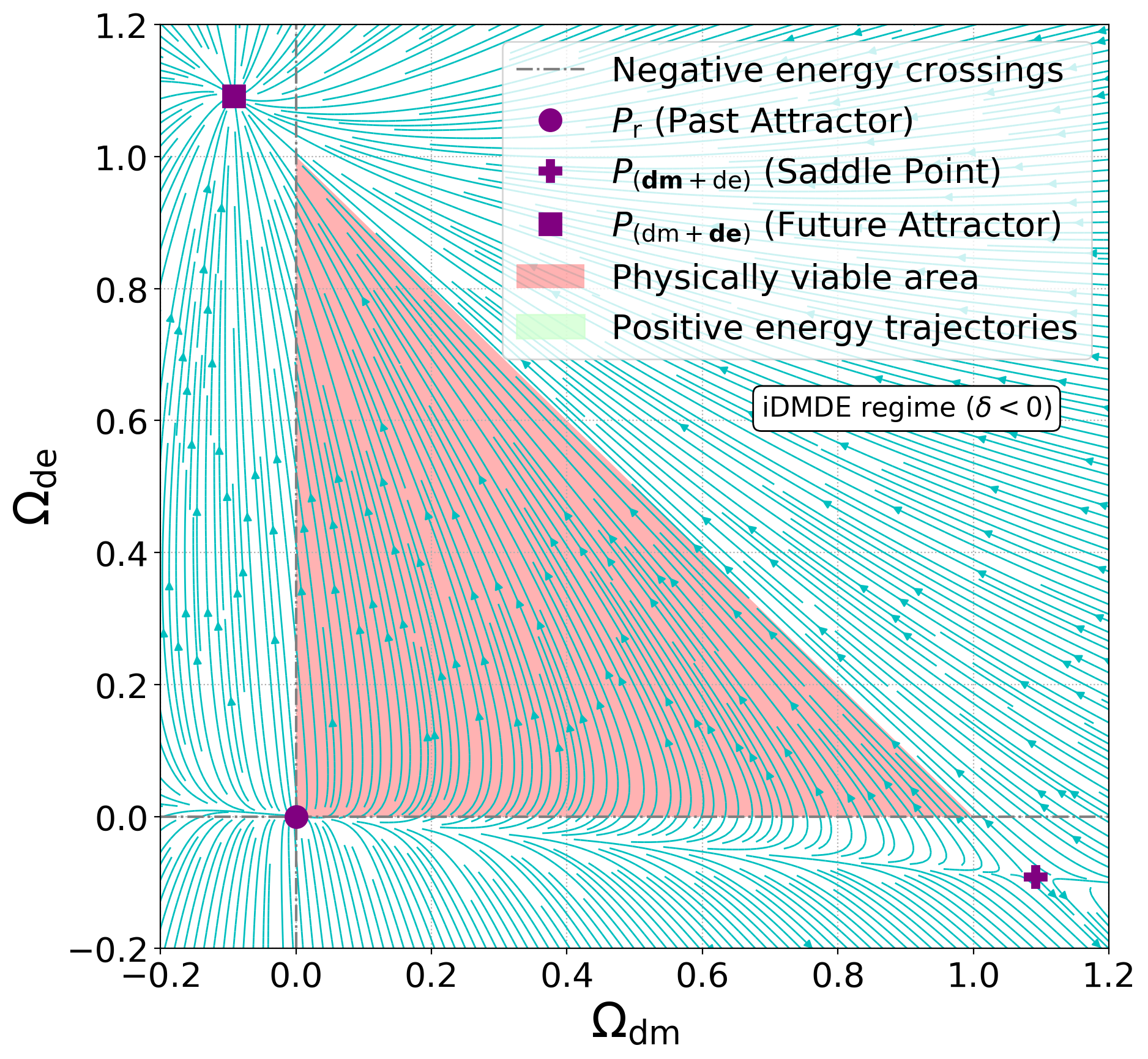}
        \label{fig:iDMDE2D}
    \end{subfigure}
\caption{2D projection of the phase portrait for $Q= 3 H (\delta_{\text{dm}} \rho_{\text{dm}} + \delta_{\text{de}}  \rho_{\text{de}})$, showing positive-energy trajectories in the iDEDM regime ($\delta_{\text{dm}}=\delta_{\text{de}}=+0.1$, left panel) and negative-energy trajectories in the iDMDE regime ($\delta_{\text{dm}}=\delta_{\text{de}}=-0.1$, right panel). }
    \label{fig:2D_Q_general_Phase_portrait_boundaries}
\end{figure}

\medskip
\noindent\underline{Critical Point $P_{\text{dm+\textbf{de}}}$: dark energy hybrid dominated phase.}
\begin{gather} \label{DSA.Q.ddm+dde.16}
\Omega_{\rm{bm}} = 0, \quad 
\Omega_{\rm{dm}} = \frac{1}{2} + \frac{\delta_{\text{dm}}-\delta_{\text{de}} +\Delta}{2 w}, \quad 
\Omega_{\rm{de}} = \frac{1}{2} - \frac{\delta_{\text{dm}}-\delta_{\text{de}} +\Delta}{2 w},  
\quad \rightarrow \quad \Omega_{\rm{r}} = 0.
\end{gather}
The parameter space showing the signs of each eigenvalue at this critical point (for $w=-1$ and $r_0=0.388$) is illustrated in Figure~\ref{fig:Linear_general_CP4_eigenvalues}.
In Figure~\ref{fig:Linear_general_CP4_eigenvalues}, we see that all three eigenvalues are negative throughout the parameter space, indicating that this critical point will always act as a stable node (sink), as also illustrated by the behavior of the trajectories around $P_{\text{dm+\textbf{de}}}$ in Figure~\ref{fig:2D_Q_general_Phase_portrait_boundaries}. From the coordinates, we find that $\Omega_{\rm{dm}}+\Omega_{\rm{de}}=1$ at this critical point, again corresponding to DM and DE hybrid dominance. In this case, however, we expect DE to be the dominant component, since $\Omega_{\rm{dm}}=0$ and $\Omega_{\rm{de}}=1$ when we set $\delta_{\text{dm}}=\delta_{\text{de}}=0$ in the critical point~\eqref{DSA.Q.ddm+dde.13}.
\\ 
We now consider the conditions required for the energy densities $\Omega_{\rm{dm}}$ and $\Omega_{\rm{de}}$ to remain positive at this critical point. For DM to be positive, we find after some algebra that:
\begin{gather} \label{DSA.Q.ddm+dde.16.5}
\begin{split}
 \Omega_{\rm{dm}}=\frac{1}{2} + \frac{ \delta_{\text{dm}}-\delta_{\text{de}} +\Delta}{2 w} > 0 
 \quad \rightarrow \quad  \delta_{\text{de}} > 0.
\end{split} 
\end{gather}
Thus, DM will be positive at the future attractor if $\delta_{\text{de}} > 0$. Taken together with the result from \eqref{DSA.Q.ddm+dde.15}, we see that \textbf{if either $\delta_{\text{dm}} < 0$ or $\delta_{\text{de}} < 0$, one of the energy densities becomes negative, ruling out positive energies in the iDMDE regime for all linear interaction models of this form.} 
For DE to be positive, we require:
\begin{gather} \label{DSA.Q.ddm+dde.17}
\begin{split}
 \Omega_{\rm{de}}=\frac{1}{2} - \frac{ \delta_{\text{dm}}-\delta_{\text{de}} +\Delta}{2 w} > 0 
 \quad \rightarrow \quad \underbrace{\Delta}_{>0} > \underbrace{\delta_{\text{de}} - \delta_{\text{dm}} + w}_{<0}.
 \end{split} 
\end{gather}
From our assumption $w \pm\delta_{\text{dm}}\pm\delta_{\text{de}}<0$, the R.H.S. of \eqref{DSA.Q.ddm+dde.17} is necessarily negative, while the determinant on the L.H.S. is always positive. Therefore, \eqref{DSA.Q.ddm+dde.17} confirms that DE will always be positive at this critical point under our initial assumptions. \\ 
Finally, we must ensure that the trajectories remain within the positive-energy domain throughout their entire evolution. This is achieved by constraining the initial coordinates $({\Omega}_{\rm{(dm,0)}},{\Omega}_{\rm{(de,0)}})$ to lie within the boundaries of the invariant manifolds in the 2D projection, which forms the $(\Omega_{\rm{dm}}, \Omega_{\rm{de}})$ plane. These boundaries define a triangular region composed of: 
(i) a straight line connecting the past attractor (critical points $P_{\text{r}}$ and $P_{\text{bm}}$) to the saddle point ($P_{\text{\textbf{dm}+de}}$); 
(ii) a second line connecting the saddle point to the future attractor ($P_{\text{dm+\textbf{de}}}$); and 
(iii) a third line directly connecting the past and future attractors. 
This bounded region is shown as the green area labelled \textit{"positive energy trajectories"} in Figure~\ref{fig:2D_Q_general_Phase_portrait_boundaries}, where we see that any trajectory outside this domain will inevitably lead to negative energy densities, even if the critical points themselves satisfy positivity conditions.
\noindent We start by obtaining the equations for the three straight lines that form the triangular region. The line in the $({\Omega}_{\rm{dm}},{\Omega}_{\rm{de}})$ plane connecting the past attractor $P_{\text{r}}$ with coordinates $({\Omega}_{\rm{dm}},{\Omega}_{\rm{de}})=(0,0)$ to the saddle point $P_{\text{\textbf{dm}+de}}$ with coordinates 
$({\Omega}_{\rm{dm}},{\Omega}_{\rm{de}})=\left(\frac{1}{2} + \frac{ \delta_{\text{dm}}-\delta_{\text{de}} -\Delta}{2 w},\frac{1}{2} - \frac{ \delta_{\text{dm}}-\delta_{\text{de}} -\Delta}{2 w}\right)$ 
is described by the straight-line equation: 
\begin{gather} \label{DSA.Q.ddm+dde.17.5}
\begin{split}
\Omega_{\rm{de}} &= m \, \Omega_{\rm{dm}} + c, 
\quad m = \frac{d \Omega_{\rm{de}}}{d \Omega_{\rm{dm}}} 
= \frac{\frac{1}{2} - \frac{ \delta_{\text{dm}}-\delta_{\text{de}} -\Delta}{2 w}}{\frac{1}{2} + \frac{ \delta_{\text{dm}}-\delta_{\text{de}} -\Delta}{2 w}}
= -\frac{ \delta_{\text{dm}}-\delta_{\text{de}} -w -\Delta}{\delta_{\text{dm}}-\delta_{\text{de}}+w -\Delta}, 
\quad c = 0, \\
\Omega_{\rm{de}} &= -\frac{ \delta_{\text{dm}}-\delta_{\text{de}} -w -\Delta}{\delta_{\text{dm}}-\delta_{\text{de}}+w -\Delta} \, \Omega_{\rm{dm}}   
\quad \rightarrow \quad r = -\frac{ \delta_{\text{dm}}-\delta_{\text{de}} +w -\Delta}{\delta_{\text{dm}}-\delta_{\text{de}}-w -\Delta}.
\end{split}
\end{gather}
For the second line, which connects the saddle point $P_{\text{\textbf{dm}+de}}$ with coordinates 
$({\Omega}_{\rm{dm}},{\Omega}_{\rm{de}})=\left(\frac{1}{2} + \frac{ \delta_{\text{dm}}-\delta_{\text{de}} -\Delta}{2 w},\frac{1}{2} - \frac{ \delta_{\text{dm}}-\delta_{\text{de}} -\Delta}{2 w}\right)$ 
to the future attractor $P_{\text{dm+\textbf{de}}}$ with coordinates 
$({\Omega}_{\rm{dm}},{\Omega}_{\rm{de}})=\left(\frac{1}{2} + \frac{ \delta_{\text{dm}}-\delta_{\text{de}} +\Delta}{2 w},\frac{1}{2} - \frac{ \delta_{\text{dm}}-\delta_{\text{de}} +\Delta}{2 w}\right)$, 
the straight-line equation is:
\begin{gather} \label{DSA.Q.ddm+dde.18}
\begin{split}
\Omega_{\rm{de}} &= m \, \Omega_{\rm{dm}} + c, \quad 
m = \frac{\left(\frac{1}{2} - \frac{ \delta_{\text{dm}}-\delta_{\text{de}} +\Delta}{2 w}\right) -\left(\frac{1}{2} - \frac{ \delta_{\text{dm}}-\delta_{\text{de}} -\Delta}{2 w}\right)}{\left(\frac{1}{2} + \frac{ \delta_{\text{dm}}-\delta_{\text{de}} +\Delta}{2 w}\right) -\left(\frac{1}{2} + \frac{ \delta_{\text{dm}}-\delta_{\text{de}} -\Delta}{2 w}\right)} 
= -1, \quad c = 1, \\
\Omega_{\rm{de}} &= - \, \Omega_{\rm{dm}} + 1   
\quad \rightarrow \quad \Omega_{\rm{dm}} + \Omega_{\rm{de}} = 1.
\end{split}
\end{gather}
For the third line, which connects the past attractor $P_{\text{r}}$ with coordinates 
$({\Omega}_{\rm{dm}},{\Omega}_{\rm{de}})=(0,0)$ 
to the future attractor $P_{\text{dm+\textbf{de}}}$ with coordinates 
$({\Omega}_{\rm{dm}},{\Omega}_{\rm{de}})=\left(\frac{1}{2} + \frac{ \delta_{\text{dm}}-\delta_{\text{de}} +\Delta}{2 w},\frac{1}{2} - \frac{ \delta_{\text{dm}}-\delta_{\text{de}} +\Delta}{2 w}\right)$, 
the straight-line equation is:
\begin{gather} \label{DSA.Q.ddm+dde.19}
\begin{split}
\Omega_{\rm{de}} &= m \, \Omega_{\rm{dm}} + c, \quad 
m = \frac{\frac{1}{2} - \frac{ \delta_{\text{dm}}-\delta_{\text{de}} +\Delta}{2 w} - 0}{\frac{1}{2} + \frac{ \delta_{\text{dm}}-\delta_{\text{de}} +\Delta}{2 w} - 0} 
= -\frac{ \delta_{\text{dm}}-\delta_{\text{de}} - w + \Delta}{\delta_{\text{dm}}-\delta_{\text{de}} + w + \Delta}, \quad c = 0, \\
\Omega_{\rm{de}} &= -\frac{ \delta_{\text{dm}}-\delta_{\text{de}} - w + \Delta}{\delta_{\text{dm}}-\delta_{\text{de}} + w + \Delta} \, \Omega_{\rm{dm}}   
\quad \rightarrow \quad r = -\frac{ \delta_{\text{dm}}-\delta_{\text{de}} + w + \Delta}{\delta_{\text{dm}}-\delta_{\text{de}} - w + \Delta}.
\end{split}
\end{gather}
From equations \eqref{DSA.Q.ddm+dde.17.5}–\eqref{DSA.Q.ddm+dde.19}, we can derive the constraints that ensure our trajectories remain within the green triangle in Figure \ref{fig:2D_Q_general_Phase_portrait_boundaries}. These constraints require the initial conditions $\Omega_{\rm{(dm,0)}}$ and $\Omega_{\rm{(de,0)}}$ to lie either above or below the corresponding straight lines. Taken together, we obtain:
\begin{gather} \label{DSA.Q.ddm+dde.20}
\begin{split}
-\frac{ \delta_{\text{dm}}  - \delta_{\text{de}} + w + \Delta}{\delta_{\text{dm}}  - \delta_{\text{de}} - w + \Delta} 
< r_0 < 
-\frac{ \delta_{\text{dm}}  - \delta_{\text{de}} + w -\Delta}{\delta_{\text{dm}}  - \delta_{\text{de}} - w - \Delta}, 
\quad 
\Omega_{\rm{(dm,0)}}+ \Omega_{\rm{(de,0)}} \le 1.
\end{split}
\end{gather}
The last condition, $\Omega_{\rm{(dm,0)}}+ \Omega_{\rm{(de,0)}} \le 1$, automatically holds under our initial assumption of a flat universe, where $\Omega_{\rm{(r,0)}}+ \Omega_{\rm{(bm,0)}}+\Omega_{\rm{(dm,0)}}+ \Omega_{\rm{(de,0)}} = 1$. After more algebra and careful checking of signs, and substituting $\Delta$ from~\eqref{DSA.Q.ddm+dde.4}, the lower bound in~\eqref{DSA.Q.ddm+dde.20} can be written in terms of $\delta_{\text{dm}}$ and $\delta_{\text{de}}$:
\begin{equation}
\begin{split}
-\frac{ \delta_{\text{dm}}  - \delta_{\text{de}} + w + \Delta}{\delta_{\text{dm}}  - \delta_{\text{de}} - w + \Delta}
< r_0 
\quad \rightarrow \quad   
\delta_{\text{dm}} r_0 + \delta_{\text{de}} < -\frac{w r_0}{(1+r_0)}.\\
\end{split}
\label{eq:delta_dm+delta_de_r_constraint_++_lower}
\end{equation} 
Similarly, the upper bound in~\eqref{DSA.Q.ddm+dde.20} may be rewritten to obtain the same expression in terms of $\delta_{\text{dm}}$ and $\delta_{\text{de}}$:
\begin{equation}
\begin{split}
r_0 &< -\frac{ \delta_{\text{dm}}  - \delta_{\text{de}} + w - \Delta}{\delta_{\text{dm}}  - \delta_{\text{de}} - w - \Delta} 
\quad \rightarrow \quad 
\delta_{\text{dm}} r_0 + \delta_{\text{de}} < -\frac{w r_0}{(1+r_0)}. \\
\end{split}
\label{eq:delta_dm+delta_de_r_constraint_++_upper}
\end{equation} 

Taking the constraints from~\eqref{DSA.Q.ddm+dde.5}, \eqref{DSA.Q.ddm+dde.15}, \eqref{DSA.Q.ddm+dde.16}, \eqref{eq:delta_dm+delta_de_r_constraint_++_lower}, and \eqref{eq:delta_dm+delta_de_r_constraint_++_upper} into account, we find the following general set of constraints:
\begin{gather} \label{DSA.Q.delta_dm+delta_de.PEC_BG}
\boxed{
\begin{aligned}
&\underline{\text{Conditions for } \rho_{\rm{dm}}\ge 0,\; \rho_{\rm{de}}\ge 0 \text{ and reality throughout the entire cosmological evolution (iDEDM):}}\\
&\quad 1.\; \delta_{\text{dm}}\ge 0;\quad 
2.\; \delta_{\text{de}}\ge 0;\quad 
3.\; \delta_{\text{dm}} r_0 + \delta_{\text{de}}\le -\dfrac{w r_0}{1+r_0};\quad 
4.\; (\delta_{\text{dm}}+\delta_{\text{de}}+w)^2 \ge 4\,\delta_{\text{de}}\delta_{\text{dm}}.
\end{aligned}
}
\end{gather}
The first two constraints, requiring positive coupling parameters, derived in~\eqref{DSA.Q.delta_dm+delta_de.PEC_BG}, match the conclusions of~\cite{Caldera_Cabral_2009_DSA}, while the full set of constraints reduces to those obtained in~\cite{vanderWesthuizen:2023hcl} for the special cases where either $\delta_{\text{dm}}=0$ or $\delta_{\text{de}}=0$, as found from analytical solutions. This convergence of results further validates the approach used for the derivation of~\eqref{DSA.Q.delta_dm+delta_de.PEC_BG}. A summary of the conditions under which DM and DE densities become negative is presented in Table~\ref{tab:Qddm+dde_energy_conditions}. We may now summarize the behavior of the system by substituting the fluid densities at each critical point into Eqs.~\eqref{DSA.r}--\eqref{DSA.r_st}, resulting in Table~\ref{tab:CP_B_ddm+dde}. \\

\begin{table}[H]
\centering
\renewcommand{\arraystretch}{1.4} 
\setlength{\tabcolsep}{10pt}
\begin{tabular}{|c|c|c|c|c|}
\hline
\textbf{} &$P_{\text{r}}$ & $P_{\text{bm}}$ & $P_{\text{\textbf{dm}+de}}$ & $P_{\text{dm+\textbf{de}}}$ \\ \hline \hline
\textbf{Class} & Source & Saddle  & Saddle & Sink \\ \hline
$\Omega_{\rm{r}}$ & $1$ & $0$ & $0$ & $0$ \\ \hline
$\Omega_{\rm{bm}}$ & $0$ & $1$ & $0$  & $0$\\ \hline
$\Omega_{\rm{dm}}$ & $0$ & $0$ & $\frac{1}{2} + \frac{ \delta_{\text{dm}}-\delta_{\text{de}} -\Delta}{2 w}$  & $\frac{1}{2} + \frac{ \delta_{\text{dm}}-\delta_{\text{de}} +\Delta}{2 w}$\\ \hline
$\Omega_{\rm{de}}$ & $0$ & $0$ & $\frac{1}{2} - \frac{ \delta_{\text{dm}}-\delta_{\text{de}} -\Delta}{2 w}$ & $\frac{1}{2} - \frac{ \delta_{\text{dm}}-\delta_{\text{de}} +\Delta}{2 w}$ \\ \hline 
$\Omega_{\text{dm}}>0$ & $\forall \delta_{\text{dm/de}}$ & $\forall \delta_{\text{dm/de}}$& $\delta_{\text{dm}}+\delta_{\text{de}}< |w|$   & $\delta_{\text{de}}>0$ \\ \hline
$\Omega_{\text{de}}>0$ & $\forall \delta_{\text{dm/de}}$ & $\forall \delta_{\text{dm/de}}$& $\delta_{\text{dm}}>0$   & $\delta_{\text{dm}}+\delta_{\text{de}}< |w|$ \\ \hline
$r$ & $\infty$ & $\infty$ & $-\frac{\delta_{\text{dm}}-\delta_{\text{de}}+ w-\Delta}{\delta_{\text{dm}}-\delta_{\text{de}}- w-\Delta}$ & $-\frac{\delta_{\text{dm}}-\delta_{\text{de}}+ w+\Delta}{\delta_{\text{dm}}-\delta_{\text{de}}- w+\Delta}$ \\ \hline
$w^{\rm{eff}}_{\rm{dm}}$ & $-\delta_{\text{dm}}$ & $-\delta_{\text{dm}}$ & $-\delta_{\text{dm}}+\delta_{\text{de}}\left[\frac{\delta_{\text{dm}}-\delta_{\text{de}}- w-\Delta}{\delta_{\text{dm}}-\delta_{\text{de}}+ w-\Delta} \right]$ & $-\delta_{\text{dm}}+\delta_{\text{de}}\left[\frac{\delta_{\text{dm}}-\delta_{\text{de}}- w+\Delta}{\delta_{\text{dm}}-\delta_{\text{de}}+ w+\Delta} \right]$ \\ \hline
$w^{\rm{eff}}_{\rm{de}}$ & $\infty$ & $\infty$ & $w +\delta_{\text{de}}-\delta_{\text{dm}}\left[\frac{\delta_{\text{dm}}-\delta_{\text{de}}+ w-\Delta}{\delta_{\text{dm}}-\delta_{\text{de}}- w-\Delta} \right]$ & $w +\delta_{\text{de}}-\delta_{\text{dm}}\left[\frac{\delta_{\text{dm}}-\delta_{\text{de}}+ w+\Delta}{\delta_{\text{dm}}-\delta_{\text{de}}- w+\Delta} \right]$\\ \hline
$\zeta$ & $\infty$ & $\infty$ & $0$ \text{(solves coincidence problem)} & $0$ \text{(solves coincidence problem)} \\ \hline
$w^{\rm{eff}}_{\rm{tot}}$ & $\frac{1}{3}$ & $0$ & $\frac{1}{2}\left[ w-\delta_{\text{dm}}+\delta_{\text{de}}+\Delta\right]$ & $\frac{1}{2}\left[ w-\delta_{\text{dm}}+\delta_{\text{de}}-\Delta\right]$ \\ \hline
$q$ & $1$ & $\frac{1}{2}$ & $\frac{1}{2}(1+\frac{3}{2}[w+\delta_{\text{de}}-\delta_{\text{dm}}+\Delta] )$ & $\frac{1}{2}(1+\frac{3}{2}[w+\delta_{\text{de}}-\delta_{\text{dm}}-\Delta] )$ \\ \hline
\end{tabular}
\caption{behavior of the model at the critical points for $Q = 3H (\delta_{\text{dm}}\rho_{\rm{dm}} + \delta_{\text{de}}\rho_{\rm{de}})$, where $\Delta = \sqrt{(\delta_{\text{dm}} + \delta_{\text{de}} + w)^2 - 4\delta_{\text{de}}\delta_{\text{dm}}}$.}
\label{tab:CP_B_ddm+dde} 
\end{table}
\noindent In Table \ref{tab:CP_B_ddm+dde}, it can be shown that if $\delta_{\text{dm}} = \delta_{\text{de}} = 0$, then $\Delta = -w$, and all the obtained expressions at the critical points $P_{\text{\textbf{dm}+de}}$ and $P_{\text{dm+\textbf{de}}}$ reduce back to the non-interacting case. We note that, at each critical point for this model, we have $w^{\rm{eff}}_{\rm{dm}} = -\delta_{\text{dm}} - \frac{\delta_{\text{de}}}{r}$, $w^{\rm{eff}}_{\rm{de}} = w + \delta_{\text{de}} + \delta_{\text{dm}} r$, and $q = \frac{1}{2} \left(1 + 3 w^{\rm{eff}}_{\rm{tot}}\right)$.
Furthermore, at $P_{\text{\textbf{dm}+de}}$ and $P_{\text{dm+\textbf{de}}}$, the expressions $w^{\rm{eff}}_{\rm{dm}} = w^{\rm{eff}}_{\rm{de}} = w^{\rm{eff}}_{\rm{tot}}$ are all algebraically equivalent, but show dynamical behavior between the last two critical points, as also illustrated in Figures \ref{fig:CP+omega_dmde_Qdm+de}, \ref{fig:CP+omega_dmde_Qdm-de}, \ref{fig:CP+omega_dmde_Qdm}, and \ref{fig:CP+omega_dmde_Qde}. It is important to note that the infinities present in $P_{\text{r}}$ and $P_{\text{bm}}$ are not physical and will not occur for this model. These only arise when the ratio $r = \frac{\Omega_{\rm{dm}}}{\Omega_{\rm{de}}}$ diverges as $\Omega_{\rm{de}} \rightarrow 0$, which will not physically occur if the positive energy conditions are maintained, since these models keep the ratio $r = \text{constant}$ fixed in the past, even as $\Omega_{\rm{dm}}, \Omega_{\rm{de}} \rightarrow 0$. Therefore, in the distant past, the expressions for $r$, $w^{\rm{eff}}_{\rm{dm}}$, $w^{\rm{eff}}_{\rm{de}}$, and $\zeta$ at points $P_{\text{r}}$ and $P_{\text{bm}}$ will be the same as the expressions found at the point $P_{\text{\textbf{dm}+de}}$. This behavior can clearly be seen in Figures \ref{fig:CP+omega_dmde_Qdm+de}, \ref{fig:CP+omega_dmde_Qdm-de}, \ref{fig:CP+omega_dmde_Qdm}, and \ref{fig:CP+omega_dmde_Qde}. In order to have accelerated expansion, $w^{\rm{eff}}_{\rm{tot}} < -\frac{1}{3}$ at the last critical point $P_{\text{\textbf{dm}+de}}$, the following condition must hold:
\begin{gather} \label{DSA.Q.ddm+dde.ACC}
\begin{split}
\frac{1}{2}\left[ w - \delta_{\text{dm}} + \delta_{\text{de}} - \Delta \right] &< -\frac{1}{3}  
\quad \rightarrow \quad  
\delta_{\text{dm}} \left( 3w + 1 \right) - \delta_{\text{de}} > w + \frac{1}{3}.
\end{split}
\end{gather} 
Similarly, if the model is in the phantom DE regime ($w < -1$), a future big rip singularity may still be avoided. No big rip will occur as long as $w^{\rm{eff}}_{\rm{tot}} \ge -1$ at the last critical point $P_{\text{dm+\textbf{de}}}$, for which we require the following condition:
\begin{gather} \label{DSA.Q.ddm+dde.BR}
\begin{split}
\frac{1}{2}\left[ w - \delta_{\text{dm}} + \delta_{\text{de}} - \Delta \right] &\ge -1  
\quad \rightarrow \quad  
\delta_{\text{dm}} \left( w + 1 \right) - \delta_{\text{de}} \le w + 1. \\ 
\end{split}
\end{gather} 
The constraints obtained in \eqref{DSA.Q.ddm+dde.ACC} and \eqref{DSA.Q.ddm+dde.BR} can be applied to all other special cases of this model where the interaction is still present in the distant future. For the models studied in this paper, only the interaction $Q=3\delta H \rho_{\text{dm}}$, where $\Omega_{\text{dm}} \rightarrow 0$ at late times, will not apply, as the model converges to uncoupled behavior in the distant future where $w^{\rm{eff}}_{\rm{tot}} = w$. For all other cases, the constraints to ensure future accelerated expansion, as well as to avoid a big rip future singularity in the phantom regime, can be found in Table \ref{tab:Com_AE_BR}.

We may now consider the stability of this system from the sign of the doom factor \eqref{DSA.doom} for this interaction:
\begin{gather} \label{DSA.Qddm+dde.doom}
\begin{split}
\textbf{d}=  \frac{Q}{3H\rho_{\rm{de}}(1+w)}=\frac{3H (\delta_{\text{dm}}\rho_{\rm{dm}}+\delta_{\text{de}}\rho_{\rm{de}})}{3H\rho_{\rm{de}}(1+w)}=\frac{1}{(1+w)} \left(\delta_{\text{de}}+ \delta_{\text{dm}}\frac{\rho_{\rm{dm}}}{\rho_{\rm{de}}} \right) = \left(\frac{\delta_{\text{de}}+ \delta_{\text{dm}}r}{1+w}\right),
\end{split}
\end{gather}
where we also apply the conditions $\rho_{\rm{dm}}>0$ and $\rho_{\rm{de}}>0$, which imply $r>0$. Since we need $\textbf{d} < 0$ to guarantee a stable universe, we can see from \eqref{DSA.Qddm+dde.doom} that this will only occur if $(\delta_{\text{de}}+ \delta_{\text{dm}}r)$ and $(1+w)$ have opposite signs. Given that we require $\delta_{\text{dm}}>0$ and $\delta_{\text{de}}>0$ for positive energy \eqref{DSA.Q.delta_dm+delta_de.PEC_BG}, this implies that we need $(1+w) > 0$ for \textit{a priori} stability, corresponding to $w < -1$, which means \textit{DE needs to be in the phantom regime}. Different combinations of parameters and their effects on energy density and stability are summarized in Table \ref{tab:Qddm+dde_stability_criteria}. 
Again, it should be kept in mind that an analysis in the Parametrized Post-Friedmann Framework might avoid this problem altogether.

\begin{table}[h]
    \centering
    \renewcommand{\arraystretch}{1.1} 
        \setlength{\tabcolsep}{5pt} 
    \begin{tabular}{|c|c|c|c|c|c|c|c|c|}
        \hline 
        $\delta_{\text{dm}} ; \delta_{\text{de}}$ & Energy flow & $w$ & Dark energy & $d$ & a priori stable & $\rho_{\text{dm}} > 0$ & $\rho_{\text{de}} > 0$ & \textbf{Viable} \\
        \hline \hline
        $+$ & DE $\to$ DM & $< -1$ & Phantom & - & $\checkmark$ & $\checkmark$ & $\checkmark$ & $\checkmark$ \\
        \hline
        $+$ & DE $\to$ DM & $> -1$ & Quintessence & + & X & $\checkmark$ & $\checkmark$ & \textbf{X} \\
        \hline
        $-$ & DM $\to$ DE & $< -1$ & Phantom & + & X & X & X & \textbf{X} \\
        \hline
        $-$ & DM $\to$ DE & $> -1$ & Quintessence & - & $\checkmark$ & X & X & \textbf{X} \\
        \hline
    \end{tabular}
    \caption{Stability and positive-energy criteria for $Q = 3H (\delta_{\text{dm}}\rho_{\rm{dm}}+\delta_{\text{de}}\rho_{\rm{de}})$.}
    \label{tab:Qddm+dde_stability_criteria}
\end{table}

 \subsection{Summary of the Main Properties for Interaction $Q = 3H (\delta_{\rm{dm}}\rho_{\rm{dm}}+\delta_{\rm{de}}\rho_{\rm{de}})$} \label{DSA.summary}
 
We summarize below the main dynamical and physical properties of the model, together with the corresponding constraints derived in the previous sections.

\begin{itemize}
    \item This model may have both negative DM densities in the future ($\Omega_{\rm{dm}} < 0$) and negative DE densities in the past ($\Omega_{\rm{de}} < 0$), if inappropriate parameters are chosen. This rules out positive energies for the iDMDE regime ($\delta_{\rm{dm}} < 0$ or $\delta_{\rm{de}} < 0$) for all linear interaction models of this form, as also illustrated in Figures~\ref{fig:Omega_Linear_dm+de}, \ref{fig:Omega_Linear_dm-de}, \ref{fig:Omega_Linear_dm} and \ref{fig:Omega_Linear_de}.  
    \item Real and positive energy densities for all components, throughout the entire cosmological evolution, can be ensured with the constraints found in \eqref{DSA.Q.delta_dm+delta_de.PEC_BG} and Table~\ref{tab:Qddm+dde_energy_conditions}: 
    \begin{enumerate}
        \item  $\delta_{\text{dm}}>0$ -- Positive DE density in the distant past,
        \item  $\delta_{\text{de}}>0$ -- Positive DM density in the distant future,
        \item $\delta_{\text{dm}}r_0+ \delta_{\text{de}}<-\frac{w r_0}{(1+r_0)}$ -- Positive DM and DE densities between the distant past and the distant future,
        \item $ (\delta_{\text{dm}}+\delta_{\text{de}}+w)^2>4\delta_{\text{de}}\delta_{\text{dm}}$ -- Real energy densities.
    \end{enumerate}
    These conditions constrain these models to only small interactions in the iDEDM regime. As a consequence, both interactions $Q = 3H \delta(\rho_{\rm{dm}}-\rho_{\rm{de}})$ and $Q = 3H \delta(\rho_{\rm{de}}-\rho_{\rm{dm}})$ can be ruled out due to the presence of negative energies, as also illustrated in Figure~\ref{fig:Omega_Linear_dm-de}.
    \item Taking the doom factor into account alongside our positive-energy conditions, DE is restricted to the phantom regime ($w<-1$) to avoid early-time instabilities, as shown in Table~\ref{tab:Qddm+dde_stability_criteria}.
    \item In the iDEDM regime, this model solves the coincidence problem both in the past (if $\delta_{\text{dm}}\neq0$) and in the future (if $\delta_{\text{de}}\neq0$), as also illustrated in Figures~\ref{fig:CP+omega_dmde_Qdm+de}, \ref{fig:CP+omega_dmde_Qdm-de}, \ref{fig:CP+omega_dmde_Qdm} and \ref{fig:CP+omega_dmde_Qde}.
    \item This model is guaranteed to feature an early radiation-dominated era in the past, a baryonic matter-dominated saddle point, as well as a dark matter--dark energy hybrid-dominated saddle point and late-time era, given the positive-energy constraints above, and our initial assumptions: $w < 0$ and $\delta_{\text{dm}}+\delta_{\text{de}}< |w|$.
    \item The late-time dark matter--dark energy hybrid-dominated era will be characterized by accelerated expansion if 
    \begin{equation}
        \delta_{\text{dm}} \left(3w+1\right)-\delta_{\text{de}} > w + \frac{1}{3},
    \end{equation}
    while a big rip future singularity may be avoided, even if $w < -1$ (which is required to avoid early-time instabilities), as long as energy flows from DE to DM and the interaction strength is sufficiently positive so that 
    \begin{equation}
        \delta_{\text{dm}} \left(w+1\right)-\delta_{\text{de}} > w + 1.
    \end{equation}
    This is also illustrated in Figures~\ref{fig:eos_tot_BR_Qdm+de}, \ref{fig:eos_tot_BR_Qdm-de}, \ref{fig:eos_tot_BR_Qdm} and \ref{fig:eos_tot_BR_Qde}.
    \item This interaction model, and all derived constraints, apply to the following special cases:
        \begin{enumerate}
            \item $Q=3H \delta(\rho_{\rm{dm}}+\rho_{\rm{de}})$ when $\delta_{\text{dm}}=\delta_{\text{de}}=\delta$ \;(see Section~\ref{Q_dm+de}),
            \item $Q=3H \delta(\rho_{\rm{dm}}-\rho_{\rm{de}})$ when $\delta_{\text{dm}}=\delta$ and $\delta_{\text{de}}=-\delta$ \;(see Section~\ref{Q_dm-de}),
            \item $Q=3H \delta\rho_{\rm{dm}}$ when $\delta_{\text{dm}}=\delta$ and $\delta_{\text{de}}=0$ \;(see Section~\ref{Q_dm}),
            \item $Q=3H \delta\rho_{\rm{de}}$ when $\delta_{\text{de}}=\delta$ and $\delta_{\text{dm}}=0$ \;(see Section~\ref{Q_de}),
            \item $\Lambda$CDM when $\delta_{\text{de}}=\delta_{\text{dm}}=0$ and $w=-1$.
        \end{enumerate}
    The above constraints and their consequences for these special cases are summarized in Tables~\ref{tab:Com_PEC}, \ref{tab:Com_real}, \ref{tab:Com_AE_BR} and \ref{tab:Com_CP}.
\end{itemize}

\begin{table}[H]
    \centering
    \renewcommand{\arraystretch}{1.2} 
    \setlength{\tabcolsep}{2pt}       
    \begin{tabular}{|c|c|c|c|c|c|c|}
        \hline
        \textbf{Conditions} & Energy flow & $\rho_{\text{(dm,past)}}$  & $\rho_{\text{(dm,future)}}$  & $\rho_{\text{(de,past)}}$ & $\rho_{\text{(de,future)}}$ & \textbf{Physical} \\
        \hline         \hline
        $\underbrace{\delta_{\text{dm}}}_{>0} r_0+ \underbrace{\delta_{\text{de}}}_{>0}<-\frac{w r_0}{(1+r_0)}$ & DE $\to$ DM & $+$ & $+$ & $+$ & $+$ & $\checkmark$ \\
        \hline
        $\underbrace{\delta_{\text{dm}}}_{<0} r_0+ \underbrace{\delta_{\text{de}}}_{>0}<-\frac{w r_0}{(1+r_0)}$ & DM $\leftrightarrows$ DE & $+$ & $+$ & $-$ & $+$  & \textbf{X} \\
        \hline
        $\underbrace{\delta_{\text{dm}}}_{>0} r_0+ \underbrace{\delta_{\text{de}}}_{<0}<-\frac{w r_0}{(1+r_0)}$ & DM $\leftrightarrows$ DE & $+$ & $-$ & $+$ & $+$  & \textbf{X} \\
        \hline
         $\underbrace{\delta_{\text{dm}}}_{<0} r_0+ \underbrace{\delta_{\text{de}}}_{<0}<-\frac{w r_0}{(1+r_0)}$ & DM $\to$ DE & $+$ & $-$ & $-$ & $+$  & \textbf{X} \\
        \hline
         $\delta_{\text{dm}} r_0+ \delta_{\text{de}}>-\frac{w r_0}{(1+r_0)}$ & DM $\leftrightarrows$ DE & $-$ & $-$ & $-$ & $-$  & \textbf{X} \\
        \hline
    \end{tabular}
\caption{Positive-energy conditions for $Q = 3H (\delta_{\text{dm}}\rho_{\rm{dm}}+\delta_{\text{de}}\rho_{\rm{de}})$.}
    \label{tab:Qddm+dde_energy_conditions}
\end{table}

\section{Finding analytical solutions} \label{Finding_analytical_solutions}

\subsection{Finding expressions for $\rho_{\rm{dm}}$ and $\rho_{\rm{de}}$}

In this section, we will follow the derivation first outlined by~\cite{Pan_2015, Pan_2017}, but we will add a small section at the end of the derivation where we use boundary conditions to get expressions in a more useful and familiar format. We start with the derivation from the coupled conservation equations \eqref{eq:conservation}, while setting the total energy density of the dark components as $\rho_{\text{t}} = \rho_{\text{dm}} + \rho_{\text{de}}$ and its corresponding time derivative as $\dot{\rho}_{\text{t}} + 3H (\rho_{\text{t}} + p_{\text{t}}) = 0$. Using this, we may then add the two conservation equations in \eqref{eq:conservation} together, which yields:
\begin{gather} \label{eq:t_conservation}
\begin{split}
(\dot{\rho}_{\text{dm}}+\dot{\rho}_{\text{de}}) + 3H (\rho_{\text{dm}} + \rho_{\text{de}}+ w \rho_{\text{de}} ) &= Q-Q \quad  \\ \rightarrow \quad
\dot{\rho}_{\text{t}} + 3H (\rho_{\text{t}} + w \rho_{\text{de}}) &= 0. 
\end{split}
\end{gather}
Furthermore, to help ease calculations, the derivative may be changed to be with respect to $x=3 \ln a$, which has the consequence that for any variable $\rho$:
\begin{gather} \label{eq:variable_change}
\begin{split}
\frac{d}{dt} \rho &= 3H \frac{d\rho}{dx} \quad  \rightarrow \quad \dot{\rho} = 3H \rho'.
\end{split}
\end{gather}
Applying the change of variable technique from \eqref{eq:variable_change} to the conservation equation \eqref{eq:t_conservation} yields the following result:
\begin{gather} \label{eq:t'_conservation}
\begin{split}
3H \rho'_{\text{t}} + 3H (\rho_{\text{t}} + w \rho_{\text{de}}) &= 0 \\
\rho_{\text{de}} &= - \frac{\rho_{\text{t}} + \rho_{\text{t}}'}{w} .
\end{split}
\end{gather}
Similarly, an expression for $\rho_{\text{dm}}$ may be obtained by substituting \eqref{eq:t'_conservation} into the expression for the total energy $\rho_{\text{t}} = \rho_{\text{dm}} + \rho_{\text{de}}$, which is equivalent to:
\begin{gather} \label{eq:rhodm_expression}
\begin{split}
\rho_{\text{dm}} &= \rho_{\text{t}} -  \rho_{\text{de}}  
= \rho_{\text{t}} + \frac{\rho_{\text{t}} + \rho_{\text{t}}'}{w}  
= \frac{(1 + w) \rho_{\text{t}} + \rho_{\text{t}}'}{w}. 
\end{split}
\end{gather}

We are now in need of a differential equation for $\rho_{\text{t}}$, whose solution we can substitute back into expressions \eqref{eq:t'_conservation} and \eqref{eq:rhodm_expression}. To do this, we substitute the interaction function $Q = 3H \left( \delta_{\text{dm}} \rho_{\text{dm}} + \delta_{\text{de}} \rho_{\text{de}} \right)$ into the DE conservation equation \eqref{eq:conservation}, while also applying the change of variables $\dot{\rho}_{\text{de}} = 3H \rho'_{\text{de}}$:
\begin{gather} \label{eq:de_conservation_Q}
\begin{split}
3H \rho'_{\text{de}} + 3H (1 + w) \rho_{\text{de}} &= -3H \left( \delta_{\text{dm}} \rho_{\text{dm}} + \delta_{\text{de}} \rho_{\text{de}} \right) \;\rightarrow\;
\rho'_{\text{de}} + (1 + w + \delta_{\text{de}}) \rho_{\text{de}} + \delta_{\text{dm}} \rho_{\text{dm}} = 0 .
\end{split}
\end{gather}
To express this in terms of $\rho_{\text{t}}$ only, we need an expression for $\rho'_{\text{de}}$, obtained by differentiating $\rho_{\text{de}}$ from \eqref{eq:t'_conservation}: 
\begin{gather} \label{eq:rhode_prime}
\begin{split}
\rho_{\text{de}}' = \left( - \dfrac{\rho_{\text{t}} + \rho_{\text{t}}'}{w} \right)' = - \left( \dfrac{\rho_{\text{t}}' + \rho_{\text{t}}''}{w} - \dfrac{ (\rho_{\text{t}} + \rho_{\text{t}}') w' }{ w^2 } \right).
\end{split}
\end{gather}
We can now obtain a single second-order differential equation for $\rho_{\text{t}}$ by substituting the expressions for $\rho'_{\text{de}}$ \eqref{eq:rhode_prime}, $\rho_{\text{de}}$ \eqref{eq:t'_conservation}, and $\rho_{\text{dm}}$ \eqref{eq:rhodm_expression} into \eqref{eq:de_conservation_Q}, which, after some algebra, gives:
\begin{gather} \label{eq:rho_t_conservation}
\begin{split}
\rho_{\text{t}}'' + \left[ 2 + w + \delta_{\text{de}} - \delta_{\text{dm}} - \dfrac{ w' }{ w } \right] \rho_{\text{t}}' + \left[ (1 + w)(1 - \delta_{\text{dm}}) + \delta_{\text{de}} - \dfrac{ w' }{ w } \right] \rho_{\text{t}} = 0.
\end{split}
\end{gather}
The equation obtained in \eqref{eq:rho_t_conservation} allows for a variable equation of state $w(x)$, but for our purposes, we will assume a constant equation of state from here onwards. 
For the case where we assume a constant equation of state, the derivative $w'$ disappears, which reduces \eqref{eq:rho_t_conservation} to a linear homogeneous differential equation: 
\begin{gather} \label{eq:rho_t_constant_conservation}
\begin{split}
\rho_{\text{t}}'' + \underbrace{\left[ 2 + w + \delta_{\text{de}} - \delta_{\text{dm}} \right]}_{p} \rho_{\text{t}}' + \underbrace{\left[ (1 + w)(1 - \delta_{\text{dm}}) + \delta_{\text{de}}\right]}_{q} \rho_{\text{t}} = 0.
\end{split}
\end{gather}
The general solution to the differential equation \eqref{eq:rho_t_constant_conservation} is:
\begin{equation} \label{eq:general_solution}
\rho_{\text{t}}(x) = \rho_1 e^{\lambda_1 x} + \rho_2 e^{\lambda_2 x}   
\quad \rightarrow \quad  
\rho_{\text{t}}(a) = \rho_1 a^{3\lambda_1} + \rho_2 a^{3\lambda_2},
\end{equation}
where we have noted that since $x = 3 \ln a$, we also have $e^{\lambda (3 \ln a)} = a^{3 \lambda}$.  
The characteristic equation associated with this linear homogeneous differential equation, and its solution for $\lambda$, is given by:
\begin{equation} \label{eq:characteristic_equation}
\lambda^2 + p\, \lambda + q = 0   
\quad ; \quad 
\lambda = \frac{ -p \pm \sqrt{ p^2 - 4q } }{2}.
\end{equation}
We now need to compute the discriminant $D = p^2 - 4q$ in order to find the equations for $\lambda_1$ and $\lambda_2$. After some algebra, we find:
\begin{gather} \label{eq:p_2_4q}
\begin{split}
p^2 &=  4 + 4w + 4\delta_{\text{de}} - 4\delta_{\text{dm}} + w^2 + 2w \delta_{\text{de}} - 2w \delta_{\text{dm}} + \delta_{\text{de}}^2 - 2\delta_{\text{de}} \delta_{\text{dm}} + \delta_{\text{dm}}^2, \\ 
-4q &= - 4 - 4w + 4\delta_{\text{dm}} + 4w \delta_{\text{dm}} - 4\delta_{\text{de}}, \\
D &= \left( w + \delta_{\text{de}} + \delta_{\text{dm}} \right)^2 - 4\delta_{\text{de}} \delta_{\text{dm}}.
\end{split}
\end{gather} 
Substituting the expression from equation \eqref{eq:rho_t_constant_conservation} into equation \eqref{eq:characteristic_equation}, and keeping with the notation in~\cite{Pan_2017}, we set \( \lambda_1 = m_+ \) and \( \lambda_2 = m_- \):
\begin{equation}
\boxed{
m_{\pm} = \frac{ -\left( 2 + w + \delta_{\text{de}} - \delta_{\text{dm}} \right) \pm \sqrt{ \left( w + \delta_{\text{de}} + \delta_{\text{dm}} \right)^2 - 4\, \delta_{\text{de}}\, \delta_{\text{dm}} } }{2}
}.
\label{eq:m_pm}
\end{equation}
To find expressions for $\rho_{\text{dm}}$ and $\rho_{\text{de}}$ in terms of $\rho_{\text{t}}$, we first need an additional expression for $\rho'_{\text{t}}$. This can be obtained by taking the derivative of equation \eqref{eq:general_solution}, while letting \( \lambda_1 = m_+ \) and \( \lambda_2 = m_- \):
\begin{equation}\label{eq:rho_t_prime_x}
\rho_{\text{t}}'(x) = \rho_1\, m_+\, e^{m_+ x} + \rho_2\, m_-\, e^{m_- x} 
\quad \rightarrow \quad 
\rho_{\text{t}}'(a) = \rho_1\, m_+\, a^{3 m_{+}} + \rho_2\, m_-\, a^{3 m_{-}},
\end{equation}
where we again used the relation $e^{\lambda(3 \ln a)} = a^{3 \lambda}$.  
We may now substitute $\rho'_{\text{t}}(a)$ from \eqref{eq:rho_t_prime_x} and $\rho_{\text{t}}(a)$ from \eqref{eq:general_solution} into $\rho_{\text{dm}}$ from \eqref{eq:rhodm_expression}:
\begin{gather} \label{eq:rhodm_a_final}
\begin{split}
\rho_{\text{dm}}(a) &= \frac{(1 + w)\, \rho_{\text{t}}(a) + \rho_{\text{t}}'(a)}{w} = \rho_1 \left[ \frac{ 1 + w + m_+ }{w} \right] a^{3 m_+} 
+ \rho_2 \left[ \frac{ 1 + w + m_- }{w} \right] a^{3 m_-}.
\end{split}
\end{gather}
Similarly, substituting \eqref{eq:rho_t_prime_x} and \eqref{eq:general_solution} into \eqref{eq:t'_conservation}, we find:
\begin{gather} \label{eq:rho_de_a_final}
\begin{split}
\rho_{\text{de}}(a) &= - \frac{\rho_{\text{t}}(a) + \rho_{\text{t}}'(a)}{w}= - \rho_1 \left[ \frac{1 + m_+}{w} \right] a^{3 m_+}
- \rho_2 \left[ \frac{1 + m_-}{w} \right] a^{3 m_-}.
\end{split}
\end{gather}
Equations \eqref{eq:rhodm_a_final} and \eqref{eq:rho_de_a_final} are the final expressions obtained in~\cite{Pan_2017}.
The following section marks the first new part of the derivation.  
Using the boundary conditions from the present-day values of the energy densities, we have $\rho_{\text{dm}}(a=1) = \rho_{\text{(dm,0)}}$ and $\rho_{\text{de}}(a=1) = \rho_{\text{(de,0)}}$. Substituting these conditions into \eqref{eq:rhodm_a_final} and \eqref{eq:rho_de_a_final}, we obtain expressions for the current DM and DE densities in these models:
\begin{equation}
\rho_{\text{(dm,0)}} = \rho_1 \left[\frac{ 1 + w + m_+ }{w} \right] + \rho_2 \left[ \frac{ 1 + w + m_- }{w} \right],  
\quad 
\rho_{\text{(de,0)}} = - \rho_1 \left[\frac{ 1  + m_+ }{w} \right] - \rho_2 \left[ \frac{ 1  + m_- }{w} \right].
\label{eq:rhodm0_de_0_c}
\end{equation}
We may note that $\rho_{\text{(dm,0)}} + \rho_{\text{(de,0)}} = \rho_1 + \rho_2$, which is straightforward to show. 
This relation also implies that $\rho_2 = \rho_{\text{(dm,0)}} + \rho_{\text{(de,0)}} - \rho_1$, which can be substituted into \eqref{eq:rhodm0_de_0_c} and solved for $\rho_1$. After some algebra, we find:
\begin{gather} \label{eq:rho_1_derivation}
\begin{split}
\rho_1 &= \frac{\rho_{\text{(de,0)}}\left(m_- +  w + 1 \right) + \rho_{\text{(dm,0)}} \left( m_- + 1 \right)}{m_- - m_+}.
\end{split}
\end{gather}
Substituting $\rho_1$ from \eqref{eq:rho_1_derivation} into $\rho_2 = \rho_{\text{(dm,0)}} + \rho_{\text{(de,0)}} - \rho_1$, we obtain $\rho_2$, which after some algebra is given by:
\begin{gather} \label{eq:rho_2_derivation}
\begin{split}
\rho_2 &=- \frac{\rho_{\text{(de,0)}}\left(m_+ +  w + 1 \right) + \rho_{\text{(dm,0)}} \left( m_+ + 1 \right)}{m_- - m_+}.
\end{split}
\end{gather}
Now we can substitute the integration constants \eqref{eq:rho_1_derivation} and \eqref{eq:rho_2_derivation} back into the expression for the DM energy density \eqref{eq:rhodm_a_final} to obtain the general solution:
\begin{empheq}[box=\fbox]{equation}
\begin{split}
\rho_{\text{dm}} = +& \left[\frac{m_++w+1}{w (m_- - m_+)} \right] \left[\rho_{\text{(de,0)}} (m_- + w + 1) + \rho_{\text{(dm,0)}} (m_- + 1) \right]  a^{3m_+}  \\
- & \left[\frac{m_-+w+1}{w  (m_- - m_+)} \right]\left[\rho_{\text{(de,0)}} (m_+ + w + 1) + \rho_{\text{(dm,0)}} (m_+ + 1)  \right]  a^{3m_-}
\end{split}
\label{eq:rhodm_general}
\end{empheq}
Similarly, we can substitute the integration constants \eqref{eq:rho_1_derivation} and \eqref{eq:rho_2_derivation} back into the expression for the DE density \eqref{eq:rho_de_a_final} to obtain the general solution:
\begin{empheq}[box=\fbox]{equation}
\begin{split}
\rho_{\text{de}} =\; & -\left[\frac{m_+ + 1}{w (m_- - m_+)} \right] \left[ \rho_{\text{(de,0)}} (m_- + w + 1) + \rho_{\text{(dm,0)}} (m_- + 1) \right] a^{3m_+}  \\
& + \left[\frac{m_- + 1}{w (m_- - m_+)} \right] \left[\rho_{\text{(de,0)}} (m_+ + w + 1) + \rho_{\text{(dm,0)}} (m_+ + 1)  \right]  a^{3m_-}
\end{split}
\label{eq:rhode_general}
\end{empheq}

\subsection{More analytical relations} \label{More_analytical_relations}

In order to show that our derived expressions are correct, we should be able to obtain expressions that converge to those obtained in the dynamical systems analysis in Section \ref{Sec.DSA}. The most interesting results are those for the last two hybrid-dominated critical points in Table \ref{tab:CP_B_ddm+dde}. The last point $P_{\text{(dm+\textbf{de})}}$ is obtained in the asymptotic limit $a\rightarrow \infty$, while $P_{\text{(\textbf{dm}+de)}}$ is obtained when $a\rightarrow 0$ and only DM and DE are relevant; therefore, we neglect the baryon contribution. In either case, the Hubble parameter can be approximated as being dominated by the DM and DE terms, such that:
\begin{equation}
\begin{split}
H^2= \frac{8\pi G}{3 } \left(\rho_{\text{r}}+\rho_{\text{bm}}+\rho_{\text{dm}}+\rho_{\text{de}}\right) \approx \frac{8\pi G}{3 } \left(\rho_{\text{dm}}+\rho_{\text{de}}\right) .
\end{split}\label{eq:frac_den_general_2}
\end{equation}
We assume that for both $\rho_{\text{dm}}$ in \eqref{eq:rhodm_general} and $\rho_{\text{de}}$ in \eqref{eq:rhode_general}, in the distant future $(a\rightarrow\infty)$ the terms with $a^{3m_-}$ become negligible, while in the past $(a\rightarrow 0)$ the terms with $a^{3m_+}$ can be ignored. After some algebra, the Hubble function \eqref{eq:frac_den_general_2} for the future and past approximates to:
\begin{equation}
\begin{split}
H^2_{\text{future}} &\approx H_0^2 \left(\frac{\Omega_{\text{(de,0)}} (m_- + w + 1) + \Omega_{\text{(dm,0)}} (m_- + 1) }{ (m_- - m_+)}    \right) a^{3m_+}, \\
H^2_{\text{past}} &\approx - H_0^2 \left(\frac{\Omega_{\text{(de,0)}} (m_+ + w + 1) + \Omega_{\text{(dm,0)}} (m_+ + 1) }{ (m_- - m_+)}    \right) a^{3m_-}.
\end{split}\label{eq:frac_den_general_3}
\end{equation}
In \eqref{eq:frac_den_general_3}, we have also used the relation $\Omega=\frac{8\pi G \rho}{3H^2}$. Substituting $H^2_{\text{future}}$ from \eqref{eq:frac_den_general_3}, together with $\rho_{\text{(de,future)}}$ and $\rho_{\text{(dm,future)}}$ (where the $a^{3m_-}$ terms in \eqref{eq:rhodm_general} and \eqref{eq:rhode_general} are neglected), we obtain the future fractional DM and DE densities after some algebra, corresponding to the critical point $P_{\text{(dm+\textbf{de})}}$: 
\begin{equation}
\begin{split}
\Omega_{\text{(de,future)}}=\frac{8\pi G \rho_{\text{(de,future)}}}{3H^2_{\text{future}}}= -\frac{m_++1}{w}  \quad ; \quad   \Omega_{\text{(dm,future)}}=\frac{8\pi G \rho_{\text{(dm,future)}}}{3H^2_{\text{future}}}= \frac{m_++w+1}{w}. 
\end{split}\label{eq:frac_den_general_future_dmde}
\end{equation}
Similarly, substituting $H^2_{\text{past}}$ from \eqref{eq:frac_den_general_3}, together with $\rho_{\text{(de,past)}}$ and $\rho_{\text{(dm,past)}}$ (where the $a^{3m_+}$ terms in \eqref{eq:rhodm_general} and \eqref{eq:rhode_general} are neglected), we obtain the past fractional DM and DE densities, corresponding to the critical point $P_{\text{(\textbf{dm}+de)}}$: 
\begin{equation}
\begin{split}
\Omega_{\text{(de,past)}}=\frac{8\pi G \rho_{\text{(de,past)}}}{3H^2_{\text{past}}}&= -\frac{m_-+1}{w} \quad ; \quad \Omega_{\text{(dm,past)}}=\frac{8\pi G \rho_{\text{(dm,past)}}}{3H^2_{\text{past}}}= \frac{m_-+w+1}{w}.   
\end{split}\label{eq:frac_den_general_past_dmde}
\end{equation}
It can be seen that $\Omega_{\text{de}}+\Omega_{\text{dm}}=1$ for both \eqref{eq:frac_den_general_future_dmde} and \eqref{eq:frac_den_general_past_dmde}, corresponding to hybrid dominance at both points.
As previously noted, the DM and DE densities in these models can become negative in either the past or the future. It is therefore of interest to derive expressions for the predicted redshifts $z_{\text{(dm=0)}}$ and $z_{\text{(de=0)}}$, where $\rho_{\text{dm}}$ in \eqref{eq:rhodm_general} or $\rho_{\text{de}}$ in \eqref{eq:rhode_general} becomes exactly zero, respectively, before crossing into the non-physical negative energy domain. Noting the transformation $a=(1+z)^{-1}$, we find: 
\begin{equation}
\begin{split}
z_{\text{(dm=0)}}&=\left[\frac{\left[m_++w+1\right] \left[\Omega_{\text{(de,0)}} (m_- + w + 1) + \Omega_{\text{(dm,0)}} (m_- + 1) \right]}{\left[m_-+w+1\right]\left[\Omega_{\text{(de,0)}} (m_+ + w + 1) + \Omega_{\text{(dm,0)}} (m_+ + 1)  \right]}\right]^{\frac{1}{-3(m_--m_+)}}-1, \\
z_{\text{(de=0)}}&=\left[\frac{\left[m_++1 \right] \left[ \Omega_{\text{(de,0)}} (m_- + w + 1) + \Omega_{\text{(dm,0)}} (m_- + 1) \right]}{\left[m_-+1 \right] \left[\Omega_{\text{(de,0)}} (m_+ + w + 1) + \Omega_{\text{(dm,0)}} (m_+ + 1)  \right]}\right]^{\frac{1}{-3(m_--m_+)}}-1 .
\end{split}
\label{eq:rhode_general_z_min}
\end{equation}
Some models may experience a switch in the direction of energy flow after $Q=0$.  
This sign change occurs at the redshift where $\delta_{\text{dm}}\rho_{\text{dm}}+\delta_{\text{de}}\rho_{\text{de}}=0$.  
Substituting \eqref{eq:rhodm_general} and \eqref{eq:rhode_general} into this condition and solving for $z$ gives:
\begin{equation}
\begin{split}
z_{\rm{(Q=0)}}  = \left[\frac{\left[\delta_{\text{dm}}(m_-+w+1)-\delta_{\text{de}}(m_-+1) \right] \left[\Omega_{\text{(de,0)}} (m_+ + w + 1) + \Omega_{\text{(dm,0)}} (m_+ + 1)  \right]}{\left[\delta_{\text{dm}}(m_++w+1)-\delta_{\text{de}}(m_++1) \right] \left[\Omega_{\text{(de,0)}} (m_- + w + 1) + \Omega_{\text{(dm,0)}} (m_- + 1) \right] }\right]^{\frac{1}{3(m_--m_+)}}-1.
\end{split}
\label{eq:Q=0_general}
\end{equation}
To find the redshift at which DM and DE have equal densities, we set $\rho_{\text{dm}}=\rho_{\text{de}}$ in \eqref{eq:rhodm_general} and \eqref{eq:rhode_general}, apply $a=(1+z)^{-1}$, and solve for $z$. This yields:
\begin{equation}
\begin{split}
 z_{\text{(dm=de)}}  = \left[\frac{\left[(m_-+w+1)+(m_-+1) \right] \left[\Omega_{\text{(de,0)}} (m_+ + w + 1) + \Omega_{\text{(dm,0)}} (m_+ + 1)  \right]}{\left[(m_++w+1)+(m_++1) \right] \left[\Omega_{\text{(de,0)}} (m_- + w + 1) + \Omega_{\text{(dm,0)}} (m_- + 1) \right] }\right]^{\frac{1}{3(m_--m_+)}}-1.
\end{split}
\label{eq:rhodm=de_general}
\end{equation}
To understand how these interaction functions address the coincidence problem, we analyze the ratio 
$r=\frac{\rho_{\text{dm}}}{\rho_{\text{de}}}$ in both the distant past ($a\rightarrow 0$, implying $a^{3m_+} \rightarrow 0$)  
and the distant future ($a\rightarrow \infty$, implying $a^{3m_-} \rightarrow 0$):
\begin{equation}
\begin{split}
r_{\text{past}} = \frac{\rho_{\text{dm}}}{\rho_{\text{de}}} \approx  
& \frac{-\left[\frac{m_-+w+1}{w  (m_- - m_+)} \right]\left[\rho_{\text{(de,0)}} (m_+ + w + 1) + \rho_{\text{(dm,0)}} (m_+ + 1)  \right]  a^{3m_-}}
{+\left[\frac{m_-+1}{w  (m_- - m_+)} \right]\left[\rho_{\text{(de,0)}} (m_+ + w + 1) + \rho_{\text{(dm,0)}} (m_+ + 1)  \right]  a^{3m_-}} 
= - \frac{m_-+w+1}{m_-+1}, \\[6pt]
r_{\text{future}} = \frac{\rho_{\text{dm}}}{\rho_{\text{de}}} \approx 
& \frac{+\left[\frac{m_++w+1}{w (m_- - m_+)} \right] \left[\rho_{\text{(de,0)}} (m_- + w + 1) + \rho_{\text{(dm,0)}} (m_- + 1) \right]  a^{3m_+}}
{-\left[\frac{m_++1}{w (m_- - m_+)} \right] \left[\rho_{\text{(de,0)}} (m_- + w + 1) + \rho_{\text{(dm,0)}} (m_- + 1) \right]  a^{3m_+}} 
= - \frac{m_++w+1}{m_++1}.
\end{split}
\label{eq:r_general_past_future}
\end{equation}
The redshift at which $w^{\rm{eff}}_{\rm{de}}$ crosses the phantom divide ($w^{\rm eff}_{\rm de} = -1$) 
is obtained from \eqref{eq.general.omega_eff_dm_de} as:
\begin{equation}
\begin{split}
w^{\rm{eff}}_{\rm{de}} = -1 
&= w + \delta_{\text{de}} +  \delta_{\text{dm}}\, r   
\quad \rightarrow \quad 
r = -\, \frac{1 + w + \delta_{\text{de}}}{\delta_{\text{dm}}}.
\end{split}
\label{eq:z_pc_general_1}
\end{equation}
Substituting $\rho_{\text{dm}}$ from \eqref{eq:rhodm_general} and $\rho_{\text{de}}$ from \eqref{eq:rhode_general} 
into $r$ in \eqref{eq:z_pc_general_1}, applying the transformation $a=(1+z)^{-1}$, and solving for $z$ after some algebra 
gives the redshift of the phantom crossing (pc):
\begin{equation}
\begin{split}
z_{\text{pc}} = & \Bigg[
    \frac{
      \bigl(\,[1+w+\delta_{\mathrm{de}}](m_++1)
           -\delta_{\mathrm{dm}}(m_++w+1)\bigr)
      }{
      \bigl(\,[1+w+\delta_{\mathrm{de}}](m_-+1)
           -\delta_{\mathrm{dm}}(m_-+w+1)\bigr)
      }
    \times
    \frac{
      \bigl[\Omega_{\mathrm{(de,0)}}(m_-+w+1)
           +\Omega_{\mathrm{(dm,0)}}(m_-+1)\bigr]
      }{
      \bigl[\Omega_{\mathrm{(de,0)}}(m_++w+1)
           +\Omega_{\mathrm{(dm,0)}}(m_++1)\bigr]
      }
    \Bigg]^{-\frac{1}{3(m_--m_+)}} - 1.
\end{split}
\label{eq:z_pc_general_4}
\end{equation}
The future asymptotic expression for the total effective equation of state $w_{\text{tot}}^{\text{eff}}$ will determine the final fate of the universe model. In the distant future, the universe will be dominated by DM and DE, such that $w_{\text{tot}}^{\text{eff}}$ from \eqref{DSA.omega_eff_tot} becomes: 
\begin{gather} \label{omega_eff_tot_general_1}
\begin{split}
w^{\rm{eff}}_{\rm{(tot,future)}}   
&\approx \frac{ w_{\rm{de}}\rho_{\rm{(de,future)}}}{\rho_{\rm{(dm,future)}}+\rho_{\rm{(de,future)}}} 
\approx - \left[m_+ + 1 \right],
\end{split}
\end{gather}
where we have assumed that terms with $a^{3m_-}$ become negligible for both $\rho_{\text{dm}}$ in \eqref{eq:rhodm_general} and $\rho_{\text{de}}$ in \eqref{eq:rhode_general} in the distant future $(a\rightarrow\infty)$. 
If $w^{\rm{eff}}_{\rm{tot}} < -1$, implying $m_+ > 0$ in \eqref{omega_eff_tot_general_1}, the model will experience a future \textit{big rip} singularity. The big rip is characterized by both the energy density and scale factor becoming infinite within a finite time, such that $\rho\rightarrow\infty$ and $a\rightarrow \infty$ within $t_{\text{rip}}$. 
If we assume that in the distant future $(a\rightarrow\infty)$, $\rho_{\text{dm}}$ and $\rho_{\text{de}}$ are given by their leading terms, then the Hubble function is given by \eqref{eq:frac_den_general_3}, and can be integrated to give:
\begin{equation}
\begin{split}
\left(\frac{\dot{a}}{a} \right)&\approx H_0 \sqrt{\frac{\Omega_{\text{(de,0)}} (m_- + w + 1) + \Omega_{\text{(dm,0)}} (m_- + 1) }{ (m_- - m_+)}  } \, a^{\frac{3}{2}m_+} \\
t_{\text{rip}}-t_0&\approx \frac{1}{ H_0} \sqrt{\frac{ (m_- - m_+)}{\Omega_{\text{(de,0)}} (m_- + w + 1) + \Omega_{\text{(dm,0)}} (m_- + 1) }  } \left[ -\frac{2}{3 m_+} \right] a^{-\frac{3}{2}m_+} \Bigg|_{a=1}^{\infty}. \\
\end{split}\label{eq:Big_Rip_general_2}
\end{equation}
If $m_+>0$, implying $-\frac{3}{2}m_+<0$, then the term $a^{-\frac{3}{2}m_+}$ tends to zero at $a=\infty$, and \eqref{eq:Big_Rip_general_2} becomes:
\begin{equation}
\begin{split}
t_{\text{rip}}-t_0&\approx \frac{2}{3 H_0 m_+}\sqrt{\frac{ (m_- - m_+)}{\Omega_{\text{(de,0)}} (m_- + w + 1) + \Omega_{\text{(dm,0)}} (m_- + 1) }  }. \\
\end{split}\label{eq:Big_Rip_general_4}
\end{equation}
Additionally, substituting \eqref{eq:rhodm_general} and \eqref{eq:rhode_general} into \eqref{DSA.H}, while applying the transformations $a=(1+z)^{-1}$ and $ \frac{8 \pi G }{3H_0^2} \rho_{\rm{(i,0)}}= \Omega_{\rm{(i,0)}}$, results in the Hubble function for the linear IDE models studied in this paper:
\begin{gather}
\begin{split} \label{hz_Q_linear}
h(z)^2= +& \left[\frac{1}{(m_- - m_+)} \right] \left[\Omega_{\text{(de,0)}} (m_- + w + 1) + \Omega_{\text{(dm,0)}} (m_- + 1) \right]  (1+z)^{-3m_+}\\[1mm]
-& \left[\frac{1}{  (m_- - m_+)} \right]\left[\Omega_{\text{(de,0)}} (m_+ + w + 1) + \Omega_{\text{(dm,0)}} (m_+ + 1)  \right]  (1+z)^{-3m_-} 
+\Omega_{\rm{bm,0}}(1+z)^3+\Omega_{\rm{r,0}}(1+z)^4. 
\end{split}   
\end{gather}
Finally, to make the above relations in this section directly applicable, we summarize in Table~\ref{tab:m+-} the explicit forms of $m_+$, $m_-$, and $\Delta$ for each of the interaction models considered, as well as for the uncoupled $w$CDM and $\Lambda$CDM cases. This allows for straightforward substitution into the general formulae derived above.
\begin{table}[h!]
\renewcommand{\arraystretch}{1.4} % Increases row height for better readability
\setlength{\tabcolsep}{10pt} % Adjust column spacing\centering
\begin{tabular}{|c|c|c|c|}
\hline 
$\textbf{Model}$ & $m_+$  & $m_-$ & $\Delta$ \\ \hline \hline

$Q=3H(\delta_{\text{dm}} \rho_{\text{dm}}+\delta_{\text{de}} \rho_{\text{de}})$ &$\frac{\delta_{\text{dm}}  - \delta_{\text{de}} - w- 2 + \Delta}{2} $ & $\frac{\delta_{\text{dm}}  - \delta_{\text{de}} - w- 2 - \Delta}{2}$ &$ \sqrt{(\delta_{\text{dm}}  + \delta_{\text{de}}+ w)^2 - 4 \delta_{\text{dm}} \delta_{\text{de}}}$   \\ \hline

$Q=3H\delta( \rho_{\text{dm}}+\rho_{\text{de}})$ & $\frac{- w - 2 + \Delta}{2}$ & $\frac{- w - 2 - \Delta}{2}$  & $\sqrt{ 4 \delta^2 +w^2}$ \\ \hline
$Q=3H\delta( \rho_{\text{dm}}-\rho_{\text{de}})$ & $ \frac{2\delta - w- 2 + \Delta}{2}$ &$  \frac{2\delta - w- 2 - \Delta}{2}$ & $\sqrt{w( 4  \delta+w)}$  \\ \hline

$Q=3H \delta \rho_{\text{dm}}$ & $-w-1$ & $\delta-1$ & $-(\delta+w)$ \\ \hline
$Q=3H\delta \rho_{\text{de}}$ &  $-(\delta+w+1)$ & $-1$  & $-(\delta+w)$  \\ \hline
$w$CDM & $-w-1$ & $1$ & $-w$ \\ \hline

$\Lambda$CDM & $0$ & $1$ & $1$ \\ \hline

\end{tabular}
\caption{Expressions for $m_+$, $m_-$, and $\Delta$ for each interaction kernel considered, including the uncoupled $w$CDM and $\Lambda$CDM cases.}
\label{tab:m+-}
\end{table}

\section{Background Cosmology for each interaction kernel} \label{Background_cosmology}

\subsection{Linear IDE model 1: \(Q= 3 H (\delta_{\text{dm}} \rho_{\text{dm}} + \delta_{\text{de}}  \rho_{\text{de}})\)} \label{Q_General}

In this subsection, we derive the expressions that determine the background cosmology for the most general linear interaction model studied, while the special cases are discussed in the subsequent four subsections. We substitute the expressions for $m_+$, $m_-$, and $\Delta$ for this interaction from Table \ref{tab:m+-} into the general expressions for $\rho_{\text{dm}}$ \eqref{eq:rhodm_general} and $\rho_{\text{de}}$ \eqref{eq:rhode_general} to obtain:
\begin{gather}
\begin{split} \label{eq:rho_dm_Q_linear}
\rho_{\text{dm}} = -& \frac{\delta_{\text{dm}} - \delta_{\text{de}} + w + \Delta}{4w \Delta} 
\Bigl[\rho_{\text{(de,0)}} \bigl(\delta_{\text{dm}} - \delta_{\text{de}} + w - \Delta \bigr)
+ \rho_{\text{(dm,0)}} \bigl(\delta_{\text{dm}} - \delta_{\text{de}} - w - \Delta \bigr) \Bigr] 
a^{\frac{3}{2} \bigl( \delta_{\text{dm}} - \delta_{\text{de}} - w - 2 + \Delta \bigr)} \\[1mm]
+& \frac{\delta_{\text{dm}} - \delta_{\text{de}} + w - \Delta}{4w \Delta} 
\Bigl[\rho_{\text{(de,0)}} \bigl(\delta_{\text{dm}} - \delta_{\text{de}} + w + \Delta \bigr)
+ \rho_{\text{(dm,0)}} \bigl(\delta_{\text{dm}} - \delta_{\text{de}} - w + \Delta \bigr) \Bigr] 
a^{\frac{3}{2} \bigl( \delta_{\text{dm}} - \delta_{\text{de}} - w - 2 - \Delta \bigr)} ,
\end{split}   
\end{gather}
\begin{gather}
\begin{split} \label{eq:rho_de_Q_linear}
\rho_{\text{de}} = +& \frac{\delta_{\text{dm}} - \delta_{\text{de}} - w + \Delta}{4w \Delta} 
\Bigl[\rho_{\text{(de,0)}} \bigl(\delta_{\text{dm}} - \delta_{\text{de}} + w - \Delta \bigr)
+ \rho_{\text{(dm,0)}} \bigl(\delta_{\text{dm}} - \delta_{\text{de}} - w - \Delta \bigr) \Bigr] 
a^{\frac{3}{2} \bigl( \delta_{\text{dm}} - \delta_{\text{de}} - w - 2 + \Delta \bigr)} \\[1mm]
-& \frac{\delta_{\text{dm}} - \delta_{\text{de}} - w - \Delta}{4w \Delta} 
\Bigl[\rho_{\text{(de,0)}} \bigl(\delta_{\text{dm}} - \delta_{\text{de}} + w + \Delta \bigr)
+ \rho_{\text{(dm,0)}} \bigl(\delta_{\text{dm}} - \delta_{\text{de}} - w + \Delta \bigr) \Bigr] 
a^{\frac{3}{2} \bigl( \delta_{\text{dm}} - \delta_{\text{de}} - w - 2 - \Delta \bigr)} ,
\end{split}   
\end{gather}
where $\Delta$ is the determinant:
\begin{gather}
\begin{split} \label{eq:determinant_Q_linear}
\Delta = \sqrt{ \bigl( \delta_{\text{dm}} + \delta_{\text{de}} + w \bigr)^2 - 4\, \delta_{\text{de}} \delta_{\text{dm}} } \,.
\end{split}   
\end{gather}
\begin{gather} \label{DSA.Q.delta_dm+delta_de.PEC_A}
\boxed{
\begin{aligned}
&\underline{\text{Conditions for} \; \rho_{\rm{dm}} \ge 0 \; ; \; \rho_{\rm{de}} \ge 0 \; \text{and real throughout the cosmological evolution:}} \quad \text{iDEDM with} \\
&1.\; \delta_{\text{dm}} \ge 0 \;; \quad 
  2.\; \delta_{\text{de}} \ge 0 \;; \quad 
  3.\; \delta_{\text{dm}} r_0 + \delta_{\text{de}} \le -\frac{w r_0}{1+r_0} \;; \quad 
  4.\; (\delta_{\text{dm}}+\delta_{\text{de}}+w)^2 \ge 4\delta_{\text{de}}\delta_{\text{dm}} \quad \text{(reality)}.
\end{aligned}
}
\end{gather}
We also note that, for $\rho_{\rm{dm}}$ in \eqref{eq:rho_dm_Q_linear} and $\rho_{\rm{de}}$ in \eqref{eq:rho_de_Q_linear} to be defined, we require $w \ne 0$ and $(\delta_{\text{dm}}+\delta_{\text{de}}+w)^2 - 4\delta_{\text{de}}\delta_{\text{dm}} \ne 0$. Similarly, substituting the expressions for $m_+$, $m_-$, and $\Delta$ for the interaction from Table \ref{tab:m+-} into the relevant expressions in Section \ref{More_analytical_relations} gives the corresponding results below. The fractional densities of both DM and DE will converge asymptotically in the future and past to:
\begin{equation}
\begin{split}
\Omega_{\text{(dm,past)}} &= \frac{1}{2} + \frac{\delta_{\text{dm}} - \delta_{\text{de}} - \Delta}{2w}, 
\quad \Omega_{\text{(de,past)}} = \frac{1}{2} - \frac{\delta_{\text{dm}} - \delta_{\text{de}} - \Delta}{2w}, \\
\Omega_{\text{(dm,future)}} &= \frac{1}{2} + \frac{\delta_{\text{dm}} - \delta_{\text{de}} + \Delta}{2w}, 
\quad \Omega_{\text{(de,future)}} = \frac{1}{2} - \frac{\delta_{\text{dm}} - \delta_{\text{de}} + \Delta}{2w}.
\end{split}
\label{eq:frac_den_ddm+dde_past}
\end{equation}
The DM and DE densities become zero at redshifts $z_{\text{(dm=0)}}$ and $z_{\text{(de=0)}}$, respectively:
\begin{equation}
\begin{split}
z_{\text{(dm=0)}} = \left[ 
\frac{\left[\delta_{\text{dm}} - \delta_{\text{de}} + w + \Delta\right] 
\left[\Omega_{\text{(de,0)}} (\delta_{\text{dm}} - \delta_{\text{de}} + w - \Delta) 
+ \Omega_{\text{(dm,0)}} (\delta_{\text{dm}} - \delta_{\text{de}} - w - \Delta) \right]}
{\left[\delta_{\text{dm}} - \delta_{\text{de}} + w - \Delta\right]
\left[\Omega_{\text{(de,0)}} (\delta_{\text{dm}} - \delta_{\text{de}} + w + \Delta) 
+ \Omega_{\text{(dm,0)}} (\delta_{\text{dm}} - \delta_{\text{de}} - w + \Delta) \right]}
\right]^{\frac{1}{3\Delta}} - 1.
\end{split}
\label{eq:rhodm_z_min_Q_linear_BG}
\end{equation}
\begin{equation}
\begin{split}
z_{\text{(de=0)}} &= \left[
\frac{
\left[\delta_{\text{dm}} - \delta_{\text{de}} - w + \Delta \right] 
\left[ \Omega_{\text{(de,0)}} (\delta_{\text{dm}} - \delta_{\text{de}} + w - \Delta) 
+ \Omega_{\text{(dm,0)}} (\delta_{\text{dm}} - \delta_{\text{de}} - w - \Delta) \right]
}{
\left[\delta_{\text{dm}} - \delta_{\text{de}} - w - \Delta \right] 
\left[ \Omega_{\text{(de,0)}} (\delta_{\text{dm}} - \delta_{\text{de}} + w + \Delta) 
+ \Omega_{\text{(dm,0)}} (\delta_{\text{dm}} - \delta_{\text{de}} - w + \Delta) \right]
}
\right]^{\frac{1}{3\Delta}} - 1.
\end{split}
\label{eq:rhode_z_min_Q_linear_BG}
\end{equation}
If $\delta_{\rm{dm}}$ and $\delta_{\rm{de}}$ have opposite signs, the direction of the energy flow can switch when $Q=0$, as illustrated in Figure~\ref{fig:Q_Linear_dm+de}. This sign switch occurs at the following redshift:
\begin{equation}
\begin{split}
z_{\rm{(Q=0)}} &= \Biggl[
\frac{
\delta_{\text{dm}}\bigl(\delta_{\text{dm}} - \delta_{\text{de}} + w - \Delta\bigr)
- \delta_{\text{de}}\bigl(\delta_{\text{dm}} - \delta_{\text{de}} - w - \Delta\bigr)
}{
\delta_{\text{dm}}\bigl(\delta_{\text{dm}} - \delta_{\text{de}} + w + \Delta\bigr)
- \delta_{\text{de}}\bigl(\delta_{\text{dm}} - \delta_{\text{de}} - w + \Delta\bigr)
}
\\[-2pt]
& \quad \times
\frac{
\Omega_{\text{(de,0)}}\bigl(\delta_{\text{dm}} - \delta_{\text{de}} + w + \Delta\bigr)
+ \Omega_{\text{(dm,0)}}\bigl(\delta_{\text{dm}} - \delta_{\text{de}} - w + \Delta\bigr)
}{
\Omega_{\text{(de,0)}}\bigl(\delta_{\text{dm}} - \delta_{\text{de}} + w - \Delta\bigr)
+ \Omega_{\text{(dm,0)}}\bigl(\delta_{\text{dm}} - \delta_{\text{de}} - w - \Delta\bigr)
}
\Biggr]^{-\frac{1}{3\Delta}} - 1.
\end{split}
\label{eq:Q=0_ddm+dde}
\end{equation}

The redshift $z_{\text{(dm=de)}}$ at which dark matter–dark energy equality occurs is given by:
\begin{equation}
\begin{split}
z_{\text{(dm=de)}} &= \left[
\frac{
\left[\delta_{\text{dm}} - \delta_{\text{de}} - \Delta \right]
\left[ \Omega_{\text{(de,0)}} (\delta_{\text{dm}} - \delta_{\text{de}} + w + \Delta)
+ \Omega_{\text{(dm,0)}} (\delta_{\text{dm}} - \delta_{\text{de}} - w + \Delta) \right]
}{
\left[\delta_{\text{dm}} - \delta_{\text{de}} + \Delta\right]
\left[ \Omega_{\text{(de,0)}} (\delta_{\text{dm}} - \delta_{\text{de}} + w - \Delta)
+ \Omega_{\text{(dm,0)}} (\delta_{\text{dm}} - \delta_{\text{de}} - w - \Delta) \right]
}
\right]^{-\frac{1}{3\Delta}} - 1.
\end{split}
\label{eq:rhodm=de_ddm+dde}
\end{equation}

The coincidence problem is addressed by allowing $r$ to converge to the following constants in the past and future:
\begin{equation}
\begin{split}
r_{\text{past}} \;(a \rightarrow 0) &
= -\frac{\delta_{\text{dm}} - \delta_{\text{de}} + w - \Delta}
{\delta_{\text{dm}} - \delta_{\text{de}} - w - \Delta} \; \; \;, \; \; \;   
r_{\text{future}} \;(a \rightarrow \infty) 
= -\frac{\delta_{\text{dm}} - \delta_{\text{de}} + w + \Delta}
{\delta_{\text{dm}} - \delta_{\text{de}} - w + \Delta}.
\end{split}
\label{eq:r_general_subbed_past_future}
\end{equation}
The DM and DE effective equations of state \eqref{DSA.omega_eff_dm_de} for this interaction kernel are:
\begin{gather}
\begin{split}
w^{\rm{eff}}_{\rm{dm}} &= - \frac{Q}{3 H \rho_{\rm{dm}}}
= - \frac{3 H (\delta_{\text{dm}} \rho_{\text{dm}} + \delta_{\text{de}} \rho_{\rm{de}})}
{3 H \rho_{\rm{dm}}}
= -\delta_{\text{dm}} - \frac{\delta_{\text{de}} \rho_{\rm{de}}}{\rho_{\rm{dm}}}
= -\delta_{\text{dm}} - \frac{\delta_{\text{de}}}{r}, \\[3pt]
w^{\rm{eff}}_{\rm{de}} &= w + \frac{Q}{3 H \rho_{\rm{de}}}
= w + \frac{3 H (\delta_{\text{dm}} \rho_{\text{dm}} + \delta_{\text{de}} \rho_{\rm{de}})}
{3 H \rho_{\rm{de}}}
= w + \delta_{\text{de}} + \frac{\delta_{\text{dm}} \rho_{\rm{dm}}}{\rho_{\rm{de}}}
= w + \delta_{\text{de}} + \delta_{\text{dm}} r .
\end{split}
\label{eq.general.omega_eff_dm_de}
\end{gather}
Expressions for $w^{\rm{eff}}_{\rm{dm}}$ and $w^{\rm{eff}}_{\rm{de}}$ in the asymptotic distant past and future are obtained by substituting \eqref{eq:r_general_subbed_past_future} into \eqref{eq.general.omega_eff_dm_de}. These expressions are then used to calculate $\zeta=3(w^{\rm{eff}}_{\rm{dm}}-w^{\rm{eff}}_{\rm{de}})$ from \eqref{DSA.r}:
\begin{equation}
\begin{split}
w^{\rm{eff}}_{\rm{(dm,past)}} &\approx -\delta_{\text{dm}} + \delta_{\text{de}} \left[ \frac{\delta_{\text{dm}} - \delta_{\text{de}} - w - \Delta} {\delta_{\text{dm}} - \delta_{\text{de}} + w - \Delta} \right]=\frac{1}{2} \left[ \delta_{\text{de}} - \delta_{\text{dm}} + w + \Delta \right], \\[2pt]
w^{\rm{eff}}_{\rm{(de,past)}} &\approx w + \delta_{\text{de}} - \delta_{\text{dm}} \left[ \frac{\delta_{\text{dm}} - \delta_{\text{de}} + w - \Delta}{\delta_{\text{dm}} - \delta_{\text{de}} - w - \Delta} \right]=\frac{1}{2} \left[ \delta_{\text{de}} - \delta_{\text{dm}} + w +\Delta \right], \\[2pt]
&\rightarrow w^{\rm{eff}}_{\rm{(dm,past)}}=w^{\rm{eff}}_{\rm{(de,past)}} \quad \rightarrow  \quad \zeta_{\text{past}} \approx 0 \quad \text{(solves coincidence problem)}, \\
w^{\rm{eff}}_{\rm{(dm,future)}} &\approx -\delta_{\text{dm}} + \delta_{\text{de}} \left[ \frac{\delta_{\text{dm}} - \delta_{\text{de}} - w + \Delta}{\delta_{\text{dm}} - \delta_{\text{de}} + w + \Delta} \right]=\frac{1}{2} \left[ \delta_{\text{de}} - \delta_{\text{dm}} + w - \Delta \right], \\[4pt]
w^{\rm{eff}}_{\rm{(de,future)}} &\approx w + \delta_{\text{de}} 
- \delta_{\text{dm}} \left[ \frac{\delta_{\text{dm}} - \delta_{\text{de}} + w + \Delta}
{\delta_{\text{dm}} - \delta_{\text{de}} - w + \Delta} \right]=\frac{1}{2} \left[ \delta_{\text{de}} - \delta_{\text{dm}} + w - \Delta \right], \\[4pt]
& \rightarrow w^{\rm{eff}}_{\rm{(dm,future)}}=w^{\rm{eff}}_{\rm{(de,future)}} \quad \rightarrow \quad  \zeta_{\text{future}} \approx 0 \quad \text{(solves coincidence problem)}. \\
\end{split}
\label{eq:omega_eff_dm_de_general_subbed_past_future}
\end{equation}
This interaction model will therefore \textbf{always solve the coincidence problem in both the past and the future, provided that $\delta_{\text{dm}}\neq 0$ or $\delta_{\text{de}}\neq 0$}. The effective DE equation of state is dynamic and, under the appropriate conditions, will undergo a phantom crossing $(w^{\rm{eff}}_{\rm{de}}=-1)$ at redshift $z_{\text{pc}}$: 
\begin{equation}
\begin{split}
z_{\text{pc}} &= \Biggl[
\frac{
\bigl(
[1 + w + \delta_{\text{de}}]
[\delta_{\text{dm}} - \delta_{\text{de}} - w + \Delta]
- \delta_{\text{dm}}
[\delta_{\text{dm}} - \delta_{\text{de}} + w + \Delta]
\bigr)
}{
\bigl(
[1 + w + \delta_{\text{de}}]
[\delta_{\text{dm}} - \delta_{\text{de}} - w - \Delta]
- \delta_{\text{dm}}
[\delta_{\text{dm}} - \delta_{\text{de}} + w - \Delta]
\bigr)
}
\\[2pt]
&\quad \times 
\frac{
\Omega_{\text{(de,0)}} (\delta_{\text{dm}} - \delta_{\text{de}} + w - \Delta)
+ \Omega_{\text{(dm,0)}} (\delta_{\text{dm}} - \delta_{\text{de}} - w - \Delta)
}{
\Omega_{\text{(de,0)}} (\delta_{\text{dm}} - \delta_{\text{de}} + w + \Delta)
+ \Omega_{\text{(dm,0)}} (\delta_{\text{dm}} - \delta_{\text{de}} - w + \Delta)
}
\Biggr]^{\frac{1}{3\Delta}} - 1.
\end{split}
\label{eq:z_pc_general_BG}
\end{equation}
From \eqref{eq.general.omega_eff_dm_de}, it can be seen that $w^{\rm{eff}}_{\rm{dm}}$ and $w^{\rm{eff}}_{\rm{de}}$ will exhibit a divergent phantom crossing (pc) in the iDMDE regime when $\rho_{\rm{dm}} = 0$ at $z_{\text{(dm=0)}}$ \eqref{eq:rhodm_z_min_Q_linear_BG} and $\rho_{\rm{de}} = 0$ at $z_{\text{(de=0)}}$ \eqref{eq:rhode_z_min_Q_linear_BG}, respectively.  
In general, for small couplings, since $w^{\rm{eff}}_{\rm{de}}$ nearly mimics the behavior of dark matter in the past $(w^{\rm{eff}}_{\rm{(de,past)}} > -1)$, a non-divergent phantom crossing can occur in the iDEDM regime if $w^{\rm{eff}}_{\rm{(de,future)}} < -1$.  
In summary:
\begin{equation}
\begin{split}
\text{pc direction} \; \begin{cases}
\text{iDMDE}: & \text{Divergent pc for $w^{\rm{eff}}_{\rm{dm}}$ at $z_{\text{(dm=0)}}$ \eqref{eq:rhodm_z_min_Q_linear_BG} and $w^{\rm{eff}}_{\rm{de}}$ at $z_{\text{(de=0)}}$ \eqref{eq:rhode_z_min_Q_linear_BG}, with $\rho_{\rm{dm/de}} < 0$.} \\[2pt]
\text{iDEDM}: & \text{Quintessence $(w^{\rm{eff}}_{\rm{(de,past)}} > -1) \rightarrow$ Phantom $(w^{\rm{eff}}_{\rm{(de,future)}} < -1)$, with $\rho_{\rm{dm/de}} > 0$.}
\end{cases}
\end{split}
\end{equation}
We also have $q = \frac{1}{2} \left( 1 + 3 w^{\rm{eff}}_{\rm{tot}} \right)$ and the algebraically equivalent expressions  
$w^{\rm{eff}}_{\rm{tot}} = w^{\rm{eff}}_{\rm{de}} = w^{\rm{eff}}_{\rm{dm}} = \frac{1}{2} \left[ \delta_{\text{de}} - \delta_{\text{dm}} + w - \Delta \right]$ in the distant future.  
This implies that, in the distant future, even if $w < -1$, both the deceleration parameter and effective equations of state can be larger, such that $w^{\rm{eff}}_{\rm{(tot,future)}} > -1$ (in the iDEDM regime), which may cause the model to avoid a future big rip singularity.  
This leads to the following condition for a big rip to occur:
\begin{gather} \label{omega_eff_tot_general_BG}
\boxed{
w^{\rm{eff}}_{\rm{(tot,future)}} \approx \frac{1}{2} \left[ \delta_{\text{de}} - \delta_{\text{dm}} + w - \Delta \right] < -1  
\quad \Longrightarrow \quad  
\delta_{\text{dm}} (w + 1) - \delta_{\text{de}} > w + 1
}
\end{gather}
In the case where $w^{\rm{eff}}_{\rm{(tot,future)}} < -1$, the universe will experience a big rip future singularity at time $t_{\text{rip}}$:
\begin{equation}
\begin{split}
t_{\text{rip}} &\approx \frac{4}{3 H_0 \left( \delta_{\text{dm}} - \delta_{\text{de}} - w - 2 + \Delta \right)}
\sqrt{
\frac{- 2\Delta}{
\Omega_{\text{(de,0)}} \left( \delta_{\text{dm}} - \delta_{\text{de}} + w - \Delta \right) 
+ \Omega_{\text{(dm,0)}} \left( \delta_{\text{dm}} - \delta_{\text{de}} - w - \Delta \right) 
}
}.
\end{split}
\label{eq:Big_Rip_ddm+dde_BG}
\end{equation}
No big rip will occur if condition \ref{omega_eff_tot_general_BG} is not met, or equivalently if $\delta_{\text{dm}} \left(w+1\right)-\delta_{\text{de}} \le w + 1$. The expressions obtained in this section are consistent with those from the dynamical system analysis in section \ref{DSA.Q.general}, as summarized in Table \ref{tab:CP_B_ddm+dde}, and both cases reduce back to $\Lambda$CDM when $\delta=0$ and $w=-1$, thereby validating the results found in both sections.

\subsection{Linear IDE model 2: \(Q= 3 H \delta(\rho_{\text{dm}} + \rho_{\text{de}})\)} \label{Q_dm+de}

\begin{figure}
    \centering
    \includegraphics[width=0.85\linewidth]{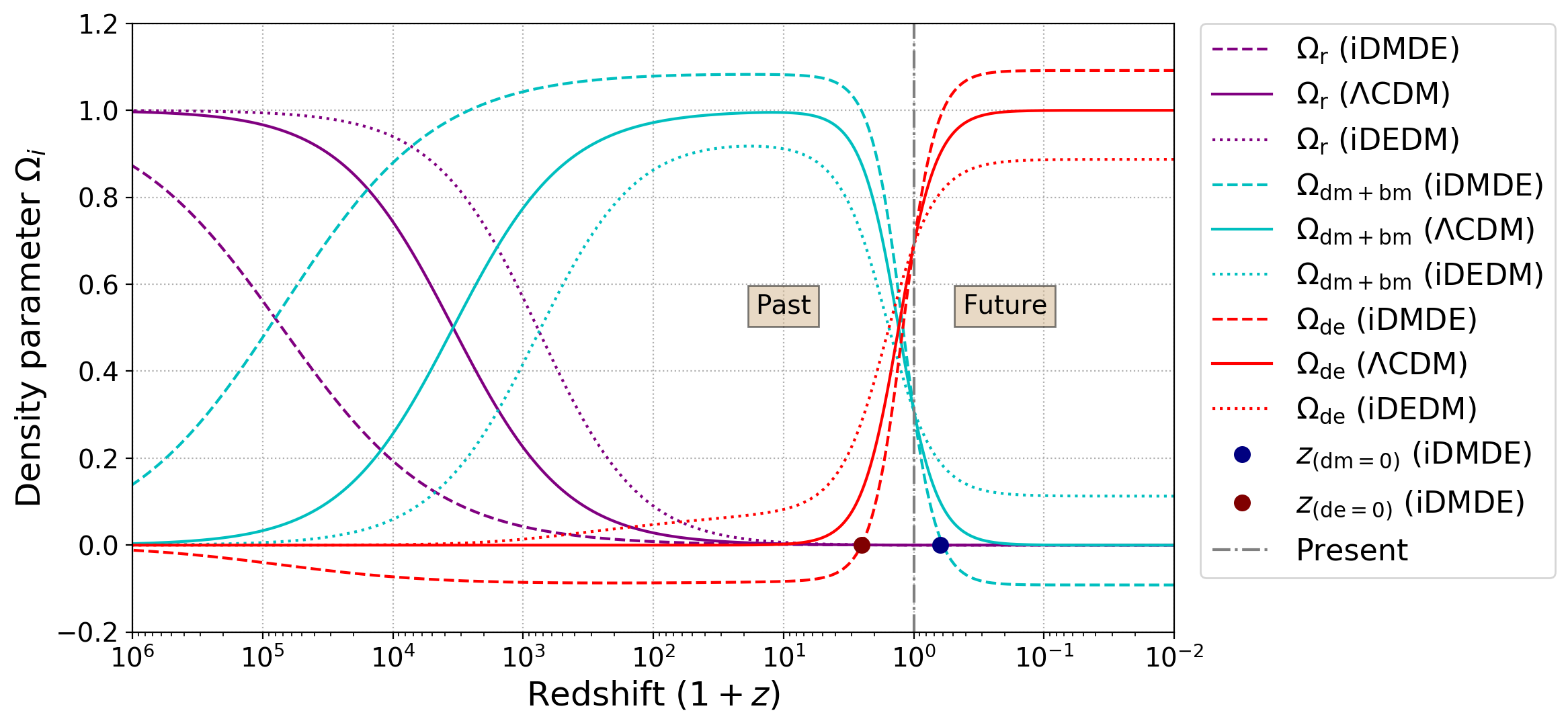}
    \caption{Density parameters vs. redshift for $Q = 3\delta H (\rho_{\text{dm}}+\rho_{\text{de}})$, with positive energy densities occurring only in the iDEDM regime ($\delta=+0.1$), while negative DE densities (in the past) and negative DM densities (in the future) are always present in the iDMDE regime ($\delta=-0.1$).}
    \label{fig:Omega_Linear_dm+de}
\end{figure}

This interaction changes the dynamics in both the distant past and future during DM and DE domination, respectively, as seen in Figure \ref{fig:Q_Linear_dm+de} and the other figures in this section. Since $Q\neq0$ when either $\rho_{\text{dm}}$ or $\rho_{\text{de}}$ becomes zero, there is no mechanism to avoid negative energies for this interaction. This model exhibits both negative DM and DE densities in the iDMDE regime, as also observed in~\cite{He_2008}, but this can be avoided with a sufficiently small interaction in the iDEDM regime, given by the conditions in \eqref{DSA.Q.dm+de.19} and illustrated in Figure \ref{fig:Omega_Linear_dm+de}. Substituting the expressions for $m_+$, $m_-$, and $\Delta$ for the interaction from Table \ref{tab:m+-} into the general expressions for $\rho_{\text{dm}}$ \eqref{eq:rhodm_general} and $\rho_{\text{de}}$ \eqref{eq:rhode_general} gives:
\begin{gather}
\begin{split} \label{eq:rho_dm_Q_dm+de}
\rho_{\text{dm}} = &-\frac{w+\Delta}{4w\Delta}\Bigl[
\rho_{\text{(de,0)}}(w-\Delta) + \rho_{\text{(dm,0)}}(-w-\Delta)
\Bigr]a^{\frac{3}{2}(-w-2+\Delta)}\\
&+\frac{w-\Delta}{4w\Delta}\Bigl[
\rho_{\text{(de,0)}}(w+\Delta) + \rho_{\text{(dm,0)}}(-w+\Delta)
\Bigr]a^{\frac{3}{2}(-w-2-\Delta)}, 
\end{split}   
\end{gather}
\begin{gather}
\begin{split} \label{eq:rho_de_Q_dm+de}
\rho_{\text{de}} = &+\frac{-w+\Delta}{4w\Delta}\Bigl[
\rho_{\text{(de,0)}}(w-\Delta) + \rho_{\text{(dm,0)}}(-w-\Delta)
\Bigr]a^{\frac{3}{2}(-w-2+\Delta)}\\
&-\frac{-w-\Delta}{4w\Delta}\Bigl[
\rho_{\text{(de,0)}}(w+\Delta) + \rho_{\text{(dm,0)}}(-w+\Delta)
\Bigr]a^{\frac{3}{2}(-w-2-\Delta)},
\end{split}   
\end{gather}
where $\Delta$ is the determinant:
\begin{gather}
\begin{split} \label{eq:determinant_Q_dm+de}
\Delta=\sqrt{w\,(4\delta+w)}\,.
\end{split}   
\end{gather}

\begin{gather} \label{DSA.Q.dm+de.19}
\boxed{
\begin{split}
&\underline{\text{Conditions for}  \text{ $\rho_{\rm{dm}}\ge0 \; ; \; \rho_{\rm{de}} \ge0 $ and real for the entire cosmological evolution:}}
 \\ &\quad\quad\quad\quad \text{iDEDM with } \
 0\le\delta \le -\frac{w r_0}{(1+r_0)^2}  \quad ; \quad \delta \le -\frac{w}{4}  \quad \text{(reality).}
\end{split}}
\end{gather}
We also note that, for $\rho_{\rm{dm}}$ \eqref{eq:rho_dm_Q_dm+de} and $\rho_{\rm{de}}$ \eqref{eq:rho_de_Q_dm+de} to be defined, we require $w\ne 0$ and $\delta \ne -\frac{w}{4}$. The reality condition in \eqref{DSA.Q.dm+de.19} was also found in~\cite{Caldera_Cabral_2009_DSA, He_2011, Pan_2020}. 
Similarly, substituting the expressions for $m_+$, $m_-$ and $\Delta$ for the interaction from Table \ref{tab:m+-} into the relevant expressions in section \ref{More_analytical_relations} gives the corresponding results below:
\begin{equation}
\begin{split}
\Omega_{\text{(dm,past)}} =\frac{1}{2}-\frac{\Delta}{2w} \quad &; \quad \Omega_{\text{(de,past)}}=  \frac{1}{2}+ \frac{\Delta}{2w},\\
\Omega_{\text{(dm,future)}} =\frac{1}{2}+\frac{\Delta}{2w} \quad &; \quad \Omega_{\text{(de,future)}}=\frac{1}{2}- \frac{\Delta}{2w}.
\end{split}
\label{eq:frac_den_dm+de_past}
\end{equation}
The DM and DE densities become zero at redshifts $z_{\text{(dm=0)}}$ and $z_{\text{(de=0)}}$:
\begin{equation}
\begin{split}
z_{\text{(dm=0)}} = \left[ 
\frac{
\left[ w + \Delta \right] \left[ \Omega_{\text{(de,0)}} ( w - \Delta) + \Omega_{\text{(dm,0)}} ( - w - \Delta) \right]
}{
\left[ w - \Delta \right] \left[ \Omega_{\text{(de,0)}} ( w + \Delta) + \Omega_{\text{(dm,0)}} ( -w + \Delta) \right]
}
\right]^{\frac{1}{3\Delta}} - 1,
\end{split}
\label{eq:rhodm_z_min_Q_dm+de_BG}
\end{equation}
\begin{equation}
\begin{split}
z_{\text{(de=0)}} = \left[ 
\frac{
\left[ -w + \Delta \right] \left[ \Omega_{\text{(de,0)}} ( w - \Delta) + \Omega_{\text{(dm,0)}} ( - w - \Delta) \right]
}{
\left[ -w - \Delta \right] \left[ \Omega_{\text{(de,0)}} ( w + \Delta) + \Omega_{\text{(dm,0)}} ( -w + \Delta) \right]
}
\right]^{\frac{1}{3\Delta}} - 1.
\end{split}
\label{eq:rhode_z_min_Q_dm+de_BG}
\end{equation}
The redshift at which the dark matter–dark energy equality occurs is given by:
\begin{equation}
\begin{split}
z_{\text{(dm=de)}} 
= \left[ 
\frac{
\Omega_{\text{(de,0)}} ( w + \Delta) + \Omega_{\text{(dm,0)}} ( - w + \Delta)
}{
\Omega_{\text{(de,0)}} ( w - \Delta) + \Omega_{\text{(dm,0)}} ( - w - \Delta)
}
\right]^{-\frac{1}{3 \Delta}} - 1.
\end{split}
\label{eq:rhodm=de_dm+de}
\end{equation}
\begin{figure}
    \centering
    \begin{subfigure}[b]{0.502\linewidth}
        \centering
        \includegraphics[width=\linewidth]{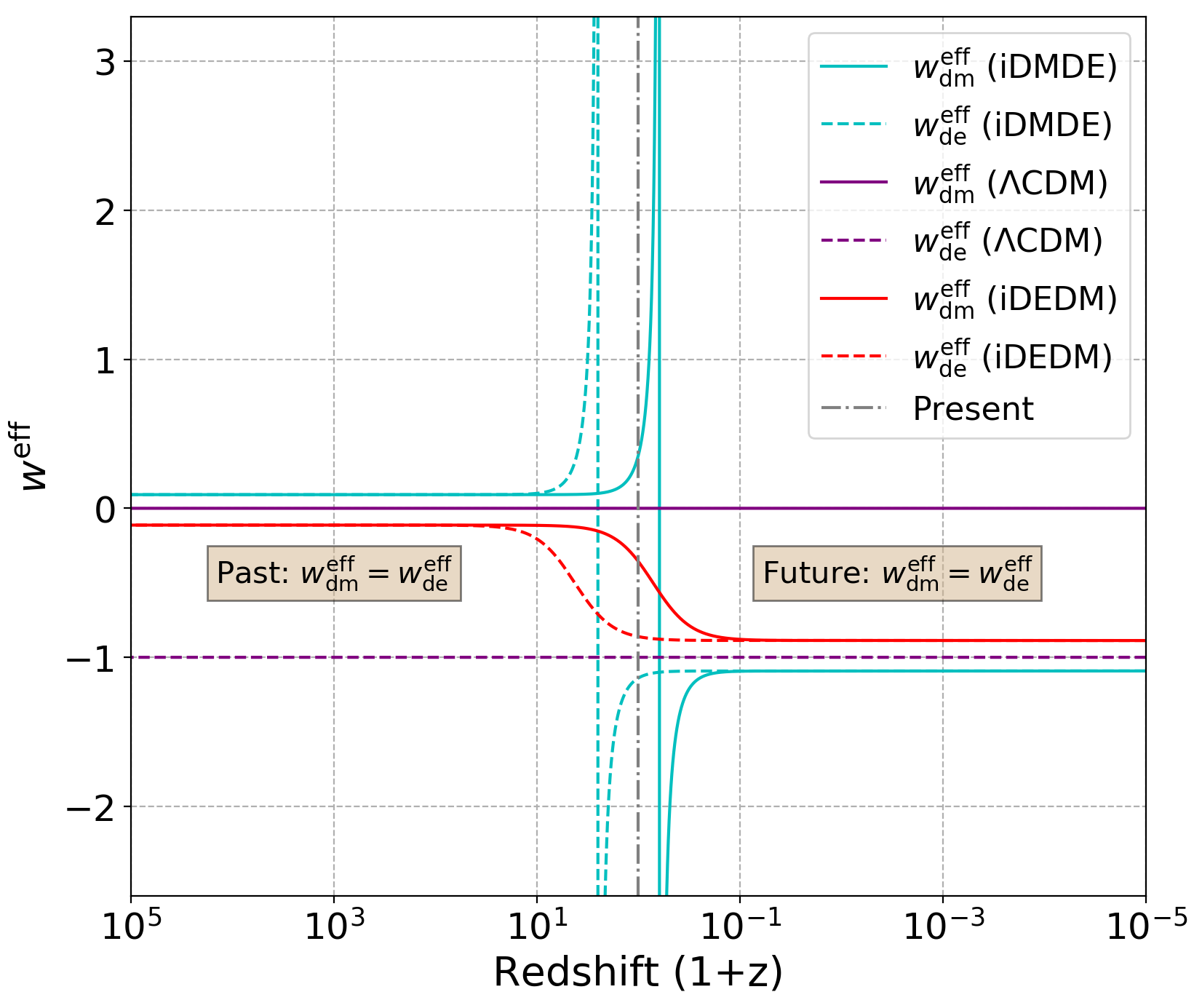}
        \label{fig:omega_dmde_Qdm+de}
    \end{subfigure}    
    \hspace{0pt} % No extra space between subfigures
    \begin{subfigure}[b]{0.483\linewidth}
        \centering
        \includegraphics[width=\linewidth]{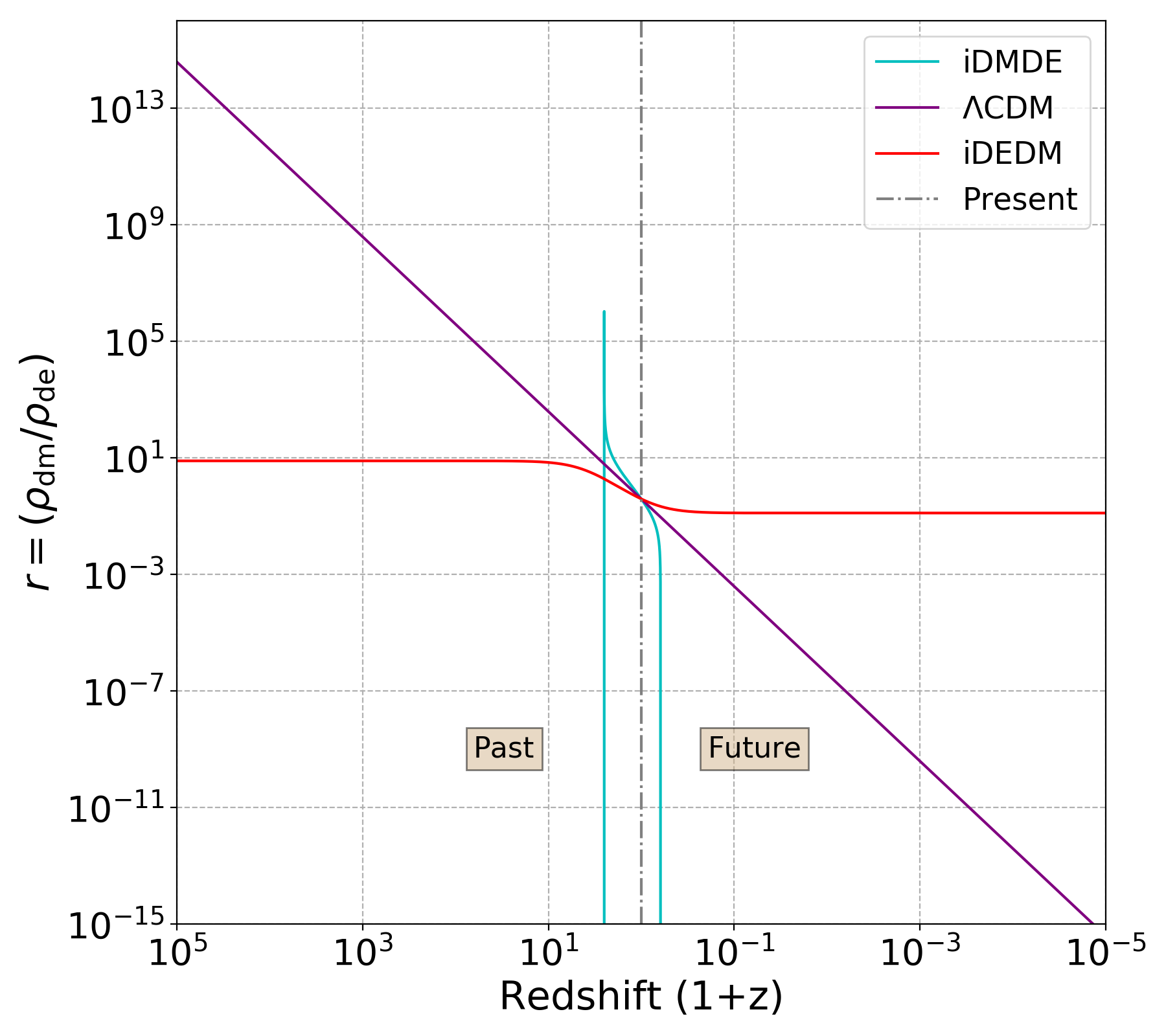}
        \label{fig:CP_Qdm+de}
    \end{subfigure}
    \caption{Effective equations of state and Coincidence Problem (CP) vs.\ redshift — $Q = 3H \delta \left(\rho_{\rm{dm}} + \rho_{\rm{de}}\right)$, with $w^{\rm{eff}}_{\rm{dm}} = w^{\rm{eff}}_{\rm{de}}$ in both the past and future, thus solving the CP ($r = \text{constant}$) in the iDEDM regime ($\delta = +0.1$). In the iDMDE regime ($\delta = -0.1$), negative DE densities (in the past) and DM densities (in the future) are always present, which also cause divergent behavior for $w^{\rm{eff}}_{\rm{dm}}$ and $w^{\rm{eff}}_{\rm{de}}$.}
    \label{fig:CP+omega_dmde_Qdm+de}
\end{figure} 
This model addresses the coincidence problem by letting $r$ converge to the following constants in the past and future:
\begin{equation}
\begin{split}
r_{\text{past}} (a \rightarrow 0) = \frac{w - \Delta}{w + \Delta} 
\quad ; \quad 
r_{\text{future}} (a \rightarrow \infty) = \frac{w + \Delta}{w - \Delta}.
\end{split}
\label{eq:r_dm+de_subbed_past_future}
\end{equation}
The DM and DE effective equations of state \eqref{DSA.omega_eff_dm_de} for this interaction kernel are given by:
\begin{gather} \label{eq.Qdm+de.omega_eff_dm_de}
\begin{split}
w^{\rm{eff}}_{\rm{dm}} &= - \delta \left( 1 + \frac{\rho_{\text{de}}}{\rho_{\text{dm}}} \right)
= - \delta \left( 1 + \frac{1}{r} \right) \quad , \quad w^{\rm{eff}}_{\rm{de}} = w + \delta \left( 1 + \frac{\rho_{\text{dm}}}{\rho_{\text{de}}} \right)
= w + \delta \left( 1 + r \right).
\end{split}
\end{gather}
Expressions for $w^{\rm{eff}}_{\rm{dm}}$ and $w^{\rm{eff}}_{\rm{de}}$ in the asymptotic distant past and future are obtained by substituting \eqref{eq:r_dm+de_subbed_past_future} into \eqref{eq.Qdm+de.omega_eff_dm_de} and after simplification yields the following results:
\begin{equation}
\begin{split}
w^{\rm{eff}}_{\rm{(dm,past)}}&=w^{\rm{eff}}_{\rm{(de,past)}}=\frac{1}{2} \left[ w + \Delta \right], \quad \rightarrow \quad \;  \zeta_{\text{past}} = 0 \; \text{(solves the coincidence problem)}.\\
w^{\rm{eff}}_{\rm{(dm,future)}}&=w^{\rm{eff}}_{\rm{(de,future)}}=\frac{1}{2} \left[ w - \Delta \right], \;\rightarrow \quad  \zeta_{\text{future}} = 0 \;\text{(solves the coincidence problem)}.
\end{split}
\label{eq:omega_eff_dm_de_dm+de_subbed_past_future}
\end{equation}
This interaction model will therefore \textbf{always solve the coincidence problem in both the past and the future} while maintaining positive energy densities in the iDEDM regime.  
The results obtained in equations \eqref{eq:r_dm+de_subbed_past_future}–\eqref{eq:omega_eff_dm_de_dm+de_subbed_past_future} are plotted in Figure~\ref{fig:CP+omega_dmde_Qdm+de}.

\begin{figure}[htbp]    \centering
\begin{subfigure}[b]{0.505\linewidth}
        \centering
        \includegraphics[width=\linewidth]{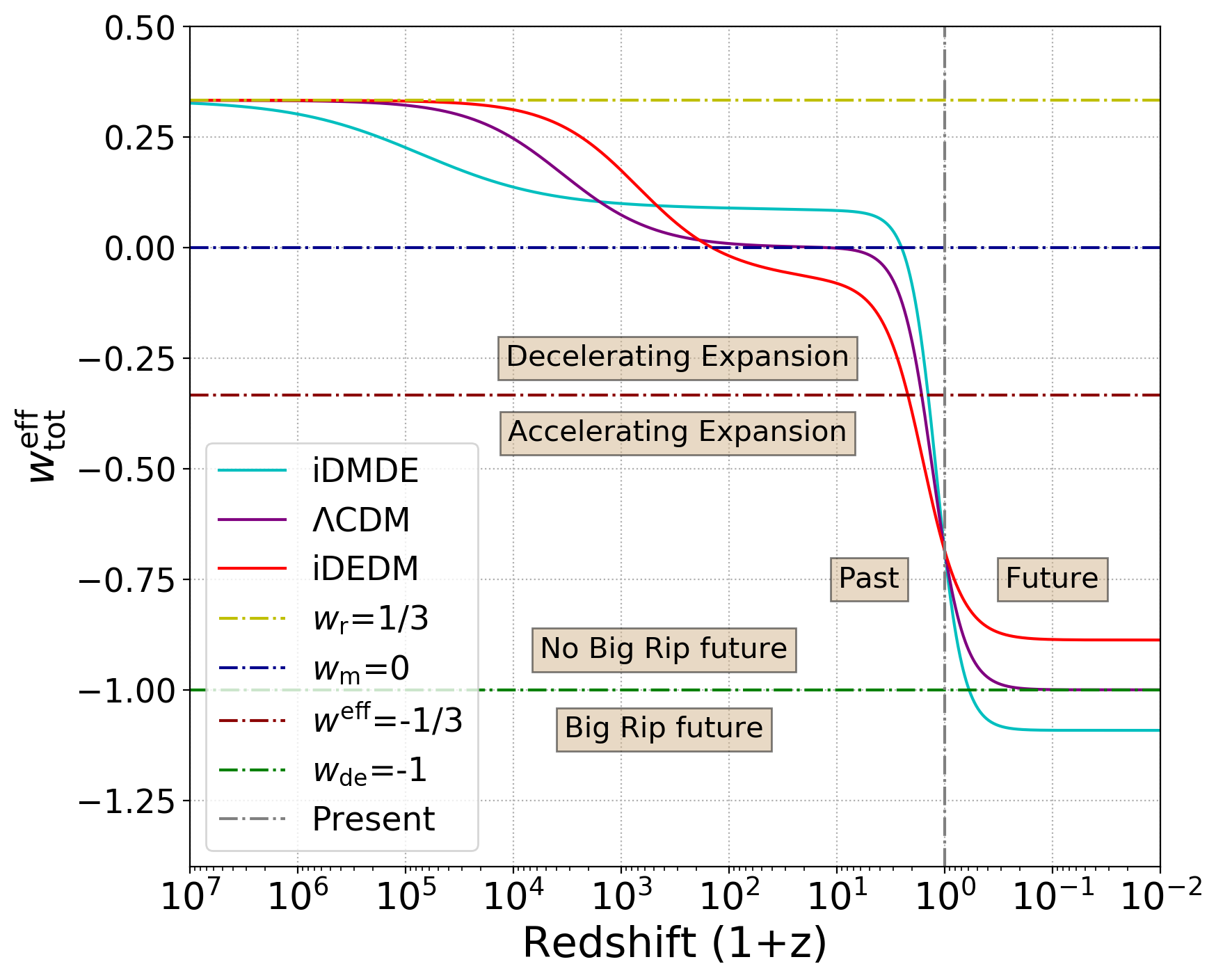}
        \label{fig:eos_tot_Qdm+de}
    \end{subfigure}    
    \begin{subfigure}[b]{0.485\linewidth}
        \centering
        \includegraphics[width=\linewidth]{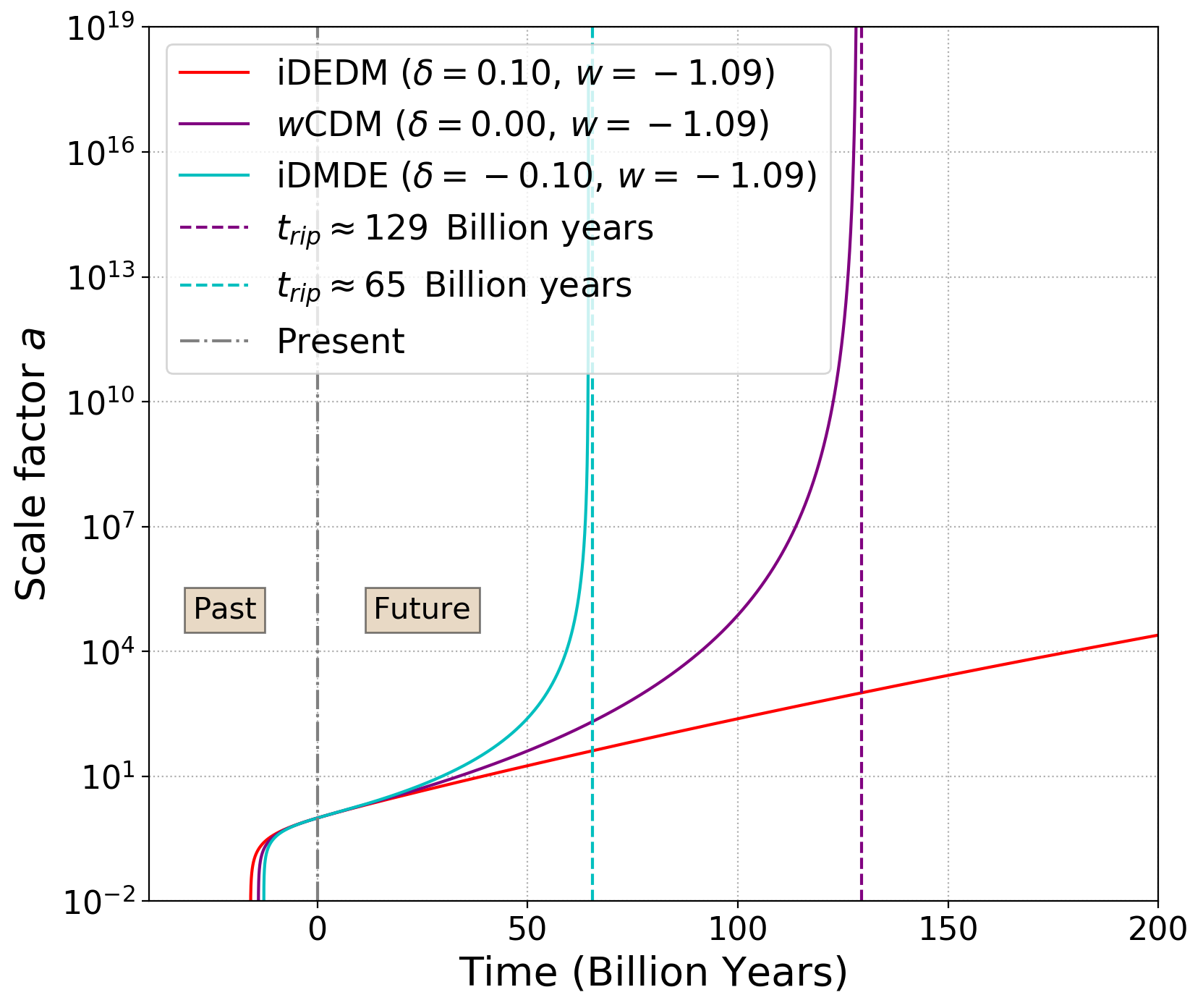}
        \label{fig:Big_rip_Qdm+de}
    \end{subfigure}    
    \hspace{0pt} % No extra space between subfigures
    \caption{Total effective equation of state $w^{\rm{eff}}_{\rm{tot}}$ and big rip future singularities — $Q=3H \delta \left(\rho_{\rm{dm}}+\rho_{\rm{de}}\right)$, with $w=-1$ (left panel) and $w=-1.09$ (right panel). In the iDEDM regime ($\delta=+0.1$), the asymptotic future may have $w^{\rm{eff}}_{\rm{tot}}>-1$ even if $w<-1$, thus avoiding a big rip. In the iDMDE regime ($\delta=-0.1$), the asymptotic future will always have $w^{\rm{eff}}_{\rm{tot}}<-1$ if $w<-1$, thus guaranteeing a future big rip singularity.}
    \label{fig:eos_tot_BR_Qdm+de}
\end{figure}

The redshift at which the DE phantom-crossing occurs is given by: 
\begin{equation}
\begin{split}
z_{\text{pc}} &= \left[
\frac{
\left( \left[ 1 + w + \delta \right] \left[ -w + \Delta \right] - \delta \left[ w + \Delta \right] \right)
\left[ \Omega_{\text{(de,0)}} (w - \Delta) + \Omega_{\text{(dm,0)}} (-w - \Delta) \right]
}{
\left( \left[ 1 + w + \delta \right] \left[ -w - \Delta \right] - \delta \left[ w - \Delta \right] \right)
\left[ \Omega_{\text{(de,0)}} (w + \Delta) + \Omega_{\text{(dm,0)}} (-w + \Delta) \right]
}
\right]^{\frac{1}{3 \Delta}} - 1.
\end{split}
\label{eq:z_pc_dm+de_BG}
\end{equation}
As seen from \eqref{eq.Qdm+de.omega_eff_dm_de}, \eqref{eq:omega_eff_dm_de_dm+de_subbed_past_future} and Figure~\ref{fig:CP+omega_dmde_Qdm+de}, for given values of $w$ and small $\delta$, there are two possible directions for the phantom-crossing (pc), with one case plagued by divergent behavior and negative energies:
\begin{equation}
\begin{split}
\text{pc direction} \begin{cases}
\text{iDMDE: Divergent pc for $w^{\rm{eff}}_{\rm{dm}}$ at $z_{\text{(dm=0)}}$ \eqref{eq:rhodm_z_min_Q_dm+de_BG} and $w^{\rm{eff}}_{\rm{de}}$ at $z_{\text{(de=0)}}$ \eqref{eq:rhode_z_min_Q_dm+de_BG}; with } \rho_{\rm{dm/de}}<0. \\
\text{iDEDM: Quintessence } (w^{\rm{eff}}_{\rm{(de,past)}} > -1) \; \rightarrow \; \text{Phantom } (w^{\rm{eff}}_{\rm{(de,future)}} < -1)\text{; with } \rho_{\rm{dm/de}} > 0.
\end{cases}
\end{split}
\end{equation}
We also have $q = \frac{1}{2} \left( 1 + 3 w^{\rm{eff}}_{\rm{tot}} \right)$, with the algebraic equivalent expressions $w^{\rm{eff}}_{\rm{tot}} = w^{\rm{eff}}_{\rm{dm}} = w^{\rm{eff}}_{\rm{de}} = \frac{1}{2} \left[ w - \Delta \right]$ in the distant future. The following condition is required for a big rip to occur:
\begin{gather} \label{omega_eff_tot_dm+de_BG}
\begin{split}
\text{Big rip condition:} \quad w^{\rm{eff}}_{\rm{(tot,future)}} &\approx \frac{1}{2} \left[ w - \Delta \right] < -1 
\quad \rightarrow \quad \delta < 1 + \frac{1}{w}.
\end{split}
\end{gather}
If $w^{\rm{eff}}_{\rm{(tot,future)}} < -1$, the universe will experience a big rip future singularity at time $t_{\text{rip}}$ given by:
\begin{equation}
\begin{split}
t_{\text{rip}} - t_0 &\approx \frac{4}{3 H_0 \left( -w - 2 + \Delta \right)}
\sqrt{ \frac{-2\Delta}{ \Omega_{\text{(de,0)}} \left( w - \Delta \right) + \Omega_{\text{(dm,0)}} \left( -w - \Delta \right) } }.
\end{split}
\label{eq:Big_Rip_dm+de_BG}
\end{equation}
The effect of the coupling on $w^{\rm{eff}}_{\rm{tot}}$, and how the interaction may cause or avoid a big rip, can be seen in Figure \ref{fig:eos_tot_BR_Qdm+de}. The time of the big rip predicted using \eqref{eq:Big_Rip_dm+de_BG} is indicated by the dashed lines in Figure \ref{fig:eos_tot_BR_Qdm+de}, which agrees with the point where the scale factor $a \rightarrow \infty$ within a finite time. We note that for phantom DE $(w < -1)$, a big rip can only be avoided in the iDEDM regime if condition \eqref{omega_eff_tot_dm+de_BG} is not satisfied, i.e., $\delta > 1 + w$.

\subsection{Linear IDE model 3: \(Q= 3 H \delta(\rho_{\text{dm}} - \rho_{\text{de}})\)} \label{Q_dm-de}

\begin{figure}
    \centering
    \includegraphics[width=0.85\linewidth]{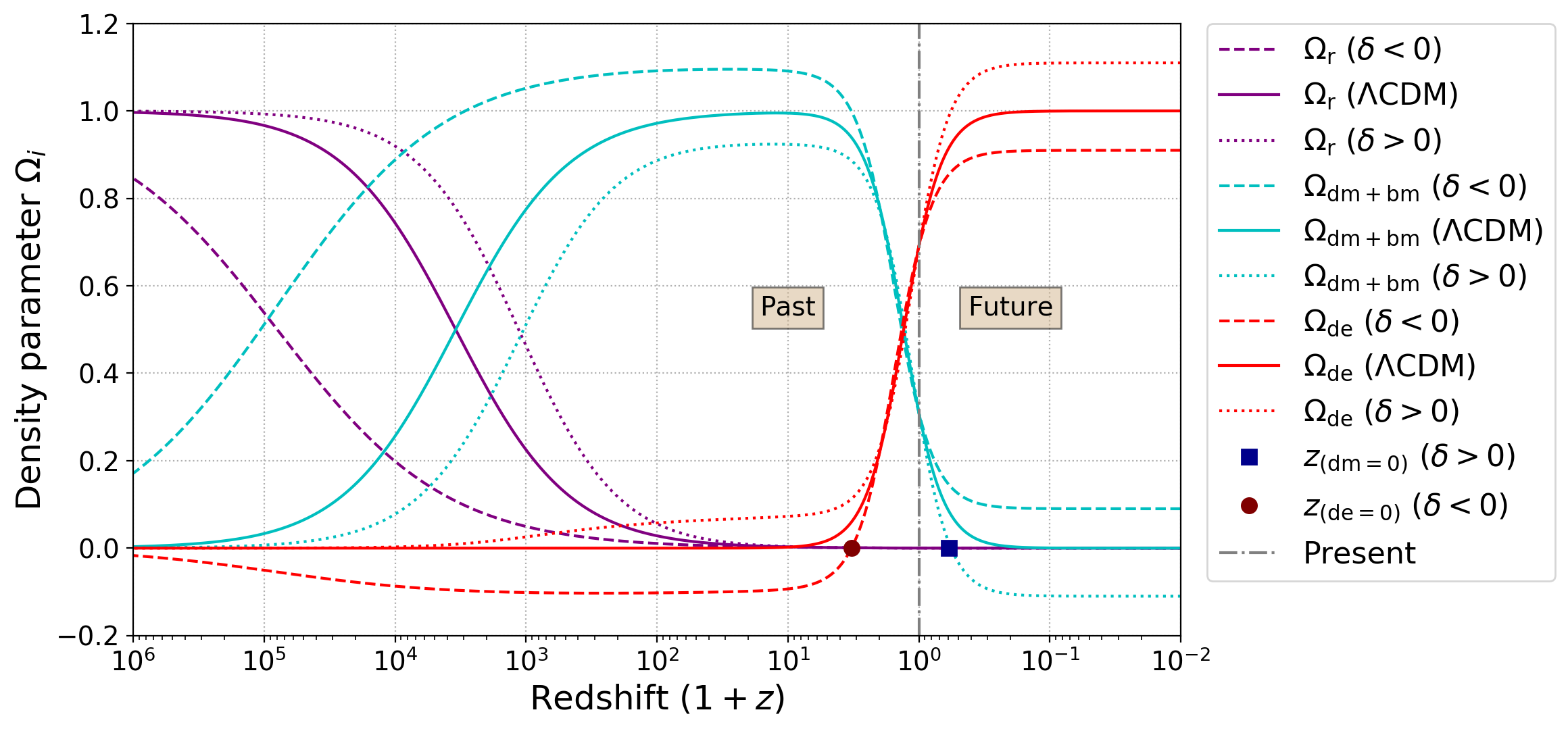}
    \caption{Density parameters vs redshift — $Q = 3\delta H (\rho_{\text{dm}}-\rho_{\text{de}})$, with negative energies being inevitable across the entire parameter space. Negative DE densities (in the past) always occur for negative couplings ($\delta=-0.1$), while negative DM densities (in the future) always occur for positive couplings ($\delta=+0.1$).}
    \label{fig:Omega_Linear_dm-de}
\end{figure}

This interaction is a sign-changing interaction; therefore, the sign of $\delta$ determines only the initial direction of energy transfer, which will switch at the dark matter–dark energy equality $z_{\rm{(dm=de)}}$ given by \eqref{eq:rhodm=de_dm-de}, as seen in Figure \ref{fig:Q_Linear_dm+de}. Sign-switching behavior is hinted at by reconstructions of $Q$ from data~\cite{guedezounme2025phantomcrossingdarkinteraction}. Regardless of the choice of $\delta$, this model will always yield either negative DM or DE densities, as seen in Figure \ref{fig:Omega_Linear_dm-de}. For small $|\delta|$, if $\delta<0$ we have $\rho_{\text{de}}<0$ in the past, while if $\delta>0$ we have $\rho_{\text{dm}}<0$ in the future. Substituting the expressions for $m_+$, $m_-$, and $\Delta$ for this interaction from Table \ref{tab:m+-} into the general expressions for $\rho_{\text{dm}}$ \eqref{eq:rhodm_general} and $\rho_{\text{de}}$ \eqref{eq:rhode_general} gives:
\begin{gather}
\begin{split} \label{eq:rho_dm_Q_dm-de}
\rho_{\text{dm}} = -& \frac{2\delta+w+\Delta}{4w \Delta} 
\Bigl[\rho_{\text{(de,0)}}\bigl(2\delta+w-\Delta\bigr)
+\rho_{\text{(dm,0)}}\bigl(2\delta-w-\Delta\bigr)\Bigr] 
a^{\frac{3}{2}\bigl(2\delta-w-2+\Delta\bigr)}\\[1mm]
&+ \frac{2\delta+w-\Delta}{4w \Delta} 
\Bigl[\rho_{\text{(de,0)}}\bigl(2\delta+w+\Delta\bigr)
+\rho_{\text{(dm,0)}}\bigl(2\delta-w+\Delta\bigr)\Bigr] 
a^{\frac{3}{2}\bigl(2\delta-w-2-\Delta\bigr)} , 
\end{split}   
\end{gather}

\begin{gather}
\begin{split} \label{eq:rho_de_Q_dm-de}
\rho_{\text{de}} = +& \frac{2\delta-w+\Delta}{4w \Delta} 
\Bigl[\rho_{\text{(de,0)}}\bigl(2\delta+w-\Delta\bigr)
+\rho_{\text{(dm,0)}}\bigl(2\delta-w-\Delta\bigr)\Bigr] 
a^{\frac{3}{2}\bigl(2\delta-w-2+\Delta\bigr)}\\[1mm]
&- \frac{2\delta-w-\Delta}{4w \Delta} 
\Bigl[\rho_{\text{(de,0)}}\bigl(2\delta+w+\Delta\bigr)
+\rho_{\text{(dm,0)}}\bigl(2\delta-w+\Delta\bigr)\Bigr] 
a^{\frac{3}{2}\bigl(2\delta-w-2-\Delta\bigr)} ,
\end{split}   
\end{gather}
where $\Delta$ is the determinant:
\begin{gather}
\begin{split} \label{eq:determinant_Q_dm-de}
\Delta = \sqrt{4\delta^2 + w^2}\,.
\end{split}   
\end{gather}

\begin{gather} \label{DSA.Q.dm-de.19}
\boxed{
\begin{split}
&\underline{\text{Conditions for $\rho_{\rm{dm}}\ge0 \; ; \; \rho_{\rm{de}} \ge0$ over the entire cosmological evolution:}} \\
&\quad\quad\quad\quad\quad\quad\text{\textbf{No viable domain exists.}}\quad\quad\quad\quad\quad\quad\quad\quad\quad\quad
\end{split}
}
\end{gather}
We also note that for $\rho_{\rm{dm}}$ \eqref{eq:rho_dm_Q_dm-de} and $\rho_{\rm{de}}$ \eqref{eq:rho_de_Q_dm-de} to be defined, we require $w\ne 0$.
Similarly, substituting the expressions for $m_+$, $m_-$ and $\Delta$ for this interaction from Table \ref{tab:m+-} into the relevant expressions in Section \ref{More_analytical_relations} gives the corresponding results below:
\begin{equation}
\begin{split}
\Omega_{\text{(dm,past)}} =\frac{1}{2}+\frac{2\delta -\Delta}{2w} \quad &; \quad \Omega_{\text{(de,past)}}=  \frac{1}{2}- \frac{2\delta -\Delta}{2w},\\
\Omega_{\text{(dm,future)}} =\frac{1}{2}+\frac{2\delta+\Delta}{2w} \quad &; \quad \Omega_{\text{(de,future)}}=\frac{1}{2}- \frac{2\delta+\Delta}{2w}.
\end{split}
\label{eq:frac_den_dm-de_past}
\end{equation}
It may be noted that for the model \(Q= 3 H \delta(\rho_{\text{de}} - \rho_{\text{dm}})\), we set $\delta \rightarrow -\delta$ in the above equations, thereby obtaining the new functions for $\rho_{\text{dm}}$ and $\rho_{\text{de}}$.
The DM and DE densities become zero at redshifts $z_{\text{(dm=0)}}$ and $z_{\text{(de=0)}}$:
\begin{equation}
\begin{split}
z_{\text{(dm=0)}}=\left[\frac{\left[2\delta + w + \Delta \right] \left[ \Omega_{\text{(de,0)}} (2\delta + w - \Delta) + \Omega_{\text{(dm,0)}} (2\delta - w - \Delta) \right]}{\left[2\delta +w - \Delta \right] \left[\Omega_{\text{(de,0)}} (2\delta + w + \Delta) + \Omega_{\text{(dm,0)}} (2\delta - w + \Delta)  \right]}\right]^{\frac{1}{3\Delta}}-1,
\end{split}
\label{eq:rhodm_z_min_Q_dm-de_BG}
\end{equation}
\begin{equation}
\begin{split}
z_{\text{(de=0)}}=\left[\frac{\left[2\delta - w + \Delta \right] \left[ \Omega_{\text{(de,0)}} (2\delta + w - \Delta) + \Omega_{\text{(dm,0)}} (2\delta - w - \Delta) \right]}{\left[2\delta - w - \Delta \right] \left[\Omega_{\text{(de,0)}} (2\delta + w + \Delta) + \Omega_{\text{(dm,0)}} (2\delta - w + \Delta) \right]}\right]^{\frac{1}{3\Delta}}-1.
\end{split}
\label{eq:rhode_z_min_Q_dm-de_BG}
\end{equation}
The redshift at which both the dark matter–dark energy equality and the change in the direction of the interaction $(Q=0)$ occur, as illustrated in Figure \ref{fig:Q_Linear_dm+de}, is given by:
\begin{equation}
\begin{split}
z_{\text{(dm=de)}} = z_{\text{(Q=0)}} &= 
\left[\frac{\left[2\delta - \Delta \right] \left[\Omega_{\text{(de,0)}} (2\delta + w + \Delta) + \Omega_{\text{(dm,0)}} (2\delta - w + \Delta) \right]}{\left[2\delta + \Delta \right] \left[\Omega_{\text{(de,0)}} (2\delta + w - \Delta) + \Omega_{\text{(dm,0)}} (2\delta - w - \Delta) \right]}\right]^{-\frac{1}{3\Delta}} - 1.
\end{split}
\label{eq:rhodm=de_dm-de}
\end{equation}
\begin{figure}[htbp]
    \centering
    \begin{subfigure}[b]{0.502\linewidth}
        \centering
        \includegraphics[width=\linewidth]{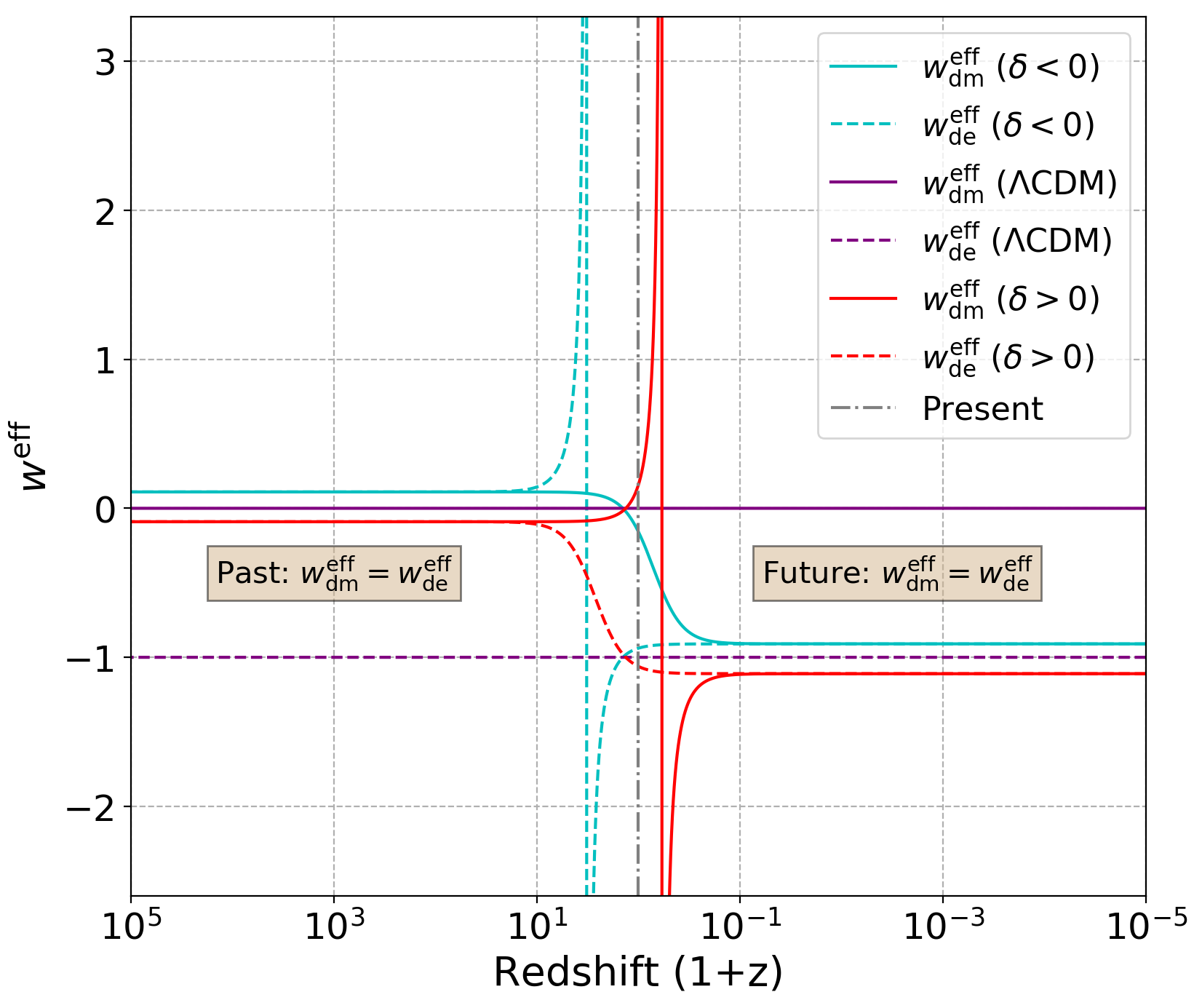}
        %\label{fig:omega_dmde_Qdm+de}
    \end{subfigure}    
    \hspace{0pt} % No extra space between subfigures
    \begin{subfigure}[b]{0.483\linewidth}
        \centering
        \includegraphics[width=\linewidth]{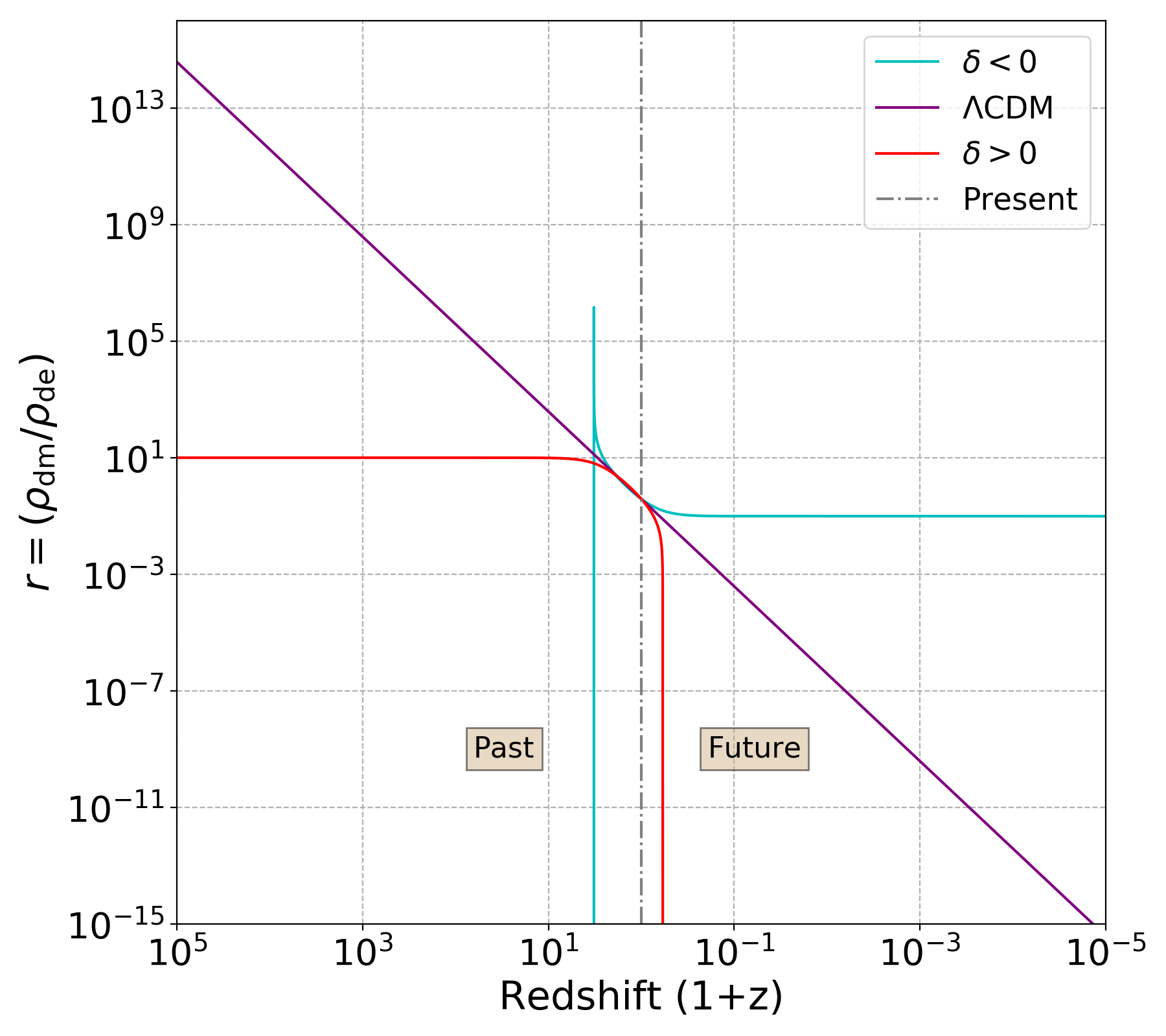}
        %\label{fig:CP_Qdm+de}
    \end{subfigure}%
    \caption{Effective equations of state and Coincidence Problem (CP) vs.\ redshift — $Q=3H \delta \left(\rho_{\rm{dm}}-\rho_{\rm{de}}\right)$, with $w^{\rm{eff}}_{\rm{dm}}=w^{\rm{eff}}_{\rm{de}}$ in both the past and future, thus solving the CP ($r=\text{constant}$). This, however, comes at the cost of negative DE densities and divergent $w^{\rm{eff}}_{\rm{de}}$ (in the past) for negative couplings ($\delta=-0.1$), and negative DM densities and divergent $w^{\rm{eff}}_{\rm{dm}}$ (in the future) for positive couplings ($\delta=+0.1$).}
    \label{fig:CP+omega_dmde_Qdm-de}
\end{figure}
This model addresses the coincidence problem by allowing $r$ to converge to the following constants in the past and future:
\begin{equation}
\begin{split}
r_{\text{past}} (a\rightarrow0) & = -\frac{2\delta+w-\Delta}{2\delta-w-\Delta} \quad , \quad 
r_{\text{future}} (a\rightarrow\infty) =  -\frac{2\delta+w+\Delta}{2\delta-w+\Delta}.
\end{split}%
\label{eq:r_dm-de_subbed_past_future}
\end{equation}
The DM and DE effective equations of state \eqref{DSA.omega_eff_dm_de} for this interaction kernel are:
\begin{gather} \label{eq.Qdm-de.omega_eff_dm_de}
\begin{split}
w^{\rm{eff}}_{\rm{dm}} &= \delta \left(\frac{\rho_{\text{de}}}{\rho_{\text{dm}}} - 1 \right) 
= \delta \left(\frac{1}{r} - 1 \right) \quad ,  \quad
w^{\rm{eff}}_{\rm{de}} = w + \delta \left(\frac{\rho_{\text{dm}}}{\rho_{\text{de}}} - 1 \right) 
= w + \delta (r - 1).
\end{split}
\end{gather}
Expressions for $w^{\rm{eff}}_{\rm{dm}}$ and $w^{\rm{eff}}_{\rm{de}}$ in the asymptotic distant past and future are obtained by substituting \eqref{eq:r_dm-de_subbed_past_future} into \eqref{eq.Qdm-de.omega_eff_dm_de} and after simplification yields the following results:
\begin{equation}
\begin{split}
w^{\rm{eff}}_{\rm{(dm,past)}}&=w^{\rm{eff}}_{\rm{(de,past)}}=\frac{1}{2} \left[2\delta+ w + \Delta \right], \quad \rightarrow \quad \;  \zeta_{\text{past}} = 0 \; \text{(solves the coincidence problem with $\rho_{\text{dm/de}}<0$)}.\\
w^{\rm{eff}}_{\rm{(dm,future)}}&=w^{\rm{eff}}_{\rm{(de,future)}}=\frac{1}{2} \left[2\delta+ w - \Delta \right], \;\rightarrow \quad  \zeta_{\text{future}} = 0 \;\text{(solves the coincidence problem with $\rho_{\text{dm/de}}<0$)}.
\end{split}
\label{eq:omega_eff_dm_de_Qdm-de_subbed_past_future}
\end{equation}
This interaction model will therefore always \textbf{solve the coincidence problem in both the past and future, but at the cost of having either negative $\rho_{\text{dm}}$ or $\rho_{\text{de}}$}. The results obtained in equations \eqref{eq:r_dm-de_subbed_past_future} to \eqref{eq:omega_eff_dm_de_Qdm-de_subbed_past_future} are plotted in Figure~\ref{fig:CP+omega_dmde_Qdm-de}.

The phantom-crossing redshift is given by:
\begin{equation}
\begin{split}
z_{\text{pc}} &= \left[\frac{\left(\left[1 + w - \delta \right]\left[2\delta - w + \Delta \right] - \delta \left[2\delta + w + \Delta\right] \right)\left[\Omega_{\text{(de,0)}} (2\delta + w - \Delta) + \Omega_{\text{(dm,0)}} (2\delta - w - \Delta) \right]}{\left(\left[1 + w - \delta \right]\left[2\delta - w - \Delta \right] - \delta \left[2\delta + w - \Delta \right] \right)\left[\Omega_{\text{(de,0)}} (2\delta + w + \Delta) + \Omega_{\text{(dm,0)}} (2\delta - w + \Delta) \right]}\right]^{\frac{1}{3 \Delta}} - 1.
\end{split}%
\label{eq:z_pc_dm-de_BG}
\end{equation}
As seen from \eqref{eq.Qdm-de.omega_eff_dm_de}, \eqref{eq:omega_eff_dm_de_Qdm-de_subbed_past_future}, and Figure~\ref{fig:CP+omega_dmde_Qdm-de}, for given values of $w$ and $\delta$, there are two possible phantom-crossing behaviors, both accompanied by divergent behavior and negative energies:
\begin{equation}
\begin{split} 
\text{pc behavior} \begin{cases}
\delta < 0 : \text{Divergent pc for $w^{\rm{eff}}_{\rm{de}}$ at $z_{\text{(de=0)}}$ \eqref{eq:rhode_z_min_Q_dm-de_BG}; with $\rho_{\rm{de}}<0$.} \\    
\delta > 0 : \text{Divergent pc for $w^{\rm{eff}}_{\rm{dm}}$ at $z_{\text{(dm=0)}}$ \eqref{eq:rhodm_z_min_Q_dm-de_BG}; with $\rho_{\rm{dm}}<0$.} \\
\end{cases}
\end{split}
\end{equation}
We also have $q = \frac{1}{2} \left(1 + 3 w^{\rm{eff}}_{\rm{tot}}\right)$, with 
$w^{\rm{eff}}_{\rm{tot}} = w^{\rm{eff}}_{\rm{dm}} = w^{\rm{eff}}_{\rm{de}} = \frac{1}{2} \left[ 2\delta + w - \Delta \right]$ in the distant future. 
This leads to the following condition for a big rip to occur:
\begin{gather} \label{omega_eff_tot_dm-de_BG}
\begin{split}
\text{Big rip condition:} \quad  
w^{\rm{eff}}_{\rm{(tot,future)}} &\approx \frac{1}{2} \left[ 2\delta + w - \Delta \right] < -1 
\quad \Rightarrow \quad 
\delta > \frac{1 + w}{2 + w}.
\end{split}
\end{gather}
If $w^{\rm{eff}}_{\rm{tot}} < -1$, the universe will experience a big rip future singularity at time $t_{\text{rip}}$:
\begin{equation}
\begin{split}
t_{\text{rip}} - t_0 &\approx 
\frac{4}{3 H_0 \left( 2\delta - w - 2 + \Delta \right)}
\sqrt{\frac{- 2\Delta}{\Omega_{\text{(de,0)}} \left( 2\delta + w - \Delta \right) 
+ \Omega_{\text{(dm,0)}} \left( 2\delta - w - \Delta \right) }}.
\end{split}
\label{eq:Big_Rip_dm-de_BG}
\end{equation}

\begin{figure}[htbp]    \centering
\begin{subfigure}[b]{0.505\linewidth}
        \centering
        \includegraphics[width=\linewidth]{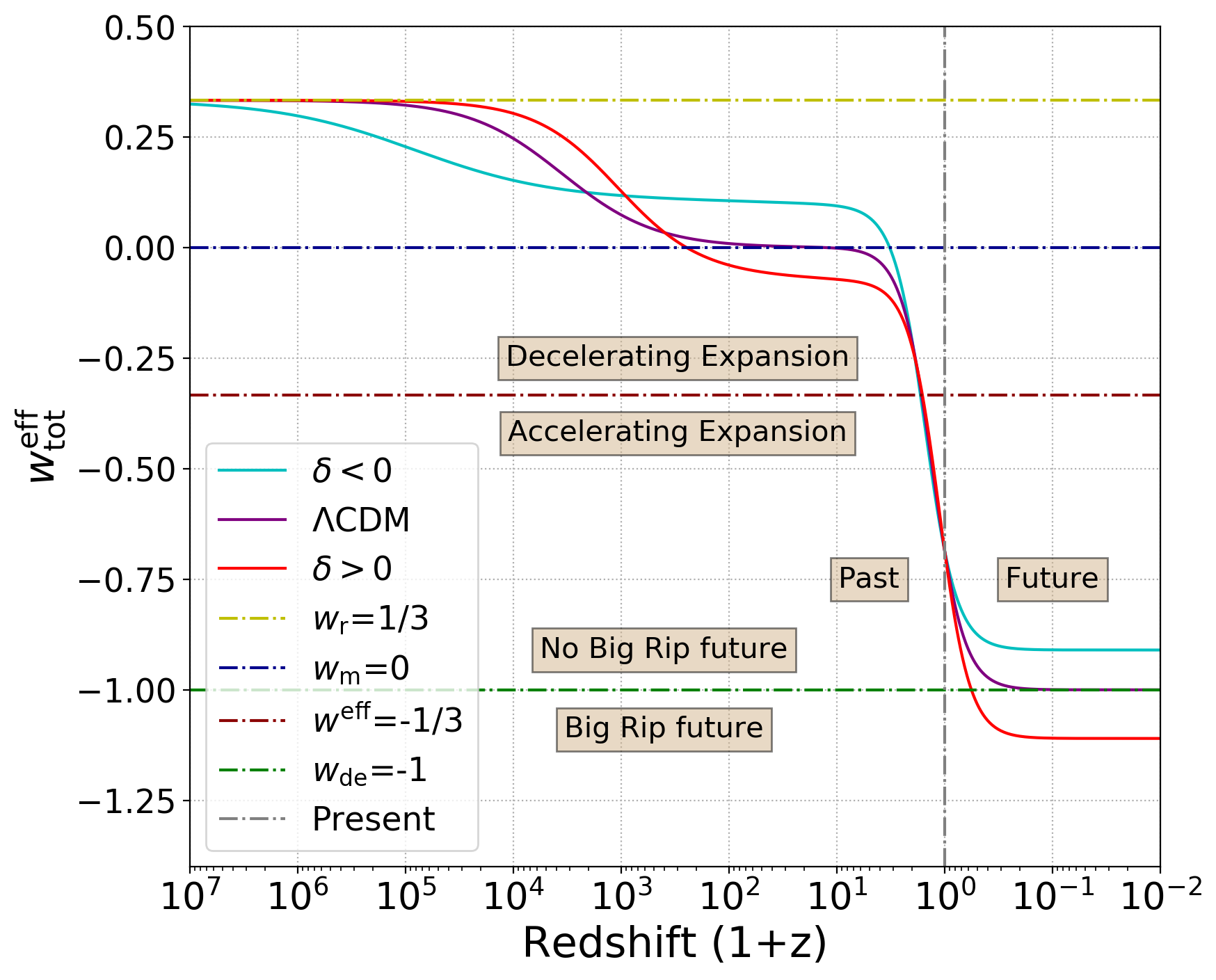}
        \label{fig:eos_tot_Qdm-de}
    \end{subfigure}    
    \begin{subfigure}[b]{0.485\linewidth}
        \centering
        \includegraphics[width=\linewidth]{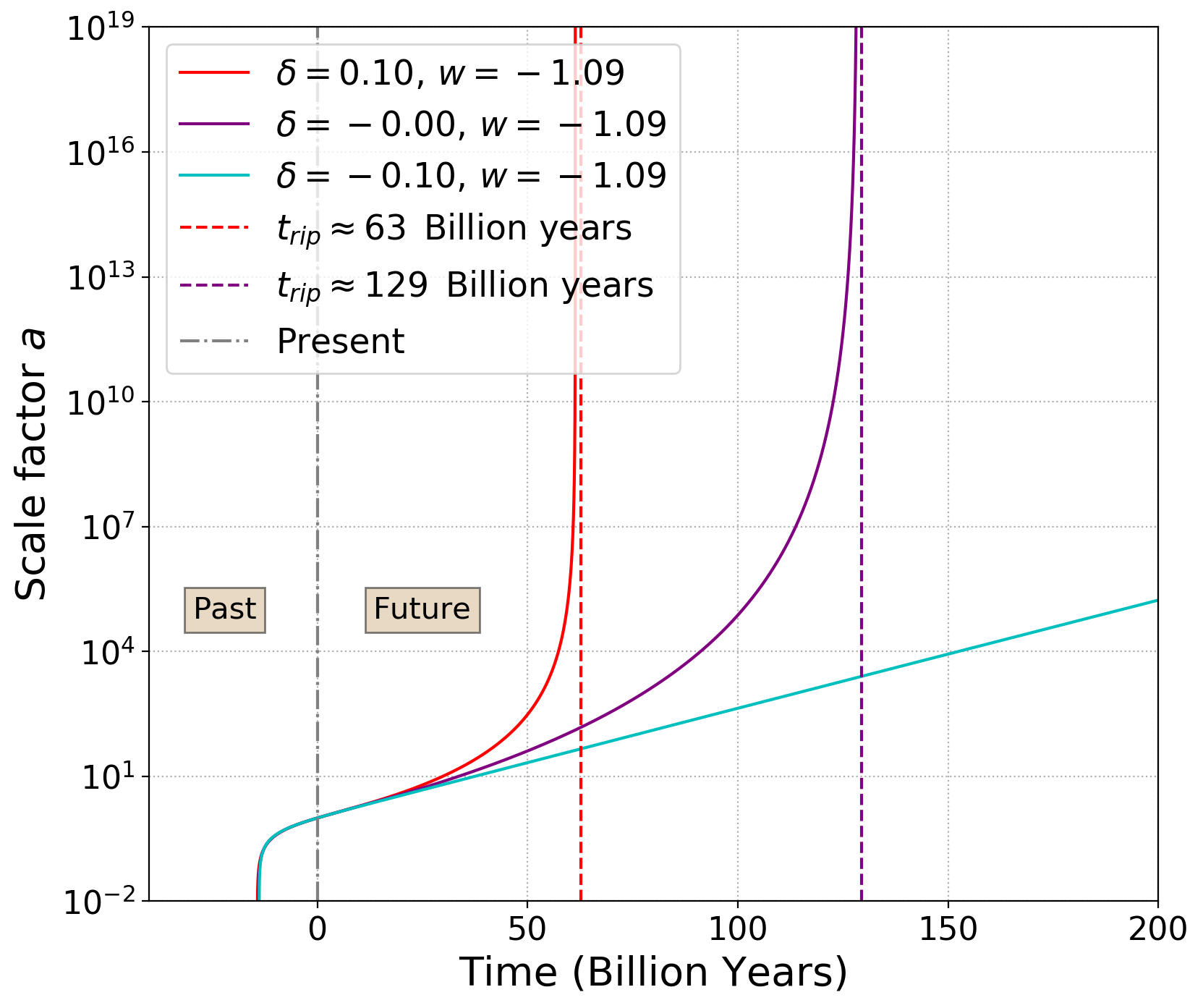}
        \label{fig:Big_rip_Qdm-de}
    \end{subfigure}    
    \hspace{0pt} % No extra space between subfigures
    \caption{Total effective equation of state $w^{\rm{eff}}_{\rm{tot}}$ and big rip future singularities for the interaction $Q=3H \delta \left(\rho_{\rm{dm}}-\rho_{\rm{de}}\right)$, with $w=-1$ (left panel) and $w=-1.09$ (right panel). For negative couplings ($\delta=-0.1$), the asymptotic future may have $w^{\rm{eff}}_{\rm{tot}} > -1$ even when $w < -1$, thereby avoiding a big rip. For positive couplings ($\delta=+0.1$), the asymptotic future will always have $w^{\rm{eff}}_{\rm{tot}} < -1$ if $w < -1$, thus guaranteeing a future big rip singularity.}
    \label{fig:eos_tot_BR_Qdm-de}
\end{figure}

The effect of the coupling on $w^{\rm{eff}}_{\rm{tot}}$, and how the interaction may cause or avoid a big rip, can be seen in Figure~\ref{fig:eos_tot_BR_Qdm-de}. 
The time of the big rip predicted using Eq.~\eqref{eq:Big_Rip_dm-de_BG} is indicated by the dashed lines in Figure~\ref{fig:eos_tot_BR_Qdm-de}, in agreement with the point where the scale factor satisfies $a \rightarrow \infty$ within a finite time. 
We note that, for phantom DE ($w < -1$), a big rip can only be avoided when DE decays into DM at late times, which in this model corresponds to $\delta < 0$.

\subsection{Linear IDE model 4: \(Q= 3 H \delta \rho_{\text{dm}}\)} \label{Q_dm}

\begin{figure}
    \centering
    \includegraphics[width=0.85\linewidth]{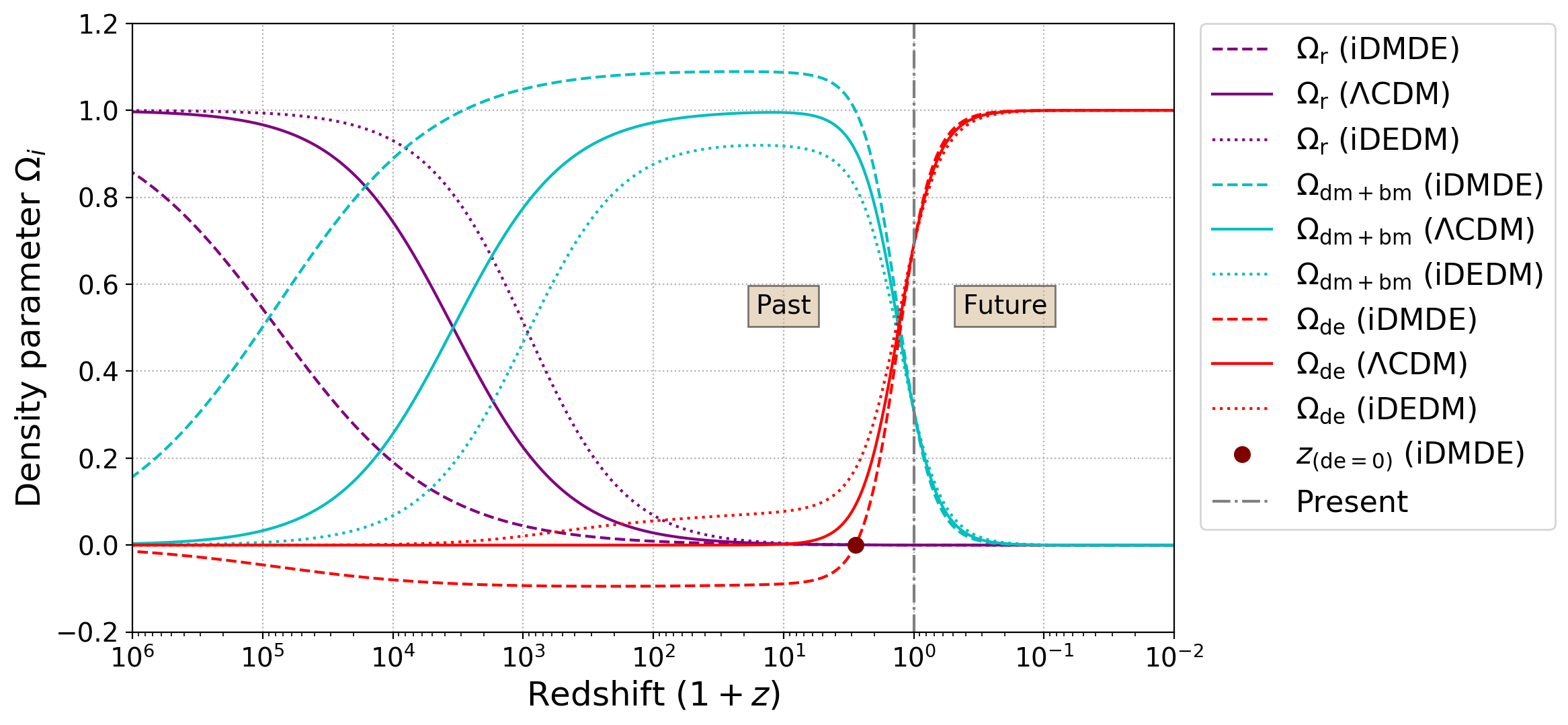}
    \caption{Density parameters vs redshift -- $Q = 3\delta H \rho_{\text{dm}}$, with positive energy densities only found in the iDEDM regime ($\delta=+0.1$), while negative DE densities (in the past) are always present in the iDMDE regime ($\delta=-0.1$).}
    \label{fig:Omega_Linear_dm}
\end{figure}

This interaction mostly changes the dynamics in the distant past during DM domination, as seen in Figure~\ref{fig:Q_Linear_dm+de} and the other figures in this section. For this interaction, $Q=0$ if $\rho_{\text{dm}}=0$, therefore we can guarantee that $\rho_{\text{dm}}\ge 0$ at all times. In contrast, this interaction will have past negative DE densities in the iDMDE regime, but this can be avoided with a sufficiently small interaction in the iDEDM regime, given by the conditions in \eqref{DSA.Q.dm.19}, as seen in Figure~\ref{fig:Omega_Linear_dm}. The results in this section were briefly discussed in~\cite{vanderWesthuizen:2023hcl} and are given to show convergence with other literature.  
Substituting the expressions for $m_+$, $m_-$, and $\Delta$ for the interaction from Table~\ref{tab:m+-} into the general expressions for $\rho_{\text{dm}}$~\eqref{eq:rhodm_general} and $\rho_{\text{de}}$~\eqref{eq:rhode_general} gives the familiar expressions:
\begin{gather}
\begin{split} \label{eq:rho_dm_Q_dm}
\rho_{\text{dm}} = \rho_{\text{(dm,0)}} a^{-3(1 -\delta)},
\end{split}   
\end{gather}
\begin{gather}
\begin{split} \label{eq:rho_de_Q_dm}
\rho_{\text{de}} = \left(\rho_{\text{(de,0)}} + \rho_{\text{(dm,0)}} \left( \frac{\delta}{\delta + w} \right) \left[  1 - a^{3 (\delta+ w)} \right] \right) a^{-3(w + 1)}.
\end{split}   
\end{gather}

\begin{gather} \label{DSA.Q.dm.19}
\boxed{
\begin{split}
&\underline{\text{Conditions for} \ \rho_{\rm{dm}}\ge0 \ ; \ \rho_{\rm{de}} \ge0 \ \text{for entire cosmological evolution:}}
 \\ &\quad \quad \quad \quad \quad \quad \text{iDEDM with} \ 
0\le\delta \le -\frac{w}{1+r_0} .
\end{split}}
\end{gather} 
We also note that, for $\rho_{\rm{de}}$ in \eqref{eq:rho_de_Q_dm} to be defined, we require $\delta \ne -w$. Similarly, substituting the expressions for $m_+$, $m_-$, and $\Delta$ for the interaction from Table \ref{tab:m+-} into the relevant expressions in Section \ref{More_analytical_relations} gives the corresponding expressions below.
\begin{equation}
\begin{split}
\Omega_{\text{(dm,past)}}=1+\frac{\delta}{w} \quad &; \quad \Omega_{\text{(de,past)}}=- \frac{\delta}{w},\\
\Omega_{\text{(dm,future)}}=0 \quad &; \quad \Omega_{\text{(de,future)}}=1.\\
\end{split}\label{eq:frac_den_dm_past}
\end{equation}
The DM density always remains positive for this model, while the DE density becomes zero at redshift $z_{\text{(de=0)}}$:
\begin{equation}
\begin{split}
z_{\text{(de=0)}}&=\left[1+\frac{\Omega_{\text{(de,0)}}}{\Omega_{\text{(dm,0)}}}\frac{\delta+w}{\delta}\right]^{-\frac{1}{3(\delta+w)}}-1.
\end{split}
\label{eq:rhode_z_min_Q_dm_BG}
\end{equation}  
The redshift at which the dark matter–dark energy equality occurs is given by:
\begin{equation}
\begin{split}
& z_{\text{(dm=de)}} = \left[\frac{\Omega_{\text{(dm,0)}} \left(2\delta+w \right)}{\Omega_{\text{(de,0)}} (\delta+w) + \Omega_{\text{(dm,0)}} \delta} \right]^{\frac{1}{3 (\delta+w)}} - 1.
\end{split}
\label{eq:rhodm=de_dm}
\end{equation}
\begin{figure}[htbp]
    \centering
    \begin{subfigure}[b]{0.503\linewidth}
        \centering
        \includegraphics[width=\linewidth]{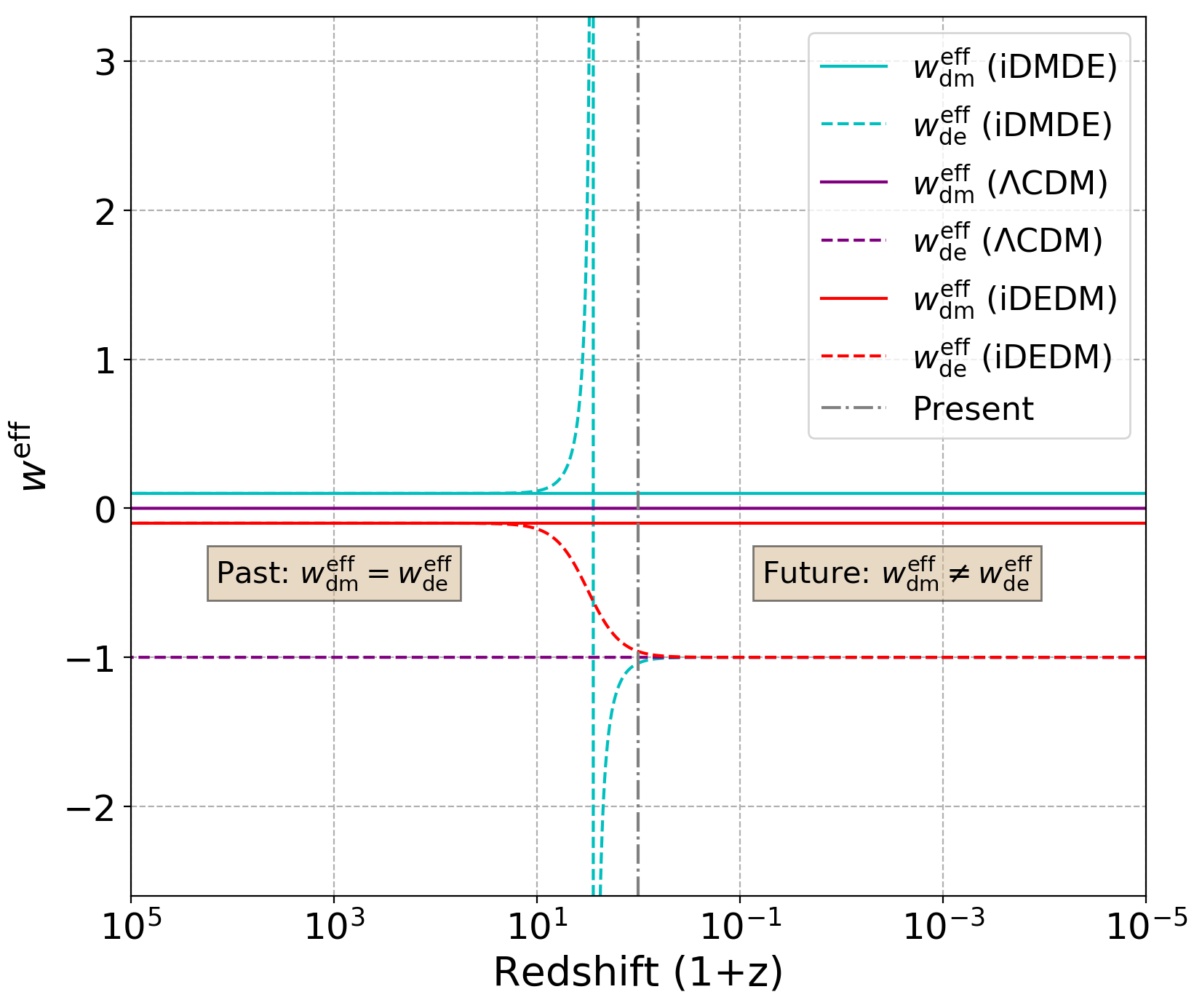}
        \label{fig:omega_dmde_Qdm}
    \end{subfigure}    
    \hspace{0pt} % No extra space between subfigures
    \begin{subfigure}[b]{0.482\linewidth}
        \centering
        \includegraphics[width=\linewidth]{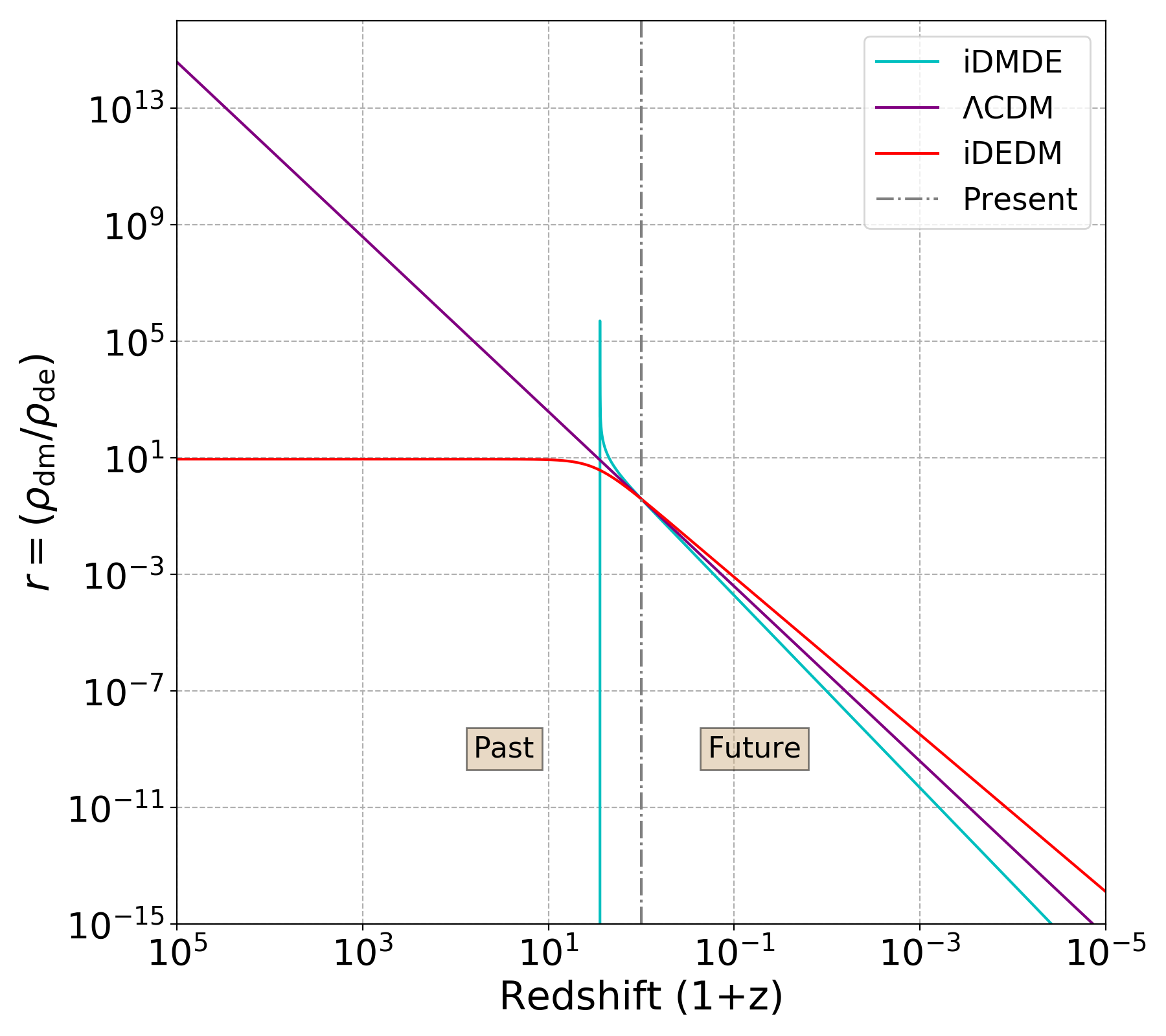}
        \label{fig:CP_Qdm+de}
    \end{subfigure}%
    \caption{Effective equations of state and Coincidence Problem (CP) vs redshift — $Q=3H \delta \rho_{\rm{dm}}$, with $w^{\rm{eff}}_{\rm{dm}}=w^{\rm{eff}}_{\rm{de}}$ in the past, thus solving the CP ($r=\text{constant}$) in the past and alleviating it in the future for the iDEDM regime ($\delta=+0.1$). In the iDMDE regime ($\delta=-0.1$), negative DE densities and divergent $w^{\rm{eff}}_{\rm{de}}$ (in the past) are always present, and the CP is worsened in the future.}
    \label{fig:CP+omega_dmde_Qdm}
\end{figure}
This model addresses the coincidence problem by letting $r$ converge to the following values in the past and future:
\begin{equation}
\begin{split}
r_{\text{past}} (a \rightarrow 0) = -1 - \frac{w}{\delta} \quad ; \quad 
r_{\text{future}} (a \rightarrow \infty) \approx 0.
\end{split}
\label{eq:r_dm_subbed_past_future}
\end{equation}
The DM and DE effective equations of state \eqref{DSA.omega_eff_dm_de} for this interaction kernel are given by:
\begin{gather} \label{eq.Qdm.omega_eff_dm_de}
\begin{split}
w^{\rm{eff}}_{\rm{dm}} &= -\delta  \quad ; \quad w^{\rm{eff}}_{\rm{de}} = w + \delta \frac{\rho_{\text{dm}}}{\rho_{\rm{de}}} = w + \delta r. 
\end{split}
\end{gather}
Expressions for $w^{\rm{eff}}_{\rm{dm}}$ and $w^{\rm{eff}}_{\rm{de}}$ in the asymptotic distant past and future are obtained by substituting \eqref{eq:r_dm_subbed_past_future} into \eqref{eq.Qdm.omega_eff_dm_de}:
\begin{equation}
\begin{split}
w^{\rm{eff}}_{\rm{(dm,past)}} &=w^{\rm{eff}}_{\rm{(de,past)}}  =  -\delta, \quad \quad \quad  \; \rightarrow \quad  \; \; \zeta_{\text{past}} = 0 \;\text{(solves the coincidence problem)}.\\
w^{\rm{eff}}_{\rm{(dm,future)}}  &= -\delta, 
\; \;w^{\rm{eff}}_{\rm{(de,future)}}   = w,  \quad \rightarrow \quad \zeta_{\text{future}}   \approx -3(w+\delta) \;\text{(alleviates coincidence problem)}. \\
\end{split}
\label{eq:omega_eff_dm_de_Qdm_subbed_past_future}
\end{equation}
This interaction model will therefore \textbf{always solve the coincidence problem in the past, while alleviating it in the future if $\delta>0$ (iDEDM regime)}. The results obtained in equations \eqref{eq:r_dm_subbed_past_future} to \eqref{eq:omega_eff_dm_de_Qdm_subbed_past_future} are plotted in Figure \ref{fig:CP+omega_dmde_Qdm}.

\begin{figure}[htbp]    \centering
\begin{subfigure}[b]{0.505\linewidth}
        \centering
        \includegraphics[width=\linewidth]{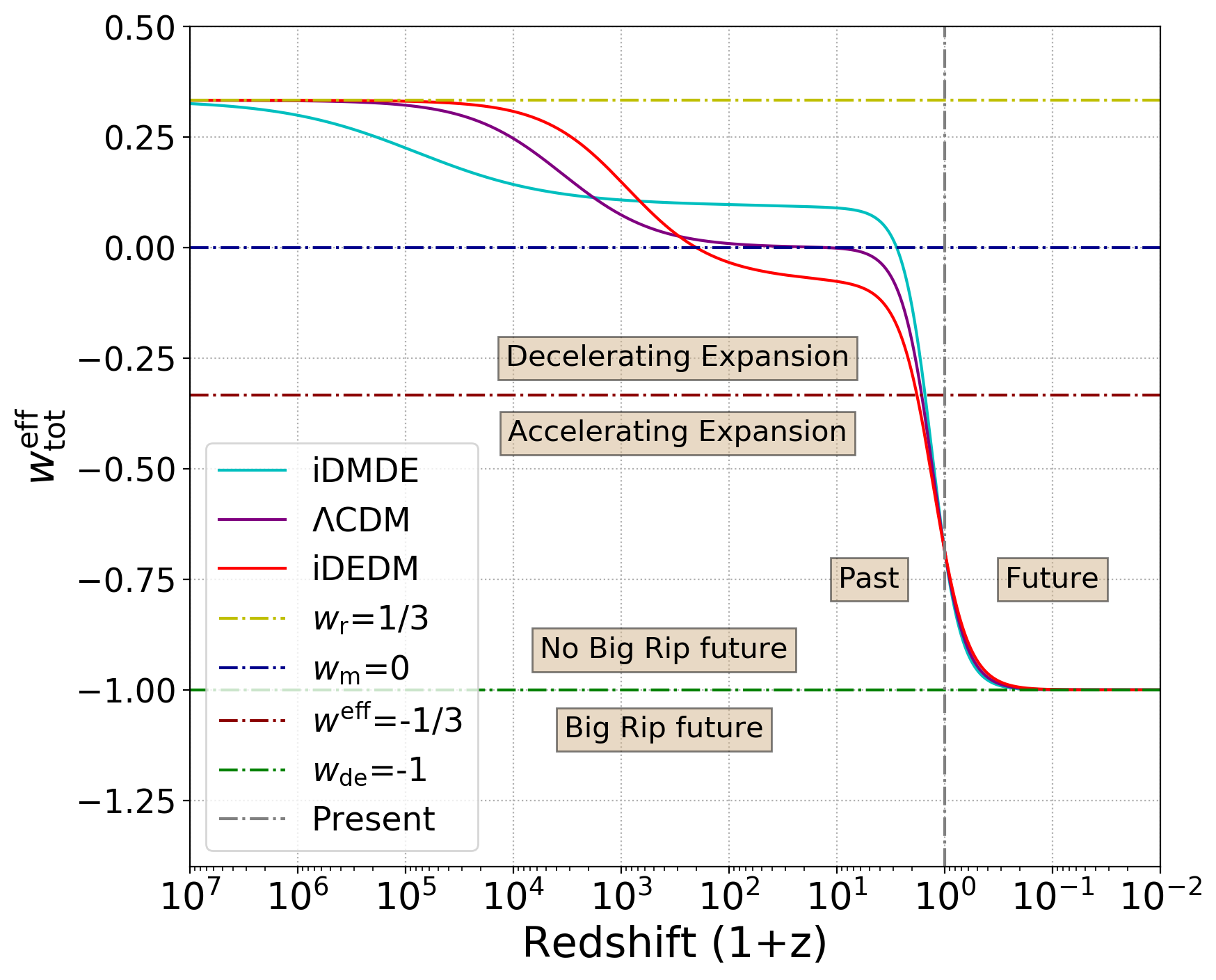}
        \label{fig:eos_tot_Qdm}
    \end{subfigure}    
    \begin{subfigure}[b]{0.485\linewidth}
        \centering
        \includegraphics[width=\linewidth]{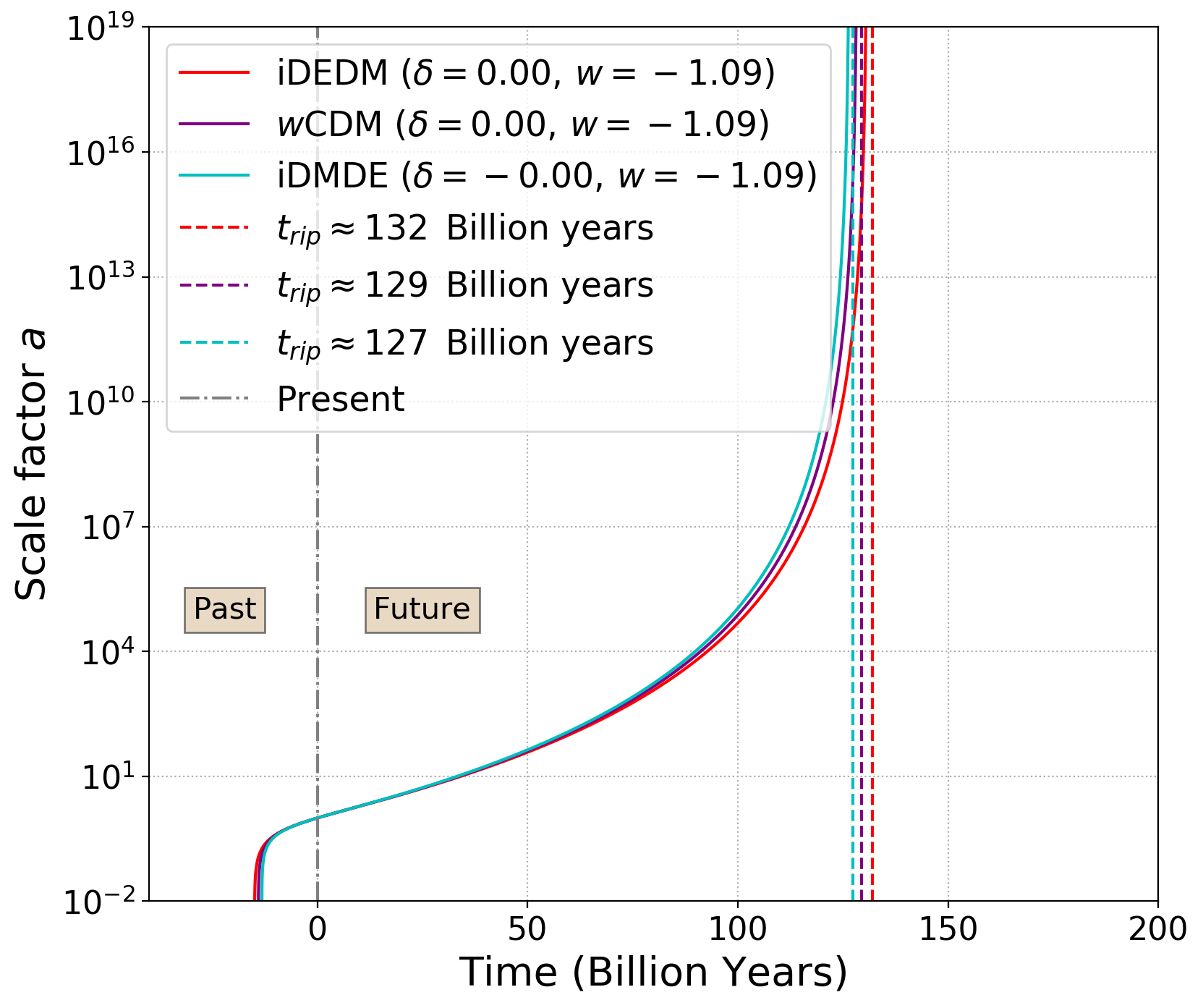}
        \label{fig:Big_rip_Qdm}
    \end{subfigure}    
    \hspace{0pt} 
    \caption{Total effective equation of state $w^{\rm{eff}}_{\rm{tot}}$ and Big Rip future singularities — $Q = 3H \delta \rho_{\rm{dm}}$, with $w = -1$ (left panel) and $w = -1.09$ (right panel). In both the iDEDM regime ($\delta = +0.1$) and iDMDE regime ($\delta = -0.1$), the effect of the interaction diminishes in the asymptotic future and $w^{\rm{eff}}_{\rm{tot}} = w$, which implies that we will always have $w^{\rm{eff}}_{\rm{tot}} < -1$ if $w < -1$, thus guaranteeing a future Big Rip singularity for both cases.}
    \label{fig:eos_tot_BR_Qdm}
\end{figure}

The phantom-crossing redshift is given by: 
\begin{equation}
\begin{split}
z_{\text{pc}}&= \left[\frac{\delta \left(1 - \delta \right) \Omega_{\text{(dm,0)}}  }{\left( 1 +w  \right)\left[\Omega_{\text{(de,0)}} (\delta+ w) + \Omega_{\text{(dm,0)}} \delta \right]}\right]^{\frac{1}{3(\delta+w)}}-1. \\
\end{split}
\label{eq:z_pc_dm_BG}
\end{equation}
As seen from \eqref{eq.Qdm.omega_eff_dm_de}, \eqref{eq:omega_eff_dm_de_Qdm_subbed_past_future} and Figure \ref{fig:CP+omega_dmde_Qdm}, given specific values of $w$ and small coupling $\delta$, we have two possibilities for the direction of the phantom-crossing (pc), with one case plagued by negative energies:
\begin{equation}
\begin{split}
 \text{pc direction} \begin{cases}
   \text{iDMDE: Divergent pc for  $w^{\rm{eff}}_{\rm{de}}$ at $z_{\text{(de=0)}}$ \eqref{eq:rhode_z_min_Q_dm_BG}; with } \rho_{\rm{de}}<0. \\
   \text{iDEDM: Quintessence } (w^{\rm{eff}}_{\rm{(de,past)}}>-1)  \rightarrow   \text{Phantom }   (w^{\rm{eff}}_{\rm{(de,future)}}<-1)\text{; with } \rho_{\rm{dm/de}}>0.
\end{cases}
\end{split}
\end{equation}
We also have $q=\frac{1}{2} \left(1+3w^{\rm{eff}}_{\rm{tot}} \right)$ and $w^{\rm{eff}}_{\rm{tot}}=w^{\rm{eff}}_{\rm{de}}=w$ in the distant future, which leads to the following condition needed for a big rip to occur:
\begin{gather} \label{omega_eff_tot_dm_BG}
\begin{split}
\text{Big rip condition: \quad} w^{\rm{eff}}_{\rm{(tot,future)}} &\approx w <-1.
\end{split}
\end{gather}
This implies that the effective equations of state will be the same as in any uncoupled model in the distant future, as seen in Figure \ref{fig:eos_tot_BR_Qdm}. In the case where $w^{\rm{eff}}_{\rm{tot}}=w<-1$, the universe will experience a big rip future singularity at time $t_{\text{rip}}$:
\begin{equation}
\begin{split}
t_{\text{rip}}-t_0 &\approx -\frac{2}{3 H_0 (w+1)}\frac{1}{\sqrt{\Omega_{\text{(de,0)}} + \Omega_{\text{(dm,0)}} \frac{\delta}{(\delta+w)} }}.
\end{split}
\label{eq:Big_Rip_dm_BG}
\end{equation}

The time of the big rip predicted using \eqref{eq:Big_Rip_dm_BG} is indicated by the dashed lines in Figure~\ref{fig:eos_tot_BR_Qdm}, which is in agreement with the point where the scale factor $a \rightarrow \infty$ within a finite time. We may note that for phantom DE $(w<-1)$, a big rip is inevitable.

\subsection{Linear IDE model 5: \(Q= 3 H \delta\rho_{\text{de}}\)} \label{Q_de}

\begin{figure}
    \centering
    \includegraphics[width=0.85\linewidth]{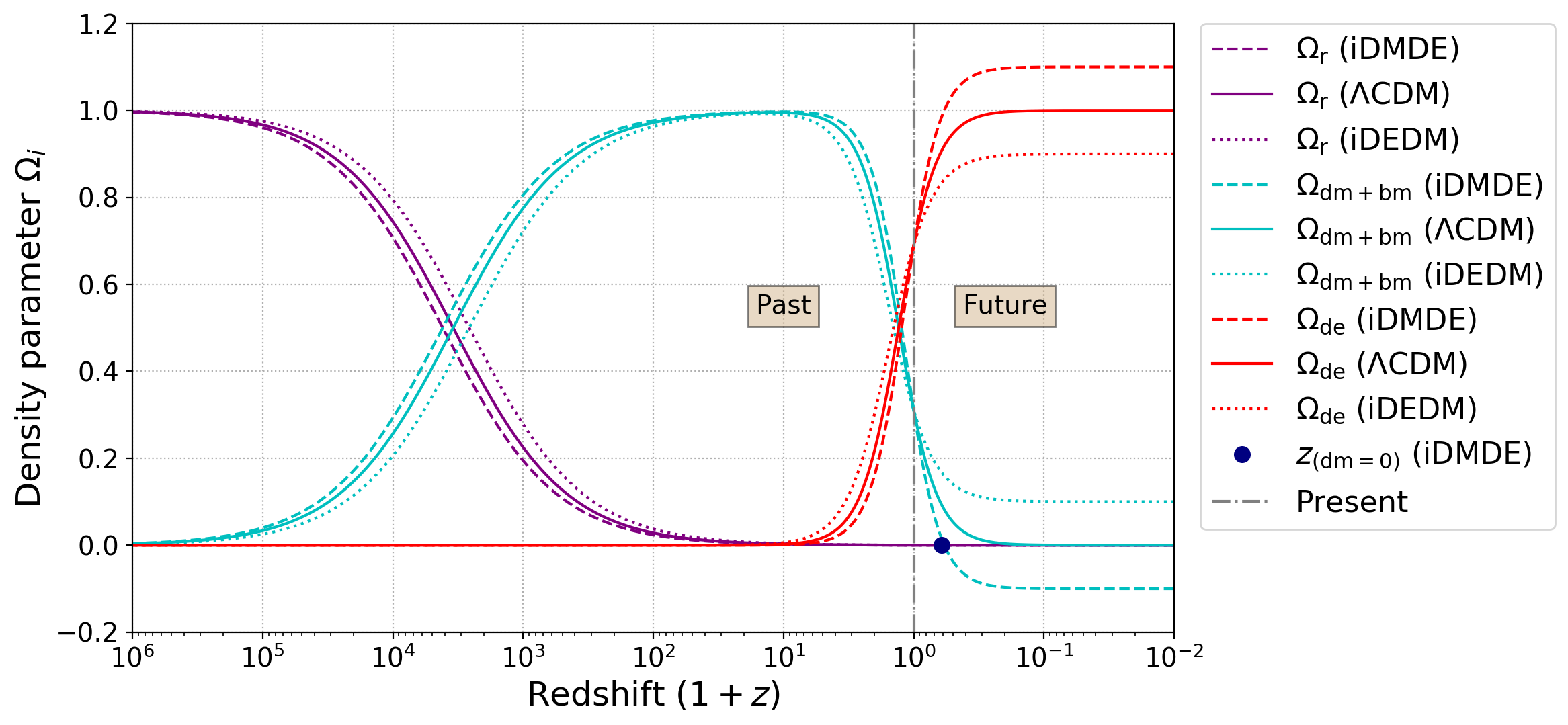}
    \caption{Density parameters vs. redshift — $Q = 3\delta H \rho_{\text{de}}$, with positive energy densities only found in the iDEDM regime ($\delta = +0.1$), while negative DM densities (in the future) are always present in the iDMDE regime ($\delta = -0.1$).}
    \label{fig:Omega_Linear_de}
\end{figure}

This interaction mostly changes the dynamics for the late-time expansion and the distant future during DE domination, as seen in Figure \ref{fig:Q_Linear_dm+de} and the other figures in this section. For this interaction $Q = 0$ if $\rho_{\text{de}} = 0$, therefore we can guarantee that $\rho_{\text{de}} \ge 0$ at all times. In contrast, this interaction will have future negative DM densities in the iDMDE regime, but this can be avoided with a sufficiently small interaction in the iDEDM regime, given by the conditions in \eqref{DSA.Q.de.confirmed}, as seen in Figure \ref{fig:Omega_Linear_de}. The results in this section were discussed in detail in~\cite{vanderWesthuizen:2023hcl} and are presented here to show convergence with published results.  
Substituting the expressions for $m_+$, $m_-$, and $\Delta$ for the interaction from Table \ref{tab:m+-} into the general expressions for $\rho_{\text{dm}}$ \eqref{eq:rhodm_general} and $\rho_{\text{de}}$ \eqref{eq:rhode_general} gives the familiar expressions:
\begin{gather}
\begin{split} \label{eq:rho_dm_Q_de}
\rho_{\text{dm}} =  \left(\rho_{\text{(dm,0)}}  + \rho_{\text{(de,0)}} \left(\frac{\delta}{\delta+w} \right) \left[1-  a^{-3(\delta + w)} \right] \right) a^{-3},
\end{split}   
\end{gather}
\begin{gather}
\begin{split} \label{eq:rho_de_Q_de}
\rho_{\text{de}} = \rho_{\text{(de,0)}} a^{-3(\delta + w + 1)}.
\end{split}   
\end{gather}

\begin{gather} \label{DSA.Q.de.confirmed}
\boxed{
\begin{split}
&\underline{\text{Conditions for}  \text{ $\rho_{\rm{dm}}\ge0 \; ; \; \rho_{\rm{de}} \ge0 $ for the entire cosmological evolution:}}
 \\& \quad  \quad \quad \quad \quad \quad \text{iDEDM with } \
0\le\delta \le -\frac{w }{\left(1+\frac{1}{r_0}\right)} .
\end{split} }
\end{gather}
We also note that for $\rho_{\rm{dm}}$ \eqref{eq:rho_dm_Q_de} to be defined, we require $\delta \ne -w$. Similarly, substituting the expressions for $m_+$, $m_-$, and $\Delta$ for the interaction from Table \ref{tab:m+-} into the relevant expressions in Section \ref{More_analytical_relations} gives the corresponding expressions below.
\begin{equation}
\begin{split}
\Omega_{\text{(dm,past)}}=1 \quad &; \quad \Omega_{\text{(de,past)}}=0,\\
\Omega_{\text{(dm,future)}}=-\frac{\delta}{w} \quad &; \quad \Omega_{\text{(de,future)}}=  1+\frac{\delta}{w}.\\
\end{split}\label{eq:frac_den_de_past}
\end{equation}
The DE always remains positive for this model, while the DM density becomes zero at redshift $z_{\text{(dm=0)}}$:
\begin{equation}
\begin{split}
z_{\text{(dm=0)}}&=\left[1+\frac{\Omega_{\text{(dm,0)}}}{\Omega_{\text{(de,0)}}} \left(\frac{ \delta+w  }{\delta } \right)\right]^{\frac{1}{3(\delta+w)}}-1. \\
\end{split}
\label{eq:rhodm_z_min_Q_de_BG}
\end{equation}
The redshift at which the dark matter–dark energy equality will occur is given by:
\begin{equation}
\begin{split}
&   z _{\text{(dm=de)}}  = \left[\frac{ \Omega_{\text{(de,0)}}(\delta) +\Omega_{\text{(dm,0)}} (\delta+w)}{\Omega_{\text{(de,0)}} (2\delta+w )    } \right]^{\frac{1}{3 (\delta+w)}}-1. \\
\end{split}
\label{eq:rhodm=de_de}
\end{equation}
This model addresses the coincidence problem by letting $r$ converge to the following in the past and future:
\begin{equation}
\begin{split}
r_{\text{past}} (a\rightarrow0)\approx\infty\quad ; \quad 
r_{\text{future}} (a\rightarrow\infty)\approx -\frac{\delta}{(\delta+w)}.
\end{split}
\label{eq:r_de_subbed_past_future}
\end{equation}
The DM and DE effective equations of state \eqref{DSA.omega_eff_dm_de} for this interaction kernel are given by:
\begin{gather} \label{eq.Qde.omega_eff_dm_de}
\begin{split}
w^{\rm{eff}}_{\rm{dm}} &= - \frac{\delta  \rho_{\text{de}}}{\rho_{\rm{dm}}}= -\frac{\delta}{r} \quad ; \quad 
w^{\rm{eff}}_{\rm{de}} =    w+\delta. 
\end{split}
\end{gather}
Expressions for $w^{\rm{eff}}_{\rm{dm}}$ and $w^{\rm{eff}}_{\rm{de}}$ in the asymptotic distant past and future are obtained by substituting \eqref{eq:r_dm+de_subbed_past_future} into \eqref{eq.Qdm+de.omega_eff_dm_de}:
\begin{equation}
\begin{split}
w^{\rm{eff}}_{\rm{(dm,past)}}  &= 0, \; \;w^{\rm{eff}}_{\rm{(de,past)}}   = w+\delta, \;  \quad \rightarrow \quad \zeta_{\text{past}} \approx -3(w+\delta)\;\text{(alleviates coincidence problem)}. \\
w^{\rm{eff}}_{\rm{(dm,future)}} &=w^{\rm{eff}}_{\rm{(de,future)}}  =  w+\delta , \quad \quad \rightarrow \quad  \zeta_{\text{future}} = 0 \; \text{(solves the coincidence problem)}. 
\end{split}
\label{eq:omega_eff_dm_de_Qde_subbed_past_future}
\end{equation}
This interaction model will therefore always \textbf{solve the coincidence problem in the future, while alleviating it in the past if $\delta>0$ (iDEDM regime)}. The results obtained in equations \eqref{eq:r_de_subbed_past_future} to \eqref{eq:omega_eff_dm_de_Qde_subbed_past_future} are plotted in the panels of Figure \ref{fig:CP+omega_dmde_Qde} below.

\begin{figure}[htbp]
    \centering
    \begin{subfigure}[b]{0.503\linewidth}
        \centering
        \includegraphics[width=\linewidth]{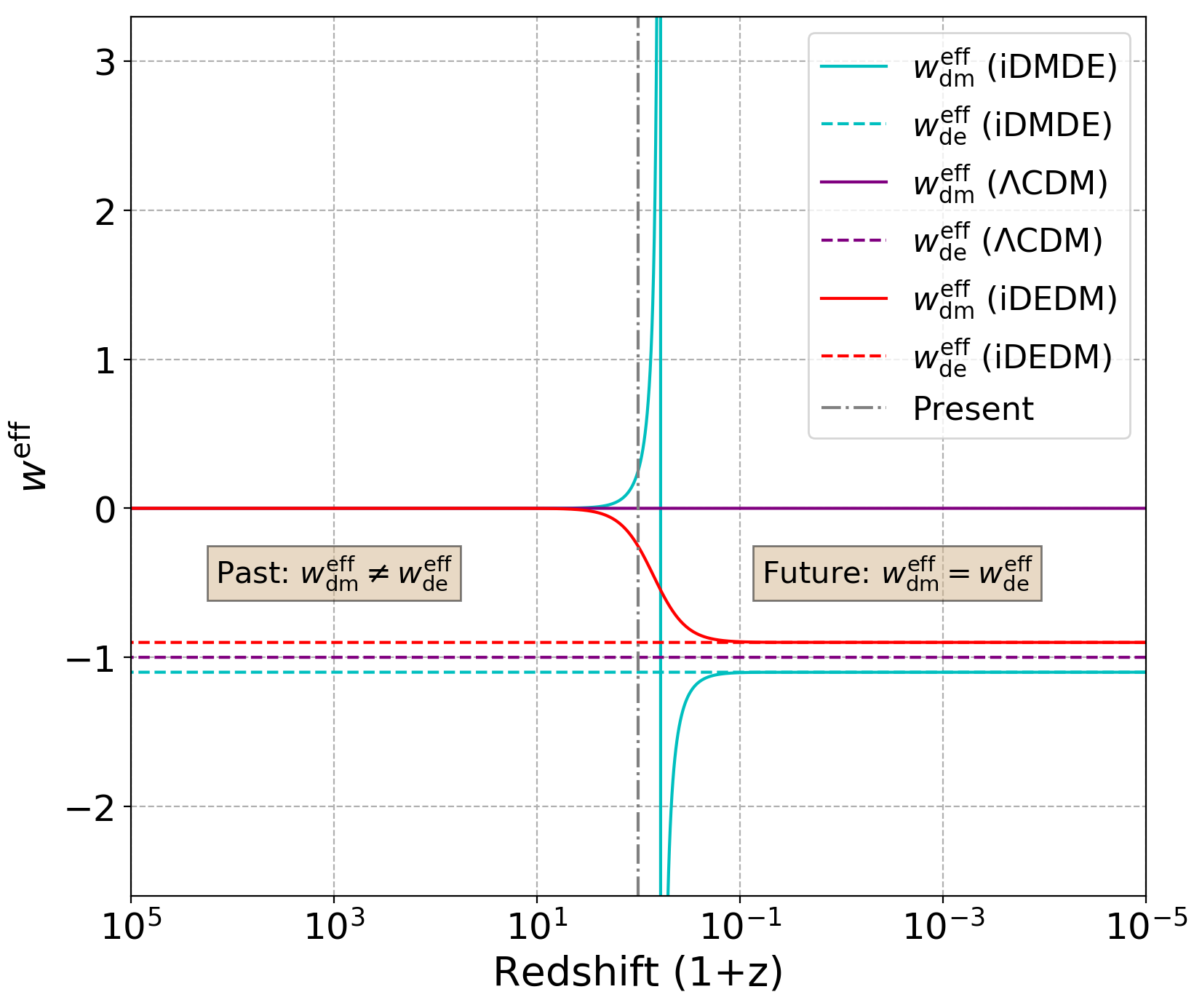}
        \label{fig:omega_dmde_Qde}
    \end{subfigure}    
    \hspace{0pt} % No extra space between subfigures
    \begin{subfigure}[b]{0.482\linewidth}
        \centering
        \includegraphics[width=\linewidth]{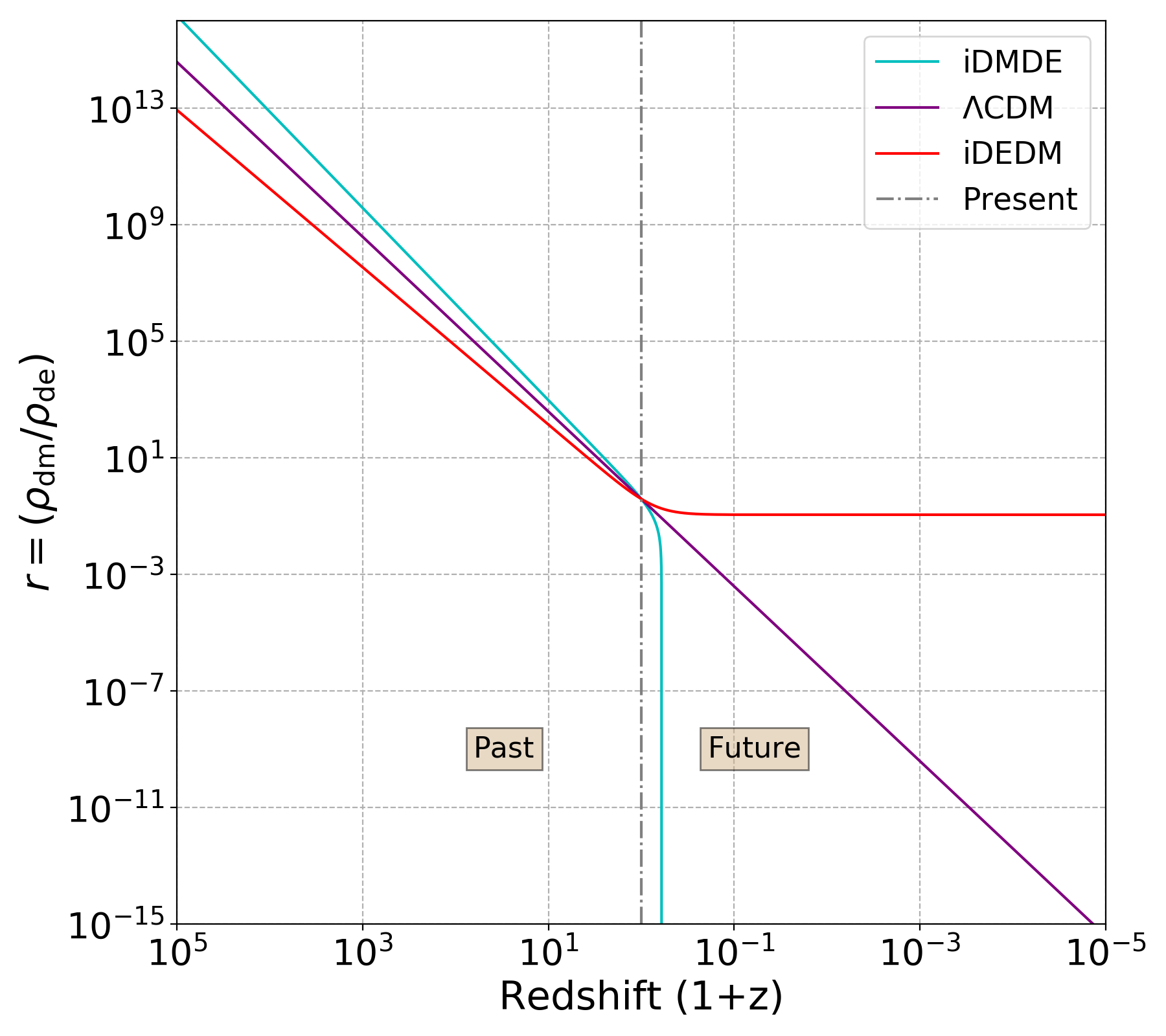}
        \label{fig:CP_Qde}
    \end{subfigure}%   
    \caption{Effective equations of state and Coincidence Problem (CP) vs redshift - $Q=3H \delta \rho_{\rm{de}}$, with $w^{\rm{eff}}_{\rm{dm}}=w^{\rm{eff}}_{\rm{de}}$ in the future, thus solving the CP ($r=\text{constant}$) in the future and alleviating it in the past for the iDEDM regime ($\delta=+0.1$). In the iDMDE regime ($\delta=-0.1$), negative DM densities and divergent $w^{\rm{eff}}_{\rm{dm}}$ (in the future) are always present, and the CP is worsened for the past.}
    \label{fig:CP+omega_dmde_Qde}
\end{figure}

From \eqref{eq.Qde.omega_eff_dm_de} and Figure \ref{fig:CP+omega_dmde_Qde}, we can see that the DE effective equation of state is constant; therefore, no DE phantom-crossing is possible for this interaction function, while in the iDMDE regime $w^{\rm{eff}}_{\rm{dm}}$ may still experience a divergent phantom-crossing at $z_{\text{(dm=0)}}$ \eqref{eq:rhodm_z_min_Q_de_BG}.
We also have $q=\frac{1}{2} \left(1+3w^{\rm{eff}}_{\rm{tot}} \right)$ and $w^{\rm{eff}}_{\rm{tot}}=w^{\rm{eff}}_{\rm{dm}}=w^{\rm{eff}}_{\rm{de}}=w+\delta$ in the distant future. This leads to the following condition needed for a big rip to occur:
\begin{gather} \label{omega_eff_tot_de_BG}
\begin{split}
\text{Big rip condition: \quad  } w^{\rm{eff}}_{\rm{(tot,future)}} \approx w +\delta <-1 \quad \rightarrow \quad \delta<-w-1.
\end{split}
\end{gather}
In the case where $w^{\rm{eff}}_{\rm{tot}}<-1$, the universe will experience a big rip future singularity at time $t_{\text{rip}}$:
\begin{equation}
\begin{split}
t_{\text{rip}}-t_0&\approx -\frac{2}{3 H_0 (\delta+w+1)}\sqrt{\frac{ \delta+w}{\Omega_{\text{(de,0)}} w  }  }. \\
\end{split}\label{eq:Big_Rip_de_BG}
\end{equation}

\begin{figure}[htbp]    \centering
\begin{subfigure}[b]{0.505\linewidth}
        \centering
        \includegraphics[width=\linewidth]{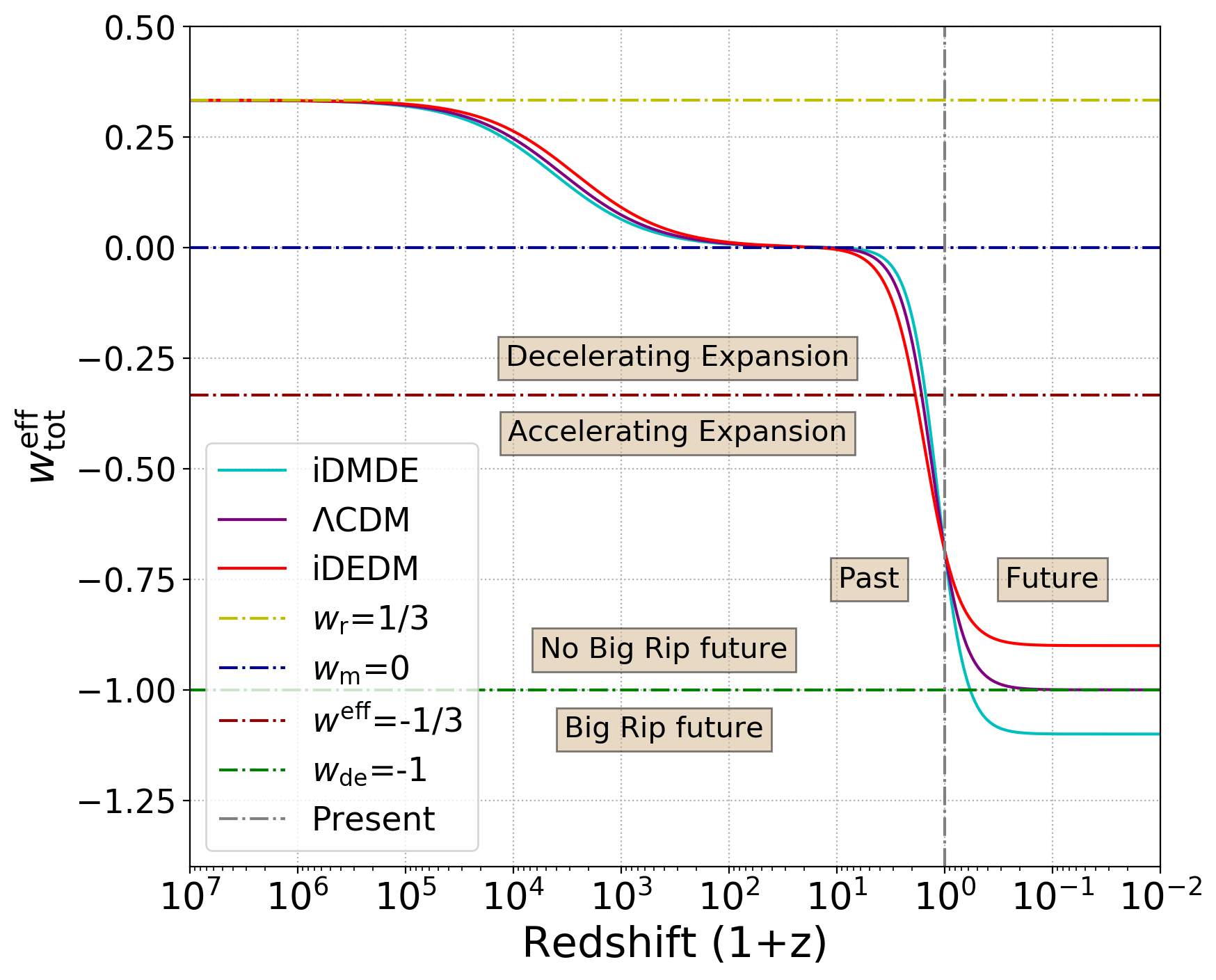}
        \label{fig:eos_tot_Qde}
    \end{subfigure}    
    \begin{subfigure}[b]{0.485\linewidth}
        \centering
        \includegraphics[width=\linewidth]{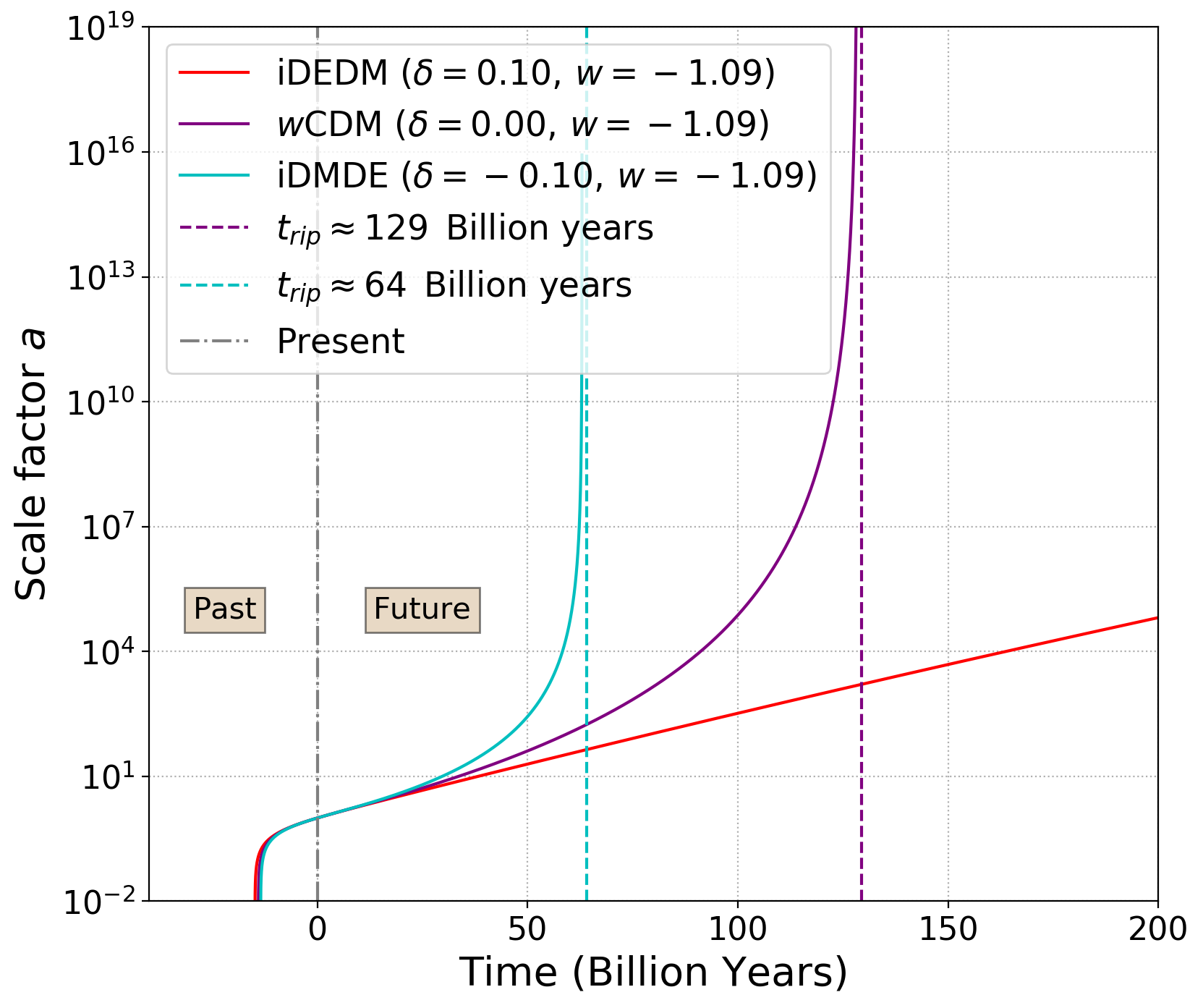}
        \label{fig:Big_rip_Qde}
    \end{subfigure}    
    \hspace{0pt} % No extra space between subfigures
    \hspace{-1cm}
    \caption{Total effective equation of state $w^{\rm{eff}}_{\rm{tot}}$ and big rip future singularities — $Q=3H \delta \rho_{\rm{de}}$, with $w=-1$ (left panel) and $w=-1.09$ (right panel). In the iDEDM regime ($\delta=+0.1$), in the asymptotic future we may have $w^{\rm{eff}}_{\rm{tot}}>-1$, even if $w<-1$, thus avoiding a big rip. In the iDMDE regime ($\delta=-0.1$), in the asymptotic future we will always have $w^{\rm{eff}}_{\rm{tot}}<-1$ if $w<-1$, thus guaranteeing a future big rip singularity.}
    \label{fig:eos_tot_BR_Qde}
\end{figure}

The effect of the coupling on $w^{\rm{eff}}_{\rm{tot}}$ and how the interaction may cause or avoid a big rip can be seen in Figure \ref{fig:eos_tot_BR_Qde}. The time of the big rip predicted using \eqref{eq:Big_Rip_de_BG} is indicated by the dashed lines in Figure \ref{fig:eos_tot_BR_Qde}, which is in agreement with where the scale factor $a\rightarrow\infty$ within a finite time. We may note that for phantom DE, a big rip can only be avoided in the iDEDM regime.

\section{Dark interactions as a dynamical dark energy equation of state $\tilde{w}(z)$} \label{reconstructed_w}
\begin{figure}
    \centering
    \begin{subfigure}[b]{0.494\linewidth}
        \centering
        \includegraphics[width=\linewidth]{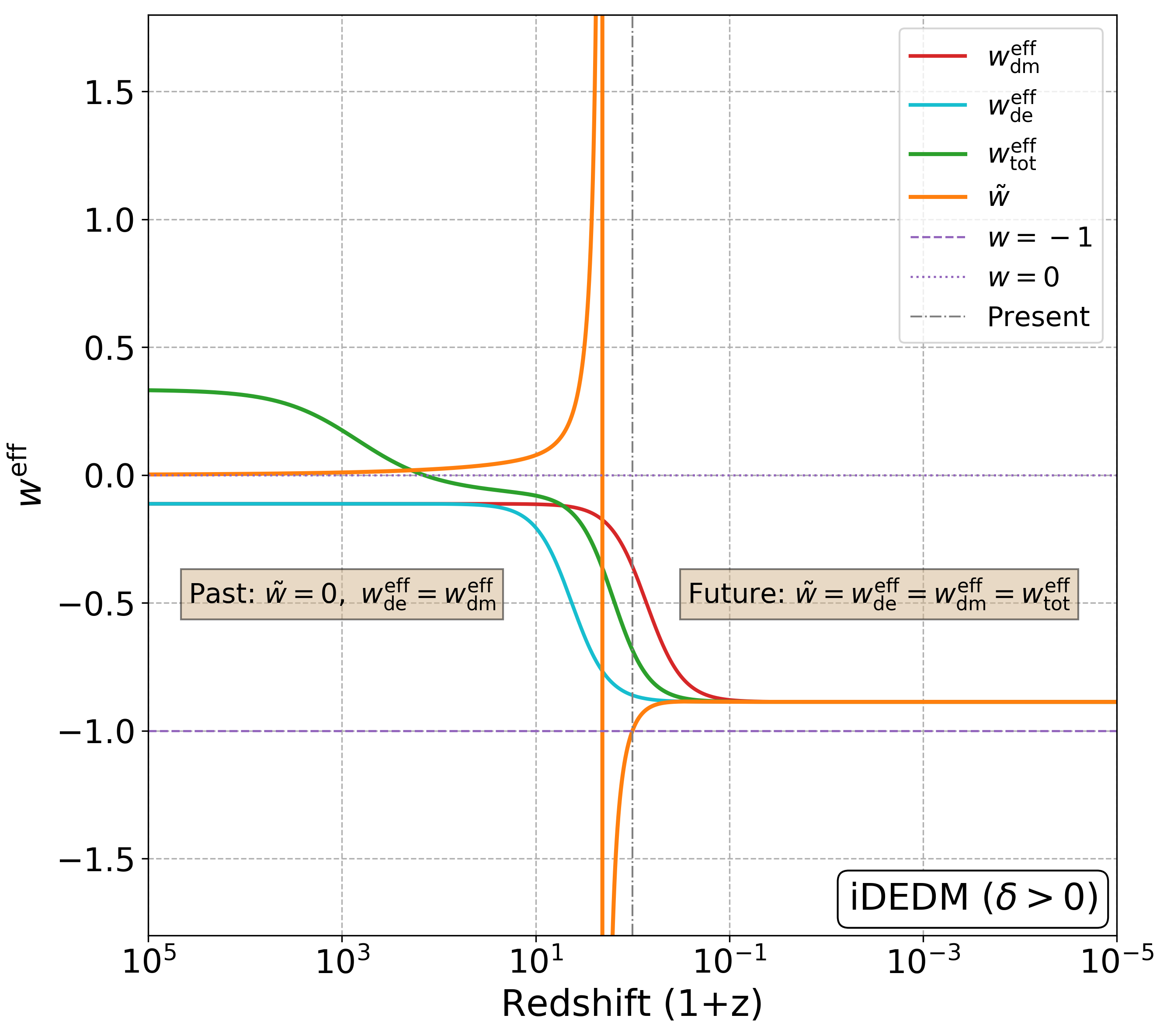}
    \end{subfigure}    
    \hspace{0pt} % No extra space between subfigures
    \begin{subfigure}[b]{0.494\linewidth}
        \centering
        \includegraphics[width=\linewidth]{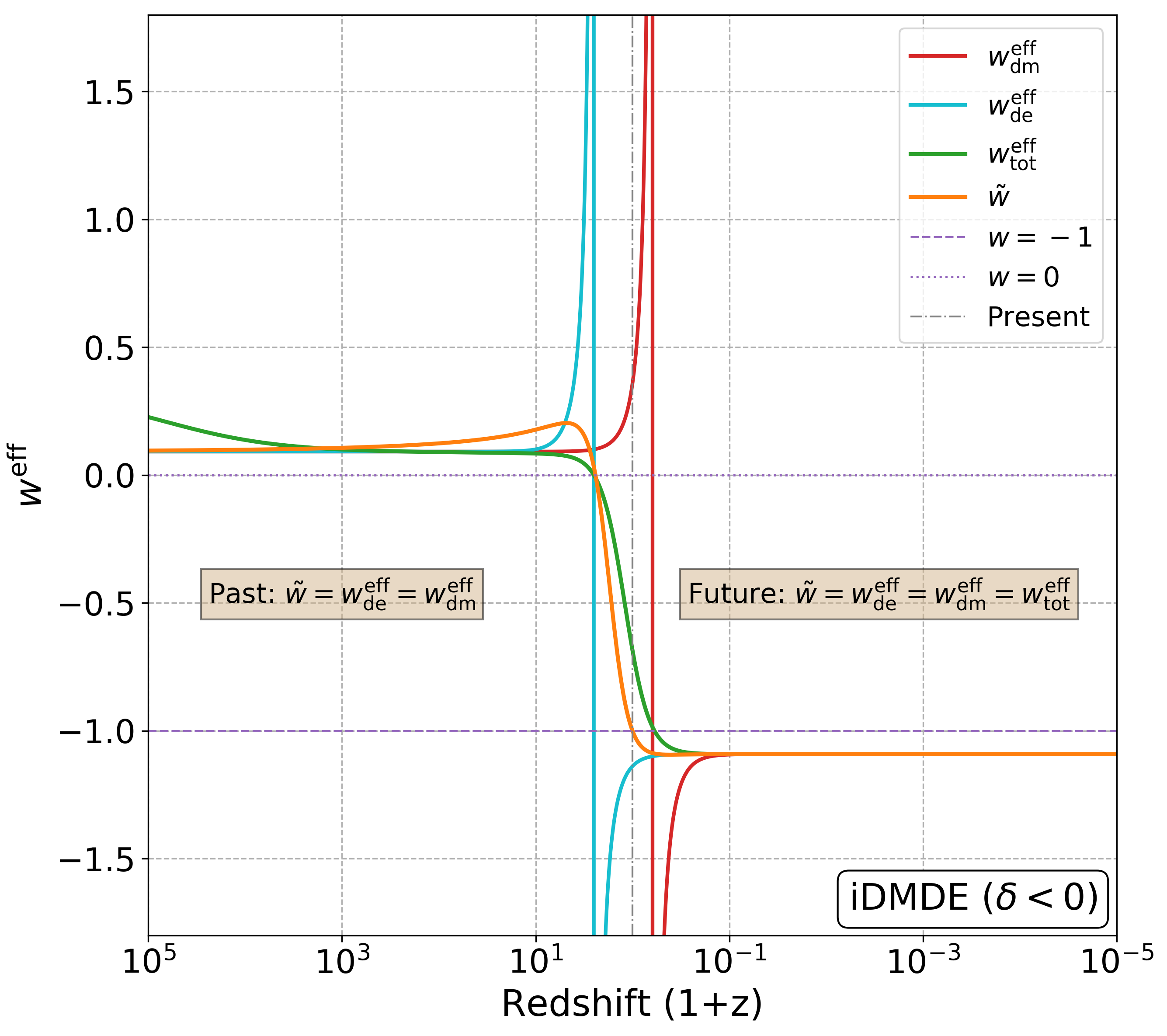}
    \end{subfigure}%
    \caption{Equations of state $\tilde{w}$, $w^{\rm eff}_{\rm de}$, $w^{\rm eff}_{\rm dm}$, $w^{\rm eff}_{\rm tot}$, and $w$ vs.\ redshift — $Q=3H\delta\left(\rho_{\rm dm}+\rho_{\rm de}\right)$. The left panel shows the iDEDM regime ($\delta=+0.1$) where only $\tilde{w}(z)$ exhibits a divergent phantom crossing. Conversely, the right panel shows the iDMDE regime ($\delta=-0.1$) where both $w^{\rm eff}_{\rm de}$ and $w^{\rm eff}_{\rm dm}$ have a divergent phantom crossing, due to $\rho_{\rm dm}$ or $\rho_{\rm de}$ becoming negative in the effective split. Additionally, for both cases we have $\tilde{w}(0)=w$, and in the asymptotic future $\tilde{w}=w^{\rm eff}_{\rm de}=w^{\rm eff}_{\rm dm}=w^{\rm eff}_{\rm tot}$. In the asymptotic past, the iDEDM regime has $\tilde{w}=0$ and $w^{\rm eff}_{\rm de}=w^{\rm eff}_{\rm dm}$, while the iDMDE regime has $\tilde{w}=w^{\rm eff}_{\rm de}=w^{\rm eff}_{\rm dm}$.}
    \label{fig:w_all_Qdm+de}
\end{figure}

\begin{figure}
    \centering
    \begin{subfigure}[b]{0.494\linewidth}
        \centering
        \includegraphics[width=\linewidth]{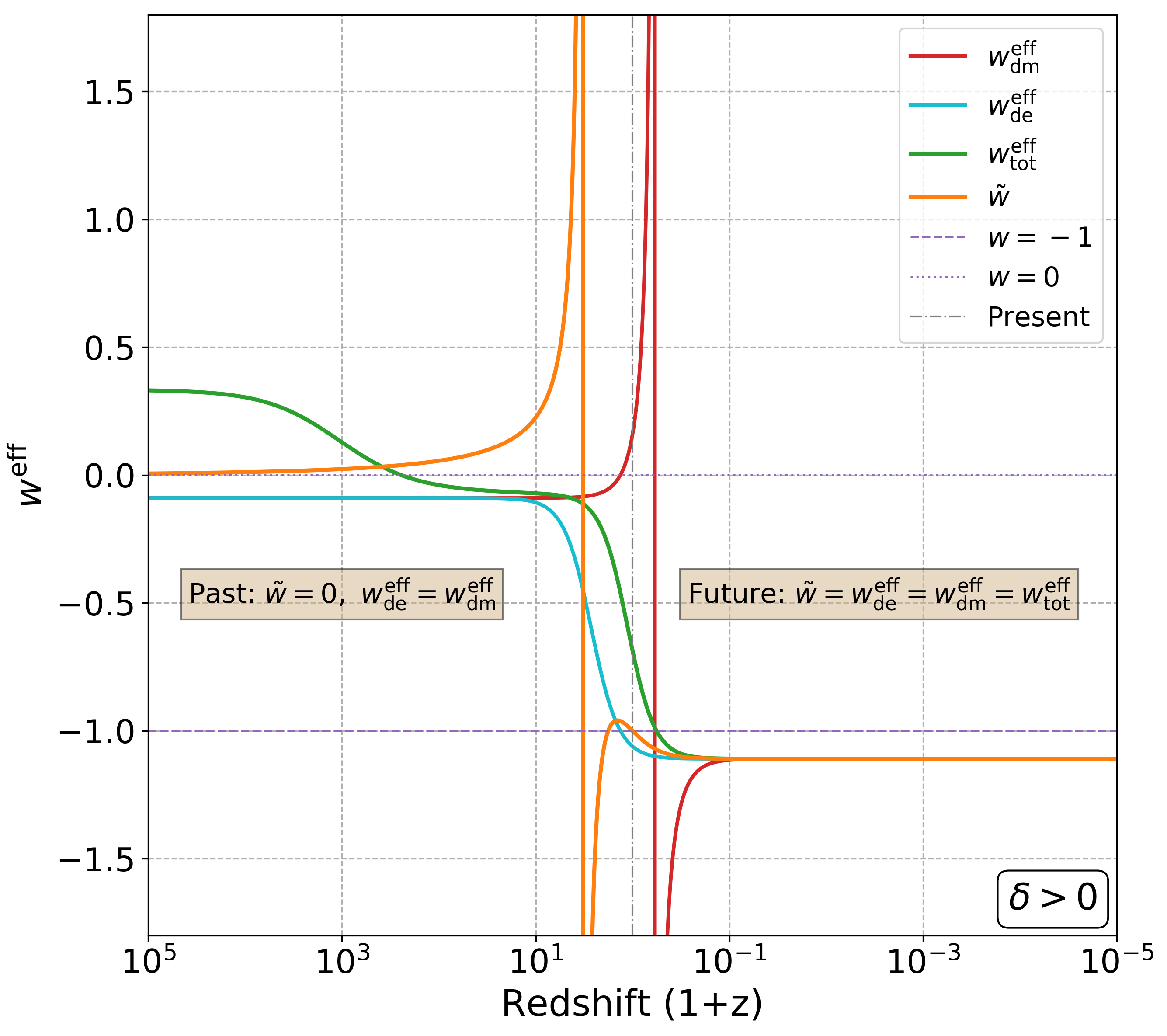}
    \end{subfigure}    
    \hspace{0pt} % No extra space between subfigures
    \begin{subfigure}[b]{0.494\linewidth}
        \centering
        \includegraphics[width=\linewidth]{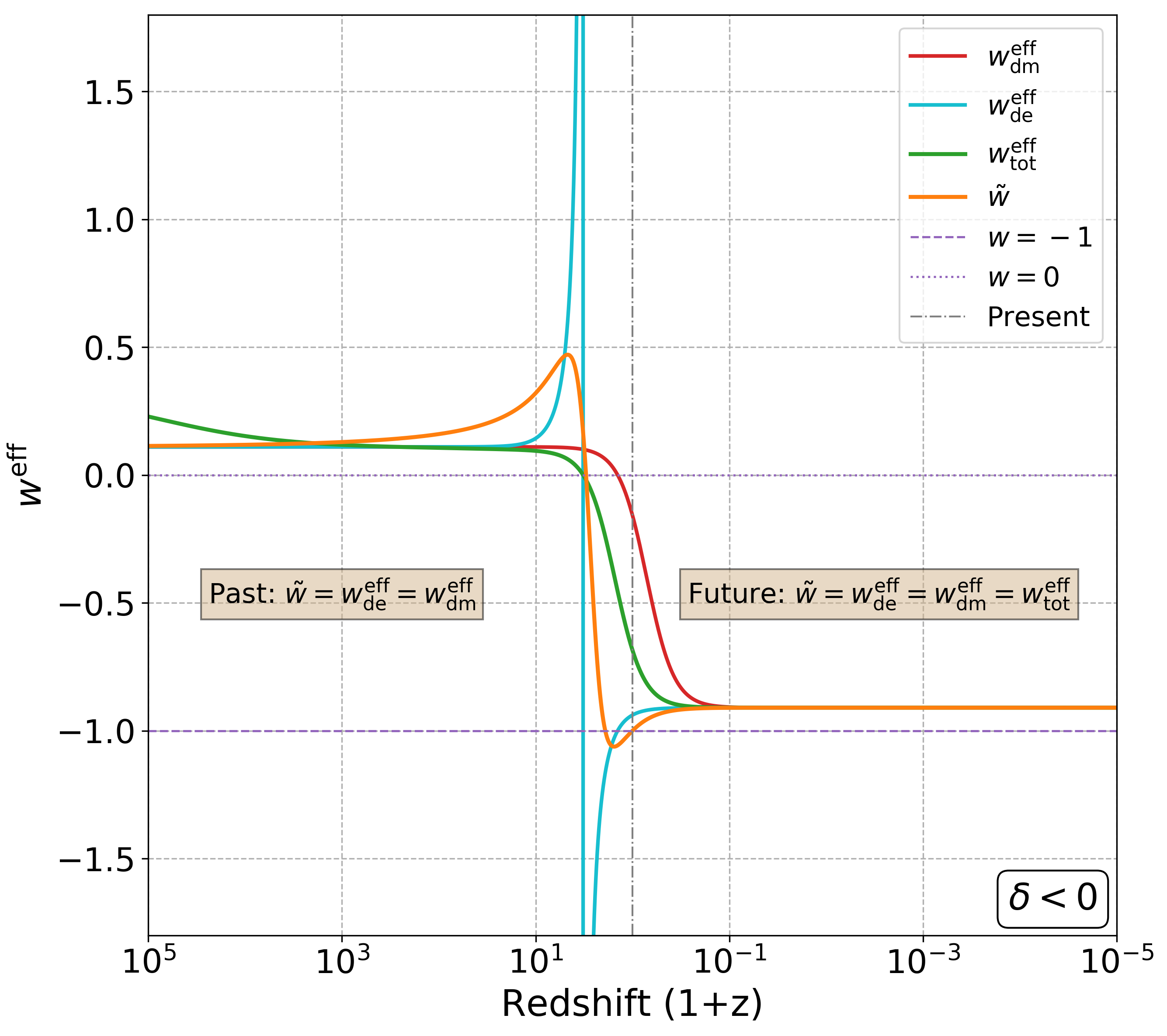}
    \end{subfigure}%
    \caption{Equations of state $\tilde{w}$, $w^{\rm eff}_{\rm de}$, $w^{\rm eff}_{\rm dm}$, $w^{\rm eff}_{\rm tot}$, and $w$ vs.\ redshift — $Q=3H\delta\left(\rho_{\rm dm}-\rho_{\rm de}\right)$. The left panel shows a positive coupling regime ($\delta=+0.1$) where both $\tilde{w}(z)$ and $w^{\rm eff}_{\rm dm}$ exhibit a divergent phantom crossing. The right panel shows a negative coupling ($\delta=-0.1$) where only $w^{\rm eff}_{\rm de}$ has a divergent phantom crossing. Additionally, for both cases we have $\tilde{w}(0)=w$, and in the asymptotic future $\tilde{w}=w^{\rm eff}_{\rm de}=w^{\rm eff}_{\rm dm}=w^{\rm eff}_{\rm tot}$. In the asymptotic past, the positive coupling has $\tilde{w}=0$ and $w^{\rm eff}_{\rm de}=w^{\rm eff}_{\rm dm}$, while the negative coupling has $\tilde{w}=w^{\rm eff}_{\rm de}=w^{\rm eff}_{\rm dm}$.}
    \label{fig:w_all_Qdm-de}
\end{figure}

\begin{figure}
    \centering
    \begin{subfigure}[b]{0.494\linewidth}
        \centering
        \includegraphics[width=\linewidth]{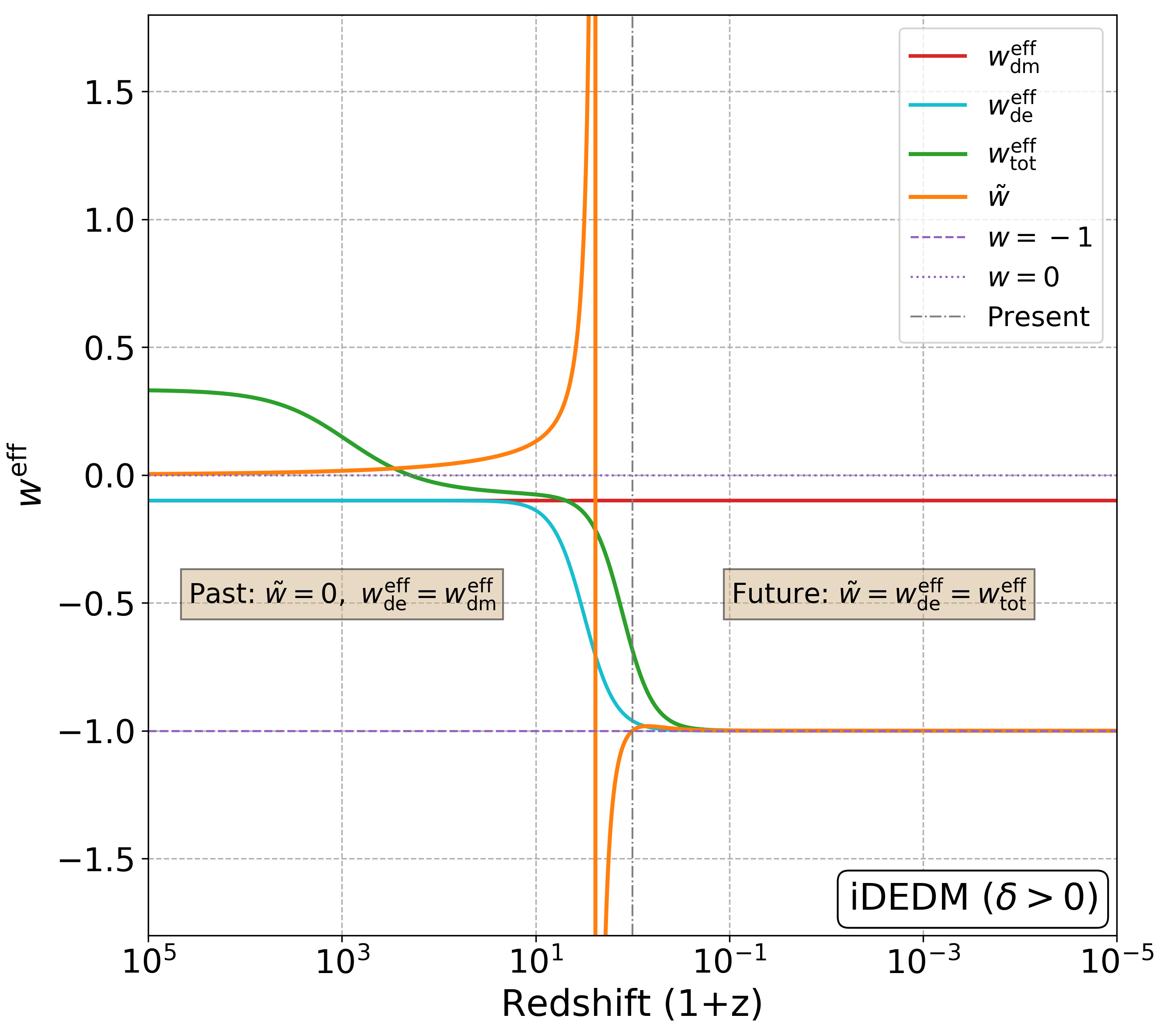}
    \end{subfigure}    
    \hspace{0pt} % No extra space between subfigures
    \begin{subfigure}[b]{0.494\linewidth}
        \centering
        \includegraphics[width=\linewidth]{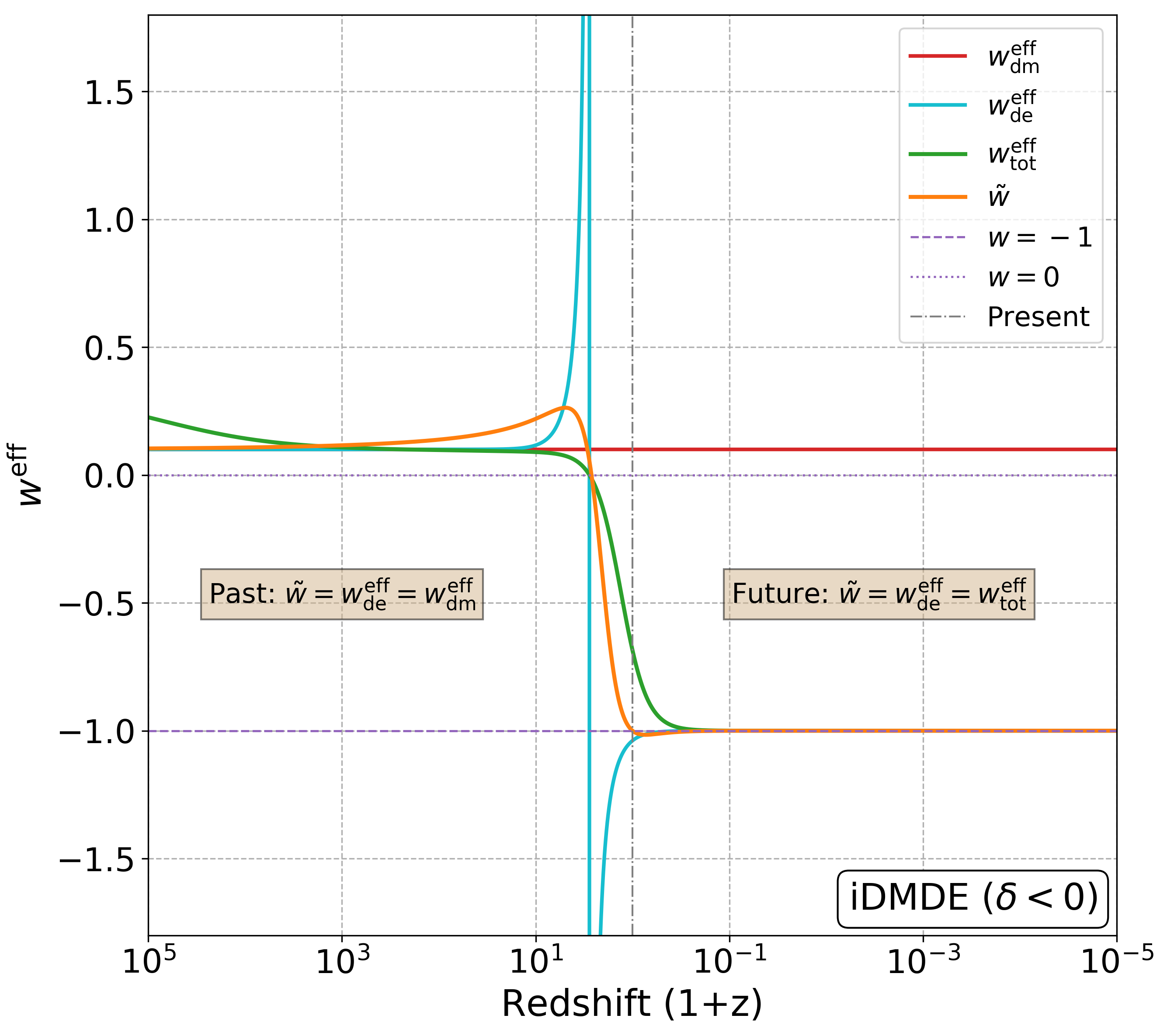}
    \end{subfigure}%
    \caption{Equations of state $\tilde{w}$, $w^{\rm eff}_{\rm de}$, $w^{\rm eff}_{\rm dm}$, $w^{\rm eff}_{\rm tot}$, and $w$ vs.\ redshift — $Q=3H\delta\rho_{\rm dm}$. The left panel shows the iDEDM regime ($\delta=+0.1$) where only $\tilde{w}(z)$ exhibits a divergent phantom crossing. Conversely, the right panel shows the iDMDE regime ($\delta=-0.1$) where only $w^{\rm eff}_{\rm de}$ has a divergent phantom crossing, due to $\rho_{\rm de}$ becoming negative in the effective split. Additionally, for both cases we have $\tilde{w}(0)=w$, and in the asymptotic future $\tilde{w}=w^{\rm eff}_{\rm de}=w^{\rm eff}_{\rm tot}$. In the asymptotic past, the iDEDM regime has $\tilde{w}=0$ and $w^{\rm eff}_{\rm de}=w^{\rm eff}_{\rm dm}$, while the iDMDE regime has $\tilde{w}=w^{\rm eff}_{\rm de}=w^{\rm eff}_{\rm dm}$.}
    \label{fig:w_all_Qdm}
\end{figure}

\begin{figure}
    \centering
    \begin{subfigure}[b]{0.494\linewidth}
        \centering
        \includegraphics[width=\linewidth]{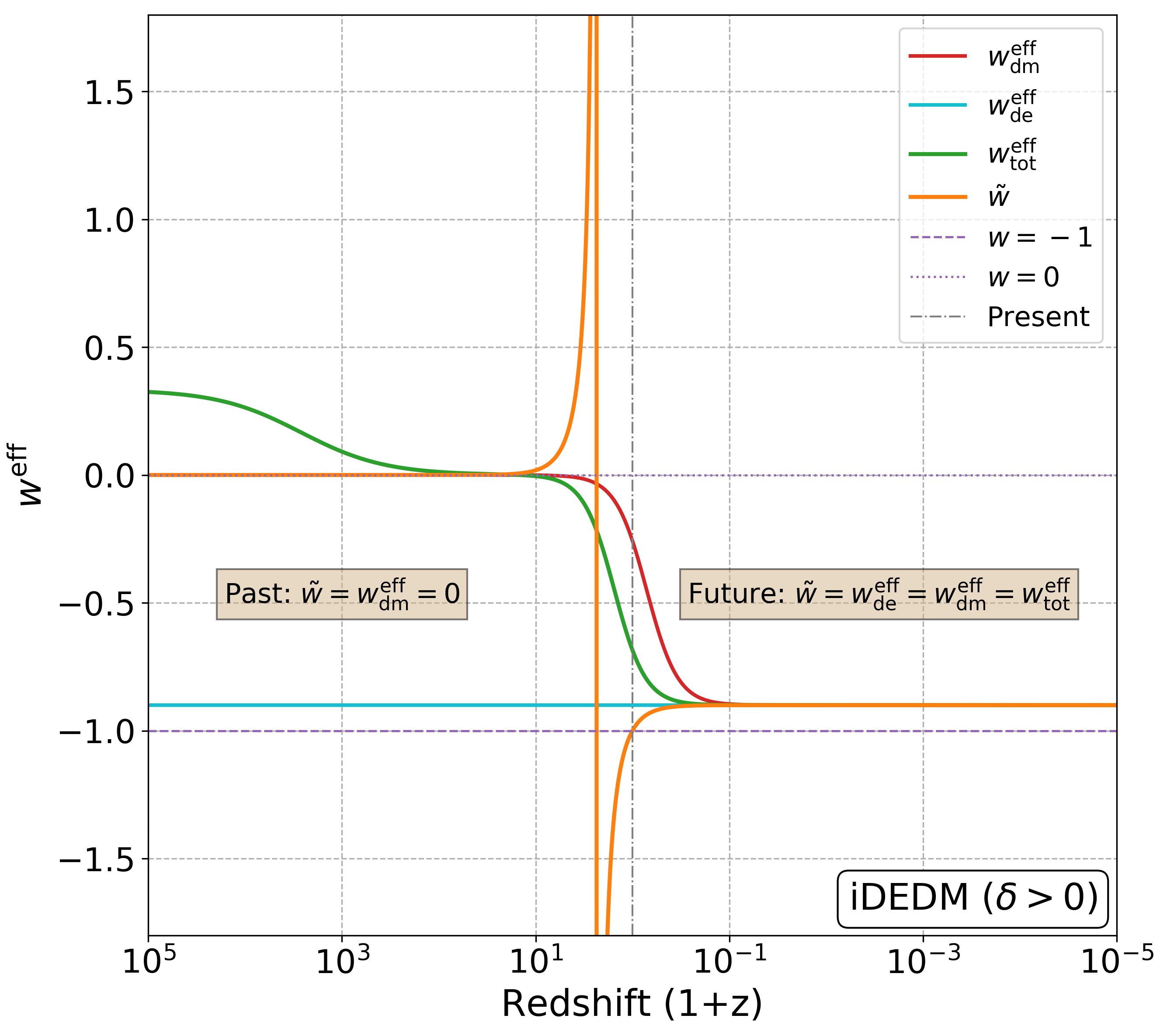}
    \end{subfigure}    
    \hspace{0pt} % No extra space between subfigures
    \begin{subfigure}[b]{0.494\linewidth}
        \centering
        \includegraphics[width=\linewidth]{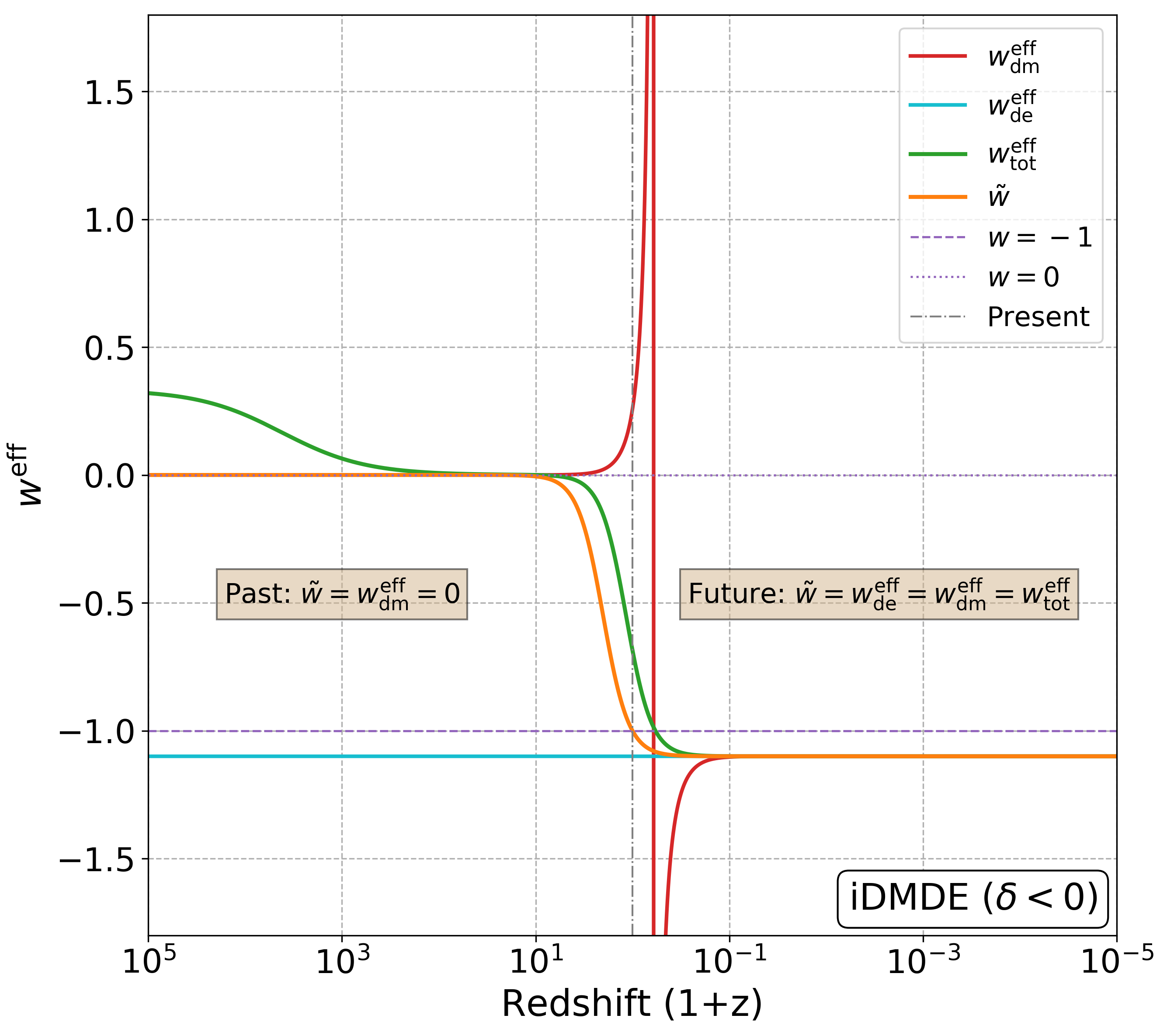}
    \end{subfigure}%
    \caption{Equations of state $\tilde{w}$, $w^{\rm eff}_{\rm de}$, $w^{\rm eff}_{\rm dm}$, $w^{\rm eff}_{\rm tot}$, and $w$ vs.\ redshift — $Q=3H\delta\rho_{\rm de}$. The left panel shows the iDEDM regime ($\delta=+0.1$) where only $\tilde{w}(z)$ exhibits a divergent phantom crossing. Conversely, the right panel shows the iDMDE regime ($\delta=-0.1$) where $w^{\rm eff}_{\rm dm}$ has a divergent phantom crossing, due to $\rho_{\rm dm}$ becoming negative in the effective split. Additionally, for both cases we have 
    $\tilde{w}(0)=w$, the asymptotic past as $\tilde{w}=w^{\rm eff}_{\rm dm}=0$ and oin the asymptotic future $\tilde{w}=w^{\rm eff}_{\rm de}=w^{\rm eff}_{\rm dm}=w^{\rm eff}_{\rm tot}$.}
    \label{fig:w_all_Qde}
\end{figure}

In section \ref{Background_cosmology}, we discussed phantom crossings for the effective DE equation of state in IDE models, but these have been accompanied by an additional effective DM equation of state, making it somewhat difficult to directly compare the behavior to alternative descriptions of DE, like the CPL parametrization, where DM is assumed to still be pressureless. 
To overcome this, we can alternatively express the dynamics of cosmological models as arising from an effective dynamical, redshift dependent equation of state $\tilde{w}(z)$. A familiar example of doing just this for IDE models is found in Section 3.2 of \cite{M.B.Gavela_2009}. We will follow this approach, but include the effect of baryons and radiation, so that we may obtain an expression of  $\tilde{w}(z)$ for each of our models considered here.  In order to reconstruct $\tilde{w}(z)$, we start by noting that the normalised Hubble parameter $h(z)$ for a dynamical dark energy model in a flat universe without any interactions in the dark sector is given by: 
\begin{gather} \label{wz_1}
\begin{split}
h^2(z)&= \Omega_{\text{(r,0)}}(1+z)^{4}+ \Omega_{\text{(bm,0)}}(1+z)^{3}+ \Omega_{\text{(dm,0)}}(1+z)^{3}+  \Omega_{\text{(de,0)}} \text{exp}\left[ 3 \int_0^z dz' \frac{1+\tilde{w}(z')}{1 + z'} \right], \\
\end{split}
\end{gather}
We can differentiate and invert $h^2(z)$ in \eqref{wz_1}, resulting in an expression for $\tilde{w}(z)$: 
\begin{gather} \label{wz_3}
\begin{split}
\tilde{w}(z)&=   \frac{(1+z) }{3 \left[ h^2(z)- \Omega_{\text{(r,0)}}(1+z)^{4}- ( \Omega_{\text{(bm,0)}}+\Omega_{\text{(dm,0)}})(1+z)^{3}\right] } \\
&\times\left[ \frac{d  h^2(z)}{dz} - 4 \Omega_{\text{(r,0)}}(1+z)^{3}-3( \Omega_{\text{(bm,0)}}+\Omega_{\text{(dm,0)}})(1+z)^{2}\right] -1.
\end{split}
\end{gather}
In general, we see in Figure \ref{fig:w_all_Qdm+de}, \ref{fig:w_all_Qdm-de}, \ref{fig:w_all_Qdm} and \ref{fig:w_all_Qde} that in the iDEDM regime $\tilde{w}(z)$ exhibits divergent behavior in the past and crosses the boundary $w(z)=w$ at $z=0$, similar to scalar-tensor theories of modified gravity, such as $f(R)$ \cite{Amendola:2007nt, Tsujikawa:2008uc, M.B.Gavela_2009}. This divergence is due to the parametrization and does not indicate a pathology, as the inherent background dynamics of both DM and DE were shown to be well behaved in the iDEDM regime. The divergence can instead be viewed as a diagnostic of IDE models in the iDEDM regime, which is discussed in more detail in Appendix C of our companion paper \cite{vanderWesthuizen:2025II}. The expressions for $\tilde{w}(z)$ given in \eqref{wz_Q_ddm+dde}, \eqref{wz_Q_dm+de}, \eqref{wz_Q_dm-de}, \eqref{wz_Q_dm} and \eqref{wz_Q_de} are obtained by substituting $h(z)$ from \eqref{hz_Q_linear}, with the corresponding $m_+$, $m_-$, and $\Delta$ for each interaction from Table \ref{tab:m+-} into \eqref{wz_3}.

\subsection{Linear IDE model 1: $Q=3 H (\delta_{\text{dm}} \rho_{\text{dm}} + \delta_{\text{de}}  \rho_{\text{de}})$}
The reconstructed dynamical DE equation of state for this interaction is given by:
\begin{gather} \label{wz_Q_ddm+dde}
\begin{split}
\tilde{w}= \Big\{ &-\left(\delta_{\text{dm}}  - \delta_{\text{de}} - w + \Delta\right) \left[\Omega_{\text{(de,0)}} \left(\delta_{\text{dm}}  - \delta_{\text{de}} + w - \Delta \right) + \Omega_{\text{(dm,0)}} \left(\delta_{\text{dm}}  - \delta_{\text{de}} - w - \Delta \right) \right]  \\
& +\left(\delta_{\text{dm}}  - \delta_{\text{de}} - w - \Delta\right)\left[\Omega_{\text{(de,0)}} \left( \delta_{\text{dm}}  - \delta_{\text{de}} + w + \Delta \right) + \Omega_{\text{(dm,0)}} \left(\delta_{\text{dm}}  - \delta_{\text{de}} - w + \Delta \right)  \right] (1+z)^{\Delta} \Big\} \\
\times\frac{1}{2}\Big\{& \left[\Omega_{\text{(de,0)}} \left(\delta_{\text{dm}}  - \delta_{\text{de}} + w - \Delta \right) + \Omega_{\text{(dm,0)}} \left(\delta_{\text{dm}}  - \delta_{\text{de}} - w - \Delta \right) \right] +2\Delta\Omega_{\rm{dm,0}}(1+z)^{\frac{3}{2}(\delta_{\text{dm}}-\delta_{\text{de}}-w+\Delta )} \\[1mm]
- &\left[ \Omega_{\text{(de,0)}} \left( \delta_{\text{dm}}  - \delta_{\text{de}} + w + \Delta \right) + \Omega_{\text{(dm,0)}} \left(\delta_{\text{dm}}  - \delta_{\text{de}} - w + \Delta \right)  \right]  (1+z)^{\Delta} 
 \Big\}^{-1}.
\end{split}
\end{gather}
We have at present that $\tilde{w}(0)=w$. We may note that both $\Delta>0$ and $\frac{3}{2}(\delta_{\text{dm}}-\delta_{\text{de}}-w+\Delta )>0$. Therefore, for the asymptotic future $(z\rightarrow-1)$, after simplification we obtain:
\begin{gather} \label{wz_Q_ddm+dde_6}
\begin{split}
\tilde{w}(z\rightarrow-1)&=\frac{1}{2}\left(\delta_{\text{de}}  - \delta_{\text{dm}} + w - \Delta\right)= w_{\rm{(de},future)}^{\rm{eff}}=w_{\rm{(dm},future)}^{\rm{eff}}=w_{\rm{(tot},future)}^{\rm{eff}}.
\end{split}
\end{gather}
For the asymptotic past, there are two possible outcomes $(z\rightarrow\infty)$, depending on which power dominates in \eqref{DSA.H}. Essentially, it can be shown that $3\Delta>\frac{3}{2}(\delta_{\text{dm}}-\delta_{\text{de}}-w+\Delta )$ if $\delta_{\rm{dm}}<0$. The two possibilities for the past are:
\begin{gather} \label{wz_Q_ddm+dde_7}
\begin{split}
\text{if } \delta_{\rm{dm}}<0:  \tilde{w}(z\rightarrow\infty)& =\frac{1}{2}\left(\delta_{\text{de}}  - \delta_{\text{dm}} + w + \Delta\right)= w_{\rm{(de},past)}^{\rm{eff}}=w_{\rm{(dm},past)}^{\rm{eff}}.\\
\text{if } \delta_{\rm{dm}}\ge0:  \tilde{w}(z\rightarrow\infty)&= 0.
\end{split}
\end{gather}

\subsection{Linear IDE model 2: $Q=3H\delta( \rho_{\text{dm}}+\rho_{\text{de}})$}

The reconstructed dynamical DE equation of state for this interaction is given by:
\begin{gather} \label{wz_Q_dm+de}
\begin{split}
\tilde{w}=& \frac{\left(    w - \Delta\right) \left[\Omega_{\text{(de,0)}} \left(   w - \Delta \right) - \Omega_{\text{(dm,0)}} \left( w + \Delta \right) \right]   -\left( w + \Delta\right)\left[\Omega_{\text{(de,0)}} \left( w + \Delta \right) - \Omega_{\text{(dm,0)}} \left(   w - \Delta \right)  \right] (1+z)^{\Delta}}{2\left[\Omega_{\text{(de,0)}} \left(   w - \Delta \right) - \Omega_{\text{(dm,0)}} \left(   w + \Delta \right)  +2\Delta\Omega_{\rm{dm,0}}(1+z)^{\frac{3}{2}(\Delta-w )} - \left[ \Omega_{\text{(de,0)}} \left(    w + \Delta \right) - \Omega_{\text{(dm,0)}} \left(   w - \Delta \right)  \right]  (1+z)^{\Delta}  \right]}  \\
\end{split}
\end{gather}
We have at present that $\tilde{w}(0)=w$. We may note that both $\Delta>0$ and $\frac{3}{2}(\Delta-w )>0$. Therefore, for the asymptotic future $(z\rightarrow-1)$ we have:
\begin{gather} \label{wz_Q_dm+de_6}
\begin{split}
\tilde{w}(z\rightarrow-1)&=\frac{1}{2}\left( w - \Delta\right)= w_{\rm{(de},future)}^{\rm{eff}}=w_{\rm{(dm},future)}^{\rm{eff}}=w_{\rm{(tot},future)}^{\rm{eff}}.
\end{split}
\end{gather}
For the asymptotic past, there are two possible outcomes $(z\rightarrow\infty)$, depending on which power dominates in \eqref{DSA.H}. Essentially, it can be shown that $3\Delta>\frac{3}{2}(\Delta -w)$ if $\delta<0$. We therefore have two possibilities for the past:
\begin{gather} \label{wz_Q_dm+de_7}
\begin{split}
\text{if } \delta<0 \text{ (iDMDE regime)}:  \tilde{w}(z\rightarrow\infty)&=\frac{1}{2}\left(w + \Delta\right)= w_{\rm{(de},past)}^{\rm{eff}}=w_{\rm{(dm},past)}^{\rm{eff}}.\\
\text{if } \delta\ge0 \text{ (iDEDM regime)}:  \tilde{w}(z\rightarrow\infty)&= 0.
\end{split}
\end{gather}

\subsection{Linear IDE model 3: $Q=3H\delta( \rho_{\text{dm}}-\rho_{\text{de}})$}

The reconstructed dynamical DE equation of state for this interaction is given by:
\begin{gather} \label{wz_Q_dm-de}
\begin{split}
\tilde{w}= \Big\{ &-\left(  2\delta - w + \Delta\right) \left[\Omega_{\text{(de,0)}} \left(  2\delta + w - \Delta \right) + \Omega_{\text{(dm,0)}} \left(  2\delta - w - \Delta \right) \right]  \\
& +\left(  2\delta - w - \Delta\right)\left[\Omega_{\text{(de,0)}} \left(   2\delta + w + \Delta \right) + \Omega_{\text{(dm,0)}} \left(  2\delta - w + \Delta \right)  \right] (1+z)^{\Delta} \Big\} \\
\times\frac{1}{2}\Big\{& \left[\Omega_{\text{(de,0)}} \left(  2\delta + w - \Delta \right) + \Omega_{\text{(dm,0)}} \left(  2\delta - w - \Delta \right) \right] +2\Delta\Omega_{\rm{dm,0}}(1+z)^{\frac{3}{2}(2\delta-w+\Delta )} \\[1mm]
- &\left[ \Omega_{\text{(de,0)}} \left(   2\delta + w + \Delta \right) + \Omega_{\text{(dm,0)}} \left(  2\delta - w + \Delta \right)  \right]  (1+z)^{\Delta} 
 \Big\}^{-1}.
\end{split}
\end{gather}
We have at present that $\tilde{w}(0)=w$. We may note that both $\Delta>0$ and $\frac{3}{2}(2\delta-w+\Delta )>0$. Therefore, for the asymptotic future $(z\rightarrow-1)$ we have:
\begin{gather} \label{wz_Q_dm-de_6}
\begin{split}
\tilde{w}(z\rightarrow-1)&=\frac{1}{2}\left(2\delta + w - \Delta\right)= w_{\rm{(de},future)}^{\rm{eff}}=w_{\rm{(dm},future)}^{\rm{eff}}=w_{\rm{(tot},future)}^{\rm{eff}}.
\end{split}
\end{gather}
For the asymptotic past, there are two possible outcomes $(z\rightarrow\infty)$, depending on which power dominates in \eqref{DSA.H}. Essentially, it can be shown that $3\Delta>\frac{3}{2}(2\delta-w+\Delta )$ if $\delta<0$. We therefore have two possibilities for the past:
\begin{gather} \label{wz_Q_dm-de_7}
\begin{split}
\text{if } \delta<0:  \tilde{w}(z\rightarrow\infty)&=\frac{1}{2}\left(2\delta + w + \Delta\right)= w_{\rm{(de},past)}^{\rm{eff}}=w_{\rm{(dm},past)}^{\rm{eff}}.\\
\text{if } \delta\ge0:  \tilde{w}(z\rightarrow\infty)&=  0.
\end{split}
\end{gather}

\subsection{Linear IDE model 4: $Q=3H\delta \rho_{\text{dm}}$}
The reconstructed dynamical DE equation of state for this interaction is given by:
\begin{gather} \label{wz_Q_dm}
\begin{split}
\tilde{w}(z)&= \frac{w \left[(\delta+w)\Omega_{\text{(de,0)}}  + \delta \Omega_{\text{(dm,0)}} \right]  -w\Omega_{\text{(dm,0)}}  (1+z)^{-3(\delta+w)}}{ \Omega_{\text{(dm,0)}}[\delta  +w  (1+z)^{-3(\delta+w)} ]+ [\delta+w]   [\Omega_{\text{(de,0)}}-\Omega_{\text{(dm,0)}}(1+z)^{-3w}]}. \\
\end{split}
\end{gather}
Expression \eqref{wz_Q_dm} is equivalent to equation (51) to what was found in \cite{M.B.Gavela_2009}, given their notation where $\delta=\frac{\xi}{3}$ and $r_0=\frac{\Omega_{\rm{(de,0)}}}{\Omega_{\rm{(dm,0)}}}$. We have at present $\tilde{w}(0)=w$. We may note that both $-(\delta+w)>0$ and $-3w>0$, given our initial assumptions. Therefore, for the asymptotic future $(z\rightarrow-1)$ we have:
\begin{gather} \label{wz_Q_dm_6}
\begin{split}
\tilde{w}(z\rightarrow-1)&= w= w_{\rm{(de},future)}^{\rm{eff}}=w_{\rm{(tot},future)}^{\rm{eff}}.
\end{split}
\end{gather}
For the asymptotic past, there are two possible outcomes $(z\rightarrow\infty)$, depending on which power dominates in \eqref{DSA.H}. Essentially, we have $-3(\delta+w)>-3w$ if $\delta<0$. We therefore have two possibilities for the past:
\begin{gather} \label{wz_Q_dm_7}
\begin{split}
\text{if } \delta<0 \text{ (iDMDE regime)}:  \tilde{w}(z\rightarrow\infty)&=-\delta= w_{\rm{(de},past)}^{\rm{eff}}=w_{\rm{(dm},past)}^{\rm{eff}}.\\
\text{if } \delta\ge0 \text{ (iDEDM regime)}:  \tilde{w}(z\rightarrow\infty)&=  0.
\end{split}
\end{gather}

\subsection{Linear IDE model 5: $Q=3H\delta \rho_{\text{de}}$}

The reconstructed dynamical DE equation of state for this interaction is given by:
\begin{gather} \label{wz_Q_de}
\begin{split}
\tilde{w}(z)&=   \frac{   w }{1-\frac{\delta}{(\delta + w )  } \left[1- (1+z)^{-3(\delta + w)}  \right]}.  \end{split}
\end{gather}
Expression \eqref{wz_Q_de} is equivalent to equation (37) in \cite{M.B.Gavela_2009}, given their notation where $\delta=\frac{\xi}{3}$. We have $\tilde{w}(0)=w$. We may note that $(\delta+w)<0$ and for the most general case, given our initial assumptions.  Therefore, for the asymptotic future $(z\rightarrow-1)$ we have:
\begin{gather} \label{wz_Q_de_6}
\begin{split}
\tilde{w}(z\rightarrow-1)&= w+\delta= w_{\rm{(de},future)}^{\rm{eff}}=w_{\rm{(dm},future)}^{\rm{eff}}=w_{\rm{(tot},future)}^{\rm{eff}}.
\end{split}
\end{gather}
For the asymptotic past, we have:
\begin{gather} \label{wz_Q_de_7}
\begin{split}
\tilde{w}(z\rightarrow\infty)&=\frac{ (\delta+w) }{  \frac{\delta}{w} (1+z)^{-3(\delta+w)}}=0=w_{\rm{(dm},past)}^{\rm{eff}}.\\
\end{split}
\end{gather}

\section{Statefinder diagnostics of IDE models} \label{statefinder}

\begin{figure}[htbp]
    \centering
    \begin{subfigure}[b]{0.49\linewidth}
        \centering
        \includegraphics[width=\linewidth]{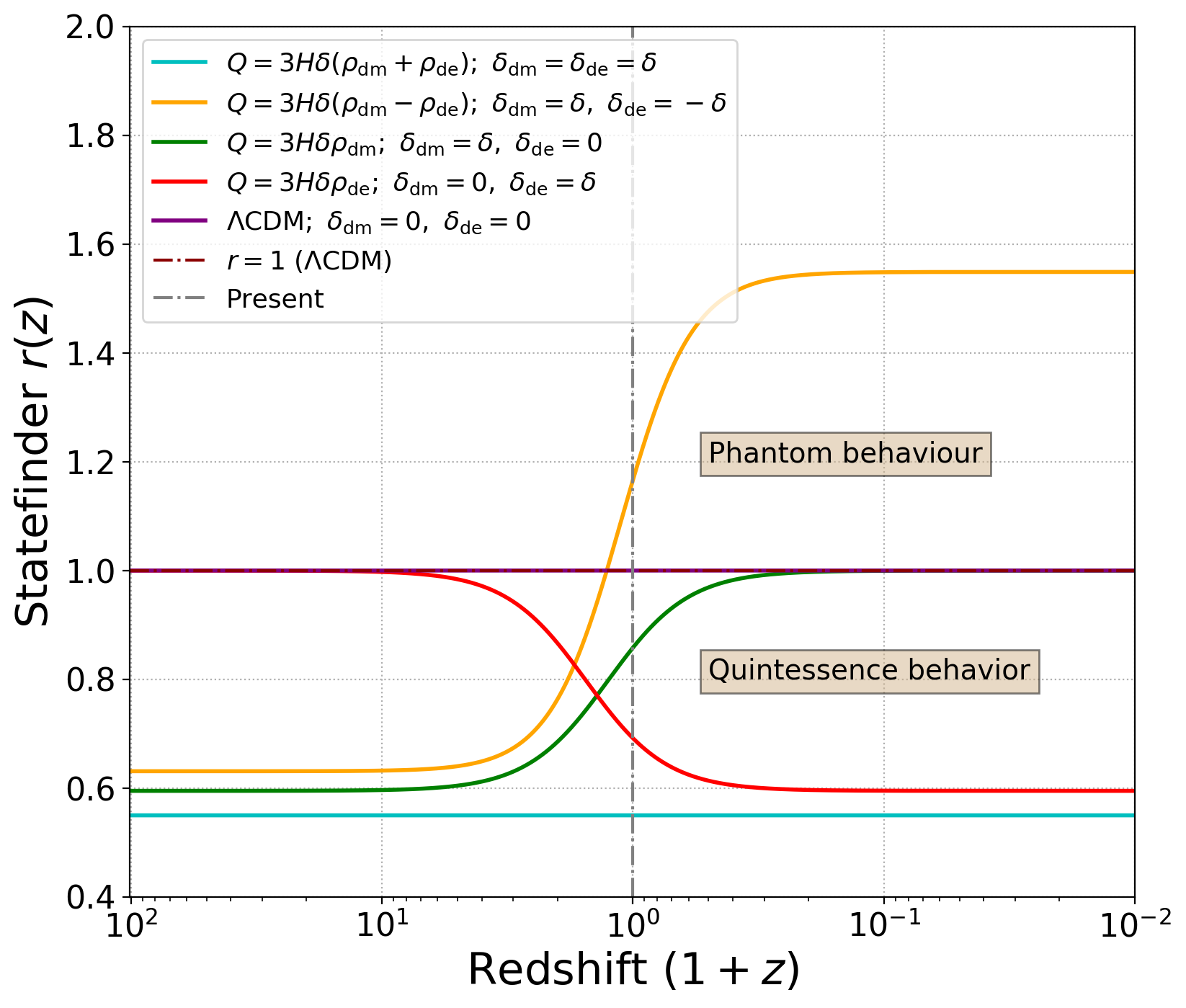}
        \label{fig:Statefinder_linear_rz_iDEDM}
    \end{subfigure}%
    \hspace{0pt} % No extra space between subfigures
    \begin{subfigure}[b]{0.49\linewidth}
        \centering
        \includegraphics[width=\linewidth]{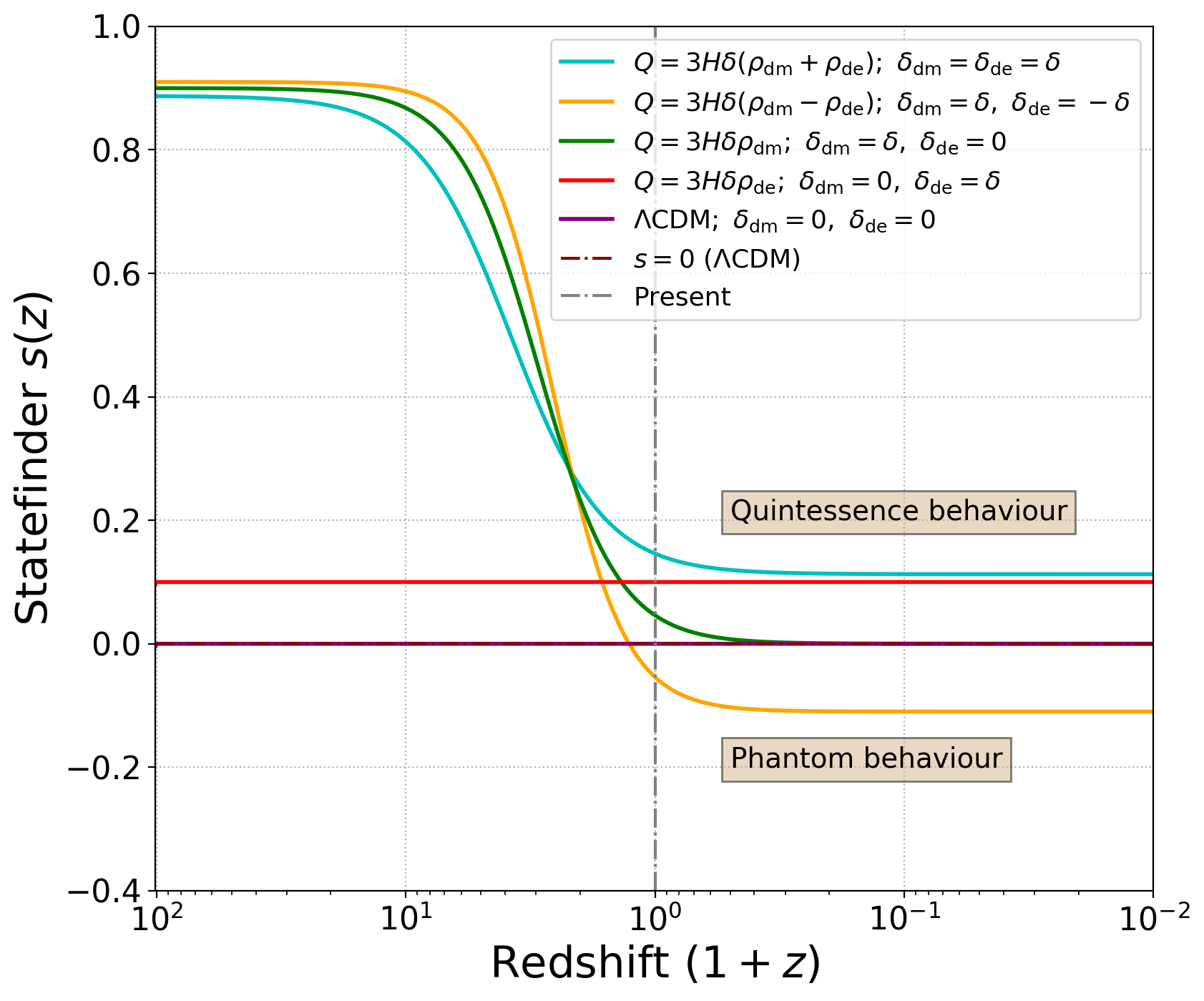}
        \label{fig:Statefinder_linear_sz_iDEDM}
    \end{subfigure}
    \vspace{0.3cm}
    \centering
    \begin{subfigure}[b]{0.49\linewidth}
        \centering
        \includegraphics[width=\linewidth]{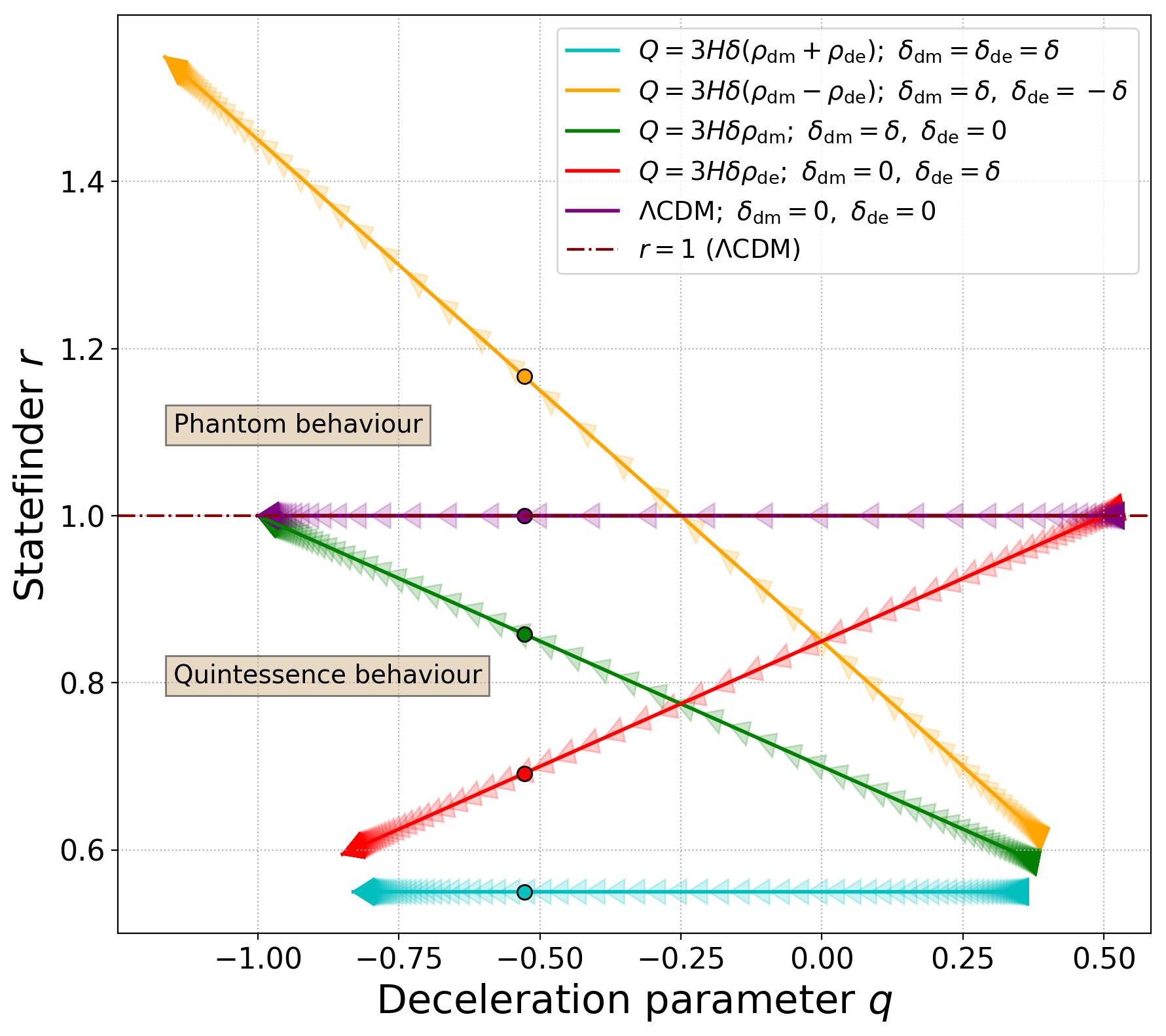}
        \label{fig:Statefinder_linear_qr_iDEDM}
    \end{subfigure}%
    \hspace{0pt} % No extra space between subfigures
    \begin{subfigure}[b]{0.49\linewidth}
        \centering
        \includegraphics[width=\linewidth]{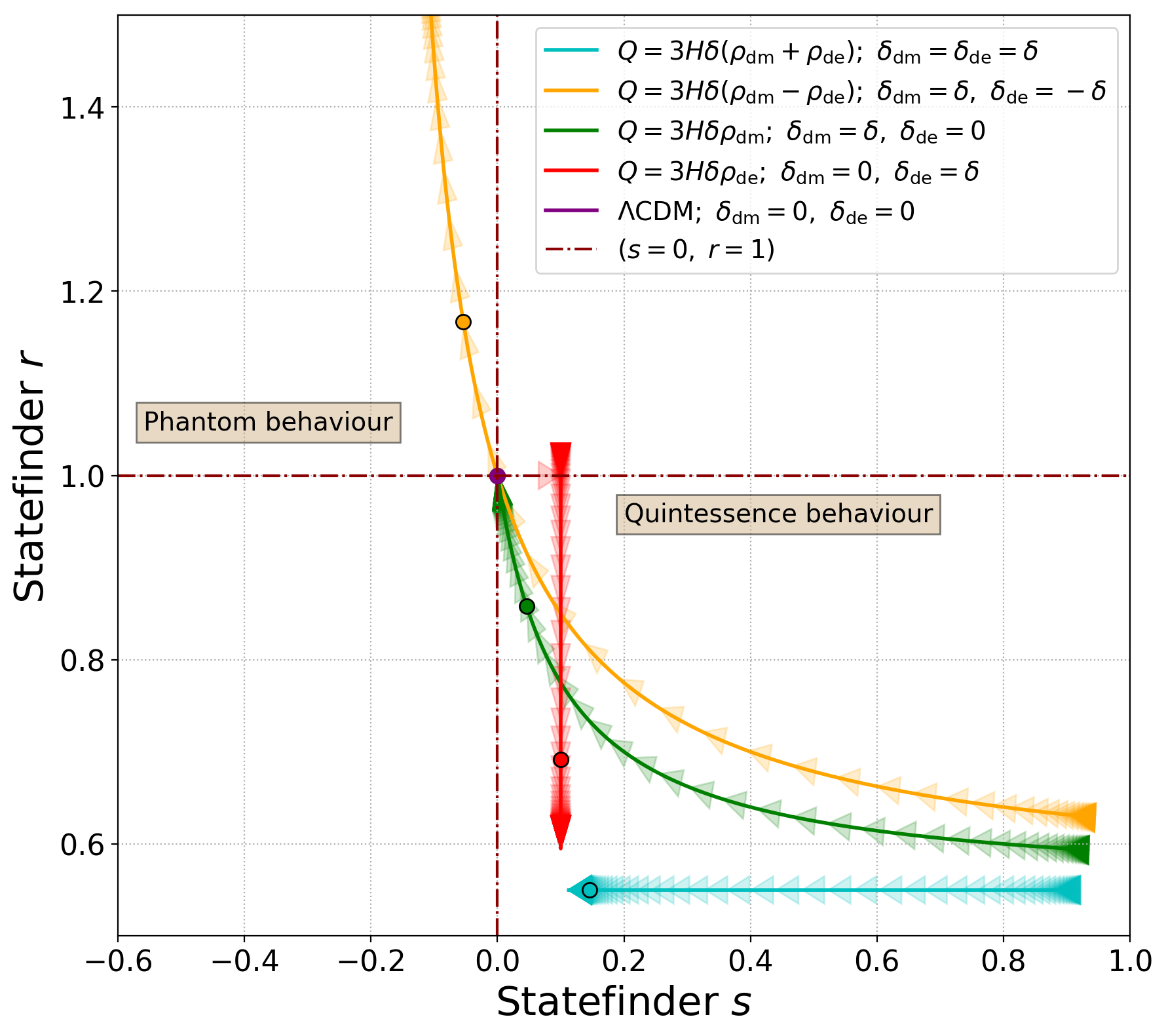}
        \label{fig:Statefinder_linear_sr_iDEDM}
    \end{subfigure}
    \caption{Statefinder parameters – $r$ vs $z$ (top left), $s$ vs $z$ (top right), $q$ vs $r$ (bottom left), and $s$ vs $r$ (bottom right) for special cases of $Q= 3 H (\delta_{\text{dm}} \rho_{\text{dm}} + \delta_{\text{de}}  \rho_{\text{de}})$ with $\delta=+0.1$ (iDEDM regime) and $w=-1$. Circles indicate the present coordinates. All trajectories end with quintessence behavior $(q>-1\; ; \; r_{\rm{sf}}<1 \; ; \;  s_{\rm{sf}}>0)$, except for the interaction $Q=3H \delta (\rho_{\text{dm}}-\rho_{\text{de}})$, which is a sign-switching interaction that changes into the iDMDE regime at late times, exhibiting phantom DE behavior and which acts similarly to Chaplygin gas or Galileon models $(q<-1\; ; \; r_{\rm{sf}}>1 \; ; \;  s_{\rm{sf}}<0)$. The interaction $Q=3H \delta \rho_{\rm{de}}$ has  a constant $s_{\rm{sf}}$,  while $r_{\rm{sf}}$ declines asymptotically, showing curves similar to those for $w$CDM or quiessence models.}
    \label{fig:Statefinder_linear_iDEDM}
\end{figure}

Using the statefinder parameters defined in \eqref{DSA.r_st} and substituting the relevant expressions from section \ref{Background_cosmology}, we can plot the evolution of $q$, $r$, and $s$ for the four linear interactions in Figure \ref{fig:Statefinder_linear_iDEDM}, which is the standard form used to differentiate between late-time cosmologies~\cite{Sahni_2003, Alam_2003, carrasco2023discriminatinginteractingdarkenergy, Bolotin_2015}. Since this diagnostic is used for late-time cosmology, or small $|t-t_0|$~\cite{Sahni_2003, Alam_2003, carrasco2023discriminatinginteractingdarkenergy}, we only consider a two-fluid system with $\rho_{\text{dm}}$ and $\rho_{\text{de}}$ up to $z=100$. It should be noted that another standard diagnostic is the O$m$-diagnostic, but it proves to be ineffective when differentiating between interaction models, since it only depends on the deceleration parameter $q$, and not on whether an interaction is present or not~\cite{Bolotin_2015}. To ease comparison, we note again that the $\Lambda$CDM model is characterized by the fixed coordinate $(r_{\rm{sf}},s_{\rm{sf}})=(1,0)$ throughout cosmic evolution, while converging during DE domination to $(q,r_{\rm{sf}})=(-1,1)$, which corresponds to a de Sitter Universe~\cite{carrasco2023discriminatinginteractingdarkenergy}. Similarly, the SCDM (Standard Cold Dark Matter) model without any DE is identified by a fixed coordinate of $(r_{\rm{sf}},s_{\rm{sf}})=(1,0)$. For ease of illustration in Figure \ref{fig:Statefinder_linear_iDEDM}, we only show the iDEDM case $\delta>0$, which is, for most interactions, the more realistic regime with positive energies.

The first conclusion we can draw from Figure \ref{fig:Statefinder_linear_iDEDM} is that for any interaction $Q \propto\rho_{\text{dm}}$ (blue, orange, and green trajectories), the trajectories differ from $\Lambda$CDM in the past, while for interactions where $Q \propto\rho_{\text{de}}$ (blue, orange, and green trajectories) the trajectories diverge in the future. Furthermore, if $\delta_{\text{dm}}=0$ (red trajectories), the trajectories start near the initial coordinates of $\Lambda$CDM, while if $\delta_{\text{de}}=0$ (green trajectories), the trajectories converge to the final coordinates of $\Lambda$CDM. Since we chose the iDEDM regime ($\delta>0$), we may see that all trajectories end with quintessence behavior $(q>-1\; ; \; r_{\rm{sf}}<1 \; ; \;  s_{\rm{sf}}>0)$~\cite{Sahni_2003, Alam_2003, carrasco2023discriminatinginteractingdarkenergy} as expected, except for $Q=3H \delta (\rho_{\text{dm}}-\rho_{\text{de}})$, which is a sign-switching interaction. Instead, this interaction changes into the iDMDE regime at late times, which exhibits phantom DE behavior and acts similarly to Chaplygin gas or Galileon models $(q<-1\; ; \; r_{\rm{sf}}>1 \; ; \;  s_{\rm{sf}}<0)$~\cite{Alam_2003, Akarsu_2014, carrasco2023discriminatinginteractingdarkenergy}. This behavior is reversed when considering the iDMDE ($\delta<0$) regime. This behavior coincides with the future asymptotic behavior of $w_{\rm{tot}}^{\rm{eff}}$ for $\delta>0$ (iDEDM regime) as illustrated in Figure \ref{fig:eos_tot_BR_Qdm+de}, \ref{fig:eos_tot_BR_Qdm-de}, \ref{fig:eos_tot_BR_Qdm} and \ref{fig:eos_tot_BR_Qde}. Another interesting case is the interaction $Q=3H \delta \rho_{\rm{de}}$, where the statefinder $s_{\rm{sf}}$ remains constant while $r_{\rm{sf}}$ declines asymptotically, showing curves in Figure \ref{fig:Statefinder_linear_iDEDM} similar to those for $w$CDM or quiessence models presented in FIG.~1 of~\cite{Alam_2003}. For $Q=3H\delta( \rho_{\text{dm}}+\rho_{\text{de}})$, it may seem as if $r_{\rm{sf}}$ is constant, but this is only true if $w=-1$, such that $r_{\rm{sf}}$ will evolve if $w\neq-1$.

All trajectories in Figure \ref{fig:Statefinder_linear_iDEDM} show the same qualitative behavior as those in~\cite{carrasco2023discriminatinginteractingdarkenergy}. Specifically, $\Lambda$CDM, $Q=3H \delta \rho_{\text{dm}}$, $Q=3H \delta \rho_{\text{de}}$, and $Q=3H \delta (\rho_{\text{dm}}-\rho_{\text{de}})$ correspond to FIG.~1, FIG.~2(b)--(c), FIG.~3(a), (c), (d), and FIG.~4(a)--(b) in~\cite{carrasco2023discriminatinginteractingdarkenergy}, respectively.
The coordinates for $q$, $r_{\text{sf}}$, and $s_{\text{sf}}$ for the general interaction $Q=3 H (\delta_{\text{dm}} \rho_{\text{dm}} + \delta_{\text{de}} \rho_{\text{de}})$ during DM domination in the past (ignoring the baryonic matter contribution for simplicity) and DE domination in the future are obtained from Table \ref{tab:CP_B_ddm+dde} by substituting $w_{\rm{de}}^{\rm{eff}}=w_{\rm{de}}^{\rm{tot}}$ in the past, present and future into \eqref{DSA.r_st},  giving the expressions found in \eqref{sr_Q_general}.
\begin{gather} \label{sr_Q_general}
\begin{split}
q_{\text{past}} &=\frac{1}{2} \left(1+\frac{3}{2}[w+\delta_{\text{de}}-\delta_{\text{dm}}+\Delta] \right), \\
r_{\text{(sf,past)}} &= 1+\frac{9}{4}\left[w+\delta_{\text{de}}-\delta_{\text{dm}}+\Delta\right]\left[1+w+\delta_{\text{de}}-\delta_{\text{dm}}+\Delta\right], \\
s_{\text{(sf,past)}} &= 1+\frac{1}{2}\left[w+\delta_{\text{de}}-\delta_{\text{dm}}+\Delta\right], \\[4pt]
q_{\text{present}} &=\frac{1}{2} \left[\Omega_{\rm{(dm,0)}}   +  \Omega_{\rm{(de,0)}}  \left(1 +3 w \right) \right] \\
r_{\text{(sf,present)}} &= 1+\frac{9}{2} \,\Omega_{\rm{(de,0)}}\, w \left[1+w+\delta_{\text{de}}+\delta_{\text{dm}}r_0\right], \\
s_{\text{(sf,present)}} &= 1+w+\delta_{\text{de}}+\delta_{\text{dm}}r_0, \\[4pt]
q_{\text{future}} &=\frac{1}{2} \left(1+\frac{3}{2}[w+\delta_{\text{de}}-\delta_{\text{dm}}-\Delta] \right), \\
r_{\text{(sf,future)}} &= 1+\frac{9}{4}\left[w+\delta_{\text{de}}-\delta_{\text{dm}}-\Delta\right]\left[1+w+\delta_{\text{de}}-\delta_{\text{dm}}-\Delta\right], \\
s_{\text{(sf,future)}} &= 1+\frac{1}{2}\left[w+\delta_{\text{de}}-\delta_{\text{dm}}-\Delta\right].
\end{split}
\end{gather}\\
Using expression \eqref{sr_Q_general} and substituting the relevant $\delta_{\rm{dm}}$ and $\delta_{\rm{de}}$  for each interaction, the starting points during DM domination, the present-day expressions, and the final expressions during DE domination for the special cases of this interaction are obtained and shown in Tables \ref{tab:Statefinder_past_linear}, \ref{tab:Statefinder_present_linear}, and \ref{tab:Statefinder_future_linear}, with $\Delta$ for the first two interactions given by \eqref{eq:determinant_Q_dm+de} and \eqref{eq:determinant_Q_dm-de}, respectively. These expressions correspond to the start and end points of the trajectories seen in Figure \ref{fig:Statefinder_linear_iDEDM}, as well as their present values indicated by the circles. It is worth noting that, in the case where $\delta=0$ and $w=-1$, the $\Lambda$CDM model is recovered.

\begin{table}[H]
\centering
\renewcommand{\arraystretch}{1.2} % Adjust row spacing
\setlength{\tabcolsep}{10pt}     % Adjust column spacing
\begin{tabular}{|c|c|c|c|}
\hline
\textbf{Model} 
&$q_{\text{(past)}}$ & $r_{\text{(sf,past)}}$
 & $s_{\text{(sf,past)}}$
\\ \hline \hline

$Q=3H\delta( \rho_{\text{dm}}+\rho_{\text{de}})$
& $\frac{1}{2}(1+\frac{3}{2}[w+\Delta] )$ & $1+\frac{9}{4}[w+\Delta][1+w+\Delta]$
 & $1+\frac{1}{2}[w+\Delta]$
\\ \hline

$Q=3H\delta( \rho_{\text{dm}}-\rho_{\text{de}})$
&$\frac{1}{2}(1+\frac{3}{2}[1+w-2\delta+\Delta] )$ & $1+\frac{9}{4}[w+2\delta+\Delta][1+w+2\delta+\Delta]$
 & $1+\frac{1}{2}[w+2\delta+\Delta]$
\\ \hline

$Q=3H \delta \rho_{\text{dm}}$
&$\frac{1}{2}(1-3 \delta)$ & $1-\frac{9}{2}\delta\left(1-\delta \right)$
 & $1-\delta$
\\ \hline

$Q=3H\delta \rho_{\text{de}}$
&$\frac{1}{2} $& $1$
 & $1+ w+\delta $

\\ \hline

$w$CDM
&$\frac{1}{2}$ & $1$
 & $1+w $
\\ \hline

$\Lambda$CDM
&$\frac{1}{2}$ & $1$
 & $0$
\\ \hline

SCDM
&$\frac{1}{2}$ & $1$
 & $1$
\\ \hline

\end{tabular}
\caption{Comparison of the deceleration parameter $q$ and the statefinder parameters $r_{\rm{sf}}$ and $s_{\rm{sf}}$ during past dark matter domination for special cases of the linear IDE model $Q = 3 H (\delta_{\text{dm}} \rho_{\text{dm}} + \delta_{\text{de}} \rho_{\text{de}})$.}
\label{tab:Statefinder_past_linear}
\end{table}

\begin{table}[H]
\centering
\renewcommand{\arraystretch}{1.2} % Adjust row spacing
\setlength{\tabcolsep}{10pt}     % Adjust column spacing
\begin{tabular}{|c|c|c|c|}
\hline
\textbf{Model} &
$q_{\rm{(present)}}$ & $r_{\text{(sf,present)}}$
 & $s_{\text{(sf,present)}}$
\\ \hline \hline

$Q=3H\delta( \rho_{\text{dm}}+\rho_{\text{de}})$
&$\frac{1}{2} \left[\Omega_{\rm{(dm,0)}}   +  \Omega_{\rm{(de,0)}}  \left(1 +3 w \right) \right]$ & $1+\frac{9}{2} \Omega_{\rm{(de,0)}}w [1+w+\delta(1+r_0)]$
 & $1+w+\delta(1+r_0)$
\\ \hline

$Q=3H\delta( \rho_{\text{dm}}-\rho_{\text{de}})$
&$\frac{1}{2} \left[\Omega_{\rm{(dm,0)}}   +  \Omega_{\rm{(de,0)}}  \left(1 +3 w \right) \right]$ & $1+\frac{9}{2} \Omega_{\rm{(de,0)}}w [1+w+\delta(r_0-1)]$
 & $1+w+\delta(r_0-1)$
\\ \hline

$Q=3H \delta \rho_{\text{dm}}$
&$\frac{1}{2} \left[\Omega_{\rm{(dm,0)}}   +  \Omega_{\rm{(de,0)}}  \left(1 +3 w \right) \right]$ & $1+\frac{9}{2} \Omega_{\rm{(de,0)}}w [1+w+\delta r_0]$
 & $1+w+\delta r_0 $
\\ \hline

$Q=3H\delta \rho_{\text{de}}$

&$\frac{1}{2} \left[\Omega_{\rm{(dm,0)}}   +  \Omega_{\rm{(de,0)}}  \left(1 +3 w \right) \right]$ & $1+\frac{9}{2} \Omega_{\rm{(de,0)}}w [1+w+\delta]$
 & $1+w+\delta$
\\ \hline

$w$CDM
&$\frac{1}{2} \left[\Omega_{\rm{(dm,0)}}   +  \Omega_{\rm{(de,0)}}  \left(1 +3 w \right) \right]$ & $1+\frac{9}{2} \Omega_{\rm{(de,0)}}w [1+w]$
 & $1+w $
\\ \hline

$\Lambda$CDM
&$\frac{1}{2} \left[\Omega_{\rm{(dm,0)}}   -2  \Omega_{\rm{(de,0)}}   \right]$ & $1$
 & $0$
\\ \hline

SCDM
&$\frac{1}{2} \Omega_{\rm{(dm,0)}}$ & $1$
 & $1$
\\ \hline

\end{tabular}
\caption{Comparison of the deceleration parameter $q$ and the statefinder parameters $r_{\rm{sf}}$ and $s_{\rm{sf}}$ at present $(a=1, \; z=0)$ for special cases of the linear IDE model $Q = 3 H (\delta_{\text{dm}} \rho_{\text{dm}} + \delta_{\text{de}} \rho_{\text{de}})$.}
\label{tab:Statefinder_present_linear}
\end{table}

\begin{table}[H]
\centering
\renewcommand{\arraystretch}{1.2} % Adjust row spacing
\setlength{\tabcolsep}{10pt}     % Adjust column spacing
\begin{tabular}{|c|c|c|c|}
\hline
\textbf{Model} &
$q_{\text{(future)}}$ & $r_{\text{(sf,future)}}$
 & $s_{\text{(sf,future)}}$
\\ \hline \hline

$Q=3H\delta( \rho_{\text{dm}}+\rho_{\text{de}})$
& $\frac{1}{2}(1+\frac{3}{2}[w-\Delta] )$ & $1+\frac{9}{4}[w-\Delta][1+w-\Delta]$
 & $1+\frac{1}{2}[w-\Delta]$
\\ \hline

$Q=3H\delta( \rho_{\text{dm}}-\rho_{\text{de}})$
& $\frac{1}{2}(1+\frac{3}{2}[w-2\delta-\Delta] )$ & $1+\frac{9}{4}[w+2\delta-\Delta][1+w+2\delta-\Delta]$
 & $1+\frac{1}{2}[w+2\delta-\Delta]$
\\ \hline

$Q=3H \delta \rho_{\text{dm}}$
&$\frac{1}{2}(1+3w)$& $1+\frac{9}{2}w \left(1+w \right)$
 & $1+w$
\\ \hline

$Q=3H\delta \rho_{\text{de}}$
 & $\frac{1}{2}(1+3[w+\delta] )$ & $1+\frac{9}{2}\left[ w+\delta \right]\left(1+\left[ w+\delta \right] \right)$
 & $1+ w+\delta $

\\ \hline

$w$CDM
&$\frac{1}{2}(1+3w)$ & $1+\frac{9}{2}w\left(1+w \right)$
 & $1+w $
\\ \hline

$\Lambda$CDM
& $-1$ & $1$
 & $0$
\\ \hline

SCDM
& $\frac{1}{2}$ & $1$
 & $1$
\\ \hline

\end{tabular}
\caption{Comparison of the deceleration parameter $q$ and the statefinder parameters $r_{\rm{sf}}$ and $s_{\rm{sf}}$ during future dark energy domination for special cases of the linear IDE model $Q = 3 H (\delta_{\text{dm}} \rho_{\text{dm}} + \delta_{\text{de}} \rho_{\text{de}})$.}
\label{tab:Statefinder_future_linear}
\end{table}

For the more realistic case where baryonic matter is included as a separate fluid, the expression for $r_{\rm{sf}}$ found in Table \ref{tab:Statefinder_past_linear} will act as a saddle point. Further into the past, $r_{\rm{sf}}$ will remain dynamic and will not converge to a specific value. This implies that all trajectories in Figure \ref{fig:Statefinder_linear_iDEDM} will also start at different points. The present expressions for $q$ in Table \ref{tab:Statefinder_present_linear} will hold after replacing $\Omega_{\rm{(dm,0)}}$ with $\Omega_{\rm{(m,0)}} = \Omega_{\rm{(bm,0)}} + \Omega_{\rm{(dm,0)}}$. Taking these exceptions into account, all other expressions found in the tables above will remain valid with the inclusion of a separate baryonic matter fluid.

\section{Summary of main results and discussions} \label{summary}

In this study, we have investigated the viability of phenomenological IDE models with a linear interaction term $Q = 3 H (\delta_{\text{dm}} \rho_{\text{dm}} + \delta_{\text{de}} \rho_{\text{de}})$, and four of its special cases. Our study focused on the background dynamics of these models, with special attention given to the often-overlooked presence of negative DM and DE densities (discussed in section \ref{BG_neg}) and future big rip singularities (discussed in section \ref{BG_rip}) that appear in certain regions of the parameter space. In section \ref{Sec.DSA}, assuming a constant DE equation of state $w$, we performed a dynamical system analysis of the model, which resulted in the four new conditions \eqref{DSA.Q.delta_dm+delta_de.PEC_BG} to avoid both imaginary and negative energy densities. From this, we found that positive energy densities are most naturally maintained when energy flows from DE to DM. Conversely, energy flow from DM to DE is typically associated with regions of parameter space where negative densities appear. While these features may signal theoretical pathologies, they could also indicate that our current phenomenological models are incomplete. Future observational constraints will be essential to clarify which scenarios remain viable. Table \ref{tab:Qddm+dde_energy_conditions} summarizes the impact of the parameter space on the positivity of DM and DE in both the past and the future. Combining these positive energy criteria with a very brief analysis of the stability of the system using the doom factor $\mathbf{d}$ led to the additional criteria for $w < -1$ to ensure stability and positive energy, as shown in Table \ref{tab:Qddm+dde_stability_criteria}. We again state that this stability analysis is limited, as other approaches are present in the literature, and that an in-depth study of these problems will be conducted in a future work.
From our analysis, we also obtained new conditions to ensure both future accelerated expansion \eqref{DSA.Q.ddm+dde.ACC} and to avoid future big rip singularities \eqref{DSA.Q.ddm+dde.BR}. A brief summary of the main results from the dynamical systems analysis is provided in section \ref{DSA.summary}.

In section \ref{Finding_analytical_solutions}, we solved the conservation equations to obtain new analytical solutions for $\rho_{\rm{dm}}$ \eqref{eq:rhodm_general} and $\rho_{\rm{de}}$ \eqref{eq:rhode_general}. The new analytical solutions were then applied in section \ref{Background_cosmology} to find a wide range of new expressions that describe the background cosmology of the model and to show convergence with the results from our dynamical system analysis. Furthermore, in this section, the behavior of four special cases of the interaction kernel was studied: $Q=3 H (\delta_{\text{dm}} \rho_{\text{dm}} + \delta_{\text{de}}  \rho_{\text{de}})$ in subsection \ref{Q_General}, $Q= 3 H \delta(\rho_{\text{dm}} + \rho_{\text{de}})$ in subsection \ref{Q_dm+de}, $Q= 3 H \delta(\rho_{\text{dm}} - \rho_{\text{de}})$ in subsection \ref{Q_dm-de}, $Q= 3 H \delta\rho_{\text{dm}}$ in subsection \ref{Q_dm}, and $Q= 3 H \delta\rho_{\text{de}}$ in subsection \ref{Q_de}. The last two interactions in this list have been widely studied before, but were included to show that the new general results we found reduce back to familiar results from the literature~\cite{vanderWesthuizen:2023hcl}, given these special cases. This analysis of the special cases also highlighted key differences in the past and future behavior of the interaction kernels, for specific values of $\delta_{\rm{dm}}$ and $\delta_{\rm{de}}$. These differences are summarized in Tables \ref{tab:Com_real}--\ref{tab:Com_CP} below. Specifically, Table \ref{tab:Com_CP} provides a comparison of how the different interactions address the coincidence problem in both the past and the future. We note that the magnitude of the problem is indicated by the size of the deviation from $\zeta=0$, as discussed in \eqref{DSA.zeta}. We show results only for the iDEDM regime, as this regime alleviates the coincidence problem, while the iDMDE regime worsens it~\cite{vanderWesthuizen:2023hcl}.   
To relate the five interactions studied to wider cosmology, we performed a statefinder analysis in section \ref{statefinder}. Here we illustrated trajectories in the statefinder phase space and found new expressions for the past, present, and future behavior of the parameters $q$, $r_{\rm{sf}}$, and $s_{\rm{sf}}$ in Tables \ref{tab:Statefinder_past_linear}--\ref{tab:Statefinder_future_linear}. A brief discussion on helpful results for observational constraints is provided below.

First, in order to constrain these models with data, it is useful to know when the DM and DE densities become imaginary and undefined (which may lead to a collapse of MCMC chains), so that this part of the parameter space can be avoided. These conditions are summarized in Table \ref{tab:Com_real}. Negative DM and DE densities may also be considered non-physical by many researchers. The derived positive energy conditions for each interaction kernel are summarized in Table \ref{tab:Com_PEC} below. To give an intuition of the range of the allowed parameters, we also include example values obtained by substituting $\Omega_{\rm{(dm,0)}}=0.266$, $\Omega_{\rm{(de,0)}}=0.685$ (which implies $r_0=0.388$), and $w=-1$. 
For any model to be realistic, it must reproduce a present era of accelerated expansion. Additionally, many researchers might consider future singularities to be unphysical and would want to avoid these in a realistic model. Conditions to ensure this, as well as example values in parentheses for $w=-1$ or $w=-1.1$, for each interaction kernel are given in Table \ref{tab:Com_AE_BR}.

\begin{table}[H]
\centering
\renewcommand{\arraystretch}{1.2} % Adjust row spacing
\setlength{\tabcolsep}{10pt}     % Adjust column spacing
\begin{tabular}{|c|c|c|}
\hline
\textbf{Interaction $Q$} 
 & Conditions to avoid imaginary $\rho_{\text{dm/de}}$
 & Conditions to avoid undefined $\rho_{\text{dm/de}}$ \\ \hline \hline

$3 H (\delta_{\text{dm}} \rho_{\text{dm}} + \delta_{\text{de}}  \rho_{\text{de}})$
 & $(\delta_{\text{dm}}+\delta_{\text{de}}+w)^2>4\delta_{\text{de}}\delta_{\text{dm}}$
 & $w\ne0$ ; $(\delta_{\text{dm}}+\delta_{\text{de}}+w)^2-4\delta_{\text{de}}\delta_{\text{dm}} \neq 0$
\\ \hline

$3H\delta( \rho_{\text{dm}}+\rho_{\text{de}})$
 & $\delta\le-\frac{w}{4}$
 & $w\ne0$ ; $\delta\ne-\frac{w}{4}$
\\ \hline

$3H\delta( \rho_{\text{dm}}-\rho_{\text{de}})$
 & $\rho_{\text{dm/de}}$ \text{ always real}
 & $w\ne0$
\\ \hline

$3H \delta \rho_{\text{dm}}$
 & $\rho_{\text{dm/de}}$ \text{ always real}
 & $\delta\ne -w$
\\ \hline

$3H\delta \rho_{\text{de}}$
 & $\rho_{\text{dm/de}}$ \text{ always real}
 & $\delta\ne -w$
\\ \hline

\end{tabular}
\caption{Conditions to avoid imaginary or undefined energy densities for different linear interaction kernels.}
\label{tab:Com_real}
\end{table}

\begin{table}[H]
\centering
\renewcommand{\arraystretch}{1.2} % Adjust row spacing
\setlength{\tabcolsep}{10pt}     % Adjust column spacing
\begin{tabular}{|c|c|c|c|}
\hline
\textbf{Interaction $Q$} 
 & \text{$\rho_{\text{dm/de}}>0$ domain} 
 & \text{$\rho_{\text{dm/de}}>0$ conditions} 
 & \text{Example values} 
\\ \hline \hline

$3 H (\delta_{\text{dm}} \rho_{\text{dm}} + \delta_{\text{de}}  \rho_{\text{de}})$
 & \text{DE $\rightarrow$ DM}
 & \multicolumn{2}{|c|}{$\; \delta_{\text{dm}}\ge0 ; \; \delta_{\text{de}}\ge0 ; \;\delta_{\text{dm}}r_0+ \delta_{\text{de}}\le-\frac{w r_0}{(1+r_0)}$}
\\ \hline

$3H\delta( \rho_{\text{dm}}+\rho_{\text{de}})$
 & \text{DE $\rightarrow$ DM}
 & $0 \le \delta \le -\frac{w r_0}{(1+r_0)^2}$
 & $0 \le \delta \le 0.201$
\\ \hline

$3H\delta( \rho_{\text{dm}}-\rho_{\text{de}})$
 & \text{No viable domain}
 & \text{No viable domain}
 & \text{No viable domain}
\\ \hline

$3H \delta \rho_{\text{dm}}$
 & \text{DE $\rightarrow$ DM}
 & $0 \leq \delta \leq -\frac{w}{(1 + r_0)}$
 & $0 \leq \delta \leq 0.720$
\\ \hline

$3H\delta \rho_{\text{de}}$
 & \text{DE $\rightarrow$ DM}
 & $0 \leq \delta \leq -\frac{w}{\left(1+\frac{1}{r_0}\right)}$
 & $0 \leq \delta \leq 0.280$
\\ \hline

\end{tabular}
\caption{Positive energy conditions for different linear interaction kernels.}
\label{tab:Com_PEC}
\end{table}

\begin{table}[H]
\centering
\renewcommand{\arraystretch}{1.2} % Adjust row spacing
\setlength{\tabcolsep}{10pt}     % Adjust column spacing
\begin{tabular}{|c|c|c|}
\hline
\textbf{Model} 
 & \text{Accelerated expansion $\left[w=-1 \right]$} 
 & \text{No big rip if $w<-1$ $\left[w=-1.1 \right]$ } 
\\ \hline \hline

$Q=3 H (\delta_{\text{dm}} \rho_{\text{dm}} + \delta_{\text{de}}  \rho_{\text{de}})$
 &   $ \delta_{\text{dm}} \left(3w+1\right)-\delta_{\text{de}}  \ge w + \frac{1}{3}$
 & $ \delta_{\text{dm}} \left(w+1\right)-\delta_{\text{de}} \le w + 1$
\\ \hline

$Q=3H\delta( \rho_{\text{dm}}+\rho_{\text{de}})$
 & $\delta\le \frac{1}{3}+\frac{1}{9 w}$ ; $\left[\delta \leq 0.222 \right]$
 & $\delta\ge 1+\frac{1}{ w}$ ; $\left[\delta \geq 0.091 \right]$
\\ \hline

$Q=3H\delta( \rho_{\text{dm}}-\rho_{\text{de}})$
 & $\delta \ge \frac{\frac{1}{3} + w}{2+3w}$ ; $\left[\delta \geq 0.666 \right]$
 & $\delta \le \frac{1 + w}{2+w}$ ; $\left[\delta \le -0.111 \right]$
\\ \hline

$Q=3H \delta \rho_{\text{dm}}$
 & $\forall \delta$ if $w\le-\frac{1}{3}$ 
 & Big rip Inevitable
\\ \hline

$Q=3H\delta \rho_{\text{de}}$
 & $\delta \leq -w-\frac{1}{3} $ ; $\left[\delta \leq 0.666\right]$
 & $\delta \geq -w-1 $ ; $\left[\delta \geq 0.1 \right]$

\\ \hline

$w$CDM 
 & $w\le-\frac{1}{3}$ 
 & Big rip Inevitable
\\ \hline

\end{tabular}
\caption{Conditions to ensure accelerated expansion and avoid a big rip for different linear interaction kernels.}
\label{tab:Com_AE_BR}
\end{table}
\begin{table}[H]
\centering
\renewcommand{\arraystretch}{1.2} % Adjust row spacing
\setlength{\tabcolsep}{8pt}      % Adjust column spacing
\begin{tabular}{|c|c|c|}
\hline 
\textbf{Model with $\delta>0$ (iDEDM)} & \text{Coindence problem (Past)} & \text{Coindence problem (Future)}  \\ \hline\hline

$Q=3 H (\delta_{\text{dm}} \rho_{\text{dm}} + \delta_{\text{de}}  \rho_{\text{de}})$  
  & \text{Solved } [$\zeta=0$]  
  & \text{Solved } [$\zeta=0$] \\ \hline

$Q=3H\delta( \rho_{\text{dm}}+\rho_{\text{de}})$  
  & \text{Solved } [$\zeta=0$]  
  & \text{Solved } [$\zeta=0$] \\ \hline

$Q=3H\delta( \rho_{\text{dm}}-\rho_{\text{de}})$  
  & \text{Solved } [$\zeta=0$]  
  & \text{Solved } [$\zeta=0$, $\rho_{\rm{dm}}<0$] \\ \hline

$Q=3H \delta \rho_{\text{dm}}$  
  & \text{Solved } [$\zeta=0$]  
  & \text{Alleviated } [$\zeta=-3(w+\delta)$]  \\ \hline

$Q=3H\delta \rho_{\text{de}}$  
  & \text{Alleviated } [$\zeta=-3(w+\delta)$]  
  & \text{Solved } [$\zeta=0$] \\ \hline

$w\text{CDM}$  
  & $\zeta=-3w$  
  & $\zeta=-3w$ \\ \hline

$\Lambda\text{CDM}$  
  & $\zeta=-3$  
  & $\zeta=-3$ \\ \hline

\end{tabular}
\caption{Potential to address the coincidence problem for different linear interaction kernels.}
\label{tab:Com_CP}
\end{table}

From Tables \ref{tab:Com_real}--\ref{tab:Com_CP}, we can see that IDE models have a complicated parameter space and many pitfalls, including undefined, imaginary, or negative energy densities (which further violate the WEC~\cite{von_Marttens_2020}) and future singularities. In most cases, however, all these issues may be avoided if there is a small energy flow from DE to DM (which is supported by thermodynamic considerations~\cite{Pav_n_2008} and alleviates the coincidence problem). IDE models still show great potential to address many long-standing and new open questions in cosmology. Nevertheless, based on the results of this and other studies, caution should be applied before making claims about IDE models solving the $H_0$ or $S_8$ discrepancies, especially in cases where energy flows from DM to DE. This study provides a guide map to avoid some of these pitfalls.

We include a brief note on how IDE may address recent observations. Recent results from DESI have suggested a DE phantom crossing and, consequently, that DE may increase at early times and then decrease at later times~\cite{DESI:2025fii}. We discussed how each of the IDE models may permit a phantom crossing for $w_{\rm{de}}^{\rm{eff}}(z)$ and $w(z)$ in Section \ref{Background_cosmology} and \ref{reconstructed_w}, but showed that this is often accompanied by both negative energy and divergent behavior. This increase in the DE density and its subsequent decrease can be replicated with a sign-switching IDE model, where energy flows from DM to DE at early times and from DE to DM at later times, as also noted by~\cite{guedezounme2025phantomcrossingdarkinteraction}. This behavior is possible with interaction kernels $Q=3 H (\delta_{\text{dm}} \rho_{\text{dm}} + \delta_{\text{de}} \rho_{\text{de}})$ with $\delta_{\text{dm}}<0$ and $\delta_{\text{de}}>0$, or for $Q=3H\delta(\rho_{\text{dm}}-\rho_{\text{de}})$ with $\delta<0$, as illustrated in Figure \ref{fig:Q_Linear_dm+de}. Unfortunately, this part of the parameter space leads to negative DE densities in the past, given the conditions in Table \ref{tab:Com_PEC} and as illustrated in Figure \ref{fig:Omega_Linear_dm-de}. It should be mentioned that negative DE in the past was also suggested by reconstructions of $Q$ in~\cite{Escamilla_2023}. 

For future work, it is important to expand our analysis to include a wider selection of interaction kernels, such as non-linear interaction kernels (studied in our companion paper \cite{vanderWesthuizen:2025II}) and, especially, interactions based on more fundamental field theory descriptions. Furthermore, to better address the possibility of a phantom crossing, we should instead map IDE models to CPL through observable fits, as was done in \cite{Wolf:2024eph, Wolf:2024stt, Wolf:2025jed}. The assumptions used throughout this paper have also limited the conclusions to only hold in cases where the Universe is always expanding. Interesting implications of the strong coupling regime (where $\delta$ is larger than the upper limit in Table~\ref{tab:Com_PEC}) will lead to negative energy densities that cause more exotic behaviors, such as a big crunch future or a non-singular bounce in the past. The implications of this part of the parameter space will be explored in future work, as this paper only focuses on the implications of IDE models with small deviations from standard cosmology. 

With a better understanding of the parameter space of IDE models, researchers may now proceed to further constrain these models with observational data, which will ultimately determine their role in describing the universe we inhabit.

\textbf{Data Availability Statements:} Data sharing is not applicable to this article as no datasets were generated or analyzed during the current study. More detailed calculations for any section can be provided by the authors on reasonable request.

\begin{acknowledgments}

EDV is supported by a Royal Society Dorothy Hodgkin Research Fellowship. This article is based upon work from the COST Action CA21136 - ``Addressing observational tensions in cosmology with systematics and fundamental physics (CosmoVerse)'', supported by COST - ``European Cooperation in Science and Technology''.

\end{acknowledgments}

\bibliographystyle{apsrev4-2}
\bibliography{References2,biblio}

\end{document}